\documentclass[12pt,a4paper]{article}
\usepackage{geometry}
\usepackage{natbib}
\usepackage{hyperref}    
\usepackage{amsthm}   
\usepackage{amsmath}
\usepackage{graphicx,psfrag,epsf}
\usepackage{enumerate}
\usepackage{mathtools}
\usepackage{amsfonts} 
\usepackage{bbm}      
\usepackage{verbatim}  
\usepackage{xcolor}   
\usepackage{booktabs} 
\usepackage{multicol}

\usepackage[noend]{algpseudocode}
\usepackage{algorithm}
\algnewcommand{\IfThenElse}[3]{
	\State \algorithmicif\ #1\ \algorithmicthen\ #2\ \algorithmicelse\ #3}
\algnewcommand{\StateFor}[2]{
	\State  #1\ \algorithmicfor\ #2}

\usepackage{subcaption} 

\usepackage{NewCommand}

\title{Root cause discovery via permutations and Cholesky decomposition} 
\author{Jinzhou Li, Benjamin B. Chu, Ines F. Scheller, \\ Julien Gagneur, Marloes H. Maathuis}
\date{\today}

\begin{document}
\maketitle

\begin{abstract}
	  This work is motivated by the following problem: Can we identify the disease-causing gene in a patient affected by a monogenic disorder?
	 This problem is an instance of root cause discovery. In particular, we aim to identify the intervened variable in one interventional sample using a set of observational samples as reference.
	 We consider a linear structural equation model where the causal ordering is unknown. 
	 We begin by examining a simple method that uses squared z-scores and characterize the conditions under which this method succeeds and fails, showing that it generally cannot identify the root cause. 
	 We then prove, without additional assumptions, that the root cause is identifiable even if the causal ordering is not. Two key ingredients of this identifiability result are the use of permutations and the Cholesky decomposition, which allow us to exploit an invariant property across different permutations to discover the root cause. 
	 Furthermore, we characterize permutations that yield the correct root cause and, based on this, propose a valid method for root cause discovery. 
	 We also adapt this approach to high-dimensional settings. Finally, we evaluate the performance of our methods through simulations and apply the high-dimensional method to discover disease-causing genes in the gene expression dataset that motivates this work.
\end{abstract}


\section{Introduction} \label{sec:intro}

\subsection{Motivation and problem statement}
Rare diseases can stem from mutations in a single gene (also known as Mendelian or monogenic disorders). 
Typically, such mutations affect the expression of the gene itself, and of further downstream genes,
ultimately leading to the disease \citep[e.g.,][]{yepez2021detection, yepez2022clinical}.
Discovering the disease-causing gene is important, as it enhances our understanding of the disease mechanism
and is a critical step towards developing potential cures.
This motivates us to consider root cause discovery, which, at a high level, is an instance of the reverse causal problem \citep[see, e.g.,][]{gelman2013ask, dawid2014fitting, Pearl2015Causes} and aims to identify the ultimate cause of an observed effect.

Specifically, we distinguish between observational samples (e.g., gene expression data from healthy individuals) and interventional samples (e.g., gene expression data from patients), 
with the former serving as a reference to detect the root cause variable in the latter.
While we assume that observational samples are independent and identically distributed, 
we do not make this assumption for interventional samples, 
since disease-causing genes often differ among rare disease patients, even among those with very similar phenotypes as in the case of mitochondrial disorders \citep{gusic2021genetic}.
Consequently, we conduct root cause discovery for rare disease patients in a personalized manner, focusing on one interventional sample at a time. 
This presents a unique challenge, distinguishing the problem from most related work where identically distributed interventional samples are available.

To formalize the problem, we consider $n$ i.i.d.\  observational samples $\bx_1, \dots, \bx_{n}$ of $X$ (e.g., data of healthy individuals) and one interventional sample $\bx^I$ of $X^I$ (e.g., data of a patient) generated from the following two linear structural equation models (SEMs), respectively:
\begin{equation} \label{model:linearSEMobs}
	X \leftarrow   b +  B X+ \err
\end{equation}
and
\begin{equation} \label{model:linearSEMint}
	X^I \leftarrow   b +  B X^I + \err + \delta,
\end{equation}
where
$X, X^I, b, \err$ and $\delta \in \bbR^p$ and $B \in \bbR^{p \times p}$.
Here, $b$ is an intercept term.
$B$ encodes the underlying causal structure. It can be visualized in a causal directed acyclic graph (DAG) on vertices $\{1,\dots,p\}$, where there is an edge from $i$ to $j$ if the edge weight $B_{ji} \neq 0$.
Further,  $\err$  is an error term that follows an arbitrary distribution with mean $0$ and diagonal covariance matrix.
Finally, $\delta = (0, \dots, 0, \delta_r, 0, \dots, 0)^T$ represents a mean-shift intervention.
It has only one nonzero entry $\delta_r$ in the $r$-th position,
indicating that the variable $X^I_r$ is intervened upon by a mean shift $\delta_r$,
modeling a mutation that results in over- or under-expression of the gene.
We refer to $r$ or $X^I_r$ as the \textit{root cause} of $X^I$.
We assume this model throughout the paper.
The assumption of a unique root cause can be seen as an extreme case of the so-called “sparse mechanism shift hypothesis” in \cite{scholkopf2021toward}.



Our goal is to discover the root cause of $X^I$ by comparing $\bx^I = (x^I_1, \dots,x^I_p)^T$ with the observational samples $\bx_1, \dots, \bx_{n}$.
Intuitively, when $|\delta_r|$ is large compared to the noise variance,
$x^I_r$ will stand out as prominently aberrant compared to the values of the $r$-th variable in the observational samples. But the intervention effect on $x^I_r$ can propagate to downstream variables, causing many other variables to appear aberrant as well.
This propagation can make it difficult to identify the root cause. 

\begin{figure}[htbp]
	\centering
	\includegraphics[width=0.7\textwidth]{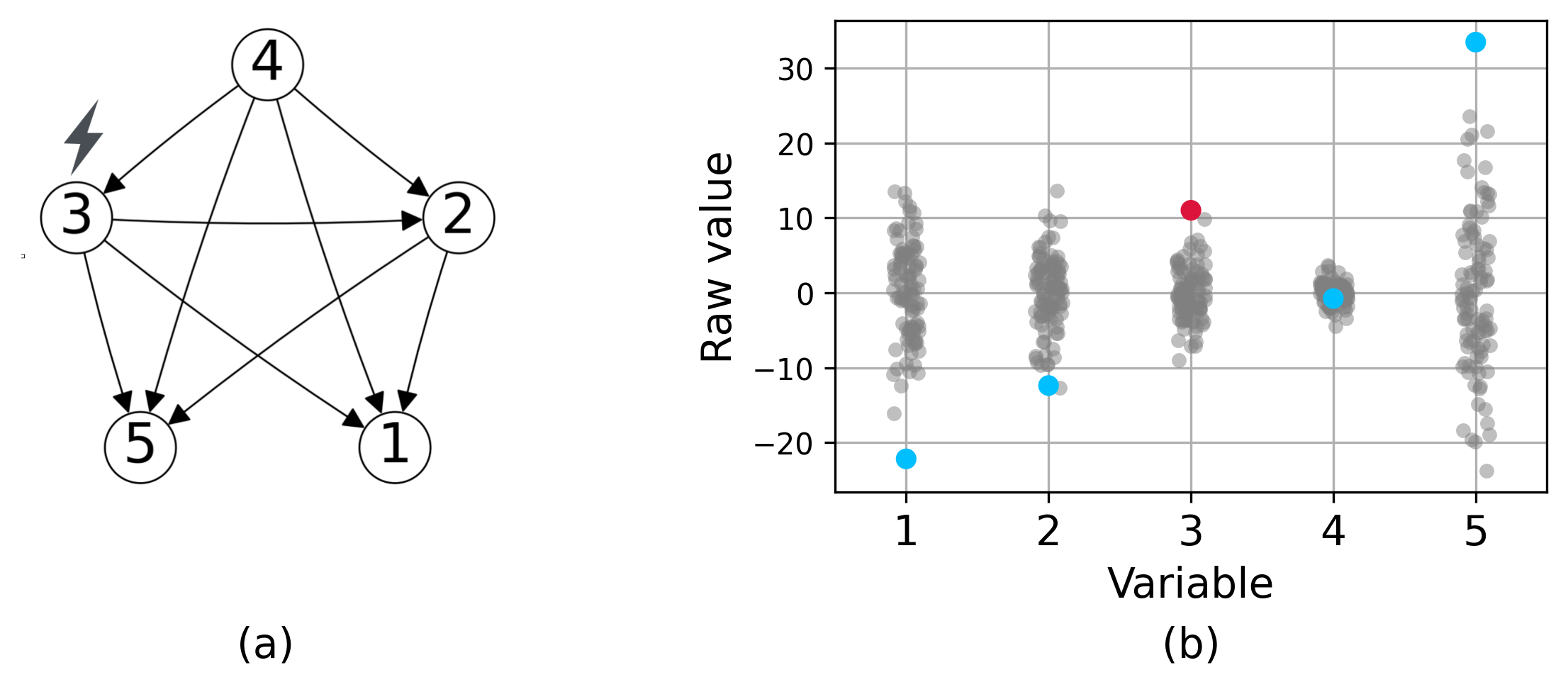}
	\caption{(a) The causal DAG that generates the samples shown in plot (b). (b) A simulated dataset based on the causal DAG in plot (a). Gray points represent $100$ observational samples, while colored points denote the data of one interventional sample.}
		\label{Fig:IntroSimuData}
\end{figure}

For a quick illustration, we look at a simulated dataset shown in Figure~\ref{Fig:IntroSimuData}(b),
where we use $b=(0,0,0,0,0)^T$,
$B=\left(\begin{smallmatrix}
	0 & 1 & -1 & -2 & 0 \\
	0 & 0 & -1 & 1 & 0 \\
	0 & 0 & 0 & -2 & 0 \\
	0 & 0 & 0 & 0 & 0 \\
	0 & -2 & 1 & 3 & 0 
\end{smallmatrix}\right),
$
$\text{Cov}(\err) = \Diag(3,2,3,2,3)$, and $\delta = (0,0,10,0,0)^T$.
The causal DAG in plot (a) represents the generating process. 
The vector $\delta$ indicates that $X^I_3$ is the root cause for the interventional sample (visualized by the lightning symbol in the DAG).
All variables except for $X^I_4$ are descendants of $X^I_3$, meaning that the intervention effect propagates to  $X^I_1$,  $X^I_2$, and $X^I_5$.
This is reflected in plot (b), where $100$ observational samples (in gray) and one interventional sample (in color) show that, except for $X^I_4$, all variables of the interventional sample appear aberrant compared to the observational samples. 

\subsection{Our contribution}
In this paper, we start by studying a commonly used quantity for detecting aberrancy: 
the squared z-score (see \eqref{formula:z-score:SampleVersion}  in Section~\ref{sec: zScore}).
We characterize when the squared z-score can and cannot successfully identify the root cause,
as the observational sample size and the intervention strength tend to infinity.
In particular, we show that the squared z-score can consistently identify the root cause if $\Var(X_k)> \alpha^2_{r \rightarrow  k} \Var(X_r)$ for all $k$, where $\alpha^2_{r \rightarrow  k}$ is the total causal effect of $X_r$ on $X_k$.
Three sufficient conditions on the generating mechanism that ensure this are: 
(1)  for any descendants of the root cause, there is no common ancestor that has a directed path to the descendant without passing through the root cause;
(2) the causal DAG is a polytree \citep[see, e.g.,][]{jakobsen2022structure, tramontano2022learning, tramontano2023learning}; and
(3) all entries of the matrix $B$ are nonnegative. 
On the other hand, if there exists a $k$ such that $\Var(X_k)< \alpha^2_{r \rightarrow  k} \Var(X_r)$, then the squared z-score cannot consistently identify the root cause.


In Section~\ref{sec: Cholesky}, 
we study the fundamental question of whether the root cause is identifiable based on the observational distribution and the first moment of the interventional distribution. 
We consider the observational distribution as it can theoretically be estimated with $n$ i.i.d.\ observational samples,
and we consider only the first moment of the interventional distribution because only one interventional sample is available.
The answer to this question is not immediately clear, as even the causal ordering may be non-identifiable in this setting.
We prove, however,
that the root cause is in fact identifiable. 
This identifiability result stems from an invariant property involving permutations of variables and the Cholesky decomposition.

The above identifiability result leads to a valid root cause discovery method which requires trying all permutations of the variables.
Considering $p!$ permutations, however, quickly becomes infeasible when the number of variables $p$ increases.
To address this problem, we characterize so-called ``sufficient" permutations that allow us to identify the root cause.
In particular, sufficient permutations are those where the parents of the root cause are positioned before it and the true descendants are positioned after it.
This leads to the second valid root cause discovery algorithm with a smaller computational burden (see Algorithm~\ref{Algo:RCD}).
At last, we develop a heuristic root cause discovery method designed for high-dimensional settings (Algorithm~\ref{Algo:RCDhighdim}),
building on our previously proposed Algorithm~\ref{Algo:RCD}. 


We examine the performance of our methods thought extensive simulations in Section~\ref{sec:simu-chap4}.
In Section~\ref{sec:realdata}, we revisit the problem of discovering disease-causing genes in the high-dimensional gene expression dataset that motivates our study. 
We apply the high-dimensional method (Algorithm~\ref{Algo:RCDhighdim}), yielding very promising results. 
We conclude the paper with discussions in Section~\ref{sec:discussion}.

\subsection{Related ideas and work} \label{sec:introRelatedIdeasAndWork}

Recall that we assume that the underlying causal structure is unknown. If the DAG or causal ordering were known, there are two natural methods for root cause discovery: (1) first identify all aberrant variables and then select the one that appears first in the causal ordering;
(2) leverage invariance: only the root cause will exhibit a changed conditional distribution given its parents
\citep[see, e.g.,][]{haavelmo1944probability, peters2016causal, janzing2019causal, li2022causal}.
See Appendix~\ref{appendix:DetailSimuSetup} for more details.

One may try to estimate the DAG or causal ordering from observational samples and then apply the ideas of the previous paragraph.
However, it is well-known that these estimation problems are highly challenging and often requires very large sample sizes \citep[see, e.g.,][]{evans2020model}. 
More importantly, causal orderings and DAGs may be unidentifiable \citep[see, e.g.,][]{spirtes2001causation,pearl2009causality},
which presents a fundamental issue that cannot be resolved regardless of sample size.
Nevertheless, in Section~\ref{sec:simu-chap4}, we implement methods based on these two ideas, using LiNGAM \citep{shimizu2006linear}, and compare their performance to that of our method.


When the DAG is unknown, there is a line of work that combines both observational and interventional samples, with the primary goal of either estimating the intervened variables (i.e., root causes) or learning the causal structure with the estimated root cause as a by-product. 
However, these methods rely on aspects of the interventional distributions like the second moments \citep{rothenhausler2015backshift, varici2021scalable, varici2022intervention}, likelihoods \citep{eaton2007exact, taeb2021perturbations}, or the entire interventional distribution \citep{squires2020permutation, jaber2020causal, ikram2022root, yang2024learning}. These approaches cannot be applied to our problem, where only a single interventional sample is available.

Our method relies on the Cholesky decomposition, which has been used before in causal structure learning \citep{ye2020optimizing, raskutti2018learning} and for 
estimating the effects of joint interventions \citep{nandy2017estimating}. 
In particular, \cite{raskutti2018learning} also combine permutations with the Cholesky decomposition to find the sparsest Cholesky factorization.
Our approach differs fundamentally from these works: we combine permutations and the Cholesky decomposition to search for an invariant property related to the root cause.

Finally, there is related work that treats root cause analysis as a causal contribution problem \citep[see, e.g.,][]{budhathoki2021did, budhathoki2022causal, okati2024root},
as well as work that focuses on root cause analysis for microservice-based applications (see \cite{hardt2023petshop} and the references therein). 
These works either require knowledge of the causal DAG or estimate it from data. One exception is the method “SCORE ORDERING” from \cite{okati2024root}, which does not require the causal graph and is based on the heuristic that small outliers are unlikely to cause larger ones. This heuristic is justified in certain scenarios but does not hold in general (see the example in Appendix~\ref{app:ZscoreExample1}).


To the best of our knowledge, no formal method currently exists in the literature to address our problem, where only a single interventional sample is available, and the causal ordering may be unidentifiable.

\section{Squared z-score is not generally valid} \label{sec: zScore}


The squared z-score is commonly used to quantify aberrancy.
Instead of examining raw values, it adjusts for each variable's marginal mean and variance. 
Intuitively, the squared z-score cannot identify the root cause in general, as it does not account for the causal relationships between variables.
In this section, we formally verify this intuition by characterizing when the squared z-score succeeds and fails.

For $j\in [p] = \{1,\dots,p\}$, the z-score of $X^{I}_{j}$ is defined as
\begin{align}\label{formula:z-score:SampleVersion}
	\whZ_{n,j} = \frac{X^{I}_{j} - \whmu_{n,j}} {\whsigma_{n,j}},
\end{align}
where $ \whmu_{n,j} = \frac{1}{n} \sum_{i=1}^{n} x_{ij}$ and $ \whsigma_{n,j} = \sqrt{ \frac{1}{n-1} \sum_{i=1}^{n} (x_{ij}-\widehat{\mu}_{n,j})^2}$ are the sample mean and standard deviation of the observational samples.


Let $r \in [p]$ be the root cause and $k \in [p]\setminus\{r\}$.
For the linear SEMs \eqref{model:linearSEMobs} and \eqref{model:linearSEMint},  we have 
\begin{align}\label{formula: simplifyfolumaXI}
	X^I_{r} = X_{r} + \delta_{r}
	\quad \text{and} \quad 
	X^I_{k} = X_{k} + \alpha_{r\rightarrow k}\delta_{r},
\end{align}
where $ \alpha_{r\rightarrow k}= (I-B)^{-1}_{kr} $ is the total causal effect of $X_r$ on $X_k$.

To gain some intuition about when the squared z-score works, we first consider the z-score $Z_{j} $ with population mean $ \mu_{j}$ and standard deviation $ \sigma_{j}$ of the observational distribution.
Then, using \eqref{formula: simplifyfolumaXI}, we have 
\begin{align*}
	Z_{r} =  \frac{ X^I_{r}- \mu_{r}} {\sigma_{r}} = \frac{X_r - \mu_r}{\sigma_{r}} +  \frac{\delta_{r}}{\sigma_{r}}
	\quad \text{and} \quad 
	Z_{k} =  \frac{ X^I_{k}- \mu_{k}} {\sigma_{k}}  =  \frac{X_k - \mu_k}{\sigma_{k}} + \frac{ \alpha_{r\rightarrow k}\delta_{r}}{\sigma_{k}}.
\end{align*}
Here the first terms on the right hand sides are standardized random variables. Hence, for $\delta_r$ large, the deterministic terms will dominant and $Z_{r}^2 > Z_{k}^2$ if $  \sigma_{k}^2 > \alpha_{r \rightarrow  k}^2 \sigma_{r}^2 $.
The corresponding sample version result is given in Theorem~\ref{thm:z-scoreMain}. 	
\begin{theorem}\label{thm:z-scoreMain}
	Let $r \in [p]$ be the root cause. For any $k \in[p] \backslash \{r\}$,  the following two statements hold:
	\begin{itemize}
		\item[(i)] If $  \sigma_{k}^2 > \alpha_{r \rightarrow  k}^2 \sigma_{r}^2 $, then 
		$\lim\limits_{\substack{n \to \infty \\ \delta_r \to \infty}} \tP\left( \whZ^2_{n,r} > \whZ^2_{n,k} \right) = 1$.
		\item[(ii)] If $  \sigma_{k}^2 < \alpha_{r \rightarrow  k}^2 \sigma_{r}^2 $, then 
		$\lim\limits_{\substack{n \to \infty \\ \delta_r \to \infty}} \tP\left( \whZ^2_{n,r} < \whZ^2_{n,k} \right) = 1$.
	\end{itemize}
\end{theorem}

In the above theorem, $n \to \infty$ allows us to obtain consistent estimators for the population mean and standard deviation of the observational distribution. 
However, since there is only one interventional sample, 
which is used to estimate $\mu^I_{j}$, the variance of this estimator does not vanish. 
To account for this, we let $\delta_r \to \infty$.
Based on our proof in Appendix~\ref{app:proofThm2.1}, one can also obtain non-asymptotic high-probability results with a more precise characterization of $\delta_r$, but we refrain from doing so here for simplicity. 
As a direct corollary, if the condition $\sigma_{k}^2 > \alpha_{r \rightarrow  k}^2 \sigma_{r}^2$ holds for all $k \in [p] \setminus {r}$, 
the squared z-score can consistently identify the root cause. 



\begin{figure}[t]
	\centering
	\begin{minipage}{0.32\textwidth}
		\centering
		\includegraphics[width=\linewidth]{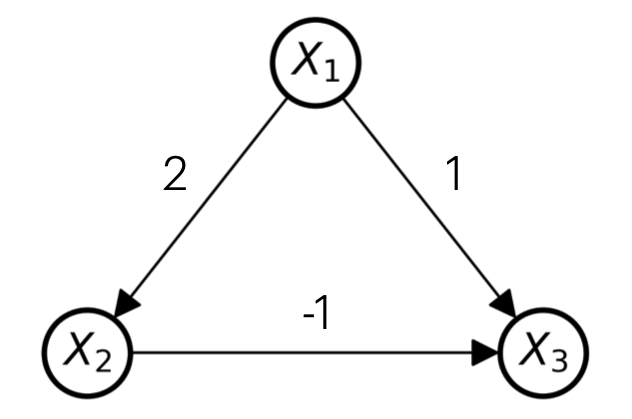}
		\\ (a)  
	\end{minipage}%
	\hspace{0.1\textwidth}  
	\begin{minipage}{0.42\textwidth}
		\centering
		\includegraphics[width=\linewidth]{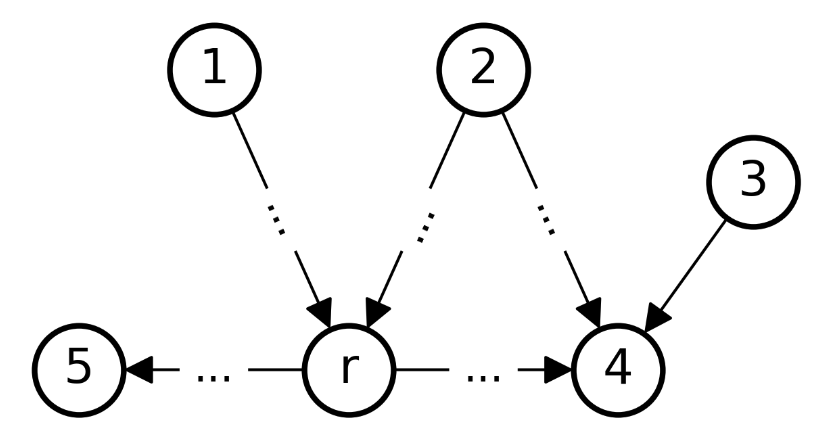}
		\\ (b)  
	\end{minipage}
	\caption{(a) The DAG corresponding to the SEM that generates data for Figure~\ref{Fig:simu1}. 
		(b) DAG used to illustrate Proposition~\ref{prop:categorizeVariablesZscore}, where $r$ denotes the root cause, and the edges with dots represent directed paths.}
	\label{Fig:DAGsSec2}
\end{figure}

\begin{figure}[htbp]
	\centering
	\includegraphics[scale=0.6]{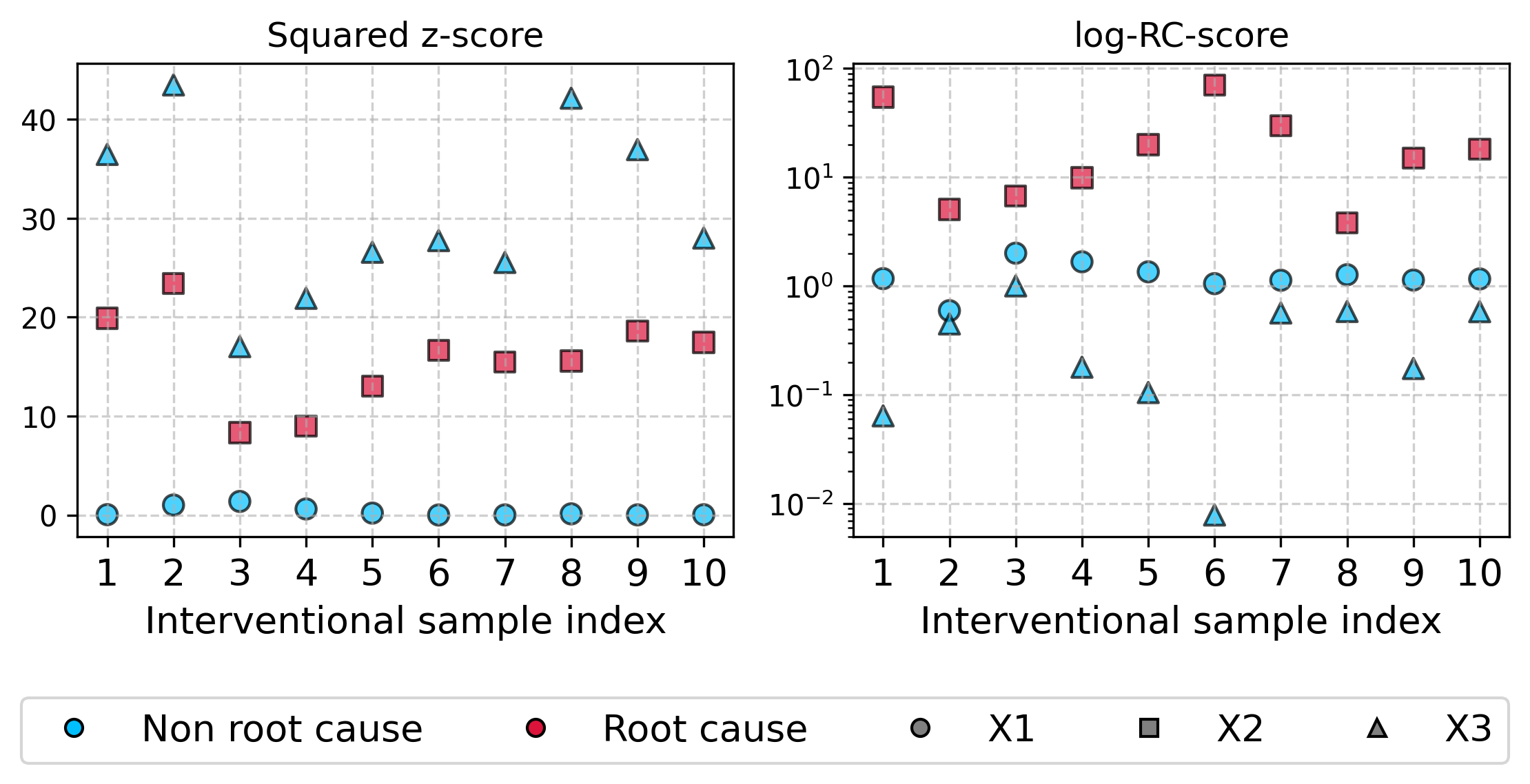}
	\caption{Simulation results based on an SEM with the DAG in Figure~\ref{Fig:DAGsSec2} (a). 
		The two plots show the squared z-scores and the RC-scores (in log scale) of $10$ interventional samples with root cause $X_2$, respectively.  For each interventional sample, red denotes the root cause while blue  represents non-root cause variables, and $X_1$, $X_2$, and $X_3$ are represented by circles, squares, and triangles, respectively.
	}
	\label{Fig:simu1}
\end{figure}

To give some intuition for conditions (i) and (ii), we rewrite $X^I_k = \alpha_{r \rightarrow k} X^I_r + N$, where \( N \) is an appropriate linear combination of the error terms of the ancestors of \( X^I_k \) excluding \( X^I_r \). Then we have $ \sigma_k^2 = \alpha_{r \rightarrow k}^2 \sigma_r^2 + \text{Var}(N) + 2\alpha_{r \rightarrow k}\text{Cov}(X^I_r, N)$.
Hence, condition (ii) holds if the term \( \alpha_{r \rightarrow k} \text{Cov}(X^I_r, N) \) is sufficiently  negative, which can only happen if there exist common ancestors of \( X^I_k \) and \( X^I_r \) (i.e.,\ confounding). This is also related to Proposition~\ref{prop:categorizeVariablesZscore} and its accompanying discussion.
To give a quick illustration, we consider a linear SEM with equal error variance $1$ and the DAG shown in Figure~\ref{Fig:DAGsSec2} (a).
For $r=2$ and $k=3$, it is easy to verify that
condition (ii) in Theorem~\ref{thm:z-scoreMain} holds,
so $X_3$ will likely have a larger squared z-score than $X_2$.
This is verified in the left plot of Figure~\ref{Fig:simu1}, which shows the squared z-scores for $10$ interventional samples where $X_2$ is the root cause.
See also \cite{ebtekar2025toward} for an explanation of this phenomenon from the perspective of the so-called conditional outlier scores.
For details of this example and further discussions, see Appendix~\ref{app:ZscoreExample1}.

A natural follow-up question is: Can we give sufficient conditions on the generating model that ensure $\sigma_{k}^2 > \alpha_{r \rightarrow  k}^2 \sigma_{r}^2$ for some or all $k \in[p] \backslash \{r\}$?
To answer this question, we first introduce some notation.
For $j \in [p]$ and a DAG $G$, let the set of its ancestors and descendants be
\begin{align*}
	\An(j) &= \{ l \in [p]\backslash\{j\}: \text{there is a directed path from $l$ to $j$ in $G$} \}, \\
\De(j) &= \{ l \in [p]\backslash\{j\}: \text{there is a directed path from $j$ to $l$ in $G$} \}.
\end{align*}
In addition, for $k \in[p] \backslash \{r\}$, define 
\[
O(r,k) = \{ j \in \An(r) \cap \An(k): \text{there is a directed path from $j$ to $k$ in $G$ that bypasses $r$} \}.
\]

Then we have the following  Proposition~\ref{prop:categorizeVariablesZscore}:
\begin{proposition}\label{prop:categorizeVariablesZscore}
	Let $r \in [p]$ be the root cause and let $k \in[p] \backslash \{r\}$. If (i) $k \not\in \De(r)$, or (ii) $k \in \De(r)$ and $O(r,k) = \emptyset$,
	then $\sigma_{k}^2 > \alpha_{r \rightarrow  k}^2 \sigma_{r}^2$.
\end{proposition}




Based on Proposition~\ref{prop:categorizeVariablesZscore}, a necessary condition for \( \sigma_{k}^2 < \alpha_{r \rightarrow k}^2 \sigma_{r}^2 \) to hold, or equivalently for a variable \( X_k \) to be problematic in the sense of having a larger asymptotic squared z-score than \( X_r \), is that \( k \in \De(r) \) and \( O(r,k) \neq \emptyset \).
For an illustration, consider Figure~\ref{Fig:DAGsSec2} (b), where $r$ is the root cause and the edges with dots represent directed paths, variables $1,2,3$ and $5$ are considered safe because $1,2,3 \not\in \De(r)$ and $5 \in \De(r)$ with $O(r, 5) = \emptyset$. The only unsafe variable is $4$, as $4 \in \De(r)$ with $O(r, 4) = \{2\} \neq \emptyset$.
Similarly, in Figure~\ref{Fig:DAGsSec2} (a), all variables are safe when $r=1$ or $r=3$. But if $r=2$, $X_3$ is unsafe as $3 \in \De(2)$ and $O(2, 3) = \{1\}$. 
See Appendix~\ref{app:ExampleProp2} for further discussion with a concrete example.

Proposition~\ref{prop:categorizeVariablesZscore} only provides a sufficient condition.
Another sufficient condition is shown in the following proposition.
\begin{proposition}\label{prop:allposSign}
	Let $r \in [p]$ be the root cause.
	If all edge weights are nonnegative, then $\sigma_{k}^2 > \alpha_{r \rightarrow  k}^2 \sigma_{r}^2$ for all $k\in[p]\backslash\{r\}$.
\end{proposition}
Corollary~\ref{coro: polytree} directly follows from Propositions~\ref{prop:categorizeVariablesZscore} and \ref{prop:allposSign}.
\begin{corollary}\label{coro: polytree}
	Let $r \in [p]$ be the root cause.
	If the underlying DAG is a polytree or if all edge weights are nonnegative, then $\lim\limits_{\substack{n \to \infty \\ \delta_r \to \infty}}  \tP \left( \whZ^2_{r} > \whZ^2_{k} \right) = 1$ for all $k\in[p]\backslash\{r\}$.
\end{corollary}

In summary, using the squared z-score identifies the root cause in some cases, but not in general.

\section{Root cause discovery via permutations and Cholesky decomposition} \label{sec: Cholesky}

\subsection{Is the root cause identifiable?}\label{sec: Cholesky-identifiability}
At this point, it is unclear whether the root cause is identifiable based solely on the observational distribution and the first moment of the interventional distribution.
As mentioned in the introduction, we focus on these two quantities because the former can be estimated from $n$ i.i.d. observational samples, while a single interventional sample only allows us to estimate the first moment of the interventional distribution.
This question is fundamental and not immediately clear, given that the causal ordering and the causal DAG are generally non-identifiable. 

In the following, we give a positive answer to this question.
In particular, we prove that the root cause is identifiable from the mean and covariance matrix of the observational distribution (denoted by $\mu_{X}$ and $\Sigma_{X}$) and the mean of the interventional distribution (denoted by $\mu_{X^{I} }$).
We focus on population quantities to address the identifiability question for now, and the sample version algorithm will be introduced in the next section.


Note that the population version of the z-score can be written as 
$\Diag(\sigma_1,\dots,\sigma_p)^{-1} (\mu_{X^{I}}  - \mu_{X})$.
Considering the generalization that takes the full covariance matrix into account, we introduce the following key term:
\begin{align}\label{formula:xi}
	\xi := L_{X}^{-1} (\mu_{X^{I}}  - \mu_{X}), 
\end{align}
where $L_{X}$ is the lower-triangular matrix with real and positive diagonals obtained by the Cholesky decomposition of $\Sigma_{X} $, i.e., $\Sigma_{X}= L_{X} L_{X}^T$.

To motivate the reason of using the Cholesky decomposition, we look at the special case where $(X_1, \dots, X_p)$ are sorted by a causal ordering. 
In this case, the matrix $B$ in model~\eqref{model:linearSEMobs} is lower triangular.
From model~\eqref{model:linearSEMobs}, we have $\Sigma_{X} =(I - B)^{-1} D_{\err} (I - B)^{-T}$, where $D_{\err} = \Diag(\sigma^2_{\err_1}, \dots, \sigma^2_{\err_p})$ is the diagonal covariance matrix of $\err$.
Hence,
$L_{X}=(I - B)^{-1} D^{1/2}_{\err}$ by the uniqueness of the Cholesky decomposition.
Additionally, since $\mu_{X^{I} } = (I - B)^{-1} (b + \delta)$ and $\mu_X = (I - B)^{-1} b$, we have 
\begin{equation} \label{formula:mu_tX_causalOrdering}
	\xi = L_{X}^{-1} (\mu_{X^{I}}  - \mu_{X}) 
	=D^{-1/2}_{\err} \delta = 
	(0, \dots, 0, \delta_r / \sigma_{\err_r}, 0, \dots, 0)^T.
\end{equation}
Thus, when $(X_1, \dots, X_p)$ are sorted by a causal ordering, one may view $\xi_i^2$ as a conditional outlier score given the true parents of $X_i$ \citep[c.f. ][]{janzing2019causal}, and we can distinguish the root cause based on $\xi$ because it is the only entry with a nonzero value.

However, since $(X_1, \dots, X_p)$ is generally not in a causal ordering,  the matrix $B$ is not lower triangular and $L_{X}$ does not equal $(I - B)^{-1} D^{1/2}_{\err}$. As a result, $\xi$ may not exhibit the informative pattern seen in~\eqref{formula:mu_tX_causalOrdering}. 
Therefore, the root cause cannot be identified using the Cholesky decomposition alone.

Remarkably, 
there is actually a (conditional) invariant property related to the root cause across different permutations of variables: conditional on $\xi(\pi)$ (see \eqref{formula:xipi}) having only one nonzero entry (we refer to this pattern as \textit{1-sparse}), the variable corresponding to this entry is invariant and must be the root cause.
This is formalized in Theorem~\ref{thm:identiOfRC}, which can be proved using linear algebra (see Lemma~\ref{lemma:identiOfRC} in Appendix~\ref{app:proofThm1}).
\begin{theorem}[Identifiability of the root cause] \label{thm:identiOfRC}
	Let $(X_1, \dots, X_p)$  be sorted in any ordering (not necessarily a causal ordering).
    The corresponding $\xi$ must have at least one nonzero element.
    Furthermore, if $\xi$ has exactly one nonzero element, then this element must be the root cause.
\end{theorem}
Theorem~\ref{thm:identiOfRC} enables us to exploit an invariant property to identify the root cause:
by trying all permutations of $(X_1, \dots, X_p)$, 
we are guaranteed to find at least one permutation (since a causal ordering must exist) for which the corresponding $\xi$ is 1-sparse, in which case the nonzero entry must correspond to the root cause.
This verifies the identifiability of the root cause. 

Notably, Theorem~\ref{thm:identiOfRC} implies that the root cause is identifiable even if the causal ordering is not.
From this perspective, discovering the root cause is a fundamentally simpler task than estimating the causal ordering or learning the causal DAG.

\subsection{Root cause discovery algorithm based on all permutations} \label{sec:RC-all-perm}
Theorem~\ref{thm:identiOfRC} inspires a root cause discovery algorithm by trying all permutations of variables.
Before presenting this algorithm, we first introduce some notation.

For a permutation $\pi=(\pi(1), \dots,\pi(p))$, let $X^{\pi} = (X_{\pi(1)}, \dots, X_{\pi(p)})^T$ and $X^{I \pi} = (X^I_{\pi(1)}, \dots, X^I_{\pi(p)})^T$,
with their corresponding means and covariance matrix denoted by $\mu_{X^{\pi}}$,  $\mu_{X^{I\pi} }$ and $\Sigma_{X^{\pi}}$, respectively.
We introduce the following notation to make the dependence of $\xi$ (see \eqref{formula:xi}) on the permutation $\pi$ explicit:
\begin{align}\label{formula:xipi}
	\xi(\pi) := L_{X^{\pi}}^{-1} (\mu_{X^{I\pi}}  - \mu_{X^{\pi}}),
\end{align}
where $L_{X^{\pi}}$ is the lower triangular matrix from the Cholesky decomposition of $\Sigma_{X^{\pi}}$. 
Note that $\xi(\pi)$ is not simply a permuted version of $\xi$ due to $L_{X^{\pi}}^{-1}$.

By Theorem~\ref{thm:identiOfRC}, if $\xi(\pi)$ is 1-sparse, the nonzero entry must be in the  $\pi^{-1}(r)$-th position, corresponding to the root cause’s position under the permutation.
We call a permutation $\pi$ \textit{sufficient} if $\xi(\pi)$ is 1-sparse, 
indicating that this permutation is sufficient to identify the root cause.
Otherwise we call it \textit{insufficient}, in which case $\xi(\pi)$ contains more than one nonzero element.
The existence of a sufficient $\pi$ is clear because a causal ordering is sufficient.
As we will show in Section~\ref{sec:characterizeSufficientPermutation}, 
non-causal orderings can also be sufficient. 

Now we introduce the sample version algorithm.
Let $\bx_1^\pi, \dots, \bx_{n}^\pi$ be $n$ i.i.d.\ observational samples of $X^{\pi}$, and let $\bx^{I\pi}$ be one interventional sample of $X^{I \pi}$.
Denote the estimators of $L_{X^\pi}$ and $\mu_{X^\pi}$ based on these observational samples by $\whL_{X^\pi}$ and $\whmu_{X^\pi}$, respectively.
Then we estimate $\xi(\pi)$ by
\begin{align} \label{formula:est_mu_X_tilde}
	\whxi(\pi)
	=  \whL_{X^\pi}^{-1} (\bx^{I\pi} - \whmu_{X^\pi}).
\end{align}


To quantify the evidence that $\whxi(\pi)$ has the 1-sparse pattern and that the corresponding largest entry is the root cause, we use
\begin{align} \label{formula:hatCpi}
	\hc(\pi) = \frac{|\whxi(\pi)|_{(1)} - |\whxi(\pi)|_{(2)}} {|\whxi(\pi)|_{(2)}},
\end{align}
where $ |\whxi(\pi)|_{(i)} $ denotes the $i$-th largest entry in $|\whxi(\pi)|$.
Note that $\hc(\pi)$ is infinite if  $\whxi(\pi) $ is 1-sparse,
so a large value of $\hc(\pi)$ indicates that $\whxi(\pi)$ is close to having the desired 1-sparse pattern.

By Theorem~\ref{thm:identiOfRC}, if $\pi$ is sufficient, 
then the root cause is likely to have the largest value in $|\whxi(\pi)|$.
Therefore, we define the set of all potential root causes as
\begin{align}\label{formula:widehatU}
	\whU = \cup_{\pi \in \Pi_{\all}}\hu(\pi), 
\end{align}
where $\hu(\pi) = \pi(\argmax_{ j \in [p]} |\whxi(\pi)|_j)$ denotes the potential root cause with respect to a permutation $\pi$ and 
$\Pi_{\all}$ is the set of all permutations of $(1,\dots,p)$.

Based on $\whU$, we assign a score to each variable $i$ as a measure of its likelihood of being the root cause. This score, which we call the \textit{RC-score}, is defined as:
\begin{align}\label{def:CholScoreAllPerm}
	\whC_i = 
	\begin{cases}
		\underset{ \pi: \hu(\pi) = i}{\max} \hc(\pi), \quad &\text{if} \quad i \in \whU \\
		\hw_i \hc_{\min}, \quad &\text{if}\quad i \in [p] \setminus \whU,
	\end{cases}
\end{align}
where $\hc_{\min} = \min_{i\in\whU} \whC_i/2$ is half of the smallest RC-score among potential root causes,
and $\hw_i = \whZ^2_{n,i} / \sum_{j\in [p] \setminus \whU} \whZ^2_{n,j}$
is the weight based on the squared z-scores.
This ensures that RC-scores for variables unlikely to be the root cause (i.e., not in $\whU$) remain smaller than the scores for all potential root causes.
A large $\whC_i$ value indicates that variable $ i$  is more likely to be the root cause.

We note that while an alternative approach that assigns a score of zero to all variables not in $ \whU $ is simpler and does not affect the theoretical consistency of the method, using $\hw_i \hc_{\min}$ has an advantage of assigning the root cause a better score in the unfavorable case where it falls outside \( \whU \). 
For example, when the root cause has no children, even if it happens to be outside $ \whU $,  using \eqref{def:CholScoreAllPerm} may still assign it a large score compared to other variables, as it might be the only aberrant variable with the largest squared z-score.

The following theorem shows that if the estimators  $\whL_{X^\pi}$ and $\whmu_{X^\pi}$ are both consistent, then using the RC-score~\eqref{def:CholScoreAllPerm} will consistently identify the root cause as the sample size and intervention strength go to infinity.
\begin{theorem} \label{thm:ConsistAlgoFullPerm}
	Let $r$ be the root cause and $\whC_i$ be obtained by ~\eqref{def:CholScoreAllPerm}.
	If $\whL_{X^\pi} \xrightarrow[n \to \infty]{p} L_{X^\pi}$ and $\whmu_{X^\pi} \xrightarrow[n \to \infty]{p} \mu_{X^\pi}$, then
	\begin{align*}
		\lim\limits_{\substack{n \to \infty \\ \delta_r \to \infty}} \tP \left( \whC_{r} > \max_{k\in[p]\backslash\{r\}} \whC_{k} \right) = 1.
	\end{align*}
\end{theorem}
We apply the RC-score~\eqref{def:CholScoreAllPerm} to the previous example in Section~\ref{sec: zScore}.
Figure~\ref{Fig:simu1} shows that our RC-score successfully identifies all root causes,
which aligns with our expectations based on the theoretical results.


\subsection{Characterization of sufficient permutations and an efficient root cause discovery algorithm} \label{sec:characterizeSufficientPermutation}

Calculating the RC-score using~\eqref{def:CholScoreAllPerm} requires evaluating $p!$ permutations, 
which is computationally infeasible for large $p$.
The purpose of considering all permutations is to ensure at least one sufficient permutation is found.
Therefore, to reduce the search space, it is crucial to understand which permutations are sufficient. 
To this end, we give a complete characterization in Theorem~\ref{thm:SufficientPermutation}.

Before presenting the theorem, we introduce two notation to be used.
For $j \in [p]$, we define its parents and real descendants as 
\begin{align*}
	\Pa(j) = \{k\in[p]\setminus\{j\}: B_{jk} \neq 0 \} 
	\quad \text{and} \quad
	\rDe(j) = \{k \in [p]\setminus\{j\}: (I-B)^{-1}_{kj} \neq 0 \},
\end{align*}
respectively. Recall that $(I-B)^{-1}_{kj} = \alpha_{j\rightarrow k}$ is the total causal effect of $X_j$ on $X_k$. 

\begin{theorem} \label{thm:SufficientPermutation}
	Let $r$ be the root cause.
	A permutation $\pi$ is sufficient if and only if the following two conditions hold:
	\begin{itemize}
		\item[(i)] $\pi^{-1}(k) < \pi^{-1}(r)$ for all $k \in \Pa(r)$,  
		\item[(ii)]  $\pi^{-1}(k) > \pi^{-1}(r)$ for all $ k \in \rDe(r)$. 
	\end{itemize}
\end{theorem}

Theorem~\ref{thm:SufficientPermutation} shows that sufficient permutations are those where the parents of the root cause are positioned before it, and the real descendants of the root cause are positioned after it.
This characterization enables us to develop a method that significantly reduces the search space of permutations.

Specifically, let $ D = \{r\} \cup \rDe(r)$ be the set comprising the root cause and its real descendants. 
If $D$ is known (though the exact root cause $r$ within $D$  does not need to be known), 
then we can generate $|D|$  permutations that are guaranteed to include a sufficient permutation.
Specifically, we  generate $|D|$ permutations by 
\begin{align}\label{formula:GeneratePermutations}
	\pi = ([p]\backslash D, i, D\backslash \{i\}), \quad i \in D,
\end{align}
where the orderings within $[p]\backslash D$ and $D\backslash \{i\}$ are arbitrary. Then, it is clear that the permutation with $i=r$ is sufficient, as it satisfies the conditions in Theorem~\ref{thm:SufficientPermutation} (see Appendix~\ref{app:SuffExample} for a concrete example).
Although we cannot target this sufficient permutation due to not knowing which variable is $r$, it suffices for our purpose that one of the $|D|$ permutations is sufficient. 

The set $D$ is unknown in practice. However, since the root cause and its real descendants are expected to be aberrant due to the intervention, 
we can estimate $D$ using the set of aberrant variables.
We use the squared z-scores $\whZ^2_{n,i}$ (see~\eqref{formula:z-score:SampleVersion}) to form an estimate $\whD = \{j \in[p]: \whZ^2_{n,j} \geq \tau \}$ for some threshold $\tau$.
It is important to note that our primary goal is to reduce the search space of permutations, and trying more permutations is generally beneficial for root cause discovery.
Therefore, we are not restricted to using just $|D|$ permutations generated from a single estimator of $D$. 
Instead, we can use multiple thresholds to obtain several sets $\whD$ and generate more permutations accordingly.
For instance, we could use all the squared z-scores $\whZ^2_{n,1}, \dots, \whZ^2_{n,p}$ as thresholds. 
Because of this, we are not facing the difficult issue of choosing an optimal threshold.

For the same reason, when generating the permutation $\pi = ([p]\backslash D, i, D\backslash \{i\})$ for a certain $i$ (see \eqref{formula:GeneratePermutations}), we can record additional permutations by using different random orderings of $[p]\backslash D$ and $D\backslash \{i\}$.
Although using more permutations does not bring any advantage from an asymptotic perspective, it tends to be beneficial for finite sample performance. The reason is that using only one permutation might result in the undesired case where the root cause is out of $\whU$ (see \eqref{formula:widehatU}), while trying more permutations increases the possibility that the root cause is included in $\widehat{U}$. 
This benefit comes at the cost of increased computational expense, which grows linearly with the number of permutations $v$. We suggest using a moderate value of $v$, depending on the available computational resources. See also Appendix~\ref{appendix:SimuForV} for empirical run-times and some more discussion.

\begin{algorithm}[htbp]
	\caption[] {\textbf{: Obtain permutations based on squared z-scores}} 
	\label{Algo:ObtainPermutations}
	\textbf{Input}:   Observational samples $\bx_1, \dots, \bx_{n}$, the interventional sample $\bx^I$, and the number of random permutations $v$.
	
	\textbf{Output}: A set of permutations.
	
	\begin{algorithmic}[1]
		\State Calculate the squared z-scores $\whZ^2_{n,1}, \dots, \whZ^2_{n,p}$ based on formula~\eqref{formula:z-score:SampleVersion} and initialize an empty set $\whPi$ to record permutations.
		\For {$i = 1, \dots, p$}
		\Statex \hspace{0.4cm} Obtain the set of aberrant variables $D= \{j \in[p]: \whZ^2_{n,j} \geq \whZ^2_{n,i} \} := \{d_1,\dots,d_u\}$.
		\For {$l = 1, \dots, u$} 
		\For {$k = 1, \dots, v$} 
		\Statex \hspace{2cm} Randomly permute $ [p]\backslash D$ and $D\backslash \{d_l\} $ to generate a permutation $\pi = ([p]\backslash D, d_l, D\backslash \{d_l\} )$, and add  $\pi$ to the permutation set $\whPi$.
		\EndFor
		\EndFor
		\EndFor
		\State \textbf{Return} $\whPi$.
	\end{algorithmic}
\end{algorithm}

We summarize the method for obtaining permutations in Algorithm~\ref{Algo:ObtainPermutations}. 
This algorithm generates a total of $vp(p+1)/2$ permutations.
In practice, to reduce computational time, one can use some reasonable thresholds such as $\{1.5, 2, 2.5, \dots, 5\}$ instead of using all squared z-scores in step 2 of this algorithm.
In addition, instead of squared z-scores, alternative approaches could be useful for estimating aberrant variable sets in Algorithm~\ref{Algo:ObtainPermutations} in different scenarios. For example, estimated tail probabilities (or the information theoretic score in \cite{budhathoki2022causal}) could be beneficial when the marginal distributions of variables exhibit strongly different tail behaviors.

In Theorem~\ref{thm:whPiContainOneSufficientPerm}, we show that as the sample size and intervention strength tend to infinity,
one of the permutations outputted by Algorithm~\ref{Algo:ObtainPermutations} is guaranteed to be sufficient with a probability tending to one.
\begin{theorem}\label{thm:whPiContainOneSufficientPerm}
	Let $\whPi$ be the set of permutations obtained by Algorithm~\ref{Algo:ObtainPermutations}, then
	\[
	\lim\limits_{\substack{n \to \infty \\ \delta_r \to \infty}}  \tP \left(\text{$\whPi$ contains at least one sufficient permutation}\right) = 1.
	\]
\end{theorem}


By applying Algorithm~\ref{Algo:ObtainPermutations}, we replace the set of all permutations in \eqref{formula:widehatU} with those generated by this algorithm. 
This significantly reduces computational expense and leads to our main root cause discovery method (Algorithm~\ref{Algo:RCD}).
As shown in the following Theorem~\ref{thm:ConsistSufficientAlgo}, if both $\whL_{X^\pi}$ and $\whmu_{X^\pi}$ are consistent, Algorithm~\ref{Algo:RCD} is guaranteed to assign the root cause the largest score with a probability approaching one, as the sample size and intervention strength increase to infinity.

\begin{algorithm}[htbp]
	\caption[] {\textbf{: Root cause discovery} } \label{Algo:RCD}
	\textbf{Input}: Observational samples $\bx_1, \dots, \bx_{n}$ and the interventional sample $\bx^I$, and the number of random permutations $v$.
	
	\textbf{Output}: RC-score $\whC_i$ for all variables.
	\begin{algorithmic}[1]
		\State Obtain $\whPi$ by implementing Algorithm~\ref{Algo:ObtainPermutations} with the same inputs as this algorithm.
		\State Initialize an empty set $\whU$.
		\For {$\pi \in \whPi$}: 
		\newline 
		\hspace{0.5cm} 
		(i) Obtain permuted samples $\bx_1^\pi, \dots, \bx_{n}^\pi$ and $\bx^{I\pi} $.
		\newline 
		\hspace{0.5cm} 
		(ii) Obtain estimators $\whL_{X^\pi}$ and $\whmu_{X^\pi}$ based on $\bx_1^\pi, \dots, \bx_{n}^\pi$.
		\newline 
		\hspace{0.5cm} 
		(iii) Get $\whxi(\pi)  =  \whL_{X^\pi}^{-1} (\bx^{I\pi} - \whmu_{X^\pi}) $,
		$\hc(\pi) = \frac{|\whxi(\pi)|_{(1)} - |\whxi(\pi)|_{(2)}} {|\whxi(\pi)|_{(2)}}$, and
		$\hu(\pi) = \pi(\argmax_{ j \in [p]} |\whxi(\pi)|_j)$.
		\newline 
		\hspace{0.5cm} 
		(iv) Add $\hu(\pi)$ into $\whU$.
		\EndFor
		\For {$i\in \whU$}: Let $\whC_i = \underset{ \pi: \pi \in \whPi \text{ and }  \hu(\pi) = i}{\max} \hc(\pi)$. 
		\EndFor
		\For {$i\in [p] \setminus \whU$}: Let $\whC_i = \hw_i \hc_{\min}$, where 
		$\hc_{\min} = \min_{j\in\whU} \whC_j/2$ and $\hw_i = \whZ^2_{n,i} / \sum_{j\in [p] \setminus \whU} \whZ^2_{n,j}$ with $\whZ^2_{n,j}$ defined by~\eqref{formula:z-score:SampleVersion}.
		\EndFor
		\State Return $\whC_i$ for all $i\in [p]$.
	\end{algorithmic}
\end{algorithm}

\begin{theorem} \label{thm:ConsistSufficientAlgo}
	Let $r$ be the root cause and $\whC_i$ be obtained by Algorithm~\ref{Algo:RCD}.
	If $\whL_{X^\pi} \xrightarrow[n \to \infty]{p} L_{X^\pi}$ and $\whmu_{X^\pi} \xrightarrow[n \to \infty]{p} \mu_{X^\pi}$, then
	\begin{align*}
		\lim\limits_{\substack{n \to \infty \\ \delta_r \to \infty}} \tP \left( \whC_{r} > \max_{k\in[p]\backslash\{r\}} \whC_{k} \right) = 1.
	\end{align*}
\end{theorem}

\subsection{An adapted root cause discovery algorithm for high-dimensional settings}

For Algorithm~\ref{Algo:RCD} to perform well, 
it is important to obtain an accurate estimate of the covariance matrix, as it is used in the Cholesky decomposition to get $\whL_{X^\pi}$. 
We found that obtaining a good covariance matrix is particularly challenging for the high-dimensional gene expression data analyzed in Section~\ref{sec:realdata}, which involves around 20,000 variables and 400 samples. 
Estimating covariance matrices in high-dimensional settings is known to be difficult and is an independent research topic from our problem (see, e.g., \cite{cai2016estimating} and \cite{fan2016overview} for reviews on this topic). 
We therefore propose a special adaptation of our method that circumvents estimating a high-dimensional covariance matrix. 
We build on Algorithm~\ref{Algo:RCD} and incorporating Lasso \citep{tibshirani1996regression} for dimension reduction in a node-wise manner.

We first present the main idea.
Consider one variable as the response and focus on the subsystem containing that variable and its Markov blanket (i.e., all variables that are conditionally dependent on it given the rest). 
Then, two key observations are: (1) If the chosen response variable is not the root cause, then it is not the root cause within the subsystem.
Specifically, if this variable is influenced by the intervention, its parents, which are included in the Markov blanket, must also be influenced, indicating it cannot be the root cause; if it is not influenced by the intervention, then it clearly cannot be the root cause. 
(2) If the chosen response variable is the root cause, it remains the root cause within the subsystem. 

Building on these two observations, we treat each variable $i$ as the response and apply cross-validated Lasso to estimate its Markov blanket.
We then implement the first three steps of Algorithm~\ref{Algo:RCD} on the subdataset corresponding to the subsystem containing variable $i$ and its estimated Markov blanket.
If $i$ is indeed the root cause, it remains the root cause within this subsystem. 
Consequently, there should exist a permutation $\pi$ for which $\xi(\pi)$ (see \eqref{formula:est_mu_X_tilde}) is 1-sparse, with $|\xi(\pi)|_{\pi^{-1}(i)}$ being the largest value. 
Hence, the corresponding $\hc(\pi)$ (see \eqref{formula:hatCpi}) is a reasonable measure of the likelihood that it is the root cause.
By considering all possible permutations, we set the root cause score for variable $i$ as the largest $\hc(\pi)$ among those $\pi$ for which $|\xi(\pi)|_{\pi^{-1}(i)}$ has the largest value, as implemented in Algorithm~\ref{Algo:RCD}.
Lastly, we assign scores to variables outside $\whU$ in a similar manner as \eqref{def:CholScoreAllPerm}.
We summarize the root cause discovery method for high-dimensional settings in Algorithm~\ref{Algo:RCDhighdim} (see Appendix~\ref{app:highdimAlgo}).

We point out a caveat regarding the high-dimensional version of the root cause discovery algorithm: after dimension reduction, latent variables may be introduced into the subsystem  before applying Algorithm~\ref{Algo:RCD}.
To assess the robustness of Algorithm~\ref{Algo:RCD} to latent variables, we conduct simulations in latent variable settings in Appendix~\ref{sec:simu-latent}.
Simulation results indicate that this algorithm is quite robust to latent variables.
We also evaluate the performance of the high-dimensional root cause discovery algorithm  in high-dimensional settings, see Appendix~\ref{appendix:SimuHighdim}.

We emphasize that this high-dimensional root cause discovery algorithm is a heuristic algorithm, and no theoretical guarantees are provided here.
We present it because it performs very well in discovering the disease-causing genes in the genetic application (see Section~\ref{sec:realdata}).
While it would be interesting to investigate its theoretical properties in high-dimensional settings, this would require studying the theoretical behavior of Algorithm~\ref{Algo:RCD} in the latent variable settings, which is beyond the scope of this paper. So we leave this for future research.

\section{Simulations} \label{sec:simu-chap4}

We now evaluate the finite sample performance of our proposed RC-score (Algorithm~\ref{Algo:RCD}). 
All simulations are carried out in Python, and the code is available at GitHub (\url{https://github.com/Jinzhou-Li/RootCauseDiscovery}).

\subsection{Simulation setup and implemented methods} \label{sec:simu-setup}

We generate observational and interventional samples according to models~\eqref{model:linearSEMobs} and \eqref{model:linearSEMint}, respectively. 
Specifically, 
for the intercept term $b$, we randomly sample its entries from the uniform distribution $U(-10,10)$.
For each component of the error term $\err$, we consider either a Gaussian distribution with mean zero or a uniform distribution $U(-a,a)$.
The variance $\sigma^2$ of each error component is independently sampled from $U(1,2)$, ensuring errors have  non-equal variances \citep{ng2024structure}.
This variance is used directly for the Gaussian distribution, and for the uniform distribution, we set  $ a = \sqrt{3\sigma^2} $ to achieve the desired variance. 
For the matrix $B$, we consider two types corresponding to the random DAG and the hub DAG.
The random DAG, commonly used in the literature, connects nodes randomly with a probability $s$.
The hub DAG is motivated by genetic interactions, where certain genes (known as hub genes) act as central connectors and significantly influence the overall network behavior.
There are $p=100$ and  $p=104$ variables for the random DAG and the hub DAG, respectively.
See Appendix~\ref{appendix:DetailSimuDAG} for more details.

To prevent the marginal variances of $X=(X_1,\dots,X_p)$ from increasing along the causal ordering \citep{reisach2021beware}.
we first sample targeted variances for each variable $X_i$ from $U(10,50)$.
We then rescale the non-zero entries of the matrix $B$ and update the error variances of source nodes (i.e., those without parents) to ensure that the final variances of $X_i$'s are close to the targeted values.
Finally, we randomly permute the ordering of the variables so that they are not sorted according to a causal ordering (with $b$, $\err$ and $B$ permuted accordingly).

For each setting,
we randomly generate $20$ matrices $B$.
For each $B$, we generate $n$ observational samples.
Moreover, we randomly choose $50$ root causes, and generate one interventional sample with intervention effect $\delta_r$ for each root cause.
The $n$ observational samples are then used for root cause discovery on each of the $50$ interventional samples.
In total, there are $m=1000$ interventional samples with possibly different root causes.

To investigate the effects of sample size $n$, the intervention strength $\delta_r$, and the sparsity level $s$, we consider the following scenarios: (i) Vary $n\in\{100, 200, 300\}$ while fixing $s=0.4$ and $\delta_r=8$;
(ii) Vary $\delta_r\in\{4, 8, 12\}$ while fixing $s=0.4$ and $n=200$;
(iii) Vary $s\in\{0.2, 0.4, 0.6\}$ while fixing $n=200$ and $\delta_r=8$.

Given $n$ observational samples and one interventional sample, we calculate the squared z-scores and our  proposed RC-scores (Algorithm~\ref{Algo:RCD}) for each variable in the interventional sample.
In addition, we implement two methods mentioned in Section~\ref{sec:introRelatedIdeasAndWork}
that require estimating the causal ordering or DAG, for which we use LiNGAM \citep{shimizu2011directlingam} (see Appendix~\ref{appendix:DetailSimuLinGam} for details on their implementation).
LiNGAM can consistently estimate a causal ordering or DAG in linear non-Gaussian settings, which are part of our simulation setup.
However, in linear Gaussian settings, which are also part of our simulation setup, the causal ordering and DAG are generally non-identifiable.
Below, we summarize the implemented methods as follows:
\begin{itemize}
	\item[1.] Squared z-score:  Based on formula~\eqref{formula:z-score:SampleVersion}. This method is denoted as ``Z-score" in the plots.
	\item[2.] The approach based on an estimated causal ordering and aberrant set: 
	The causal ordering is estimated using the Python package \textit{lingam} \citep{ikeuchi2023python}.
	We denote the three methods with the optimal threshold and thresholds of $2$ and $5$ for obtaining the aberrant variable set as ``LiNGAM-opt", ``LiNGAM-2", and ``LiNGAM-5", respectively, in the plots.
	In particular, “LiNGAM-opt” uses the squared z-score of the root cause as the threshold, which requires oracle information and is thus infeasible in practice. We present its results to illustrate the best possible performance achievable by such approaches.
	\item[3.] The approach based on an estimated DAG and residuals:
	The DAG is estimated using the same Python package \textit{lingam} as above. 
	We denote this method as “LiNGAM-Inva” in the plots.
	\item[4.] RC-score: Implemented using Algorithm~\ref{Algo:RCD} using $v=10$ random permutations and thresholds $(0.1, 0.3, \dots, 5)$ in its first step. 
	The estimator $\whL_{X^\pi}$ in step 2 is obtained by applying the Cholesky decomposition on the estimated covariance matrix, for which the sample covariance matrix is used when $n>p$, and a shrinkage estimator is used when $n<p$ (\cite{schafer2005shrinkage}, we use the Python function \textit{sklearn.covariance.ShrunkCovariance} for its implementation).  
	This method is denoted as ``RC-score" in the plots.
\end{itemize}


\subsection{Simulation results}\label{sec:simu-results}

\begin{figure}[htbp]
	\centering
	\begin{minipage}[b]{0.88\textwidth}
		\includegraphics[width=\textwidth]{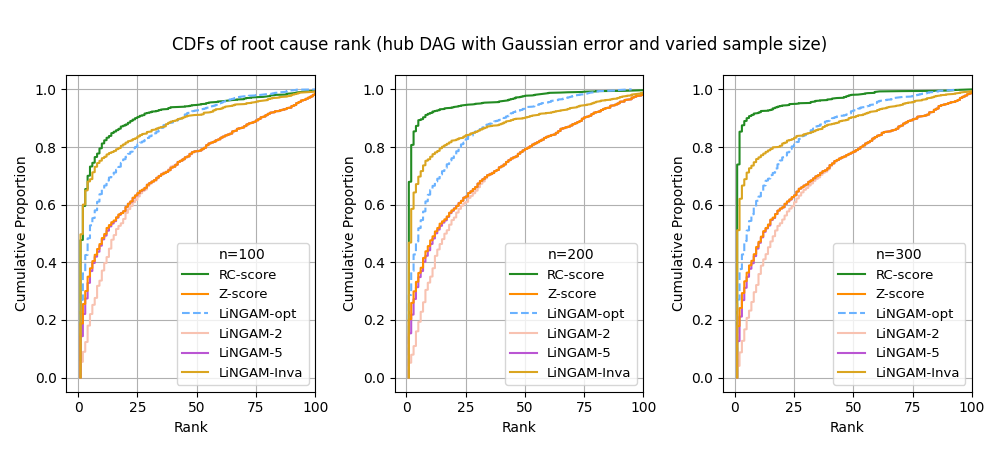}
	\end{minipage}
	\hfill
	\begin{minipage}[b]{0.88\textwidth}
		\includegraphics[width=\textwidth]{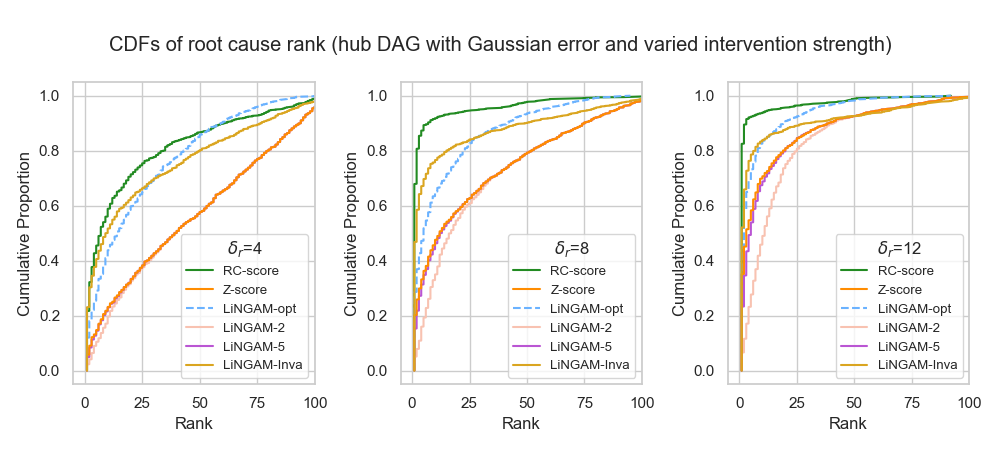}
	\end{minipage}
	\hfill
	\begin{minipage}[b]{0.88\textwidth}
		\includegraphics[width=\textwidth]{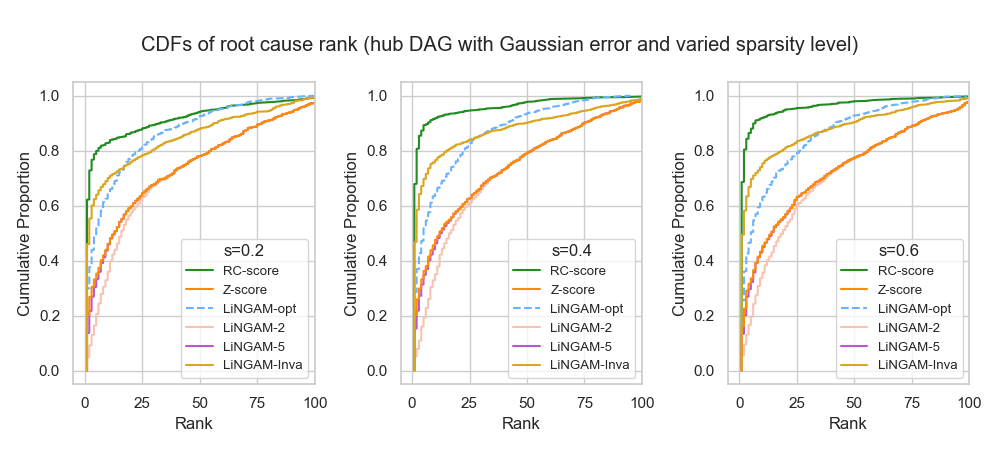}
	\end{minipage}
	\caption{CDFs of the root cause rank using the squared z-score, RC-score, and LiNGAM-based approaches in the setting with a hub DAG and Gaussian errors. The top, middle, and bottom plots display results for varying sample sizes, intervention strengths, and sparsity levels, respectively.}
	\label{Fig:MainSimuP100HubDAGGaussian}
\end{figure}

Based on the obtained scores, we calculate and record the rank of the root cause for each interventional sample. 
A smaller rank indicates a larger score for the root cause, so rank 1 is the best.
In total, we obtain $1000$ ranking values for each method, where smaller values indicate better performance.
To compare their performances, we plot the cumulative distribution function (CDF) of the root cause rank for each method. 
The value of the CDF at $x=k$ represents the percentage of times the root cause is ranked in the top $k$ among the $1000$ interventional samples. 
The results for the hub DAG with Gaussian errors are shown in Figure~\ref{Fig:MainSimuP100HubDAGGaussian}. 
The results for the hub DAG with uniform errors and the random DAG with uniform or Gaussian errors are shown in Appendix~\ref{appendix:SimuMoreMainResults}.

Our proposed RC-score outperforms the squared z-score, LiNGAM-Inva, LiNGAM-2, and LiNGAM-5 in all considered settings, and it also outperforms LiNGAM-opt in most settings. 
In particular, LiNGAM-2 and LiNGAM-5 perform worse than or similarly to the squared z-score, whereas LiNGAM-opt consistently outperforms the squared z-score, as expected. 
We plot the optimal thresholds (i.e., the squared z-score of the root cause) used in LiNGAM-opt across all settings in Appendix~\ref{appendix:SimuOptimalThre}, and these plots show significant variation in the optimal thresholds across the $1000$ interventional samples, indicating that no single fixed optimal threshold exists for all scenarios.
Even with the optimal threshold,  LiNGAM-opt does not seem to perform very well.
The performance of LiNGAM-Inva can be good if the causal DAG is well-estimated, such as in the uniform errors setting with a large sample size (see the plot with $n=300$ in Figure~\ref{Fig:MainSimuP100HubDAGUnif} in Appendix~\ref{appendix:SimuMoreMainResults}), which is particularly favorable for using LiNGAM. 
Lastly, all methods tend to perform better with larger sample sizes or stronger interventions.

The run-times for the methods are reported in Appendix~\ref{appendix:SimuCompuTime}. To give a quick idea, in the setting of a random DAG with Gaussian errors (with sample size $n = 200$, sparsity level $s = 0.4$, and intervention strength $\delta_r = 8$), one run takes on average $0.00034$ seconds for the squared z-score method, about $17$ seconds for our proposed RC score method, and about $200$ seconds for LiNGAM-based approaches.

\section{Real application on a gene expression dataset} \label{sec:realdata}

\subsection{Data description and implemented method}
In this section, we analyze the gene expression dataset mentioned in Section~\ref{sec:intro}, which motivates this paper. 
The dataset comprises gene expression data in the form of RNA-sequencing read counts from skin fibroblasts of $423$ individuals with a suspected Mendelian disorder, including $154$ non-strand-specific and $269$ strand-specific RNA-sequencing samples. The non-strand-specific dataset is available at \url{https://zenodo.org/records/4646823} \citep{yepez_2021_4646823}, and the strand-specific dataset is available at \url{https://zenodo.org/records/7510836} \citep{vicente_yepez_2023_7510836}.
These two datasets were sequenced using different protocols, yielding somewhat different distributions (17,133 differentially expressed genes out of 45,960 at an FDR of 10\% using DESeq2 \citep{love2014moderated}). Combining them increases the sample size but also introduces some confounding, leading to a bias-variance trade-off. Hence, it is generally hard to predict when combining is beneficial. In our application, however, we have a ground truth, and could therefore compare the results on the combined and separate datasets. We found that the combined dataset led to better results (see Figure~\ref{Fig:tab:realdataRankResultSSNSseparate} in Appendix~\ref{appendix:realdataSSNSseparate}). For future research, it would be interesting to explicitly model the sequencing protocol as a measured confounder and develop corresponding root cause discovery methods.
Among the $423$ patients, $58$ have (likely) known genetic mutations (see Tables S2, S3 and S4 in Additional file 1 of \cite{yepez2022clinical}).
We apply our method to identify the disease-causing gene in these patients and compare the results with the aforementioned genetic mutations, which serve as the ground truth.


When applying our method to one patient, we treat the other patients as observational samples. 
While this approach may not be ideal for detecting aberrancy and identifying the root cause, 
it is reasonable here because the aberrant genes in these rare disease patients are likely to be different. 
Better results are expected if gene expression data from healthy individuals were available as a reference.


We first apply some pre-processing and quality control steps to the raw gene expression data. 
Specifically, we filter out genes with counts less than $10$ in more than $90\%$ of the samples and remove genes that are highly correlated with others (marginal correlation greater than $0.999$). 
For the remaining genes, we follow the preprocessing procedure described by \citep{brechtmann2018outrider}, applying a log-transformation to better satisfy the linearity assumption and dividing by a size factor \citep{anders2010differential} to account for sequencing depth. This results in $p = 19736$ genes and $n=423$ samples in the pre-processed gene expression data.

We apply Algorithm~\ref{Algo:RCDhighdim} with $v=20$ random permutations for each of the $58$ patients for which we have a ground truth.
To reduce computational time, in step 2 of Algorithm~\ref{Algo:RCDhighdim}, we treat only the variables with squared z-scores greater than 1.5 as responses.
For comparison, we implement the squared z-score method.
We also implemented the LiNGAM-based methods (using \textit{DirectLiNGAM} and its high-dimensional version \textit{HighDimDirectLiNGAM}, both from the Python pacakge \textit{lingam}). However, the computation for a single patient did not complete even after 7 days on our university’s computing clusters. As a result, we do not include these methods in our comparison.

\subsection{Results}

\begin{figure}[t]
	\centering
	\begin{minipage}[b]{0.9\textwidth}
		\centering
		\includegraphics[scale=0.3]{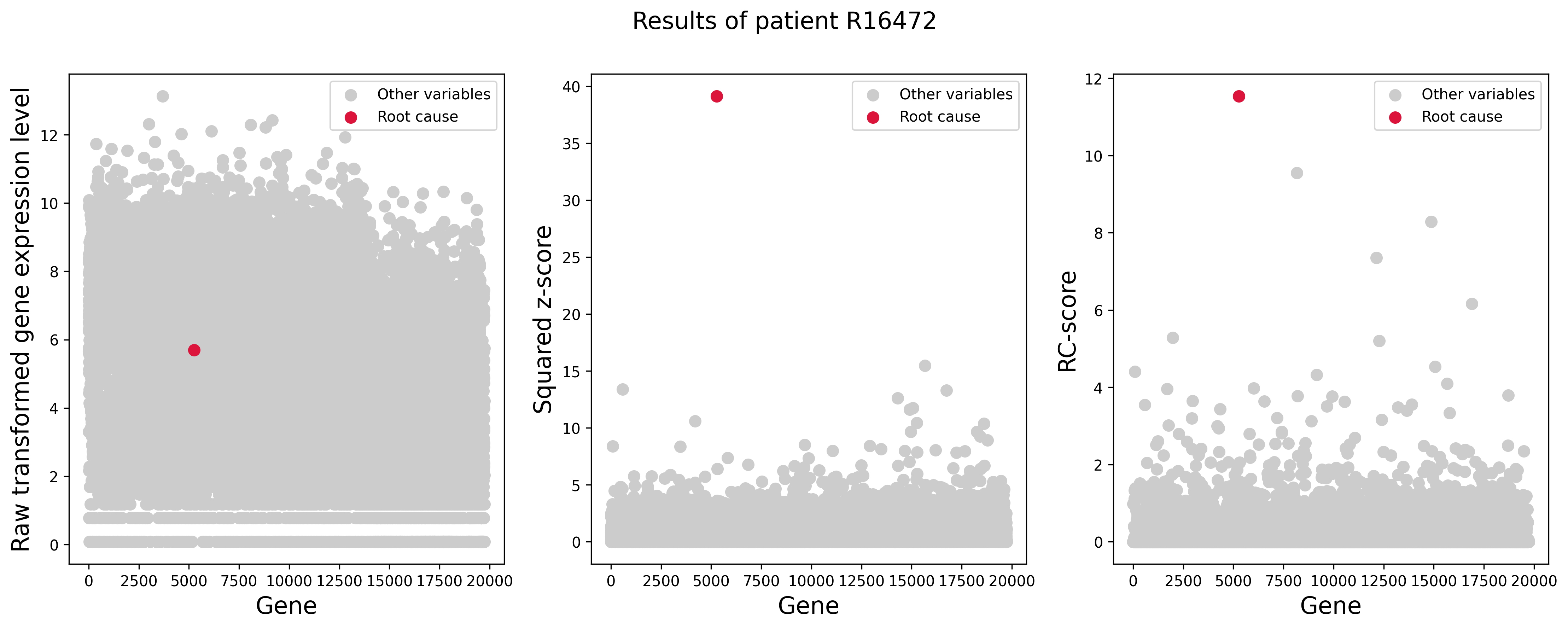}
	\end{minipage}
	\vspace{0.5cm}
	\begin{minipage}[b]{0.9\textwidth}
		\centering
		\includegraphics[scale=0.3]{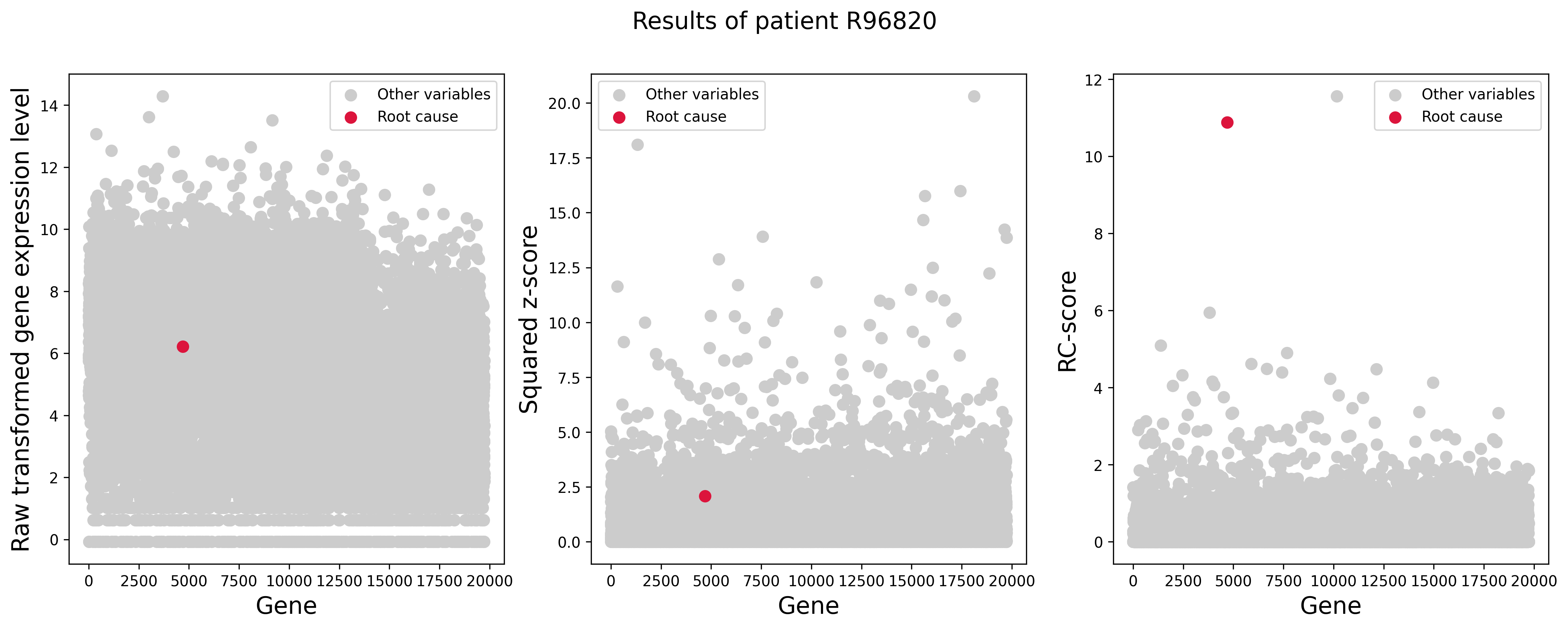}
	\end{minipage}
	\caption{The raw transformed gene expression levels, squared z-scores, and RC-scores of genes for patients $R16472$ and $R96820$.}
	\label{Fig:realdataThreeScoresPlot}
\end{figure}
Figure~\ref{Fig:realdataThreeScoresPlot} shows the raw transformed gene expression levels, squared z-scores, and RC-scores of genes for two representative patients, $R16472$ and $R96820$.
In both cases, the raw gene expression values of the root cause are obscured by other genes due to the propagation of the intervention effect, which leads to many aberrant genes.
For patient $R16472$, the squared z-score method successfully identifies the root cause. This is in line with our results in Section~\ref{sec: zScore}, which show that this method can be effective in some cases. 
The RC-score also assigns the root cause the highest score for this patient. 
However, for patient $R96820$, the squared z-score fails to distinguish the root cause while the RC-score was able to assign the root cause the second largest score. Plots for the other patients are provided in Appendix~\ref{appendix:realdata}.

Figure~\ref{Fig:tab:realdataRankResult} shows the root cause ranks based on the squared z-score and the RC-score for all $58$ patients. 
In this table, for $17$ out of $58$ patients, both methods identify the root cause correctly. For one patient, they both assign rank $2$.
For $30$ patients, the RC-score assigns a smaller rank to the root cause, indicating that it performs better than the squared z-score. 
In contrast, for $10$ patients, the squared z-score outperforms the RC-score.
It is worth noting that the RC-score often improves by a significant margin compared to the squared z-score. 
For example, for patients $R59185$ and $R34834$, the RC-score assigns the root cause a rank of 1, whereas the squared z-score gives ranks of $3057$ and $1657$, respectively. 
In cases where the squared z-score performs better, the rank differences tend to be smaller.

\begin{figure}[h!]
	\centering
	\includegraphics[scale=0.5]{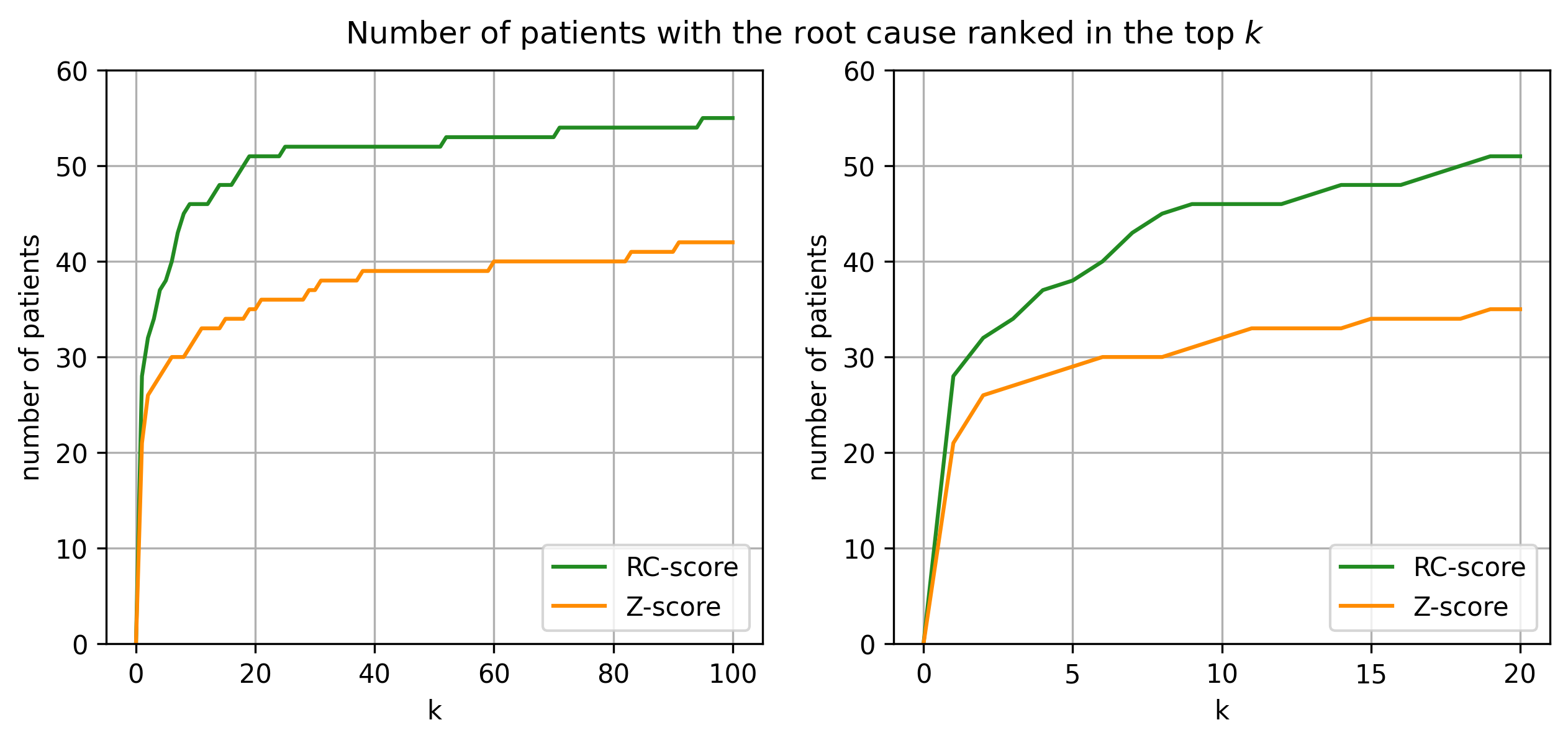}
	\caption{Number of patients with the root cause ranked in the top $k$ based on the squared z-score and the RC-score. The right plot is the zoom-in version of the left plot for $k=1,\dots,20$.}
	\label{Fig:realdataCountRank}
\end{figure}

We also count how many patients have the true root cause ranked in the top $k$ for each method, and show it in Figure~\ref{Fig:realdataCountRank}. 
It is evident that the RC-score method outperforms the squared z-score. Specifically, using the RC-score, the root cause is ranked first in $28$ out of $58$ patients, in the top 5 in $38$ patients, in the top 10 in $46$ patients, and in the top 20 in $51$ patients. 
Our method does not work well for the patients $R18626$, $R46723$, and $R12128$. 
One potential reason is that the genetic correlation structures around the root causes and their descendants in these patients are particularly complex, leading to inaccurate estimates of the associated Markov blankets and covariance matrices.

Overall, these results show that our proposed method can be useful for discovering disease-causing genes based on gene expression data.
In particular, considering that we are not in the ideal scenario where gene expression data from healthy individuals are available and with a large sample size, we expect our method to be even more effective if such a better reference is available.

\begin{figure}[h!]
	\centering
	\includegraphics[scale=0.3]{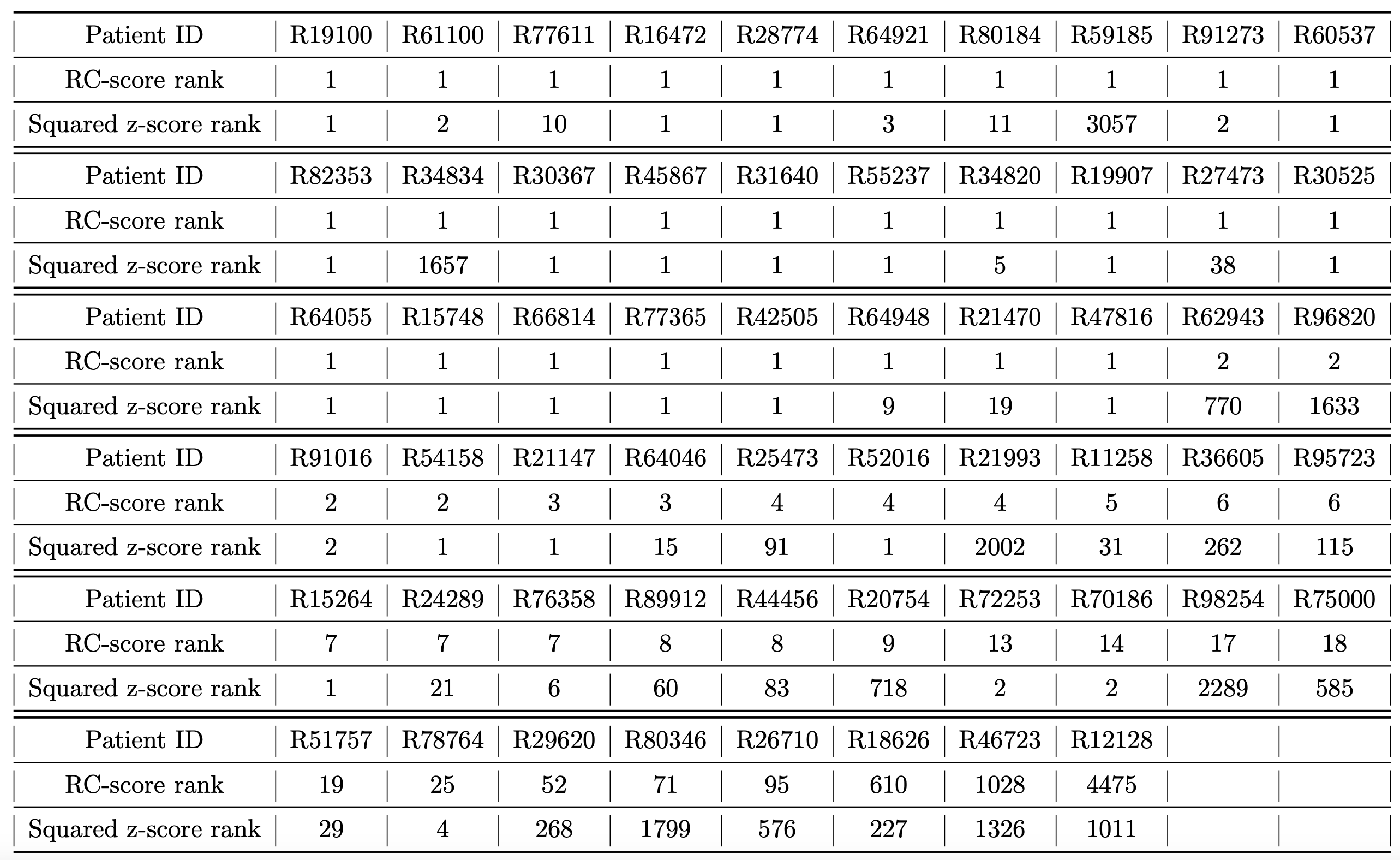}
	\caption{Table showing the rank of the root cause for $58$ patients based on the squared z-score and the RC-score. A smaller rank is better, with rank 1 being the best.}
	\label{Fig:tab:realdataRankResult}
\end{figure}

\section{Discussion} \label{sec:discussion}

There are many interesting directions for follow-up research.
One direction involves cases with latent variables. 
Although our simulations (see Appendix~\ref{sec:simu-latent}) indicate some robustness of our method to latent variables, a formal study is important and desired. This is also closely related to a deeper understanding of the theoretical properties of Algorithm~\ref{Algo:RCDhighdim}, which is designed for high-dimensional settings.

With respect to the intervention types,
we focus on mean-shift interventions in this paper because they are reasonable in genetic applications. 
There are also other types of interventions in the causal literature, such as do-intervention and variance-shift intervention \citep[see, e.g.,][]{eberhardt2007interventions, Pearl2009}. Investigating how to conduct root cause discovery in these settings is an interesting topic for future research.


Furthermore, in many applications, there may be multiple root causes, non-linear relationships between variables, or feedback loops. Thus, generalizing the current methodology to address these complexities would be important. 


Quantifying the uncertainty of our method would be very valuable. 
This seems challenging, because there are essentially two sources of uncertainty, coming from the unknown causal order and the unknown root cause.
Ideally, we would like to develop a method that captures both, so that we can
output a set of causes with a theoretical guarantee that the true root cause is inside this set with high probability. 

Finally, the main idea of using the Cholesky decomposition and permutations to exploit an invariant property opens up new possibilities of utilizing interventional samples. 
It would be interesting to leverage this idea to develop new methods in structure learning \citep[see, e.g.,][]{huang2020causal}, active causal learning, and experimental design where heterogeneous interventional samples are available.


\section*{Acknowledgments}
We are grateful to the anonymous reviewers for their constructive and valuable comments, which greatly improved the manuscript.
We thank Dominik Janzing and Daniela Schkoda for sharing with us a simplified proof of Lemma~\ref{lemma:identiOfRC}, which has been included in
Supplementary Material~\ref{app:proofThm1}.
Jinzhou Li gratefully acknowledges support by the SNSF Grant P500PT-210978.
Benjamin B. Chu gratefully acknowledges support by the grants R01MH113078, R56HG010812, R01MH123157, and the Stanford Biomedical Informatics NLM training grant T15 LM007033–40. 
Julien Gagneur and Ines F. Scheller gratefully acknowledge support by the Deutsche Forschungsgemeinschaft (DFG, German Research Foundation) via the IT Infrastructure for Computational Molecular Medicine (project \#461264291).

\bibliographystyle{apalike}
\bibliography{Reference}

\addtocontents{toc}{\vspace{.5\baselineskip}}
\addtocontents{toc}{\protect\setcounter{tocdepth}{1}}
\appendix
\newpage

\begin{center}
	{\bf Supplementary Material for ``Root cause discovery via permutations and Cholesky decomposition"}
\end{center}

The supplementary material consists of the following five appendices.
\begin{itemize}
	\item[A] Illustrations of theoretical results
	\item[B] High-dimensional version of the root cause discovery method
	\item[C] Proofs of Section~\ref{sec: zScore}
	\item[D] Proofs of Section~\ref{sec: Cholesky}
	\item[E] Supplementary materials for simulations and the real application
\end{itemize}

\section{Illustrations of theoretical results}
\subsection{A concrete example to illustrate Theorem~\ref{thm:z-scoreMain}} \label{app:ZscoreExample1}

\begin{example} \label{example:zScore}
	Consider a linear SEM 
	\begin{equation*}
		\begin{pmatrix}
			X_1 \\
			X_2 \\
			X_3
		\end{pmatrix}
		\leftarrow
		\begin{pmatrix}
			1 \\
			1 \\
			1
		\end{pmatrix}
		+
		\begin{pmatrix}
			0 & 0 & 0 \\
			2 & 0 & 0 \\
			1 & -1 & 0
		\end{pmatrix}
		\begin{pmatrix}
			X_1 \\
			X_2 \\
			X_3
		\end{pmatrix}
		+
		\begin{pmatrix}
			\err_1 \\
			\err_2 \\
			\err_3
		\end{pmatrix},
	\end{equation*}
	where $\err_1$, $\err_2$ and $\err_3$ follow standard Gaussian distribution. 
	The corresponding causal DAG with edge weights is shown in Figure~\ref{Fig:DAGsSec2}(a).
	It easily follows from this model that $\sigma_1^2 = 1$, $\sigma_2^2 = 5$, $\sigma_3^2 = 3$, $\alpha_{1 \rightarrow 2}=2$,
	$\alpha_{1 \rightarrow 3}=-1$, $\alpha_{2 \rightarrow 3}=-1$, and that all other $\alpha_{j \rightarrow k}$ equal $0$.
	Hence, when $X_1$ or $X_3$ is the root cause, condition (i) in Theorem~\ref{thm:z-scoreMain} holds for all $k \neq r$,
	so we should be able to identify the root cause.
	In contrast, when $X_2$ is the root cause, we have $\sigma_{3}^2 < \alpha_{2\rightarrow 3}^2 \sigma_2^2$,
	so $X_3$ will have a larger squared z-score than $X_2$ with high probability.
	
	We conduct simulations to verify these theoretical results. 
	We generate 100 observational samples and 30 interventional samples with an intervention strength $\delta_r = 10$. 
	Specifically, the first 10 interventional samples correspond to $r=1$ (i.e., $X_1$ being the root cause),  the next $10$ correspond to $r=2$, and the final $10$ correspond to $r=3$.
	We use the 100 observational samples to calculate the sample mean and standard deviation and obtain squared z-scores for each individual interventional sample based on formula \eqref{formula:z-score:SampleVersion}. 

	Figure~\ref{Fig:simu1} shows the results, with the first row displaying the squared z-scores for these 30 interventional samples.
	As expected from Theorem~\ref{thm:z-scoreMain}, the squared z-scores successfully identify the root cause for $X_1$ and $X_3$ (see the first and third plots containing interventional samples 1–10 and 21–30, respectively), but fail to identify the root cause for $X_2$ (see the middle plot containing interventional samples 11–20).
	In the second row, we present the performance of our RC-scores that we will introduce in Section~\ref{sec: Cholesky}. They successfully identify the root cause in all cases.
	
	\begin{figure}[htbp]
		\centering
		\includegraphics[scale=0.5]{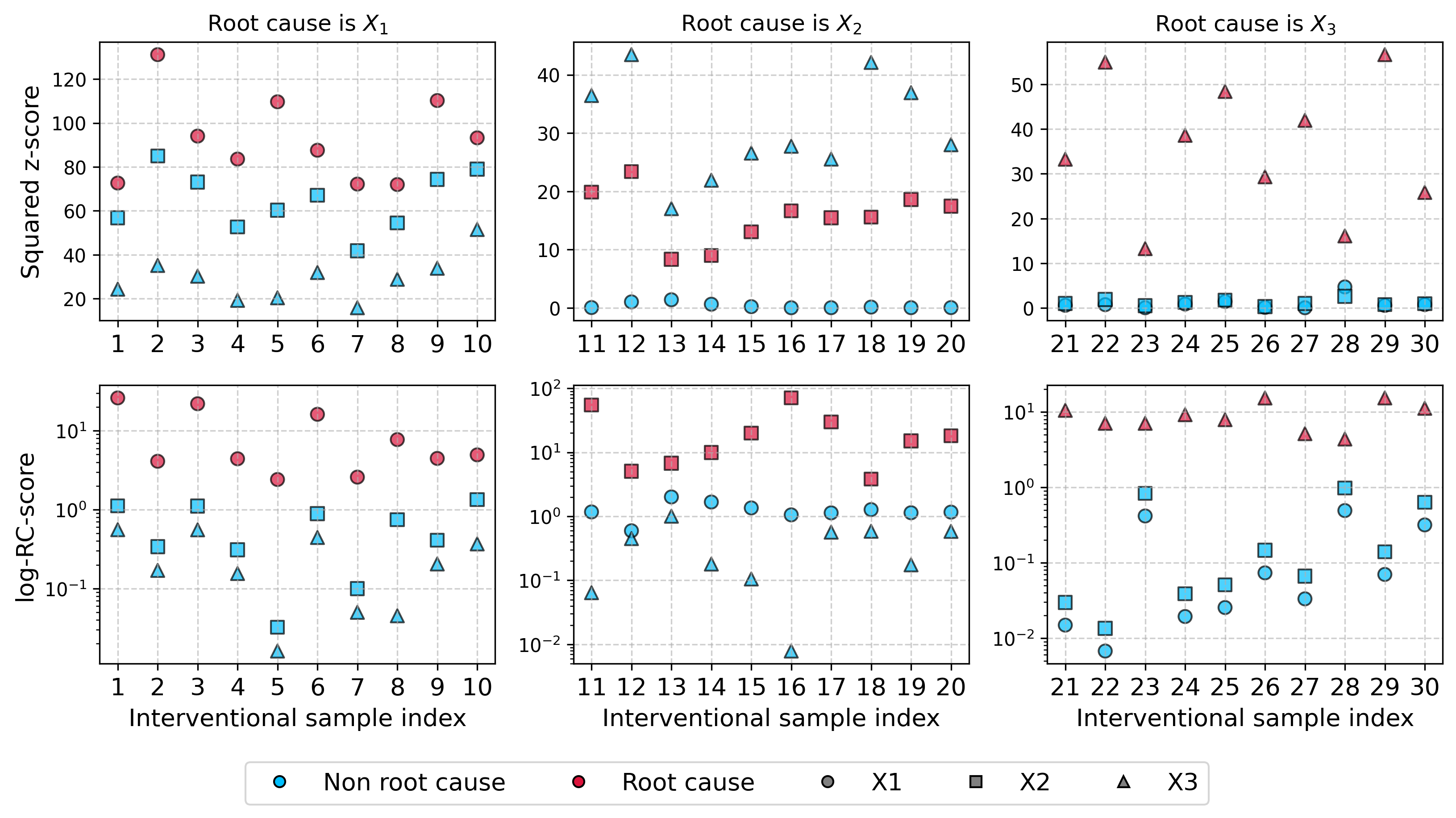}
		\caption{Simulation results based on Example~\ref{example:zScore}. 
			The three plots in the first row show the squared z-scores of the interventional samples, with root causes $X_1$, $X_2$, and $X_3$, respectively. The three plots in the second row display the RC-scores (in log scale) for the same interventional samples. For each interventional sample, red denotes the root cause while blue  represents non-root cause variables, and $X_1$, $X_2$, and $X_3$ are represented by circles, squares, and triangles, respectively.
		}
		\label{Fig:Appsimu1}
	\end{figure}

\subsection{Further discussion of Proposition~\ref{prop:categorizeVariablesZscore} with a concrete example} \label{app:ExampleProp2}
To obtain better understanding of Proposition~\ref{prop:categorizeVariablesZscore},
let us look at Example~\ref{example:zScore} a bit more closely.
In the case where $X_2$ is the root cause, if the edge $X_1 \rightarrow X_3$ were absent, the variance of $X_3$ would be larger, namely $6$,  which would result in a smaller squared z-score than that of the root cause.
However, due to the additional path $X_1 \rightarrow X_3$, which has the opposite sign of $X_1 \rightarrow X_2 \rightarrow X_3$, the variance of $X_3$ becomes smaller, namely $3$, causing its squared z-score to be larger than that of the root cause.
If we change the SEM to $X_3 = 1+X_1+X_2+\err_3$, i.e., changing the sign of the edge weight of $X_1 \rightarrow X_3$,
then $X_3$ is still unsafe according to Proposition~\ref{prop:categorizeVariablesZscore}, but $\sigma_{3}^2 > \alpha_{2 \rightarrow  3}^2 \sigma_{2}^2 $. This shows that Proposition~\ref{prop:categorizeVariablesZscore} only provides a sufficient condition.
\end{example}

\subsection{A concrete example to illustrate Theorem~\ref{thm:SufficientPermutation}} \label{app:SuffExample}
Consider the case where $p=6$, the root cause is $3$, and $D=\{1,3,6\}$.
First, we put $[p] \setminus D=\{2,4,5\}$ at the start of the permutation.
Then, we generate $3$ permutations by placing $1$, $3$, and $6$ in turn in the fourth position (i.e., the first available position after \{2,4,5\}) and randomly placing the remaining two elements.
An example of these $3$ permutations could be: $(2,4,5,1,3,6)$, $(4,5,2,3,1,6)$, and $(2,5,4,6,3,1)$.
In this case, the second permutation $(4,5,2,3,1,6)$ is sufficient.

\section{High-dimensional version of the root cause discovery method} \label{app:highdimAlgo}
We summary the high-dimensional version of the root cause discovery method in Algorithm~\ref{Algo:RCDhighdim}.
Treating all $p$ variables as responses in step 2 of Algorithm~\ref{Algo:RCDhighdim} guarantees that the root cause will be treated as a response. In practice, to reduce computational time, one may instead select a subset of variables likely to contain the root cause, for example, those whose squared z-scores exceed a certain threshold.

\begin{algorithm}[h!]
	\caption[] {\textbf{: Root cause discovery (high-dimensional version)} } \label{Algo:RCDhighdim}
	\textbf{Input}: Observational samples $\bx_1, \dots, \bx_{n}$ and the interventional sample $\bx^I$, and the number of random permutations $v$.
	
	\textbf{Output}: RC-score $\whC_i$ for all variables.
	\begin{algorithmic}[1]
		\State Initialize $\wtU=\emptyset$. 
		\For {$i=1,\dots,p$}: 
		\newline 
		\hspace{0.5cm} 
		(i) Treat $X_i$ as the response and other variables as covariates, run the cross-validated Lasso using observational samples to obtain an estimated Markov blanket $\whS$.
		Let $\bx_{\new,1}, \dots, \bx_{\new,n}$ and $\bx^I_{\new}$ be the new observational and interventional samples containing only variables in $\{ i \} \cup \whS$.
		\newline 
		\hspace{0.5cm} 
		(ii) Using $\bx_{\new,1}, \dots, \bx_{\new,n}$, $\bx^I_{\new}$, and $v$ as inputs, run the first three steps of Algorithm~\ref{Algo:RCD} to obtain $\whU$, $\whPi$, $\hc(\pi)$, and $\hu(\pi)$ for all $\pi \in \whPi$.
		\newline 
		\hspace{0.5cm} 
		(iii) If $i \in \whU$, let $\whC_i = \underset{\pi: \pi \in \whPi \text{ and } \hu(\pi)=i }{ \max} \hc(\pi)$ and include $i$ in $\wtU$.
		\EndFor
		\For {$i\in [p] \setminus \wtU$}: Let $\whC_i = \hw_i \hc_{\min}$, where 
		$\hc_{\min} = \min_{i\in\wtU} \whC_i/2$ and $\hw_i = \whZ^2_{n,i} / \sum_{j\in [p] \setminus \wtU} \whZ^2_{n,j}$ with $\whZ^2_{n,j}$ defined by~\eqref{formula:z-score:SampleVersion}.
		\EndFor
		\State Return $\whC_i$ for all $i\in [p]$.
	\end{algorithmic}
\end{algorithm}

\section{Proofs of Section~\ref{sec: zScore}} \label{app:proofSec2}

\subsection{Proof of Theorem~\ref{thm:z-scoreMain}} \label{app:proofThm2.1}
\begin{proof}(Theorem~\ref{thm:z-scoreMain})
	
	Recall that $ \alpha_{r\rightarrow j} = (I-B)^{-1}_{jr} $, 
	\begin{align*}
		X  =  (I - B)^{-1} (b + \err )
		\quad 
		\text{and}
		\quad 
		X^I= (I - B)^{-1} (b + \err + \delta ) = X + (I - B)^{-1}  \delta.
	\end{align*}
	Hence, for $j \in[p]$, we have 
	\begin{align}\label{formula:X^I_jRewrite}
		X^I_j  = X_j  + \alpha_{r\rightarrow j} \delta_r,
	\end{align}
	with $\alpha_{r\rightarrow r}=1$.

	Recall that $Z_{j} =  (X^{I}_{j} - \mu_{j})/ \sigma_{j} $ is the population version of the z-score of the interventional variable $X^I_j$, and let 
	$Z^o_j = (X_{j} - \mu_{j}) / \sigma_{j}$ be the corresponding term for the observational variable $X_j$.
	
	Let $k \in [p] \setminus \{r\}$, by \eqref{formula:X^I_jRewrite},
	we have
	\begin{align*}
		Z^2_{r} - Z^2_{k}
		=& \left( \frac{X^{I}_{r} - \mu_{r}} {\sigma_{r}} \right)^2 - \left( \frac{X^{I}_{k} - \mu_{k}} {\sigma_{k}} \right)^2 	\\
		=& \left( \frac{X_{r} + \delta_{r} - \mu_{r}} {\sigma_{r}} \right)^2 - \left( \frac{X_{k} +  \alpha_{r \rightarrow k}\delta_{r} - \mu_{k}} {\sigma_{k}} \right)^2 \\
		=&  \left( Z^o_{r} + \frac{ \delta_{r} } {\sigma_{r}} \right)^2 - \left( Z^o_{k} + \frac{  \alpha_{r \rightarrow k}\delta_{r} } {\sigma_{k}} \right)^2 \\
		=& (Z^o_{r})^2 - (Z^o_{k})^2 + \delta_r\left( \frac{2Z^o_{r}}{\sigma_r} - \frac{2Z^o_{k} \alpha_{r \rightarrow k}}{\sigma_k} \right) + \delta_r^2 \left( \frac{\sigma_{k}^2 - \alpha_{r \rightarrow  k}^2 \sigma_{r}^2}{\sigma_r^2\sigma_k^2} \right) \\
		:=& H_1 + \delta_r H_2 + \delta_r^2 h_3,
	\end{align*}
	where $H_1 = (Z^o_{r})^2 - (Z^o_{k})^2 $, $H_2 = \frac{2Z^o_{r}}{\sigma_r} - \frac{2Z^o_{k} \alpha_{r \rightarrow k}}{\sigma_k}$, and $h_3 =  \frac{\sigma_{k}^2 - \alpha_{r \rightarrow  k}^2 \sigma_{r}^2}{\sigma_r^2\sigma_k^2}$.
	Note that $\delta_r^2 h_3$ is the leading term if $\delta_r \to \infty$.
	
	In the first case of $  \sigma_{k}^2 > \alpha_{r \rightarrow  k}^2 \sigma_{r}^2 $, we have $h_3 > 0$.
	Fix a small positive $\eta$, so for a sufficiently large $\delta_r$, we have $\delta_r^2 h_3 > \eta$.
	For any $\epsilon>0$, when $\delta_r$ is large enough, we have
	\begin{equation} \label{proof:ZrMinorsZkLessThanEta}
		\begin{aligned}
			\tP\left( Z^2_{r} - Z^2_{k} \leq \eta \right) 
			&= \tP\left( H_1 + \delta_rH_2 \leq - \delta_r^2 h_3 + \eta \right) \\
			&\leq \tP\left( |H_1 + \delta_rH_2| \geq \delta_r^2 h_3 - \eta \right) \\
			&\leq \frac{\E[|H_1 + \delta_rH_2|]}{\delta_r^2 h_3 - \eta} \\
			&\leq \frac{\E[|H_1|] + \E[|\delta_rH_2|]}{\delta_r^2 h_3 - \eta} \\
			&= \frac{\E[|(Z^o_{r})^2 - (Z^o_{k})^2|] + \E\left[\left|\frac{2\delta_rZ^o_{r}}{\sigma_r} - \frac{2\delta_rZ^o_{k} \alpha_{r \rightarrow k}}{\sigma_k} \right| \right]} {\delta_r^2 h_3 - \eta} \\
			&\leq \frac{\E[(Z^o_{r})^2| + \E[(Z^o_{k})^2] + \frac{2|\delta_r|}{\sigma_r} \E[ |Z^o_{r}| ] + \frac{2|\delta_r\alpha_{r \rightarrow k}|}{\sigma_k}  \E[ |Z^o_{k}| ] } {\delta_r^2 h_3 - \eta} \\
			&\leq \frac{2}{\delta_r^2 h_3 - \eta} \left(1 + \frac{|\delta_r|}{\sigma_r} + \frac{|\delta_r\alpha_{r \rightarrow k}|}{\sigma_k}   \right) \\
			&\leq \epsilon/3,
		\end{aligned}
	\end{equation}
	where we used  Markov's inequality in the second inequality, 
	and $\E[(Z^o_{j})^2] = 1$ and $\E[|Z^o_{j}|] \leq 1$ for $j\in[p]$ in the second last inequality, as $ Z^o_j$ is a standardized random variable with mean $0$ and variance $1$.
	
	Because $\whmu_{n, j} \xrightarrow[]{p} \mu_{j}$ and $\whsigma_{n, j}
	\xrightarrow[]{p} \sigma_{j}$ as $n \to \infty$,
	we have $\whZ^2_{n, j} \xrightarrow[]{p} Z^2_{j}$ as $n \to \infty$ by the continuous mapping theorem.
	Hence, for any $\epsilon>0$ and large enough $n$, we have
	\begin{align} \label{proof:ZnZconverge}
		\tP\left( |\whZ^2_{n, r} - Z^2_{r}| > \eta/4\right) \leq \epsilon/3
		\quad \text{and} \quad 
		\tP\left( |\whZ^2_{n, k} - Z^2_{k}| > \eta/4\right) \leq \epsilon/3.
	\end{align}
	
	Therefore, for any $\epsilon>0$, when $\delta_r$ and $n$ are sufficiently large, 
	we have
	\begin{align*}
		\tP\left( \whZ^2_{n, r} > \whZ^2_{n, k} \right) 
		\geq & \tP\left( Z^2_{r} - Z^2_{k}> \eta, |\whZ^2_{n, r} - Z^2_{r}| \leq \eta/4, |\whZ^2_{n, k} - Z^2_{k}| \leq \eta/4 \right) \\
		\geq & 1 - \tP\left( Z^2_{r} - Z^2_{k} \leq \eta\right) - \tP\left( |\whZ^2_{n, r} - Z^2_{r}| >\eta/4\right)- \tP\left( |\whZ^2_{n, k} - Z^2_{k}| > \eta/4 \right) \\
		\geq &1 - \epsilon,
	\end{align*}
	which by definition proves that
	\begin{align*}
		\lim\limits_{\substack{n \to \infty \\ \delta_r \to \infty}} \tP\left( \whZ^2_{n,r} > \whZ^2_{n,k} \right) = 1.
	\end{align*}
	
	The second case of $  \sigma_{k}^2 - \alpha_{r \rightarrow  k}^2 \sigma_{r}^2 < 0$ can be proved similarly.
\end{proof}

\subsection{Proof of Propositions~\ref{prop:categorizeVariablesZscore} and \ref{prop:allposSign}}
\begin{proof}(Propositions~\ref{prop:categorizeVariablesZscore} and \ref{prop:allposSign})
	We start with the proof of Propositions~\ref{prop:categorizeVariablesZscore}.
	
	For (i), when $k \not\in \De(r)$, we have $ \alpha_{r \rightarrow k} = 0$, so
	$ \sigma_{k}^2 > \alpha_{r \rightarrow  k}^2 \sigma_{r}^2 = 0$. 
	
	For (ii), 
	recall that $X  =  (I - B)^{-1} (b + \err )$ and $ \alpha_{j\rightarrow k} = (I-B)^{-1}_{kj} $,  
	so  
	\begin{align*}
		X_k = \mu_k + \sum_{j \neq k}  \alpha_{j\rightarrow k} \err_j + \err_k,
	\end{align*}
	where $\mu_k = \E[X_k]$.
	
	From now on, without loss of generality, we assume that $X_1,\dots, X_p$ is sorted by a causal ordering. That is, 
	$i < j$ for $j \in \De(i)$, where $\De(i)$ denotes the descendants of $i$.
	Then we have 
	\begin{align*}
		X_k = \mu_k + \sum_{j =1}^{k-1}  \alpha_{j\rightarrow k} \err_j + \err_k.
	\end{align*}
	
	Denote $\theta_j^2 = \Var(\err_j)$, then
	\begin{align*}
		\sigma_{k}^2 = \Var(X_k) = \sum_{j =1}^{k-1}  \alpha^2_{j\rightarrow k} \theta^2_j + \theta^2_k.
	\end{align*}
	
	For $k \in \De(r)$, we have $k>r$.
	The condition 
	\begin{align*}
		 &\sigma_{k}^2 - \alpha_{r \rightarrow  k}^2 \sigma_{r}^2 \\
		=
		& \sum_{j =1}^{k-1}  \alpha^2_{j\rightarrow k} \theta^2_j + \theta^2_k
		- \alpha_{r \rightarrow  k}^2 \left(\sum_{j =1}^{r-1}  \alpha^2_{j\rightarrow r} \theta^2_j + \theta^2_r \right) \\
		=& \alpha^2_{1\rightarrow k} \theta^2_1 + \dots + \alpha^2_{k-1\rightarrow k} \theta^2_{k-1} + \theta^2_k
		-  \left(\alpha_{r \rightarrow  k}^2 \alpha^2_{1\rightarrow r} \theta^2_1 + \dots + \alpha_{r \rightarrow  k}^2 \alpha^2_{r-1\rightarrow r} \theta^2_{r-1} \right) - \alpha_{r \rightarrow  k}^2 \theta^2_r \\
		=& (\alpha^2_{1\rightarrow k} - \alpha_{r \rightarrow  k}^2 \alpha^2_{1\rightarrow r})\theta^2_1 + \dots
		+ (\alpha^2_{r-1\rightarrow k} - \alpha_{r \rightarrow  k}^2 \alpha^2_{r-1\rightarrow r})\theta^2_{r-1} \\
		&+ \alpha_{r \rightarrow  k}^2 \theta^2_r  - \alpha_{r \rightarrow  k}^2 \theta^2_r \\
		&+ \bbone_{\{ k \geq r+2 \}} (\alpha_{r+1 \rightarrow  k}^2 \theta^2_{r+1} +\dots + \alpha_{k-1 \rightarrow  k}^2 \theta^2_{k-1} )+ \theta^2_{k}\\
		:=& U + V,
	\end{align*}
	where
	\begin{align*}
		U = (\alpha^2_{1\rightarrow k} -  \alpha^2_{1\rightarrow r}\alpha_{r \rightarrow  k}^2)\theta^2_1 + \dots
		+ (\alpha^2_{r-1\rightarrow k} -  \alpha^2_{r-1\rightarrow r}\alpha_{r \rightarrow  k}^2)\theta^2_{r-1} 
	\end{align*}
	and
	\begin{align*}
		V = \bbone_{\{ k \geq r+2 \}} (\alpha_{r+1 \rightarrow  k}^2 \theta^2_{r+1} +\dots + \alpha_{k-1 \rightarrow  k}^2 \theta^2_{k-1}) + \theta^2_{k} > 0.
	\end{align*}
	
	Consider the terms in $U$ and note that $\alpha_{j \rightarrow  k}$ is the total causal effect of $X_j$ on $X_k$, and  $\alpha_{j \rightarrow  r} \alpha_{r \rightarrow  k}$ is the total causal effect of $X_j$ on $X_k$ \textit{through} $X_r$.
	Hence, if $O(r,k) = \emptyset$, then $\alpha_{j \rightarrow  k} = \alpha_{j \rightarrow  r} \alpha_{r \rightarrow  k}$ for all $j<r$. This implies that $U=0$ and $\sigma_{k}^2 > \alpha_{r \rightarrow  k}^2 \sigma_{r}^2 $.
	
	Next, we prove Propositions~\ref{prop:allposSign}.
	
	We only need to consider the case where $k \in \De(r)$.
	If all edge weights are nonnegative,
	then the causal effects along all paths are nonnegative, hence
	\begin{align*}
		\alpha_{j \rightarrow  k} \geq \alpha_{j \rightarrow  r} \alpha_{r \rightarrow  k}.
	\end{align*}
	This implies $U \geq 0$ and $\sigma_{k}^2 > \alpha_{r \rightarrow  k}^2 \sigma_{r}^2 $.
\end{proof}

\section{Proofs of Section~\ref{sec: Cholesky}} \label{app:proofSec3}

We first introduce some notation that will be used in the following proofs.

For a permutation $\pi$, let $P^{\pi}$ be the permutation matrix that permutes the rows of the identity matrix according to $\pi$.
Then, for the permuted $X^{\pi}$ and $X^{I{\pi}}$ (see Section~\ref{sec:RC-all-perm}), we have 
\begin{equation*}
	X^{\pi}  =  (I - B^{\pi})^{-1} (b^{\pi}+ \err^{\pi} )
	\quad 
	\text{and}
	\quad
	X^{I{\pi}}= (I - B^{\pi})^{-1} (b^{\pi} + \err^{\pi} + \delta^{\pi}),
\end{equation*}
where $b^{\pi} = P^{\pi} b$, $\err^{\pi} = P^{\pi}\err$, and $B^{\pi} = P^{\pi}B(P^{\pi})^T$.
It is then follows that $\mu_{X^{\pi}} = (I - B^{\pi})^{-1} b^{\pi}$,
$\mu_{X^{I\pi} } = (I - B^{\pi})^{-1} (b^{\pi} + \delta^{\pi})$,
and
$\Sigma_{X^{\pi}} = (I - B^{\pi})^{-1} D_{\err^{\pi}} (I - B^{\pi})^{-T}$.
Hence, for the $\xi(\pi)$ defined in Section~\ref{sec:RC-all-perm}, we have
\begin{align} \label{formula:xi(pi)}
	\xi(\pi) =  L_{X^{\pi}}^{-1} ( \mu_{X^{I\pi} } - \mu_{X^{\pi}})
	=L_{X^{\pi}}^{-1}  (I - B^{\pi})^{-1} \delta^{\pi}.
\end{align}

\subsection{Proof of Theorem~\ref{thm:identiOfRC}} \label{app:proofThm1}
Before proving Theorem~\ref{thm:identiOfRC}, we introduce the following Lemma~\ref{lemma:identiOfRC}, whose proof will be provided after the proof of Theorem~\ref{thm:identiOfRC}.
\begin{lemma} \label{lemma:identiOfRC}
	Let $A \in \bbR^{p \times p}$ be an invertible matrix for which there exists a permutation matrix $P \in \bbR^{p \times p}$ such that $PAP^T$ is lower-triangular. 
	Let $L$ be the lower-triangular matrix obtained by  the Cholesky decomposition of $A A^T$.
	For any fixed $j \in [p]$, if the $j$-th column of $L^{-1} A$ has exactly one non-zero element, then this non-zero element must be in the $j$-th position of this column.
\end{lemma}

We first prove Theorem~\ref{thm:identiOfRC}.
\begin{proof}(Theorem~\ref{thm:identiOfRC})
    Using the notation $\xi(\pi) =  L_{X^{\pi}}^{-1} ( \mu_{X^{I\pi} } - \mu_{X^{\pi}})$ (see Section~\ref{sec:RC-all-perm}),
    proving Theorem~\ref{thm:identiOfRC} is equivalent to proving that for any permutation $\pi$: 
    \begin{itemize}
    	\item[(i)] $\xi(\pi)$ must have at least one nonzero element;
    	\item[(ii)] If $\xi(\pi)$ has exactly one nonzero element, then this element must be in the $\pi^{-1}(r)$-th position, that is, the position of the root cause.
    \end{itemize}
    We prove these two statements by applying Lemma~\ref{lemma:identiOfRC}.
	
	Let $ A= (I - B^{\pi})^{-1} D^{1/2}_{\err^{\pi}} $ which satisfies the conditions in Lemma~\ref{lemma:identiOfRC}.
	Note that $\Sigma_{X^{\pi}} =A A^T$, so $ L_{X^{\pi}}$ is the lower-triangular matrix obtained by implementing the Cholesky decomposition of $A A^T$. 
	
	By \eqref{formula:xi(pi)}, we have that $\xi(\pi) =L_{X^{\pi}}^{-1}  (I - B^{\pi})^{-1} \delta^{\pi}$.
	Since $\delta^{\pi}$ has only one nonzero entry in the $\pi^{-1}(r)$-th position and $D_{\err^{\pi}}$ is a diagonal matrix with positive diagonals,
	 $\xi(\pi)$ has the same support as the $\pi^{-1}(r)$-th column of $L_{X^{\pi}}^{-1} A= L_{X^{\pi}}^{-1}  (I - B^{\pi})^{-1} D_{\err^{\pi}}^{1/2}$.
	
	Because $L_{X^{\pi}}^{-1} A$ is invertible, its $\pi^{-1}(r)$-th column must contain at least one nonzero element, hence 
	$\xi(\pi)$ must have at least one nonzero element. This proves statement (i).
	
	Statement (ii) follows by applying Lemma~\ref{lemma:identiOfRC} with $j=\pi^{-1}(r)$.
\end{proof}

Now we give the proof of Lemma~\ref{lemma:identiOfRC}.
\begin{proof}(Lemma~\ref{lemma:identiOfRC})
	Denote
	\begin{equation*}
		A=
		\begin{pmatrix}
			a_{11} & a_{12} & \dots & a_{1p} \\
			a_{21} & a_{22} & \dots & a_{2p} \\
			\vdots & \vdots & \ddots & \vdots \\
			a_{p1} & a_{p2} & \dots & a_{pp} \\
		\end{pmatrix}
		=
		\begin{pmatrix}
			a_1^T  \\
			a_2^T  \\
			\vdots \\
			a_p^T \\
		\end{pmatrix}
		= [a^{(1)}, a^{(2)}, \dots, a^{(p)}],
	\end{equation*}
	so
	\begin{equation*}
		A^T = [a_1, a_2, \dots, a_p].
	\end{equation*}
	Since $A$ is invertible and $PAP^T$ is lower-triangular, it follows that all diagonal elements of $A$ are non-zero. 
	
	Let $A^T=QR$ be the QR decomposition of $A^T$, where
	\[
	Q =  
	\begin{pmatrix}
		q_{11} & q_{12} & \dots & q_{1p} \\
		q_{21} & q_{22} & \dots & q_{2p} \\
		\vdots & \vdots & \ddots & \vdots \\
		q_{p1} & q_{p2} & \dots & q_{pp} \\
	\end{pmatrix}= [q^{(1)}, \dots, q^{(p)}]
	\]
	is an orthogonal matrix (i.e., $QQ^T=Q^TQ=I$, where $I$ is the identity matrix) and 
	\begin{equation} \label{proof:formulaOfR}
		R =
		\begin{pmatrix}
			\innP{q^{(1)}, a_1} & \innP{q^{(1)},a_2}  & \dots & \innP{q^{(1)},a_p} \\
			0 & \innP{q^{(2)},a_2}  & \dots & \innP{q^{(2)},a_p} \\
			\vdots & \vdots  & \ddots & \vdots \\
			0 & 0  & \dots & \innP{q^{(p)},a_p} \\
		\end{pmatrix}
	\end{equation}
	is an upper-triangular matrix with positive diagonal elements. We have $L L^T = A A^T = R^T Q^T Q R = R^T R$, so $L=R^T$ by the uniqueness of the Cholesky decomposition for a positive-definite matrix. 
	
	We now consider the $j$-th column of $L^{-1} A $, i.e., $L^{-1} A e_j$, where $e_j = (0, \dots, 0, 1, 0, \dots, 0)^T \in \bbR^{p}$ with the non-zero element $1$ in the $j$-th position. 
	Suppose that this column contains one non-zero element, i.e., $L^{-1} A e_j = c e_k$ for some $c \neq 0$ and $k \in [p]$. We need to show that this non-zero element is in the $j$-th position, i.e., $k=j$.
	
	First note that $L^{-1} A e_j = c e_k$, $L=R^T$ and the form of $R$ (see \eqref{proof:formulaOfR}) imply
	\begin{equation*}
		a^{(j)} = A e_j = c L e_k = c R^T e_k
		= c \begin{pmatrix}
			0 \\
			\vdots\\
			0\\
			\innP{q^{(k)},a_k} \\
			\vdots\\
			\innP{q^{(k)},a_p} \\
		\end{pmatrix}.
	\end{equation*}
	Since $a_{jj} \neq 0$, we must have $k \leq  j$.
	
	Next, note that 
	\begin{equation*}
		c e_k = L^{-1} A e_j = L^{-1} R^T Q^T e_j = L^{-1} L Q^T e_j = Q^T e_j,
	\end{equation*}
	so that
	\begin{equation*}
		e_j = c Q e_k = c q^{(k)}.
	\end{equation*}
	Since $1 = \lVert e_j \rVert_2 = \lVert c q^{(k)} \rVert_2 = |c|$, we have $c = \pm 1$ and $q^{(k)} = \pm e_j$.
	
	It is well-known by the Gram-Schmidt process that $q^{(k)}$ can be written as a linear combination of vectors $a_1, \dots, a_k$, 
	so there exists  $b_{h_1}, \dots, b_{h_m} \neq 0$ such that 
	$q^{(k)} = b_{h_1} a_{h_1} + \dots + b_{h_m} a_{h_m}$,
	where $1 \leq m\leq k$ and $h_1, \dots, h_m \leq k$.
	
	Now assume $k < j$, we will show that this leads to a contradiction, which then concludes that $k=j$.
	
	We have $h_1, \dots, h_m \leq k<j$ and hence the $h_1, \dots, h_m $-th elements of $q^{(k)}$ equal zero as  $q^{(k)} = \pm e_j$. Looking at these elements, we have 
	\begin{equation*}
		\begin{aligned}
			\begin{pmatrix}
				0 \\
				\vdots\\
				0\\
			\end{pmatrix}
			= &
			b_{h_1}
			\begin{pmatrix}
				a_{h_1 h_1 } \\
				\vdots\\
				a_{h_1 h_m } \\
			\end{pmatrix}
			+ \dots + 
			b_{h_m}
			\begin{pmatrix}
				a_{h_m h_1 } \\
				\vdots\\
				a_{h_m h_m }
			\end{pmatrix}
			= 
			\begin{pmatrix}
				a_{h_1 h_1 } & \dots & a_{h_m h_1} \\
				\vdots  & \ddots & \vdots \\
				a_{h_1 h_m } & \dots & a_{h_m h_m}
			\end{pmatrix}
			\begin{pmatrix}
				b_{h_1 } \\
				\vdots\\
				b_{h_m}
			\end{pmatrix}.
		\end{aligned}
	\end{equation*}
	Denote
	\begin{equation*}
		\tilde{A} =
		\begin{pmatrix}
			a_{h_1 h_1 } & \dots & a_{h_m h_1} \\
			\vdots & \ddots & \vdots \\
			a_{h_1 h_m }& \dots & a_{h_m h_m} \\
		\end{pmatrix},
	\end{equation*}
	then $\tilde{A}$ must be non-invertible, otherwise $b_{h_1}=\dots= b_{h_m}=0$ and $q^{(k)} =0$. However, $\tilde{A}$ is obtained by taking the $h_1$, $\dots$, $h_m$-th rows and columns of $A^T$, whose diagonal elements are non-zero and for which there exists a permutation matrix $P$ such that $P A^T P^T$ is  upper-triangular, hence $\tilde{A}$ must be invertible, which contradicts the previous conclusion. 
\end{proof}

We also provide a simplified proof of Lemma~\ref{lemma:identiOfRC}, originally due to Dominik Janzing and Daniela Schkoda.
To this end, we first show the following lemma.

\begin{lemma}\label{lemma:byDJandDS}
	Let $LL^T = AA^T$, where $L\in \bbR^{p \times p}$ is an invertible lower triangular matrix and $A\in \bbR^{p \times p}$ satisfies that there exists a permutation matrix $P \in \bbR^{p \times p}$ such that $PAP^T$ is lower-triangular. 
	If the $k$-th column of $L$ is proportional to the $j$-th column of $A$, then $k = j$.
\end{lemma}
\begin{proof}(Lemma~\ref{lemma:byDJandDS})
	Since $LL^T = AA^T$ and $L $ is invertible, it follows that $A$ is also invertible, and hence $A_{jj} \neq 0$.
	Moreover, since the first $k - 1$ entries of the $k$-th column of $L$ are zero, we must have $k \leq j$. 
	It is then sufficient to show that $k \geq j$. 
	
	By $LL^T = A A^T$, the matrix $L^{-1} A$ is orthogonal (for any two invertible matrices $A$ and $B$, the equation $AA^T = BB^T$ implies that $B^{-1}A$ is orthogonal),
	hence $A^{-1} L = (L^{-1} A)^{-1} = (L^{-1} A)^T = A^T L^{-T}$.
	Since $L e_k = c A e_j$ for some $c \neq 0$ by assumption, we have $ c e_j = A^{-1} L e_k = A^T L^{-T} e_k$. 
	Denote $\lambda := L^{-T} e_k$. Since $L^{-T}$ is upper triangular, we have $\lambda_{k+1} = \cdots = \lambda_p = 0$. Therefore,
	\[
	ce_j = A^T (L^{-T} e_k) = A^T \lambda = (A^T)_{:,1:k} \lambda_{1:k}.
	\]
	Restricting to the first $k$ entries of the equation, we obtain
	\[
	c(e_j)_{1:k} = (A^T)_{1:k,1:k} \lambda_{1:k}.
	\]
	Since $\lambda_k \neq 0$, and $(A^T)_{1:k,1:k}$ is invertible (since there exists a permutation matrix $P$ such that $P A P^T$ is lower-triangular with non-zero diagonals), we have $(e_j)_{1:k} \neq 0$, which implies that $k \geq j$.
\end{proof}

Now we give the second proof of Lemma~\ref{lemma:identiOfRC}.
\begin{proof}(Second proof of Lemma~\ref{lemma:identiOfRC})
	Let $e_j = (0, \dots, 0, 1, 0, \dots, 0)^T \in \bbR^{p}$ with the non-zero element $1$ in the $j$-th position. 
	Suppose that the $j$-th column of $L^{-1} A $ contains one non-zero element, i.e., $L^{-1} A e_j = c e_k$ for some $c \neq 0$ and $k \in [p]$. 
	We need to show that $k=j$.
	Note that $ L^{-1} A e_j = c e_k \Longleftrightarrow A e_j = c L e_k$, thus $k=j$ by Lemma~\ref{lemma:byDJandDS}.
\end{proof}

\subsection{Proof of Theorem~\ref{thm:ConsistAlgoFullPerm}}
\begin{proof}(Theorem~\ref{thm:ConsistAlgoFullPerm})
	The proof is almost the same as the proof of Theorem~\ref{thm:ConsistSufficientAlgo}, 
	with the only difference that we now directly have
	\begin{align*}
		\tP \left(\text{$ \Pi_{\all}$ contains at least one sufficient permutation}\right) = 1.
	\end{align*}
	Hence we omit the proof.
\end{proof}

\subsection{Proof of Theorem~\ref{thm:SufficientPermutation}}
Before proving Theorem~\ref{thm:SufficientPermutation}, we present Lemma~\ref{lemma:OneNoneZeroWRTA}, with its proof given after the proof of Theorem~\ref{thm:SufficientPermutation}.
\begin{lemma} \label{lemma:OneNoneZeroWRTA}
	Let $A \in \bbR^{p \times p}$ be an invertible matrix for which there exists a permutation matrix $P \in \bbR^{p \times p}$ such that $PAP^T$ is lower-triangular. 
	Let $L$ be the lower-triangular matrix obtained by  the Cholesky decomposition of $A A^T$.
	For any fixed $j \in [p]$, if
	\begin{itemize}
		\item[(i)] $A_{kj}=0$ for all $k<j$, and 
		\item[(ii)]$(A^{-1})_{jk}=0$ for all $k>j$
	\end{itemize}
	hold, then the $j$-th column of $L^{-1} A$ has exactly one non-zero element in the $j$-th position.
	If one of the above two conditions does not hold, then the $j$-th column of $L^{-1} A$ has at least two non-zero elements.
\end{lemma}

We first give the proof of Theorem~\ref{thm:SufficientPermutation}.
\begin{proof}(Theorem~\ref{thm:SufficientPermutation})
	We prove this theorem by applying Lemma~\ref{lemma:OneNoneZeroWRTA}.
	
	Let $ A= (I - B^{\pi})^{-1} D^{1/2}_{\err^{\pi}} $ which satisfies the conditions in Lemma~\ref{lemma:OneNoneZeroWRTA}.
	Note that $\Sigma_{X^{\pi}} =A A^T$, so $ L_{X^{\pi}}$ is the lower-triangular matrix obtained by implementing the Cholesky decomposition of $A A^T$. 
	
	Because $D_{\err^{\pi}} $ is a diagonal matrix with positive diagonals, $A$ has the same support as $(I - B^{\pi})^{-1} $ and 
	the off-diagonal part of $A^{-1}$ has the same support as $B^{\pi}$. 
	Hence, we have
	\begin{align*}
		\text{$\pi^{-1}(k) < \pi^{-1}(r)$ for all $k \in \Pa(r)$} 
		&\Longleftrightarrow \text{$B^{\pi}_{\pi^{-1}(r)j} = 0$ for all $j > \pi^{-1}(r)$} \\
		&\Longleftrightarrow \text{$(A^{-1})_{\pi^{-1}(r)j}=0$ for all $j>\pi^{-1}(r)$},
	\end{align*}
	\begin{align*}
		\text{ $\pi^{-1}(k) > \pi^{-1}(r)$ for all $ k \in \rDe(r)$ } 
		&\Longleftrightarrow \text{$(I-B^{\pi})^{-1}_{j\pi^{-1}(r)} = 0$ for all $j < \pi^{-1}(r)$} \\
		&\Longleftrightarrow \text{$A_{j\pi^{-1}(r)}=0$ for all $j<\pi^{-1}(r)$}.
	\end{align*}
	
	By \eqref{formula:xi(pi)}, we have that $\xi(\pi) =L_{X^{\pi}}^{-1}  (I - B^{\pi})^{-1} \delta^{\pi}$.
	Since $\delta^{\pi}$ has only one nonzero entry in the $\pi^{-1}(r)$-th position and $D_{\err^{\pi}}$ is a diagonal matrix with positive diagonals,
	$\xi(\pi)$ has the same support as the $\pi^{-1}(r)$-th column of $L_{X^{\pi}}^{-1} A= L_{X^{\pi}}^{-1}  (I - B^{\pi})^{-1} D_{\err^{\pi}}^{1/2}$.
	
	The results then follow from applying Lemma~\ref{lemma:OneNoneZeroWRTA} with $j=\pi^{-1}(r)$.
\end{proof}

Now we prove Lemma~\ref{lemma:OneNoneZeroWRTA}.
\begin{proof}(Lemma~\ref{lemma:OneNoneZeroWRTA})
	First, we prove that if conditions (i) and (ii) hold for some fixed $j\in [p]$, then the $j$-th column of $L^{-1} A$ has exactly one non-zero element in the $j$-th position.
	
	Let $c = (c_1,\dots,c_p)^T$ be the $j$-th column of $L^{-1} A$.
	It suffices to prove that $c_j \neq 0$ and $c_l=0$ for any $l \in[p]\setminus \{j\}$.
	We prove this case by case.
	
	\textbf{Case 1:} $l < j$.
	Because $L$ is lower-triangular, we have  $(L^{-1})_{ik}=0$ for all $k>i$, so
	$c_l = \sum_{k=1}^{p} (L^{-1})_{lk} A_{kj} = \sum_{k=1}^{l} (L^{-1})_{lk} A_{kj} $.
	Then, by condition (i) that $A_{kj}=0$ for all $k<j$, we have 
	$c_l = \sum_{k=1}^{l} (L^{-1})_{lk} A_{kj} = 0$ as $l < j$.
	
	\textbf{Case 2:} $l = j$.
	Following the same arguments as above, we have $c_l = c_j =  (L^{-1})_{jj} A_{jj}$.
    $A$ is invertible and $PAP^T$ is lower-triangular implies that all diagonal elements of $A$ are non-zero,
    and $L$ is lower-triangular with positive diagonals implies that all diagonal elements of $L^{-1}$ are non-zero.
    Therefore, $c_j =(L^{-1})_{jj} A_{jj} \neq 0$.
  
	\textbf{Case 3:} $l > j$.
	Note that $L^{-1} A = L^T A^{-T}$ as $A A^T = L L^T$,
	so $c_l =  \sum_{k=1}^{p} (L^T )_{lk} (A^{-T})_{kj}$.
	$L$ is lower-triangular implies that $(L^T)_{lk}=0$ for all $l>k$,
	so  $c_l =  \sum_{k=1}^{p} (L^T )_{lk} (A^{-T})_{kj} =  \sum_{k=l}^{p} (L^T )_{lk} (A^{-T})_{kj}$.
	Additionally, condition (ii) that $(A^{-1})_{jk}=0$ for all $k>j$ implies that $(A^{-T})_{kj}=0$ for all $k>j$,
	hence $c_l = \sum_{k=l}^{p} (L^T )_{lk} (A^{-T})_{kj} = 0$ as $l > j$.
	
	 The claim is then proved by combining the above three cases.
	
	Next, we prove that if conditions (i) or (ii) does not hold, then the $j$-th column of $L^{-1} A$ has at least two non-zero elements.
	
	If condition (i) does not hold, that is, there exists some $k<j$ such that $A_{kj} \neq 0$.
	Denote the smallest such $k$ by $m$, then $c_m = (L^{-1})_{mm} A_{mj} \neq 0$.
	Because $m \neq j$, there must exist another non-zero element $c_{m'}$ for some $m' \in [p] \backslash\{m\}$, otherwise it contradicts Lemma~\ref{lemma:identiOfRC}.
	
	Similarly, if condition (ii) does not hold, that is, there exits some $k>j$ such that $(A^{-1})_{jk} \neq 0$, which is equivalent to $(A^{-T})_{kj} \neq 0$.
	Denote the largest such $k$ by $u$, then $c_u = (L^{T})_{uu} (A^{-T})_{uj} \neq 0$.
	Because $u \neq j$, there must exist another non-zero element $c_{u'}$ for some $u' \in [p] \backslash\{u\}$, otherwise it contradicts Lemma~\ref{lemma:identiOfRC}.
\end{proof}

\subsection{Proof of Theorem~\ref{thm:whPiContainOneSufficientPerm}}
\begin{proof}(Theorem~\ref{thm:whPiContainOneSufficientPerm})
	If we obtain an aberrant set $D=\{r\} \cup \rDe(r)$ in step 2 of Algorithm~\ref{Algo:ObtainPermutations},
	then by steps 3 and 4 of Algorithm~\ref{Algo:ObtainPermutations} and Theorem~\ref{thm:SufficientPermutation}, 
	$\whPi$ contains at least one sufficient permutation.
	Therefore, we prove the result by showing that $D=\{r\} \cup \rDe(r)$ happens with a probability tending to one as $n$ and $\delta_r$ tend to infinity.
	
	In the case of $\{r\} \cup \rDe(r)=[p]$, $D=\{r\} \cup \rDe(r)=[p]$ occurs when $\min_{j\in [p]} \whZ^2_{n,j}$ is used as a threshold in step 2 of Algorithm~\ref{Algo:ObtainPermutations}, hence 
	\begin{align*}
		\tP \left(\text{$\whPi$ contains at least one sufficient permutation}\right) 
		\geq \tP \left(D=\{r\} \cup \rDe(r)\right) 
		= 1.
	\end{align*}

	Now consider the case where $\{r\} \cup \rDe(r) \neq [p]$.
	Let $R = \{r\} \cup \rDe(r)$ for simplicity of notation.
	We will show that for large enough  $n$ and $\delta_r$, 
	all $\whZ^2_{n,j}$ with $j \in R$ are larger than $\whZ^2_{n,k}$, $k \not \in R$, 
	hence when $\min_{j \in R} \whZ^2_{n,j}$ is used as a threshold in step 2 of Algorithm~\ref{Algo:ObtainPermutations}, we have $D=R$.
	
	We first consider the population version terms.
	For $i \in [p]$, recall that
	\begin{align*}
		Z_{i} =  \frac{X^{I}_{i} - \mu_{i}} {\sigma_{i}},
		\quad 
		Z^o_i = \frac{X_{i} - \mu_{i}} {\sigma_{i}},
		\quad \text{and} \quad
		X^I_i = X_i +  \alpha_{r \rightarrow i} \delta_r,
	\end{align*}
	where 
	$\alpha_{r \rightarrow i} $ is the total causal effect of $X_r$ on $X_i$, and
	$\mu_{i}$ and $\sigma_{i}$ are the population mean and standard deviation of $X_i$, respectively.
	Note that $Z^o_i$ is a standardized random variable with mean $0$ and variance $1$.
	
	For any $j \in R$ and $k \not \in R$,
	we have $ \alpha_{r \rightarrow j} \neq 0$ and  $ \alpha_{r \rightarrow k} = 0$ (here we use the convention that $\alpha_{r\rightarrow r}=1$).
	Hence, 
	\begin{align*}
		Z^2_{j} = (Z^o_j)^2 + \delta_r \frac{2 \alpha_{r \rightarrow j}Z^o_j}{\sigma_j} +\delta_r^2  \frac{\alpha_{r \rightarrow j}^2}{\sigma_j^2},
		\quad 
		Z^2_{k} = (Z^o_{k})^2,
	\end{align*}
	and 
	\begin{align*}
		Z^2_{j} - Z^2_{k}
		= (Z^o_j)^2 - (Z^o_{k})^2 + \delta_r \frac{2 \alpha_{r \rightarrow j}Z^o_j}{\sigma_j} +  \delta_r^2\frac{\alpha_{r \rightarrow j}^2}{\sigma_j^2} 
		:= J_1 + \delta_rJ_2 + \delta_r^2 J_3,
	\end{align*}
	where $J_1 = (Z^o_j)^2 - (Z^o_{k})^2$, $J_2 = \frac{2 \alpha_{r \rightarrow j}Z^o_j}{\sigma_j} $, and $J_3 = \frac{\alpha_{r \rightarrow j}^2}{\sigma_j^2} > 0 $.
	
	Fix a small positive $\eta$. For sufficiently large $\delta_r$, we have $- \delta_r^2 J_3 + \eta < 0$.
	Then, for any $\epsilon>0$, when $\delta_r$ is large enough, we have
	\begin{equation} \label{proof:ZRZSLessThanEta}
		\begin{aligned}
			\tP\left( Z^2_{j} - Z^2_{k} \leq \eta \right) 
			&= \tP\left( J_1 + \delta_rJ_2 \leq - \delta_r^2 J_3 + \eta \right) \\
			&\leq \tP\left( |J_1 + \delta_rJ_2| \geq \delta_r^2 J_3 - \eta\right) \\
			&\leq \frac{\E[|J_1+\delta_rJ_2 |]}{ \delta_r^2 J_3 - \eta} \\
			&\leq \frac{\E[(Z^o_j)^2]  +  \E[(Z^o_{k})^2] + \frac{|2 \alpha_{r \rightarrow j}\delta_r|}{\sigma_j}  \E[| Z^o_j |]}{ \delta_r^2  \frac{\alpha_{r \rightarrow j}^2}{\sigma_j^2} - \eta} \\
			&\leq \frac{2\sigma_j^2 + \sigma_j|2 \alpha_{r \rightarrow j}\delta_r| }{ \delta_r^2 \alpha_{r \rightarrow j}^2 - \eta\sigma_j^2} \\
			& \leq \frac{\epsilon}{3|R|(p-|R|)},
		\end{aligned}
	\end{equation}
	where we used  Markov's inequality in the second inequality. 
	Therefore, for any $\epsilon > 0$ and sufficiently large $\delta_r$, we have
	\begin{equation} \label{proof:minZRmaxZSLargerThanEta}
		\begin{aligned}
			& \tP\left( \min_{j \in R} Z^2_{j} - \max_{k \not\in R} Z^2_{k} > \eta \right) \\
			=&  \tP\left( \underset{j \in R}{\cap} \underset{k \not\in R}{\cap} \{Z^2_{j} - Z^2_{k} > \eta\} \right) \\
			\geq & \sum_{j \in R} \sum_{k \not\in R} \tP\left(Z^2_{j} - Z^2_{k} > \eta\right)  - |R| (p - |R|) + 1 \\
			\geq & \sum_{j \in R} \sum_{k \not\in R} \left(1- \frac{\epsilon}{3|R|(p - |R|)}\right) - |R| (p - |R|) + 1 \\
			= & 1-\epsilon/3,
		\end{aligned}
	\end{equation}
	where we used \eqref{proof:ZRZSLessThanEta} in the last inequality.

    Now we consider the sample version terms.
	Because $\whmu_{n, j} \xrightarrow[n \to \infty]{p} \mu_{j}$ and $\whsigma_{n, j}
	\xrightarrow[n \to \infty]{p} \sigma_{j}$,
	we have $\min_{j \in R} \whZ^2_{n, j} \xrightarrow[n \to \infty]{p} \min_{j \in R} Z^2_{j}$ 
	and $\max_{k \not\in R} \whZ^2_{n,k} \xrightarrow[n \to \infty]{p}  \max_{k \not\in R} Z^2_{k}$
	by the continuous mapping theorem.
	Hence, for any $\epsilon>0$ and large enough $n$, we have
	\begin{align} \label{proof:ZnZconvergeRS}
		\tP\left( |\min_{j \in R} \whZ^2_{n,j} - \min_{j \in R} Z^2_{j} | \leq \eta/4\right) \geq 1-\epsilon/3
		\quad \text{and} \quad 
		\tP\left( |\max_{k \not\in R} \whZ^2_{n,k}  - \max_{k \not\in R} Z^2_{k}| \leq \eta/4\right) \geq 1-\epsilon/3.
	\end{align}
	
	Therefore, by \eqref{proof:minZRmaxZSLargerThanEta} and \eqref{proof:ZnZconvergeRS}, for any $\epsilon>0$ and sufficiently large $\delta_r$ and $n$, we have
	\begin{align*}
		& \tP \left(\text{$\whPi$ contains at least one sufficient permutation}\right) \\
		\geq & \tP \left( D = R \right) \\
		\geq & \tP\left( \min_{j\in R} \whZ^2_{n,j} > \max_{k \not\in R} \whZ^2_{n,k} \right) \\
		\geq & \tP\left( \min_{j\in R} Z^2_{j} - \max_{k \not\in R} Z^2_{k} > \eta, | \min_{j\in R} \whZ^2_{n,j}- \min_{j\in R} Z^2_{j}| \leq \eta/4,  | \max_{k \not\in R} \whZ^2_{n,k}- \max_{k \not\in R} Z^2_{k}| \leq \eta/4 \right) \\
		\geq & \tP\left( \min_{j\in R} Z^2_{j} - \max_{k \not\in R} Z^2_{k} > \eta \right) +
		\tP\left( | \min_{j\in R} \whZ^2_{n,j}- \min_{j\in R} Z^2_{j}| \leq \eta/4 \right) \\
		&+ \tP\left(| \max_{k \not\in R} \whZ^2_{n,k}- \max_{k \not\in R} Z^2_{k}| \leq \eta/4 \right) - 2 \\
		\geq & 1- \epsilon,
	\end{align*}
	which by definition proves that
	\begin{align*}
		\lim\limits_{\substack{n \to \infty \\ \delta_r \to \infty}} \tP \left(\text{$\whPi$ contains at least one sufficient permutation}\right) = 1.
	\end{align*}
	
\end{proof}

\subsection{Proof of Theorem~\ref{thm:ConsistSufficientAlgo}}

Before giving the proof, we first introduce some notation and two lemmas that will be used in the proof of Theorem~\ref{thm:ConsistSufficientAlgo}.
Let 
\begin{align*}
	\wtxi(\pi)
	:=  L_{X^\pi}^{-1} (\bx^{I\pi} - \mu_{X^\pi}),
\end{align*}
which has mean $\bbE[\wtxi(\pi)] = L_{X^\pi}^{-1} (\mu_{X^{I\pi}} - \mu_{X^\pi}) = L_{X^{\pi}}^{-1}  (I - B^{\pi})^{-1} \delta^{\pi} = \xi(\pi)$ (see \eqref{formula:xi(pi)}). 
Recall that $\delta^{\pi}=(0, \dots, 0, \delta_r, 0, \dots, 0)^T$, where $\delta_r$ is in the $\pi^{-1}(r)$-th position.
Let 
\begin{align*}
	\tc(\pi) = \frac{|\wtxi(\pi)|_{(1)} - |\wtxi(\pi)|_{(2)}} {|\wtxi(\pi)|_{(2)}}
	\quad \text{and} \quad
	\tu(\pi) = \pi (\argmax_{ j \in [p]} |\wtxi(\pi)|_j),
\end{align*}
where $ |\wtxi(\pi)|_{(i)}$ denotes the $i$-th largest entry in $|\wtxi(\pi)|$.
Then we have the following two lemmas showing some properties of $\tu(\pi) $ and $\tc(\pi)$ for sufficient and insufficient $\pi$, respectively.
\begin{lemma} \label{lemma:resOfSufficientPi}
	Let $r$ be the root cause.
	For a sufficient permutation $\pi$, we have
	\begin{align}\label{proof-thm2:XtildeSufficientPiResult}
		\tc(\pi) \xrightarrow[\delta_{r} \to \infty]{p} \infty
		\quad \text{and} \quad 
		\tu(\pi) \xrightarrow[\delta_{r} \to \infty]{p} r.
	\end{align}
\end{lemma}
\begin{lemma} \label{lemma:resOfInsufficientPi}
	There exists some constant $C>0$ such that
	for any insufficient permutation $\pi$, we have
	\begin{align}\label{proof-thm2:XtildeInsufficientPiResult}
		\lim\limits_{\delta_{r} \to \infty} \tP \left( \tc(\pi)  < C\right) = 1.
	\end{align}
\end{lemma}

We now give the proof of Theorem~\ref{thm:ConsistSufficientAlgo}, followed by the proofs of the above two lemmas.
\begin{proof}(Theorem~\ref{thm:ConsistSufficientAlgo})
	We divide the proof into two parts.
	In the first part, we show some properties for sufficient and insufficient $\pi$ that will be used later.
	In the second part, we prove the main result.
	
	\textbf{Part 1:} Prove some properties related to $\pi$.
	
	First, because 
	\[
	\whL_{X^\pi} \xrightarrow[n \to \infty]{p} L_{X^\pi}
	\quad \text{and} \quad 
	\whmu_{X^\pi} \xrightarrow[n \to \infty]{p} \mu_{X^\pi},
	\]
	we have by the continuous mapping theorem that
	\[
	\whmu_{\wtxi(\pi)} = \whL_{X^\pi}^{-1} (\bx^{I\pi} - \whmu_{X^\pi})
	\xrightarrow[n \to \infty]{p} 
	L_{X^\pi}^{-1} (\bx^{I\pi} - \mu_{X^\pi})
	=\wtxi(\pi),
	\]
	\begin{align}\label{proof:hcConverTotc}
		\hc(\pi) = \frac{\left|\whmu_{\wtxi(\pi)}\right|_{(1)} - \left|\whmu_{\wtxi(\pi)}\right|_{(2)}} {\left|\whmu_{\wtxi(\pi)}\right|_{(2)}}
		\xrightarrow[n \to \infty]{p} 
		\frac{|\wtxi(\pi)|_{(1)} - |\wtxi(\pi)|_{(2)}} {|\wtxi(\pi)|_{(2)}}
		=\tc(\pi),
	\end{align}
	and
	\begin{align}\label{proof:huConverTotu}
		\hu(\pi) = \pi(\argmax_{ j \in [p]} |\whmu_{\wtxi(\pi)}|_j)
		\xrightarrow[n \to \infty]{p} 
		\pi (\argmax_{ j \in [p]} |\wtxi(\pi)|_j) = \tu(\pi).
	\end{align}
	
	\textbf{Part 1-1:} Prove some properties related to sufficient $\pi$.
	
	For a sufficient $\pi$, because
	\begin{align*}
		\tP \left(  \hu(\pi) = r \right) 
		&\geq  \tP \left(  \tu(\pi) = r, | \hu(\pi) -  \tu(\pi)| < 1/2 \right) \\
		&\geq  \tP \left(  \tu(\pi)= r \right) + \tP \left( | \hu(\pi) -  \tu(\pi)| < 1/2 \right) - 1,
	\end{align*}
	by \eqref{proof:huConverTotu} and Lemma~\ref{lemma:resOfSufficientPi},
	we have
	\begin{align}\label{proof:huConverTor}
		\hu(\pi) \xrightarrow[n, \delta_r \to \infty]{p} r.
	\end{align}
	In addition, for any $M>0$, we have
	\begin{align*}
		&\tP \left( \hc(\pi) \bbone_{\{\hu(\pi) =r \}} \geq M \right) \\
		\geq & \tP \left( \tc(\pi) \geq 2M, \tu(\pi)=r , |\hu(\pi)-\tu(\pi)|<1/2, |\hc(\pi) - \tc(\pi)| < M/2 \right) \\
		\geq & \tP \left( \tc(\pi) \geq 2M \right) + \tP \left( \tu(\pi)=r  \right) +
		\tP \left( |\hu(\pi)-\tu(\pi)|<1/2\right) + \tP \left(  |\hc(\pi) - \tc(\pi)| < M/2 \right) - 3,
	\end{align*}
	so by \eqref{proof:hcConverTotc}, \eqref{proof:huConverTotu} and  Lemma~\ref{lemma:resOfSufficientPi}, we have
	\begin{align}\label{proof:hcIndConverTor}
		\hc(\pi) \bbone_{\{\hu(\pi)=r \}} \xrightarrow[n, \delta_r \to \infty]{p} \infty.
	\end{align}
	Furthermore,  for any $M>0$ and $k \in [p]\setminus\{r\}$, we have
	\begin{align*}
		\tP \left( \hc(\pi) \bbone_{\{\hu(\pi)=k \}} < M \right) 
		\geq  &\tP \left( \hu(\pi) = r \right) \\
		\geq &\tP \left( \tu(\pi) = r, | \hu(\pi) -  \tu(\pi) |<1/2 \right) \\
		\geq & \tP \left( \tu(\pi) = r \right) + \tP \left( | \hu(\pi) -  \tu(\pi) |<1/2 \right) -1,
	\end{align*}
	so by Lemma~\ref{lemma:resOfSufficientPi} and \eqref{proof:huConverTotu}, we have
	\begin{align}\label{proof:MainThmFesibleBigThanC}
		\lim\limits_{\substack{n \to \infty \\ \delta_r \to \infty}} \tP \left( \hc(\pi) \bbone_{\{\hu(\pi)=k \}} < M \right) = 1.
	\end{align}
	
	\textbf{Part 1-2:} Prove some properties related to insufficient $\pi$.
	
	For an insufficient $\pi$ and any $j\in[p]$, let $C$ be the constant from Lemma~\ref{lemma:resOfInsufficientPi}, we have
	\begin{align*}
		\tP \left( \hc(\pi) \bbone_{\{\hu(\pi)=j \}} < 2C  \right)
		\geq & \tP \left( \hc(\pi)  < 2C  \right) \\
		\geq & \tP \left( \tc(\pi)  < C, |\hc(\pi) - \tc(\pi)| < C/2 \right) \\
		\geq & \tP \left( \tc(\pi)  < C \right) + \tP \left( |\hc(\pi) - \tc(\pi)| < C/2\right) -1,
	\end{align*}
	so by \eqref{proof:hcConverTotc} and  Lemma~\ref{lemma:resOfInsufficientPi}, we have
	\begin{align}\label{proof:MainThmInfesibleBigThanC}
		\lim\limits_{\substack{n \to \infty \\ \delta_r \to \infty}} \tP \left( \hc(\pi) \bbone_{\{ \hu(\pi)=j \}} < 2C  \right) = 1.
	\end{align}
	
	\textbf{Part 2:} Prove the main result.
	
	We first analyze $\tP \left( \whC_k < \whC_r \right) $ for $k \in [p]\setminus\{r\}$, then obtain the main result by using the union bound.
	
	Denote all permutations by $\{\pi_i: i\in[p!]\}$, and
	let $\whd_i= \bbone_{\{ \pi_i \in \whPi\}} \in\{0,1\}$ be a data-dependent random variable indicating whether a permutation is contained in $\whPi$.
	Then, for $j\in[p]$, we can rewrite $\whC_j$ (see Algorithm~\ref{Algo:RCD}) as
	\begin{align*}
		\whC_j = 
		\underset{i \in [p!]}{\max} 
		\left\{  
		\hc(\pi_i) \whd_i \bbone_{\{\hu(\pi_i) = j \}},
		\hw_j \hc_{\min} 
		\left(
		\underset{l \in [p!]}{\Pi}
		(1 - \whd_i \bbone_{\{\hu(\pi_l) = j \}} ) 
		\right)
		\right\}.
	\end{align*}
	For any $k \in [p]\setminus\{r\}$, we have
	\begin{equation} \label{proof:CjhatLessThanCrhat}
		\begin{aligned}
			\tP \left( \whC_k < \whC_r \right) 
			= & 
			\tP \left(
			\underset{i \in [p!]}{\max} 
			\left\{   \hc(\pi_i) \whd_i \bbone_{\{\hu(\pi_i) = k \}},
			\hw_k \hc_{\min} \left( \underset{l \in [p!]}{\Pi} (1 - \whd_i \bbone_{\{ \hu(\pi_l) = k \}} )  \right) \right\}
			< \whC_r \right) \\
			\geq &
			\tP \left(
			\underset{i \in [p!]}{\max} 
			\left\{   \hc(\pi_i) \bbone_{\{\hu(\pi_i) = k \}} \right\} < \whC_r,
			\hc_{\min}  < \whC_r 
			\right) \\
			\geq &
			\tP \left( \underset{i \in [p!]}{\max}  
			\left\{   \hc(\pi_i) \bbone_{\{\hu(\pi_i) = k \}} \right\} < \whC_r \right) 
			+ \tP \left( \hc_{\min}  < \whC_r  \right) - 1,
		\end{aligned}
	\end{equation}
	where we used the fact that $\hw_k \hc_{\min} \left( \underset{l \in [p!]}{\Pi} (1 - \whd_i \bbone_{\{k= \hu(\pi_l) \}} )  \right) \leq \hc_{\min}$ for the first inequality.
	
	Next, we show that 
	\begin{align*}
		\lim\limits_{\substack{n \to \infty \\ \delta_r \to \infty}} \tP \left( \underset{i \in [p!]}{\max}  
		\left\{   \hc(\pi_i) \bbone_{\{\hu(\pi_i) = k \}} \right\} < \whC_r \right) = 1
		\quad \text{and} \quad
		\lim\limits_{\substack{n \to \infty \\ \delta_r \to \infty}} \tP \left( \hc_{\min}  < \whC_r  \right)  = 1.
	\end{align*}
	
	\textbf{Part 2-1:} Prove $\lim\limits_{\substack{n \to \infty \\ \delta_r \to \infty}} \tP \left( \underset{i \in [p!]}{\max}  
		\left\{   \hc(\pi_i) \bbone_{\{\hu(\pi_i) = k \}} \right\} < \whC_r \right) = 1$.
	
	Let $C$ be the constant from Lemma~\ref{lemma:resOfInsufficientPi},
	we have 
	\begin{equation*} 
		\begin{aligned}
			&\lim\limits_{\substack{n \to \infty \\ \delta_r \to \infty}} \tP \left( \underset{i \in [p!]}{\max}  
			\left\{   \hc(\pi_i) \bbone_{\{\hu(\pi_i) = k \}} \right\} < \whC_r \right) \\
			\geq &
			\lim\limits_{\substack{n \to \infty \\ \delta_r \to \infty}} \tP \left( \underset{i \in [p!]}{\max}  
			\left\{   \hc(\pi_i)  \bbone_{\{\hu(\pi_i) = k \}} \right\} < 2C, \whC_r \geq 2C \right) \\
			\geq &
			\lim\limits_{\substack{n \to \infty \\ \delta_r \to \infty}} \tP \left( \underset{i \in [p!]}{\max}  
			\left\{   \hc(\pi_i) \bbone_{\{\hu(\pi_i) = k \}} \right\} < 2C \right) + 
			\lim\limits_{\substack{n \to \infty \\ \delta_r \to \infty}} \tP \left( \whC_r \geq 2C \right) - 1.
		\end{aligned}
	\end{equation*}
	
	By \eqref{proof:MainThmFesibleBigThanC}, \eqref{proof:MainThmInfesibleBigThanC} and using the union bound, we have
	\begin{align} \label{proof:maxLessthanB}
		\lim\limits_{\substack{n \to \infty \\ \delta_r \to \infty}} 
		\tP \left( \underset{i \in [p!]}{\max}  
		\left\{   \hc(\pi_i)  \bbone_{\{ \hu(\pi_i) = k \}} \right\} < 2C \right) = 1.
	\end{align}
	
	Let the first $u$ permutations be all sufficient permutations without loss of generality,
	then we have
	\begin{align*}
		\tP \left( \whC_r \geq 2C \right)
		\geq  & \tP \left( \underset{i \in [u]}{\max} \left\{  \hc(\pi_i) \whd_i \bbone_{\{r = \hu(\pi_i) \}} \right\} \geq 2C \right) \\
		\geq &\tP \left( \underset{i \in [u]}{\min} \left\{  \hc(\pi_i) \bbone_{\{r = \hu(\pi_i) \}} \right\} \geq 2C,
		\sum_{i=1}^{u} \whd_i \geq 1 \right) \\
		= &\tP \left( \underset{i \in [u]}{\bigcap} \left\{  \hc(\pi_i) \bbone_{\{r = \hu(\pi_i) \}} \geq 2C \right\},
		\sum_{i=1}^{u} \whd_i \geq 1 \right) \\
		\geq & \sum_{i \in [u]} \tP \left( \hc(\pi_i) \bbone_{\{r = \hu(\pi_i) \}} \geq 2C\right) +
		\tP \left(\sum_{i=1}^{u} \whd_i \geq 1 \right) - |u| \\
		= &\sum_{i \in [u]} \tP \left( \hc(\pi_i) \bbone_{\{r = \hu(\pi_i) \}} \geq 2C\right) \\
		&+ \tP \left(\text{$\whPi$ contains at least one sufficient permutation}\right) - |u| .
	\end{align*}
	Hence, by \eqref{proof:hcIndConverTor} and Theorem~\ref{thm:whPiContainOneSufficientPerm}, we have
	\begin{align}\label{proof:whCrToInfty}
		\lim\limits_{\substack{n \to \infty \\ \delta_r \to \infty}} \tP \left( \whC_r \geq 2C \right) =1.
	\end{align}
	Therefore, by combining  \eqref{proof:whCrToInfty} and \eqref{proof:maxLessthanB}, we have
	\begin{equation} \label{proof:maxLessthanhCr}
		\begin{aligned}
			\lim\limits_{\substack{n \to \infty \\ \delta_r \to \infty}} \tP \left( \underset{i \in [p!]}{\max}  
			\left\{   \hc(\pi_i) \bbone_{\{\hu(\pi_i) = k \}} \right\} < \whC_r \right) 
			= 1.
		\end{aligned}
	\end{equation}

	\textbf{Part 2-2:} Prove $\lim\limits_{\substack{n \to \infty \\ \delta_r \to \infty}} \tP \left( \hc_{\min}  < \whC_r  \right)  = 1.$
	
	Recall that we let the first $u$ permutations be all sufficient permutations without loss of generality.
	Then,
	\begin{align*}
		\tP \left( \hc_{\min}  < \whC_r  \right) 
		\geq & \tP \left( r \in \whU \right) \\
		\geq  & \tP \left( \underset{i \in [u]}{\bigcap} \left\{ \hu(\pi_i) = r \right\}, \sum_{i=1}^{u} \whd_i \geq 1  \right) \\
		\geq & \sum_{i \in [u]} \tP \left( \hu(\pi_i) = r \right) +
		\tP \left( \sum_{i=1}^{u} \whd_i \geq 1  \right)  - u\\
		= & \sum_{i \in [u]} \tP \left(  \hu(\pi_i) = r \right) +
		\tP \left(\text{$\whPi$ contains at least one sufficient permutation}\right)  - u.
	\end{align*}
	Hence, by \eqref{proof:huConverTor}  and Theorem~\ref{thm:whPiContainOneSufficientPerm}, we have
	\begin{align}\label{proof:cminLessthanhCr}
		\lim\limits_{\substack{n \to \infty \\ \delta_r \to \infty}} \tP \left( \hc_{\min} < \whC_r  \right) = 1.
	\end{align}

	\textbf{Part 2-3:} Use the union bound to show the final result.
	
	By using the union bound and combining \eqref{proof:CjhatLessThanCrhat}, \eqref{proof:maxLessthanhCr} and  \eqref{proof:cminLessthanhCr}, we have
	\begin{align*}
		\lim\limits_{\substack{n \to \infty \\ \delta_r \to \infty}} 
		\tP \left( \whC_{r} > \max_{k\in[p]\backslash\{r\}} \whC_{k} \right) 
		= 1.
	\end{align*}

\end{proof}

In the following, we give the proofs of Lemma~\ref{lemma:resOfSufficientPi} and \ref{lemma:resOfInsufficientPi}.
\begin{proof}(Lemma~\ref{lemma:resOfSufficientPi})
	We divide the proof into two parts.
	In the first part, we show some results that will be used later.
	In the second part, we prove the main result.
	
	\textbf{Part 1:} Some results that will be used later.
	
	For a sufficient permutation $\pi$, we have
	\begin{align*}
		\bbE[\wtxi(\pi)]_{\pi^{-1}(r)} = (L_{X^{\pi}}^{-1}  (I - B^{\pi})^{-1} )_{\pi^{-1}(r), \pi^{-1}(r)} \delta_{r}
		:= a\delta_{r}
		\xrightarrow[]{\delta_{r} \to \infty} \infty
	\end{align*}
		and
		\begin{align*}
		\bbE[\wtxi(\pi)]_{\pi^{-1}(k)} =  0 \quad \text{for }  k \in[p]\setminus\{r\}
	\end{align*}
	by Theorem~\ref{thm:identiOfRC},
	where $a = (L_{X^{\pi}}^{-1}  (I - B^{\pi})^{-1} )_{\pi^{-1}(r), \pi^{-1}(r)} $ does not depend on $\delta_{r}$.
	
	\textbf{Part 1-1:} Analyze the property of $\wtxi(\pi)_{\pi^{-1}(r)}$.

	For any $\epsilon>0$ and $M>0$, let $\delta_r$ be large enough such that $ \frac{|a\delta_{r}|}{2} \geq M$ and $\frac{4}{|a\delta_{r}|^2} \leq \epsilon$.
	Then, 
	\begin{equation}\label{proof:lemma1:probIneuq1}
		\begin{aligned}
			\tP \left( |\wtxi(\pi)_{\pi^{-1}(r)}| \geq M  \right) 
			&\geq \tP \left( |\wtxi(\pi)_{\pi^{-1}(r)}| \geq \frac{|a\delta_{r}|}{2}\ \right) \\
			&\geq \tP \left( |\wtxi(\pi)_{\pi^{-1}(r)} -  a\delta_{r} | \leq \frac{|a\delta_{r}|}{2}\  \right) \\
			&\geq 1- \frac{4}{|a\delta_{r}|^2}\\
			&\geq 1- \epsilon,
		\end{aligned}
	\end{equation}
	where we used Chebyshev's inequality and the fact that $\Sigma_{\wtxi(\pi)} = I$ for the second last inequality.
	
	\textbf{Part 1-2:} Analyze the property of $\max_{k\in[p]\backslash\{r\}} |\wtxi(\pi)_{\pi^{-1}(k)}| $.
	
	For any $\epsilon>0$, let $\wtM \geq \sqrt{\frac{p-1}{\epsilon}}>0$, so 
	$\tP\left( |\wtxi(\pi)_{\pi^{-1}(k)}| \geq \wtM \right)  \leq \frac{1}{\wtM^2}  \leq \frac{\epsilon}{p-1}$ by Chebyshev's inequality.
	Therefore, by the union bound, we have
	\begin{align}\label{proof:lemma1:probIneuq2}
		\tP \left(\max_{k\in[p]\backslash\{r\}} |\wtxi(\pi)_{\pi^{-1}(k)}| < \wtM \right) \geq 1 - \epsilon.
	\end{align}

	\textbf{Part 2:} Prove the main results.
	
	\textbf{Part 2-1:} Prove $\tu(\pi) \xrightarrow[\delta_{r} \to \infty]{p} r$ by definition.
	
	For any $\epsilon >0$, taking $\wtM \geq \sqrt{\frac{p-1}{\epsilon}}$ and $\delta_r$ large enough such that $ \frac{|a\delta_{r}|}{2} \geq \wtM$ and $\frac{4}{|a\delta_{r}|^2} \leq \epsilon$.
	Then by  \eqref{proof:lemma1:probIneuq1} and \eqref{proof:lemma1:probIneuq2}, we have
	\begin{align*}
		\tP\left(  \tu(\pi) = r \right)
		= &\tP\left(  \pi (\argmax_{ j \in [p]} |\wtxi(\pi)|_j ) = r \right) \\
		= &\tP \left( |\wtxi(\pi)_{\pi^{-1}(r)}| > \max_{k\in[p]\backslash\{r\}} |\wtxi(\pi)_{\pi^{-1}(k)}| \right) \\
		\geq&  \tP \left( |\wtxi(\pi)_{\pi^{-1}(r)}| \geq \wtM, \max_{k\in[p]\backslash\{r\}} |\wtxi(\pi)_{\pi^{-1}(k)}| < \wtM \right) \\
		\geq & \tP \left( |\wtxi(\pi)_{\pi^{-1}(r)}| \geq \wtM \right) + 
		\tP \left(\max_{k\in[p]\backslash\{r\}} |\wtxi(\pi)_{\pi^{-1}(k)}| < \wtM \right) - 1 \\
		\geq &1- 2 \epsilon,
	\end{align*}
	which means that 
	\begin{align} \label{proof:lemma1:res1}
		\tu(\pi) 
		\xrightarrow[\delta_{r} \to \infty]{p} r.
	\end{align}
	
	\textbf{Part 2-2:} Prove $\tc(\pi) \xrightarrow[\delta_{r} \to \infty]{p} \infty$ by definition.
	
	For any $\epsilon>0$ and $M>0$,
	taking $\wtM \geq \sqrt{\frac{p-1}{\epsilon}}$ and $\delta_r$ large enough such that $ \frac{|a\delta_{r}|}{2} \geq M \wtM + \wtM $ and $\frac{4}{|a\delta_{r}|^2} \leq \epsilon$.
	Then, by \eqref{proof:lemma1:probIneuq1}, \eqref{proof:lemma1:probIneuq2} and \eqref{proof:lemma1:res1},
	we have
	\begin{align*}
		\tP\left(  \tc(\pi) \geq M \right)
		=&\tP\left(  \frac{|\wtxi(\pi)|_{(1)} - |\wtxi(\pi)|_{(2)}} {|\wtxi(\pi)|_{(2)}} \geq M \right) \\
		\geq & \tP\left(  \frac{|\wtxi(\pi)_{\pi^{-1}(r)}| - \wtM} {\wtM} \geq M,
		\max_{k\in[p]\backslash\{r\}} |\wtxi(\pi)_{\pi^{-1}(k)}| < \wtM,
		\tu(\pi) = r \right) \\
		\geq & \tP\left( |\wtxi(\pi)_{\pi^{-1}(r)}| \geq M \wtM + \wtM \right) +
		\tP\left( \max_{k\in[p]\backslash\{r\}} |\wtxi(\pi)_{\pi^{-1}(k)}| < \wtM \right) \\ 
		&+ \tP\left(\tu(\pi) = r \right) - 2 \\
		\geq &1- 4\epsilon,
	\end{align*}
	which means that $\tc(\pi) \xrightarrow[\delta_{r} \to \infty]{p} \infty$.
	
\end{proof}

\begin{proof}(Lemma~\ref{lemma:resOfInsufficientPi})
	For an insufficient permutation $\pi$, by Theorem~\ref{thm:identiOfRC}, 
	there must exist a set $F^\pi \subseteq [p]$ with $|F^\pi|\geq 2$ such that 
	\begin{align*}
		\bbE[\wtxi(\pi)]_{j} = (L_{X^{\pi}}^{-1}  (I - B^{\pi})^{-1} )_{j, \pi^{-1}(r)} \delta_{r}
		\xrightarrow[]{\delta_{r} \to \infty} \infty \quad \text{for }  j \in F^\pi
	\end{align*}
	and 
	\begin{align*}
		\bbE[\wtxi(\pi)]_k =  0 \quad \text{for }  k \not\in F^\pi.
	\end{align*}
	Therefore, by using similar arguments based on Chebyshev's inequality and the union bound as in the Part 1 of the proof of Lemma~\ref{lemma:resOfSufficientPi}, we have
	\begin{equation}\label{proof:lemma2:limit1}
		\begin{aligned}
			&\lim\limits_{\delta_{r} \to \infty} 
			\tP \left( \text{the largest and second largest elements of $|\wtxi(\pi)|$ are both in $F^\pi$} \right) \\
			\geq &\lim\limits_{\delta_{r} \to \infty} 
			\tP \left( \min_{j \in F^\pi} |\wtxi(\pi)_{j}| > \max_{k \not\in F^\pi } |\wtxi(\pi)_{k}| \right) \\ =& 1.
		\end{aligned}
	\end{equation}
	In particular, the above holds trivially in the case of $F^\pi = [p]$.
	
	Then, denote
	\begin{align*}
		\wtxi(\pi)
		= L_{X^\pi}^{-1} ((I - B^{\pi})^{-1} (b^{\pi} + \err^{\pi})  - \mu_{X^\pi}) + L_{X^\pi}^{-1}(I - B^{\pi})^{-1} \delta^{\pi}
		:= u + \delta_r v,
	\end{align*}
	where $u=L_{X^\pi}^{-1} ((I - B^{\pi})^{-1} (b^{\pi} + \err^{\pi})  - \mu_{X^\pi}) $ is a random vector 
	and $v = L_{X^\pi}^{-1}(I - B^{\pi})^{-1} \delta^{\pi} / \delta_r$ is a deterministic vector whose support is $F^\pi$.
	Note that both terms do not depend on $\delta_r$.
	Hence, for any $i,j \in F^\pi$,
	\begin{align}\label{proof:lemma2:limit2}
		\frac{|\wtxi(\pi)_{i}| - |\wtxi(\pi)_{j}|} {|\wtxi(\pi)_{j}|}
		= \frac{|u_{i} + \delta_r v_{i}| - |u_{j} + \delta_r v_{j}| } {|u_{j} + \delta_r v_{j}|}
		\xrightarrow[\delta_{r} \to \infty]{p}
		\frac{|v_{i}| - | v_{j}| } {| v_{j}|}. 
	\end{align}
	Let 
	\[
	C = \max_{\pi \text{ is insufficient }} \max_{i,j\in F^\pi} \frac{|v_{i}| - | v_{j}| } {| v_{j}|} + 1.
	\]
	Because of \eqref{proof:lemma2:limit2}, we have
	\begin{align*}
		\lim\limits_{\delta_{r} \to \infty} \tP \left( \frac{|\wtxi(\pi)_{i}| - |\wtxi(\pi)_{j}|} {|\wtxi(\pi)_{j}|} \geq C \right) = 0
	\end{align*}
	by definition. 
	Then, since
	\begin{align*}
		&\tP \left( \max_{i,j\in F^\pi}\frac{|\wtxi(\pi)_{i}| - |\wtxi(\pi)_{j}|} {|\wtxi(\pi)_{j}|} \geq C \right) \\
		= &\tP \left( \bigcup_{i,j\in F^\pi} \left\{\frac{|\wtxi(\pi)_{i}| - |\wtxi(\pi)_{j}|} {|\wtxi(\pi)_{j}|} \geq C \right\}  \right) \\
		\leq &\sum_{i,j\in F^\pi} \tP \left( \frac{|\wtxi(\pi)_{i}| - |\wtxi(\pi)_{j}|} {|\wtxi(\pi)_{j}|} \geq C \right),
	\end{align*}
	we have
	\begin{align}\label{proof:lemma2:limit3}
		\lim\limits_{\delta_{r} \to \infty} \tP \left( \max_{i,j\in F^\pi}\frac{|\wtxi(\pi)_{i}| - |\wtxi(\pi)_{j}|} {|\wtxi(\pi)_{j}|} < C \right) 
		= 1.
	\end{align}
	
	Finally, because
	\small{
	\begin{align*}
		&\tP \left( \tc(\pi) <  C \right) \\
		= & \tP \left(\frac{|\wtxi(\pi)|_{(1)} - |\wtxi(\pi)|_{(2)}} {|\wtxi(\pi)|_{(2)}} < C \right) \\
		\geq & \tP \left( \text{the largest and second largest elements of $|\wtxi(\pi)|$ are both in $F^\pi$}, 
		\max_{i,j\in F^\pi}\frac{|\wtxi(\pi)_{i}| - |\wtxi(\pi)_{j}|} {|\wtxi(\pi)_{j}|} < C \right) \\
		\geq & \tP \left( \text{the largest and second largest elements of $|\wtxi(\pi)|$ are both in $F^\pi$}\right) \\
		&+ \tP \left( \max_{i,j\in F^\pi}\frac{|\wtxi(\pi)_{i}| - |\wtxi(\pi)_{j}|} {|\wtxi(\pi)_{j}|} < C \right) - 1,
	\end{align*}
}
\normalsize
	we have $\lim\limits_{\delta_{r} \to \infty} \tP \left( \tc(\pi)< C \right) = 1$ by \eqref{proof:lemma2:limit1} and \eqref{proof:lemma2:limit3}.
\end{proof}

\section{Supplementary materials for simulations and the real application}

\subsection{Simulation results illustrating the effect of $v$ in Algorithm~\ref{Algo:RCD}} \label{appendix:SimuForV}

We illustrate the benefit and cost in run-time of using a large number of random permutations $v$ in Algorithm~\ref{Algo:RCD} through simulations.

We consider a random DAG and uniform errors as described in Section~\ref{sec:simu-setup}. 
The number of variables is $p=50$, the sample size is $n=200$, the intervention strength is $\delta_r=4$, and the sparsity level is $s=0.4$.
We generate $m=1000$ interventional samples (see Section~\ref{sec:simu-setup} for more details) and apply Algorithm~\ref{Algo:RCD} with $v=1$ and $v=10$ to these samples.

\begin{figure}[t]
	\centering
	\begin{minipage}{0.37\textwidth}
		\centering
		\includegraphics[width=\linewidth]{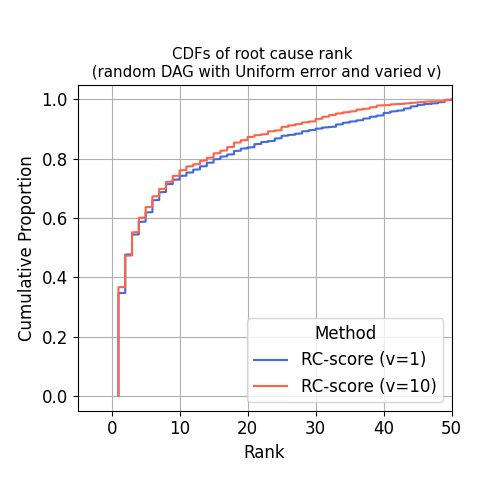}
			\\ (a) 
	\end{minipage}%
	\hspace{0.1\textwidth}  
	\begin{minipage}{0.47\textwidth}
		\centering
		\includegraphics[width=\linewidth]{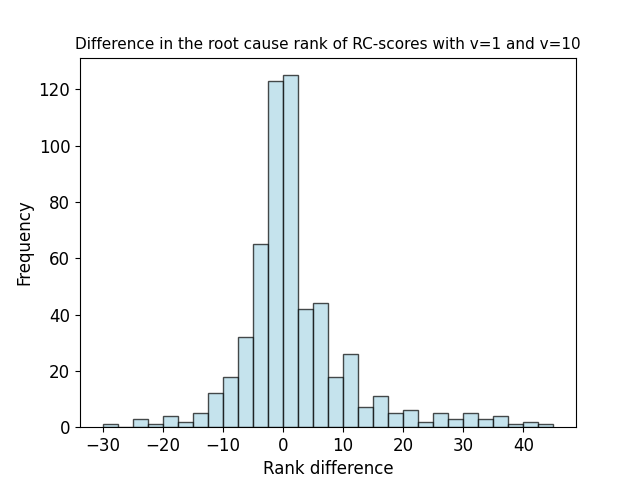}
			\\ (b) 
	\end{minipage}
	\caption{(a) CDFs of the root cause ranks using the RC-score with $v=1$ and $v=10$ in the setting with a random DAG and uniform errors. (b) Histogram of the differences in the root cause ranks between $v=1$ and $v=10$ for the $589$ samples for which these two methods lead to different ranks for the root cause.}
	\label{Fig:effectV}
\end{figure}

The CDFs of the root cause rank of these two methods (see Section~\ref{sec:simu-results} for more details) are shown in Figure~\ref{Fig:effectV} (a). 
We can see that using $v=10$ yields better results than using $v=1$.

We investigate the results a bit more closely.
Among the $1000$ interventional samples, the root cause ranks differ between $v=1$ and $v=10$ in $589$ samples.
We then calculate the rank difference between $v=1$ and $v=10$ for these $576$ samples. 
A large positive difference indicates that using $v=1$ yields a much worse rank than using $v=10$,
while a small negative difference suggests the opposite.
Figure~\ref{Fig:effectV} (b) shows the histogram of the rank differences, which is clearly right-skewed,
indicating that it is more likely to obtain a much worse score with $v=1$ than with $v=10$.
For example, there are $31$ samples where the rank difference exceeds $20$, but only $5$ sample where the rank difference is less than $-20$.
This is largely due to the fact that using only one permutation leads to more cases where the root cause lies outside $\whU$, resulting in a lower score.
In particular, when using $v=1$, the root cause lies outside $\whU$ in $137$ samples, while this happens in $32$ samples when using $v=10$.

\begin{figure}[htbp]
	\centering
	\includegraphics[scale=0.6]{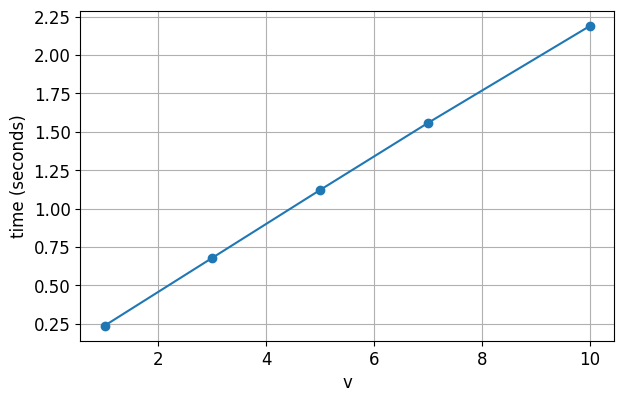}
	\caption{Average run-time of Algorithm 2 with different values of $v$.
	}
	\label{Fig:runtime}
\end{figure}

As mentioned in Section~\ref{sec:characterizeSufficientPermutation}, the benefit of using a larger $v$ comes at the cost of increased computational expense, which grows linearly with $v$.
This is because using $v = k$ results in $k$ times more permutations than using $v = 1$ in step 1 of Algorithm~\ref{Algo:RCD}. For each permutation, we compute $\widehat{\xi}(\pi)$ in step 3 of Algorithm~\ref{Algo:RCD}, which is the most computationally expensive step. Consequently, the overall run-time increases linearly with $v$.
We verify this in Figure~\ref{Fig:runtime}, which shows the average run-time of Algorithm~2 for $v = 1, 3, 5, 7, 10$. A clear linear trend is observed.


\subsection{More details about the simulation setup and implemented methods in Section~\ref{sec:simu-setup}} \label{appendix:DetailSimuSetup}
\subsubsection{Details of the simulation setup} \label{appendix:DetailSimuDAG}

For the matrix $B$, we consider two types corresponding to the following DAG types:
\begin{itemize}
	\item Random DAG: This is a commonly used DAG where nodes are connected in a completely random manner. 
	Specifically, to ensure acyclicity, we generate a lower-triangular matrix $B$ of size $p=100$ with sparsity level $s \in (0,1)$ (i.e., $s \cdot 100\%$ of the edges are expected to be present).
	Each entry in the lower-triangular part of $B$ has a probability $s$ of being non-zero, and each non-zero entry is assigned a random value drawn from $U(-1,1)$ as the edge weight.
	\item Hub DAG: This DAG is motivated by genetic interactions, where certain genes (known as hub genes) act as central connectors and significantly influence the overall network behavior. 
	In our simulations, we consider $4$ hub nodes, each with an upper node block of size $15$, where the nodes point to the hub node, and a lower node block of size $10$, to which the hub node points. 
	Additionally, each hub node is pointed to by $4$ nodes from other upper node blocks and points to $3$ nodes from other lower node blocks. 
	Nodes within the upper and lower node blocks form a random DAG, as described above, with sparsity level $s$.
	In total, there are $p=104$ variables.
	Based on the description, the corresponding matrix $B$ is constructed as a lower-triangular matrix to ensure acyclicity, 
	and each non-zero entry of $B$ is assigned a random value drawn from $U(-1,1)$ as the edge weight.
\end{itemize}

\subsubsection{Implementation details for the LiNGAM-based methods} \label{appendix:DetailSimuLinGam}


The first LiNGAM-based method estimates a causal ordering and an aberrant variable set, and then ranks the variables based on these estimates.
Specifically, we estimate the aberrant variable set based on squared z-scores. 
Then, based on the aberrant variable set and the estimated causal ordering, we assign ranks to the aberrant variables: an aberrant variable receives a smaller rank if it appears earlier in the estimated causal ordering. 
Variables not in the aberrant set are ranked after the aberrant ones based on their squared z-scores.
A threshold is required to obtain this set. 
The optimal threshold would be the squared z-score of the root cause, because any other threshold would yield a root cause rank that is at most as good as using the squared z-score of the root cause.
However, determining this optimal threshold in a data-dependent manner is challenging. 
In our simulations, we use this optimal threshold to assess the best possible performance of the approach, 
though it is important to note that this is infeasible in practice due to the requirement of prior knowledge of the root cause. 
We also use two fixed thresholds $2$ and $5$ for comparison.

The second LiNGAM-based method uses an estimated DAG and residuals.
For each variable $X_i$, 
let $\widehat{\Pa}_i$ denote its estimated parents.
We then apply cross-validated Lasso to fit a linear model $\hat{f}(X_{\widehat{\Pa}_i})$ based on observational samples, with $X_i$ as the response variable and $X_{\widehat{\Pa}_i}$ as the covariates.
Next, for the interventional sample $ X^I$, we compute the standardized absolute residual for each variable $X^I_i$ as \(\frac{|X_i^I - \hat{f}(X_{\widehat{\Pa}_i}^I)|}{\hat{\sigma}_{\text{obs}}}\), where $\hat{\sigma}_{\text{obs}}$ is the sample standard deviation of the residuals obtained from the fitted model $\hat{f}(X_{\widehat{\Pa}_i})$ on the observational samples.
The logic behind it is that if $X^I_i$ is not the root cause (i.e., it is not intervened upon), the distribution of $X^I_i$ given its parents remains invariant across different environments, so the  standardized residual should be small.
On the other hand, if  $X^I_i$ is the root cause (i.e., it is intervened upon), its value should be quite different from $\hat{f}(X^I_{\widehat{\Pa}_i})$, leading to a large  standardized residual.

\subsection{More simulation results for Section~\ref{sec:simu-results}} \label{appendix:SimuMoreMainResults}
\begin{figure}[htbp]
	\centering
	\begin{minipage}[b]{0.88\textwidth}
		\includegraphics[width=\textwidth]{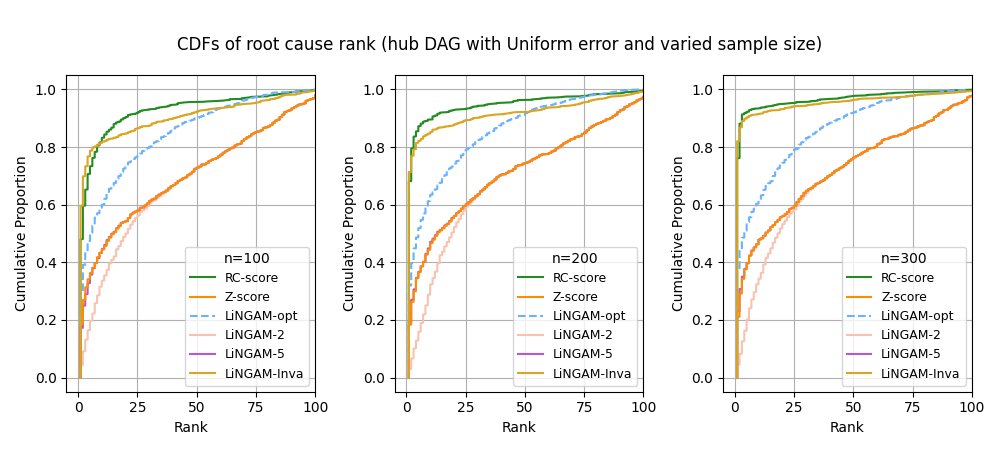}
	\end{minipage}
	\hfill
	\begin{minipage}[b]{0.88\textwidth}
		\includegraphics[width=\textwidth]{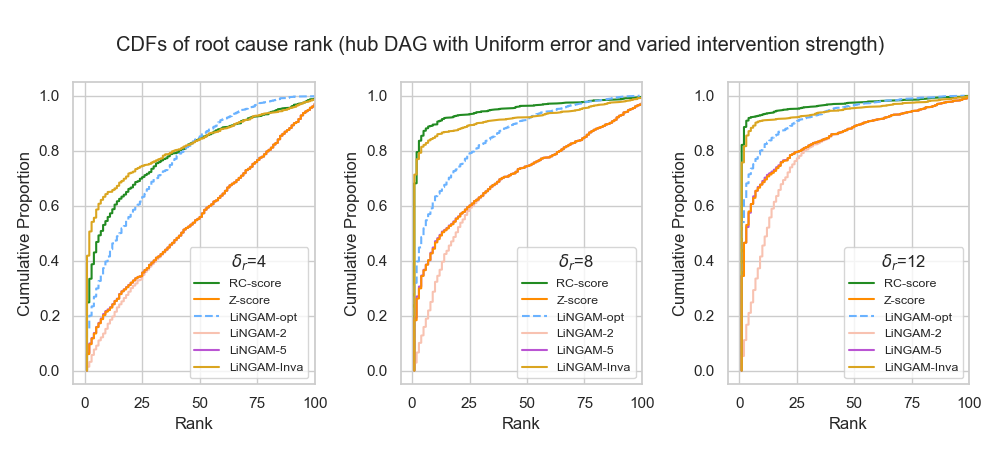}
	\end{minipage}
	\hfill
	\begin{minipage}[b]{0.88\textwidth}
		\includegraphics[width=\textwidth]{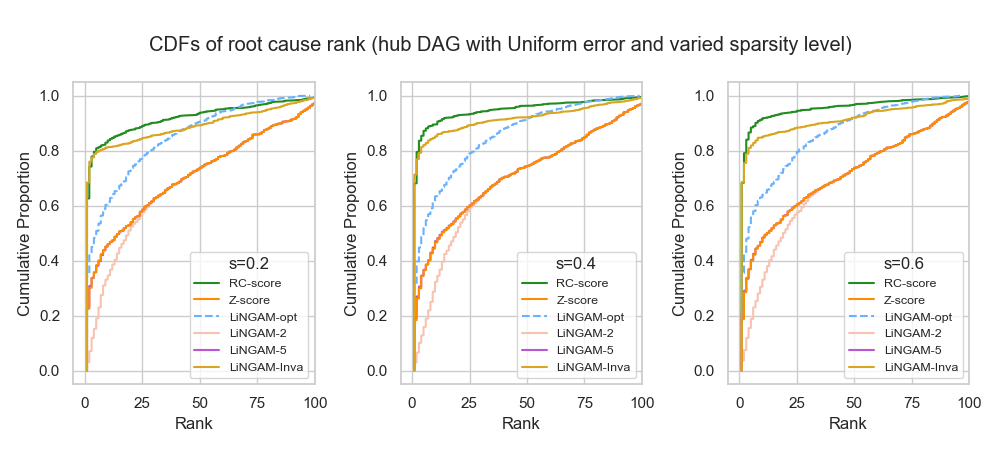}
	\end{minipage}
	\caption{CDFs of the root cause rank using the squared z-score, RC-score, and LiNGAM-based approaches in the setting with a hub DAG and uniform errors. The top, middle, and bottom plots display results for varying sample sizes, intervention strengths, and sparsity levels, respectively.}
	\label{Fig:MainSimuP100HubDAGUnif}
\end{figure}

\begin{figure}[h!]
	\centering
	\begin{minipage}[b]{0.88\textwidth}
		\includegraphics[width=\textwidth]{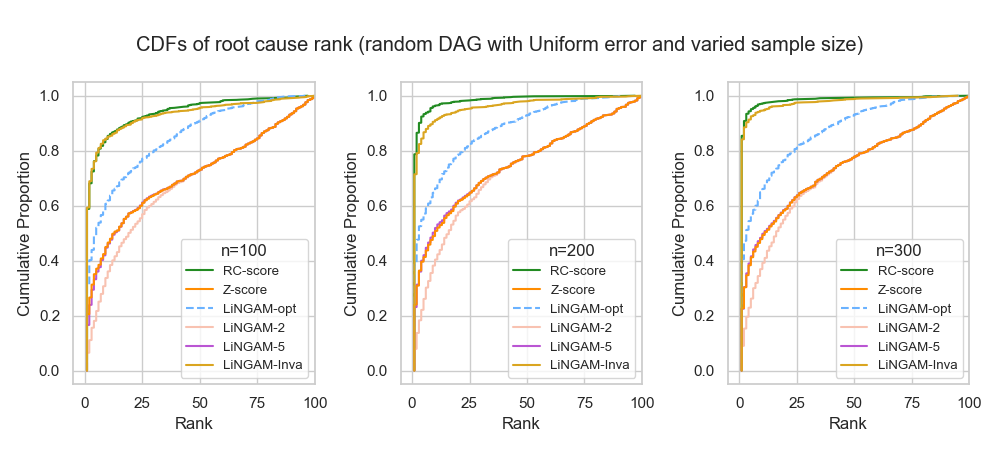}
	\end{minipage}
	\hfill
	\begin{minipage}[b]{0.88\textwidth}
		\includegraphics[width=\textwidth]{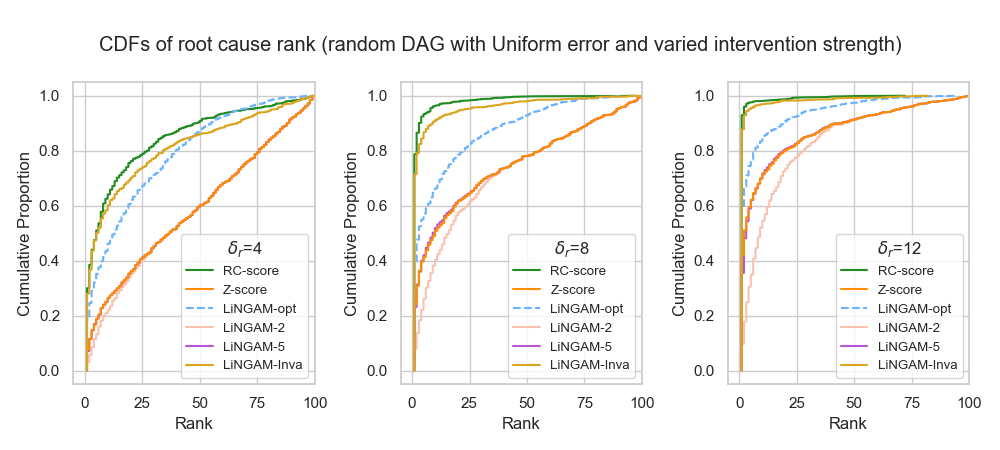}
	\end{minipage}
	\hfill
	\begin{minipage}[b]{0.88\textwidth}
		\includegraphics[width=\textwidth]{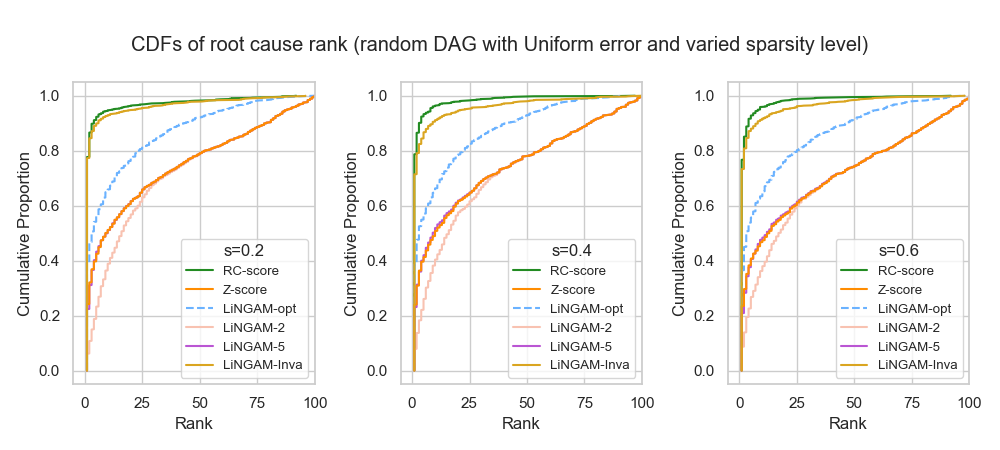}
	\end{minipage}
	\caption{CDFs of the root cause rank using the squared z-score, RC-score, and LiNGAM-based approaches in the setting with a random DAG and uniform errors. The top, middle, and bottom plots display results for varying sample sizes, intervention strengths, and sparsity levels, respectively.}
	\label{Fig:MainSimuP100RandomDAGUnif}
\end{figure}

\begin{figure}[h!]
	\centering
	\begin{minipage}[b]{0.88\textwidth}
		\includegraphics[width=\textwidth]{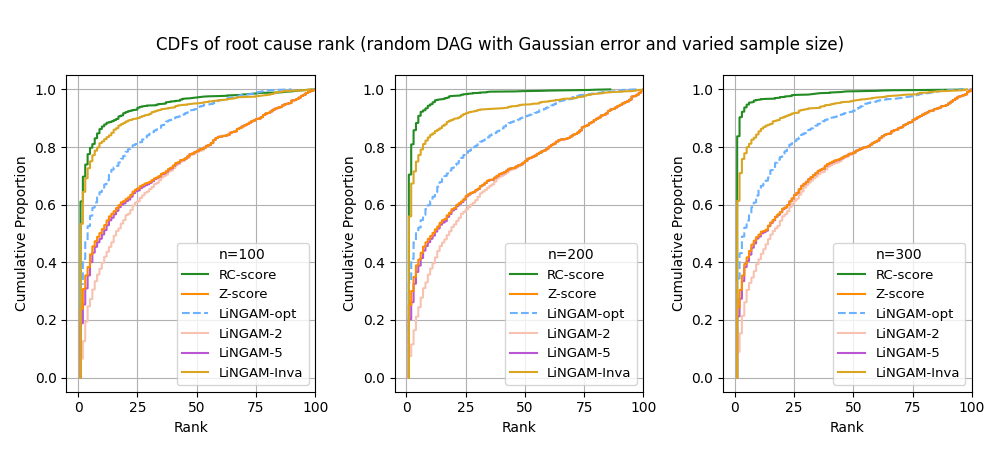}
	\end{minipage}
	\hfill
	\begin{minipage}[b]{0.88\textwidth}
		\includegraphics[width=\textwidth]{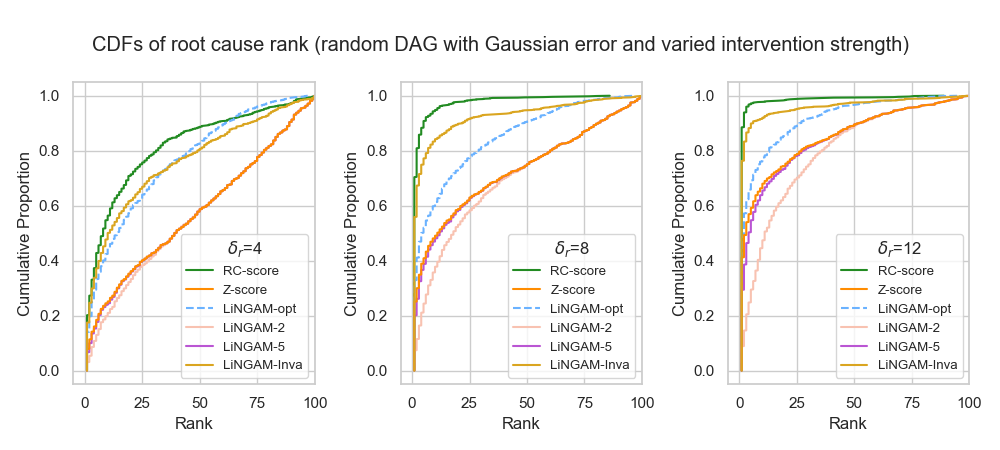}
	\end{minipage}
	\hfill
	\begin{minipage}[b]{0.88\textwidth}
		\includegraphics[width=\textwidth]{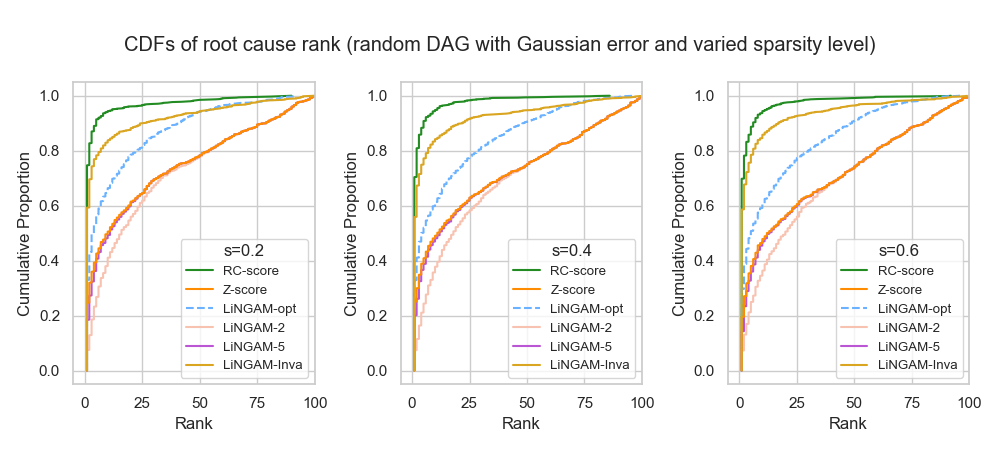}
	\end{minipage}
	\caption{CDFs of the root cause rank using the squared z-score, RC-score, and LiNGAM-based approaches in the setting with a random DAG and Gaussian errors. The top, middle, and bottom plots display results for varying sample sizes, intervention strengths, and sparsity levels, respectively.}
	\label{Fig:MainSimuP100RandomDAGGaussian}
\end{figure}

The simulation results for the hub DAG with uniform errors and random DAGs with uniform or Gaussian errors are shown in
Figures~\ref{Fig:MainSimuP100HubDAGUnif},
Figures~\ref{Fig:MainSimuP100RandomDAGUnif}, and \ref{Fig:MainSimuP100RandomDAGGaussian}, respectively. 

\subsection{The optimal thresholds for the LiNGAM-opt approach in Section~\ref{sec:simu-results}} \label{appendix:SimuOptimalThre}

\begin{figure}[ht]
	\centering
	\begin{minipage}[b]{0.49\textwidth}
		\includegraphics[width=\textwidth]{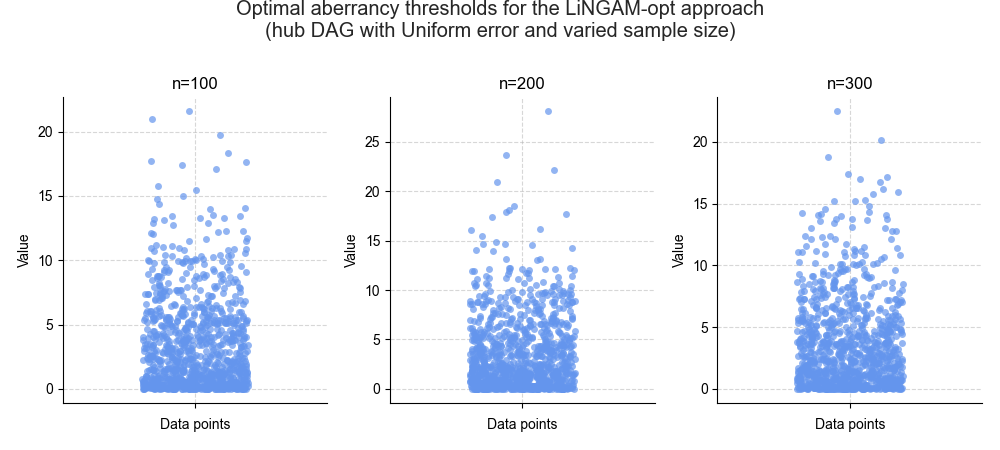}
	\end{minipage}
	\vspace{20pt}
	\begin{minipage}[b]{0.49\textwidth}
		\includegraphics[width=\textwidth]{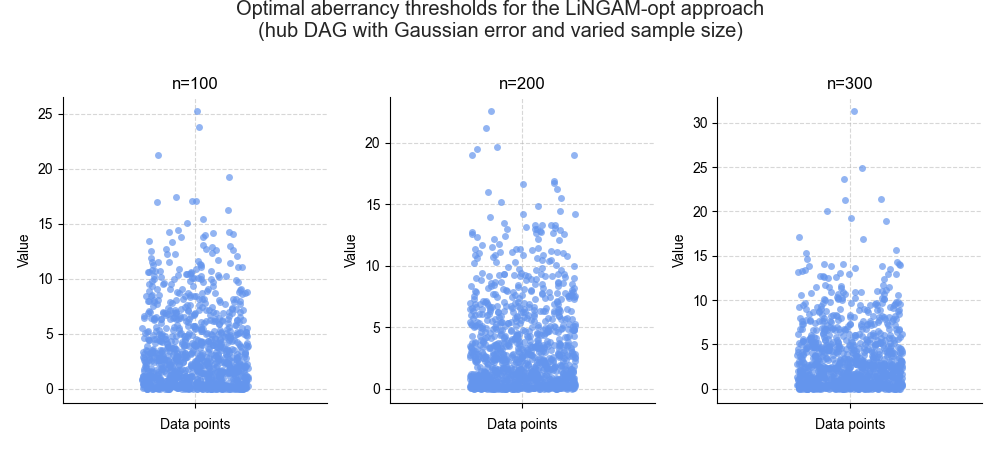}
	\end{minipage}
	\begin{minipage}[b]{0.49\textwidth}
		\includegraphics[width=\textwidth]{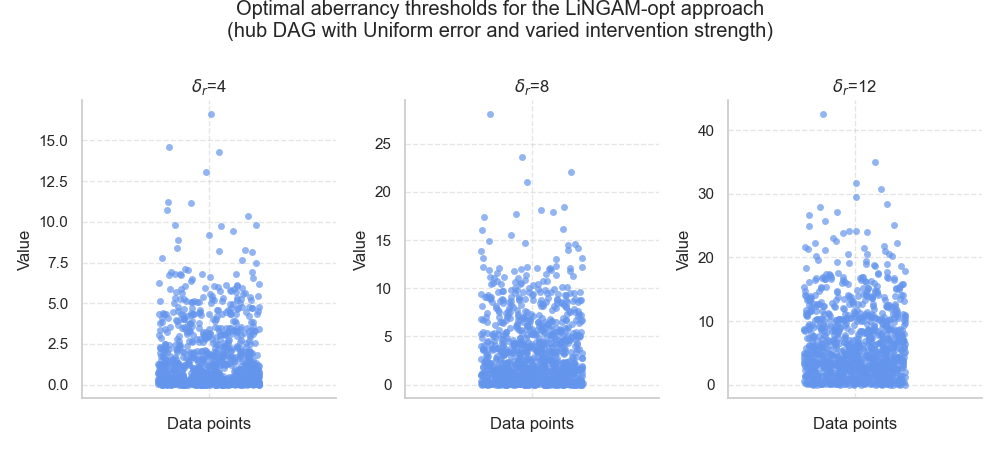}
	\end{minipage}
	\vspace{20pt}
	\begin{minipage}[b]{0.49\textwidth}
		\includegraphics[width=\textwidth]{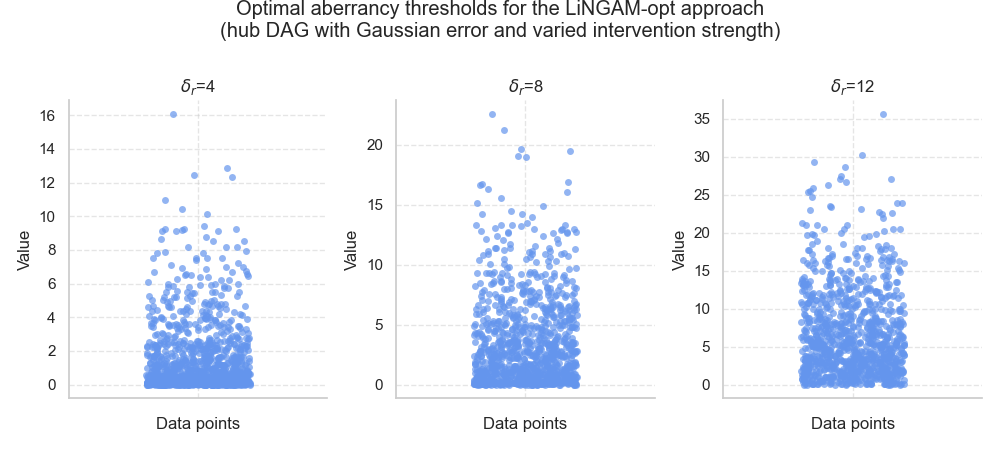}
	\end{minipage}
	\begin{minipage}[b]{0.49\textwidth}
		\includegraphics[width=\textwidth]{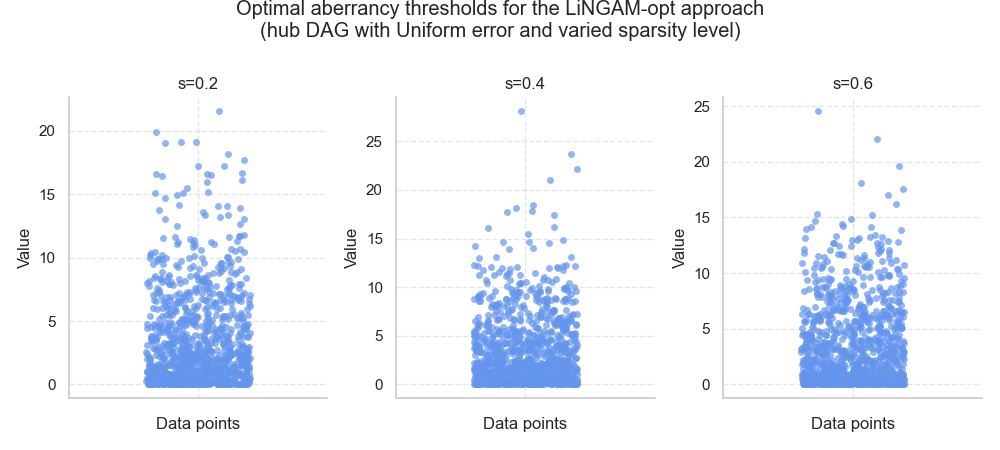}
	\end{minipage}
	\vspace{20pt}
	\begin{minipage}[b]{0.49\textwidth}
		\includegraphics[width=\textwidth]{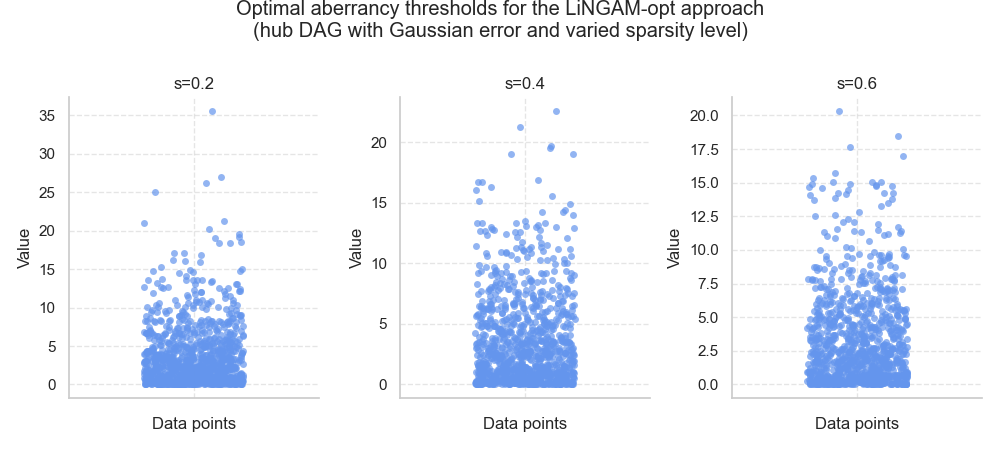}
	\end{minipage}
	\caption{The optimal thresholds used in the LiNGAM-opt approach for simulation settings with hub DAG. The error types are described in the subtitle of each plot.}
	\label{Fig:OptThreLiNGAMhub}
\end{figure}

\begin{figure}[ht]
	\centering
	\begin{minipage}[b]{0.49\textwidth}
		\includegraphics[width=\textwidth]{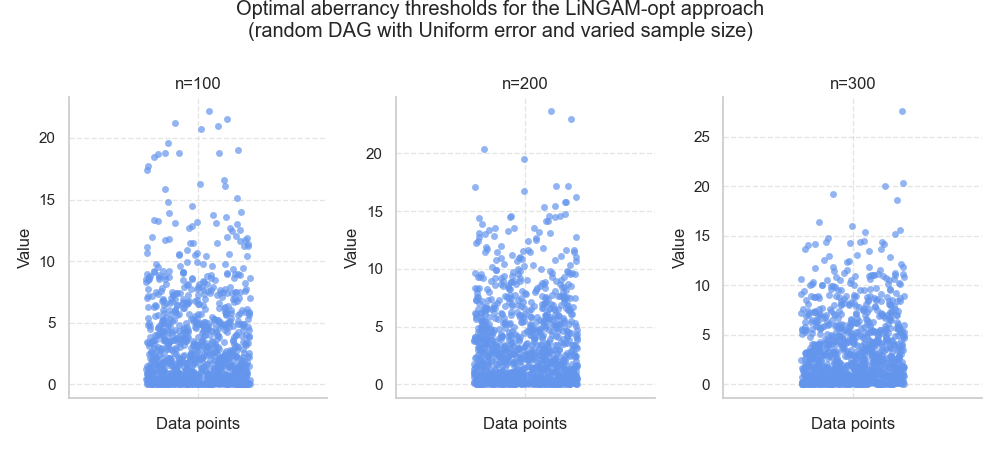}
	\end{minipage}
	\vspace{20pt}
	\begin{minipage}[b]{0.49\textwidth}
		\includegraphics[width=\textwidth]{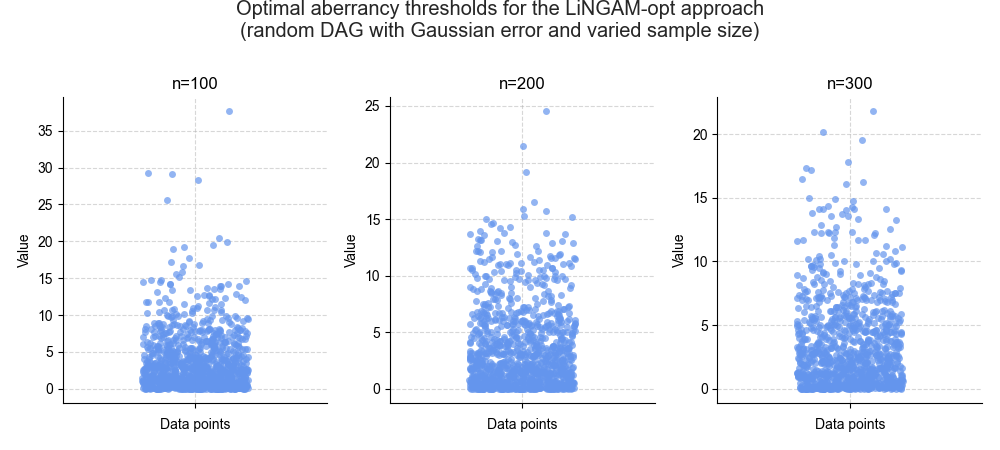}
	\end{minipage}
	\begin{minipage}[b]{0.49\textwidth}
		\includegraphics[width=\textwidth]{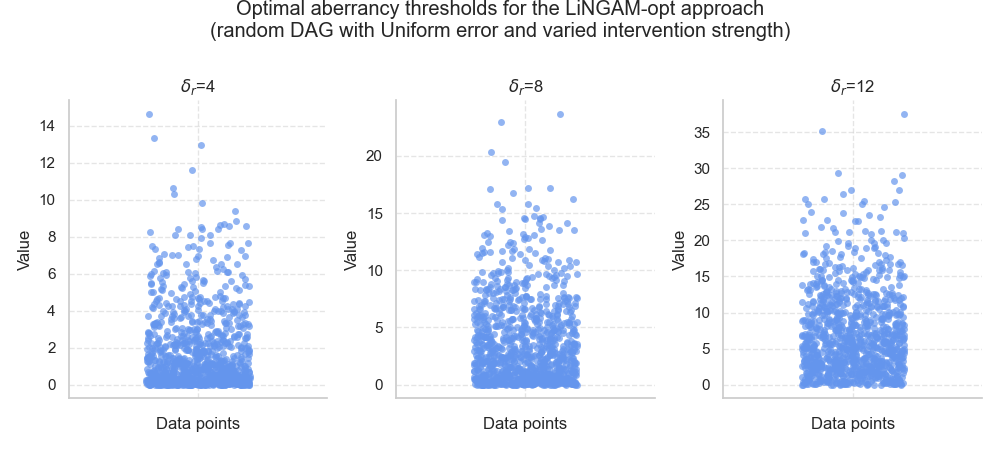}
	\end{minipage}
	\vspace{20pt}
	\begin{minipage}[b]{0.49\textwidth}
		\includegraphics[width=\textwidth]{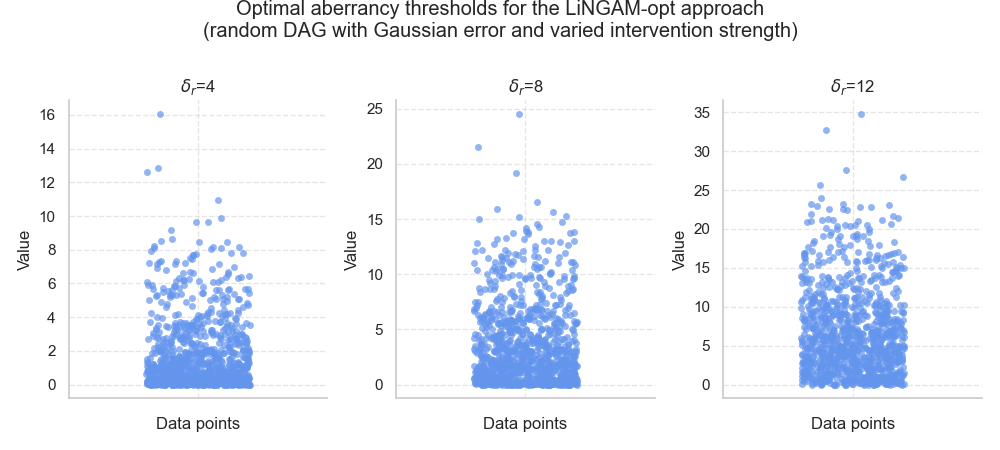}
	\end{minipage}
	\begin{minipage}[b]{0.49\textwidth}
		\includegraphics[width=\textwidth]{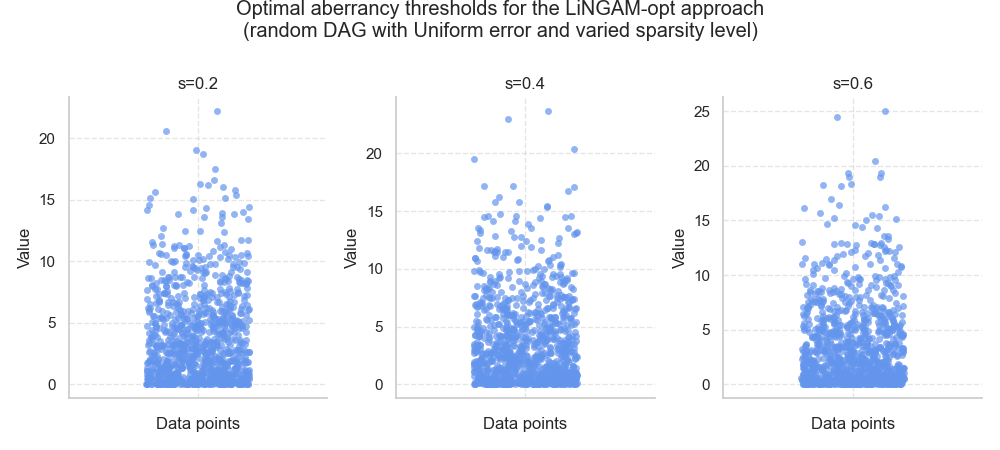}
	\end{minipage}
	\vspace{20pt}
	\begin{minipage}[b]{0.49\textwidth}
		\includegraphics[width=\textwidth]{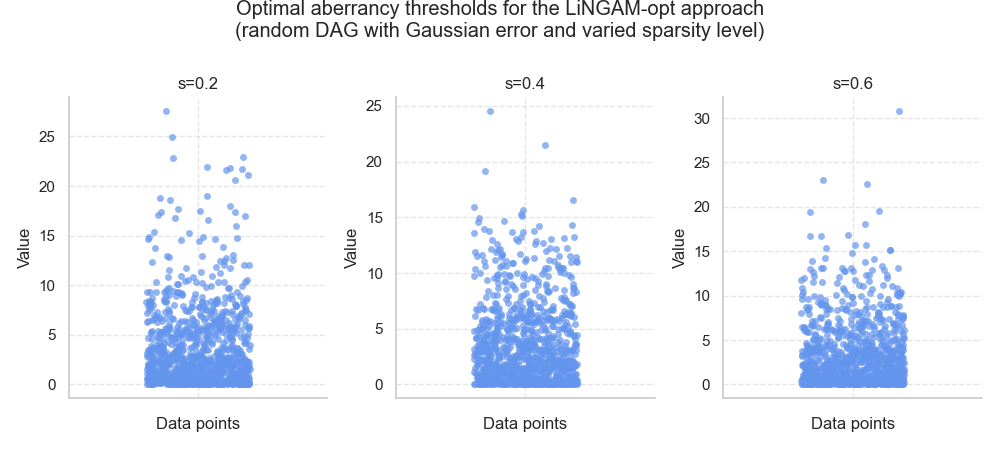}
	\end{minipage}
	\caption{The optimal thresholds used in the LiNGAM-opt approach for simulation settings with random DAG. The error types are described in the subtitle of each plot.}
	\label{Fig:OptThreLiNGAMrandom}
\end{figure}

In Figure~\ref{Fig:OptThreLiNGAMhub} and \ref{Fig:OptThreLiNGAMrandom}, we show 
the optimal thresholds (that is, the squared z-score of the root cause) used in the LiNGAM-opt approach for hub and random DAGs, respectively. 
We observe that the optimal thresholds vary significantly across the 1000 interventional samples in each setting, indicating that no single fixed optimal threshold exists.

\subsection{Run-time of different methods in Section~\ref{sec:simu-results}} \label{appendix:SimuCompuTime}

\begin{figure}[t]
	\centering
	\begin{minipage}{0.44\textwidth}
		\centering
		\includegraphics[width=\linewidth]{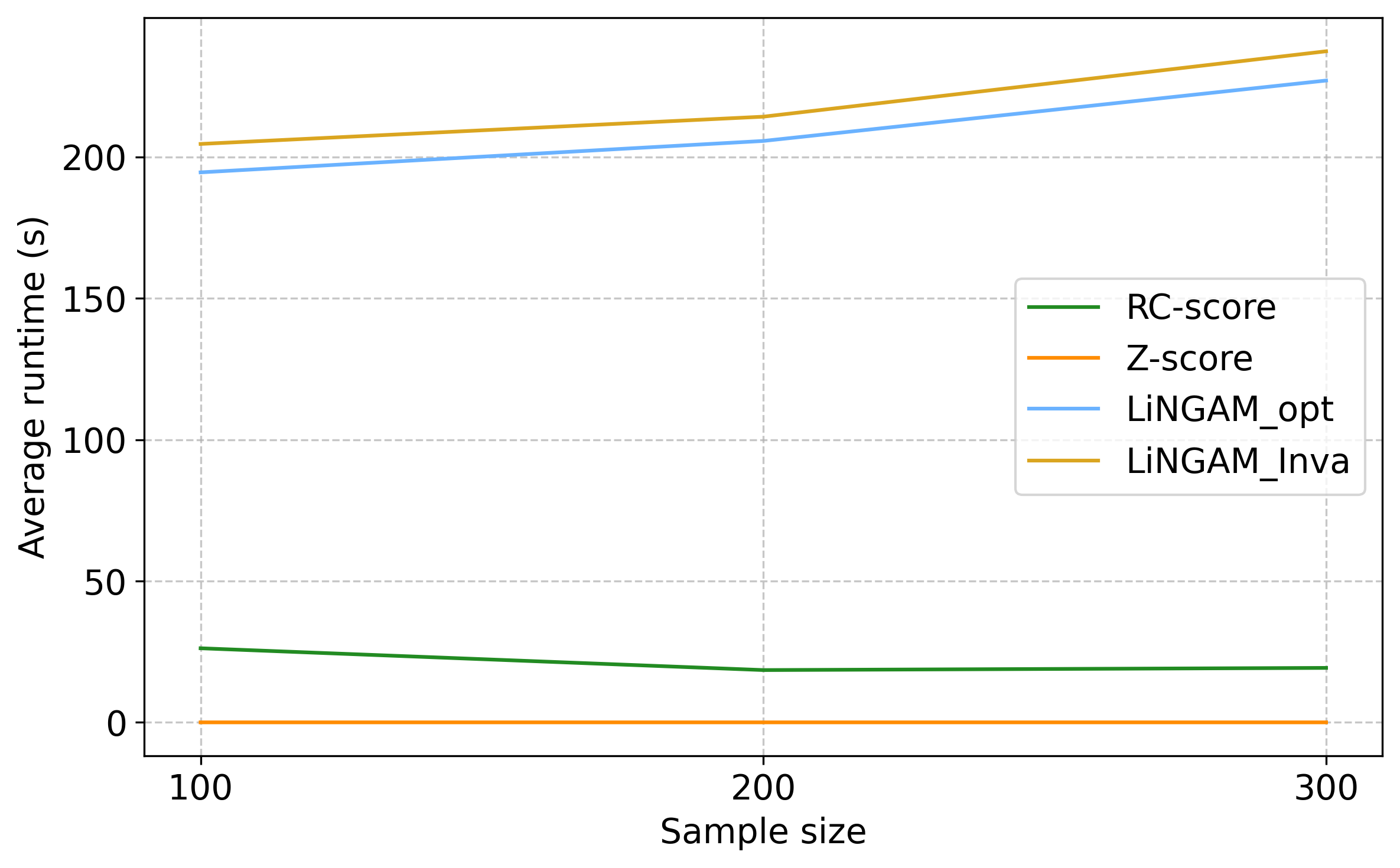}
	\end{minipage}%
	\hspace{0.1\textwidth}  
	\begin{minipage}{0.44\textwidth}
		\centering
		\includegraphics[width=\linewidth]{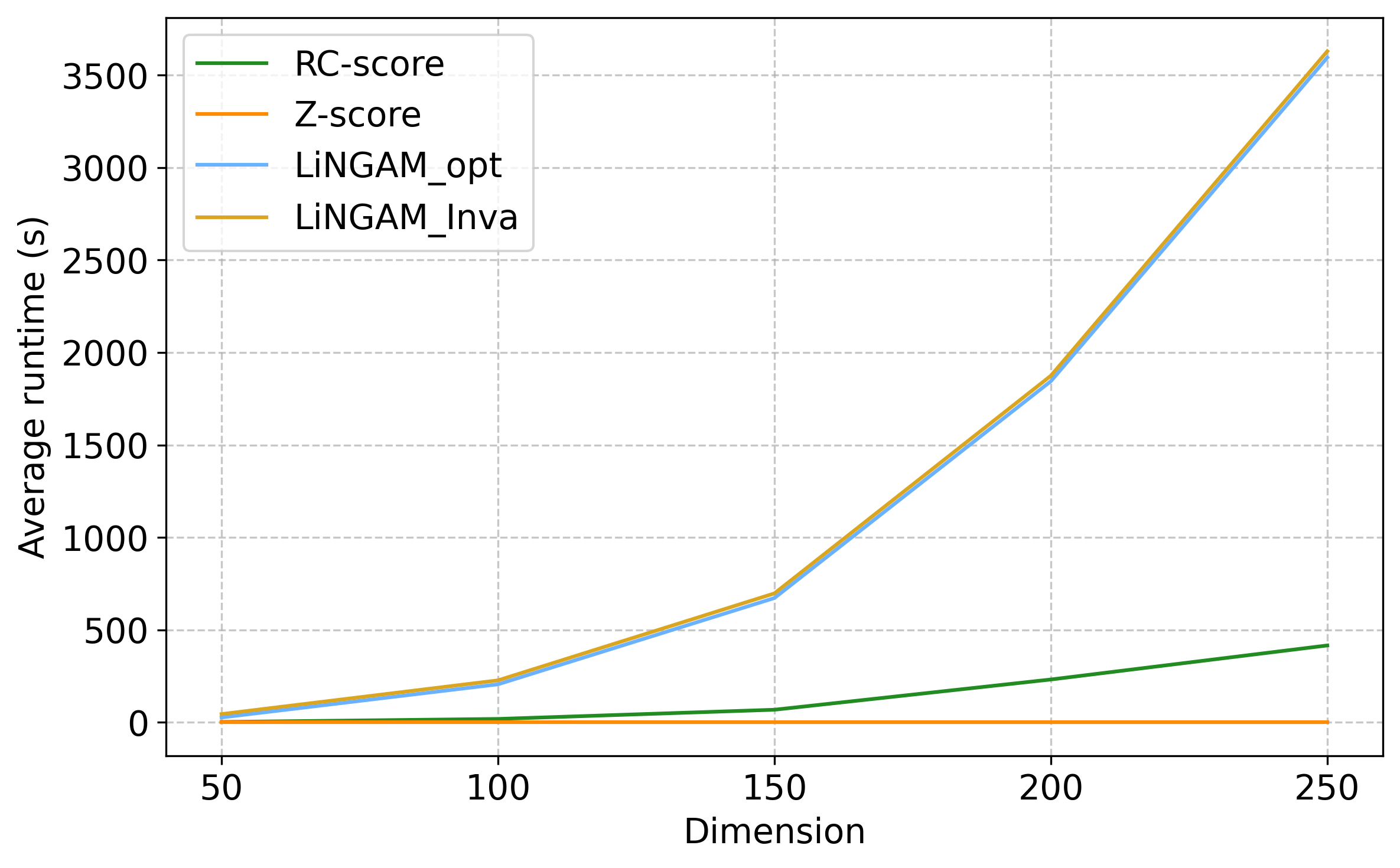}
	\end{minipage}
	\caption{Average run-time (in seconds) for different methods over 1000 interventional samples under a random DAG with Gaussian errors. The left plot shows run-times as a function of the sample size $n$ (with $p = 100, s = 0.4$, and $\delta_r = 8$), and the right plot shows run-times as a function of the dimension $p$ (with $n = 200, s = 0.4$, and $\delta_r = 8$)}
	\label{Fig:Runtime}
\end{figure}

We now consider the run-times of the different methods implemented in Section~\ref{sec:simu-results}. The run-times of LiNGAM-2, LiNGAM-5, and LiNGAM-opt are very similar, since the only difference among these approaches is the choice of threshold, while the most computationally intensive part is the estimation of the causal DAG by LiNGAM. We therefore only report the run-times for LiNGAM-opt as a representative of this class of methods.
In addition, because run-time is mainly affected by the dimension $p$ and the simulations in Section~\ref{sec:simu-results} use a fixed $p$, we report run-times only for the random DAG with Gaussian errors setting with $p = 100$, $s = 0.4$, $\delta_r = 8$, and varying sample sizes. 
We also examine the run-times for varying dimension $p$ under the random DAG with Gaussian errors setting with $n = 200$, $s = 0.4$, and $\delta_r = 8$.
See Figure~\ref{Fig:Runtime} for the run-time results.


\subsection{Simulations in latent variable settings} \label{sec:simu-latent}



In this section, we empirically examine the robustness of our proposed RC-scores (Algorithm~\ref{Algo:RCD}) to the presence of latent variables. 
Since squared z-scores are calculated marginally and are not affected by latent variables, we use them for comparison.
We consider random DAGs and hub DAGs with Gaussian errors as described in Section~\ref{sec:simu-setup},
with the variances of errors independently sampled from $U(1,5)$. 
We generate samples in the same manner as in Section~\ref{sec:simu-setup}, with the only difference that we introduce latent variables by randomly discarding $\eta$ percent of the variables in the generated samples. 
We fix the sample size at $n=200$ and the intervention strength at $\delta_r=12$,
and we vary the latent proportion $\eta$ and the sparsity level $s$ to investigate their effects.
To this end, we consider the following scenarios:
\begin{itemize}
	\item Vary $\eta \in\{0.1,0.3,0.5,0.7\}$ while fixing $s=0.2$.
	\item Vary $s \in\{0.2, 0.4, 0.6, 0.8\}$ while fixing $\eta=0.3$.
\end{itemize}

\begin{figure}[h!]
	\centering
	\begin{minipage}[b]{0.48\textwidth}
		\includegraphics[width=\textwidth]{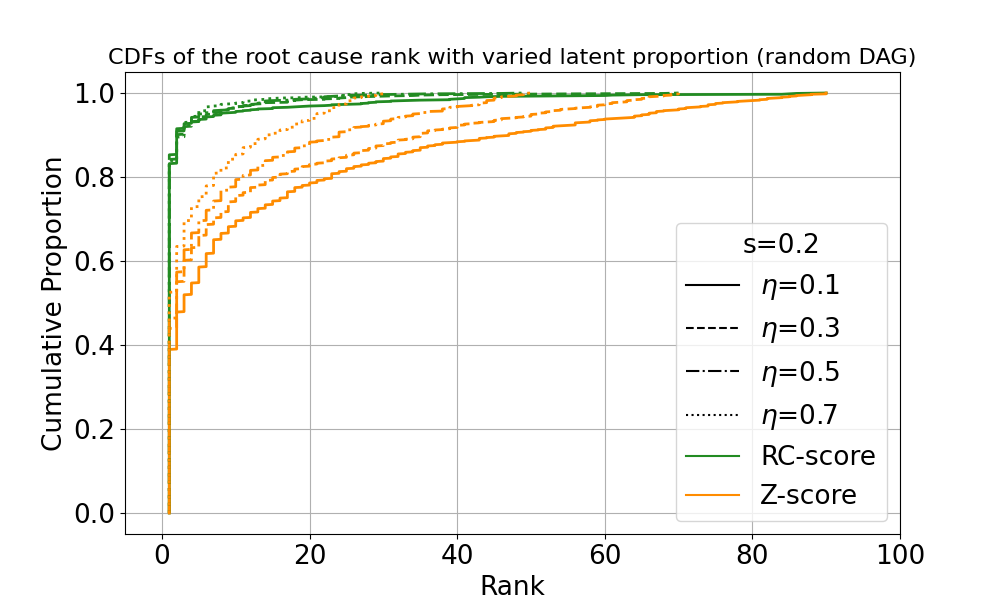}
	\end{minipage}
	\begin{minipage}[b]{0.48\textwidth}
		\includegraphics[width=\textwidth]{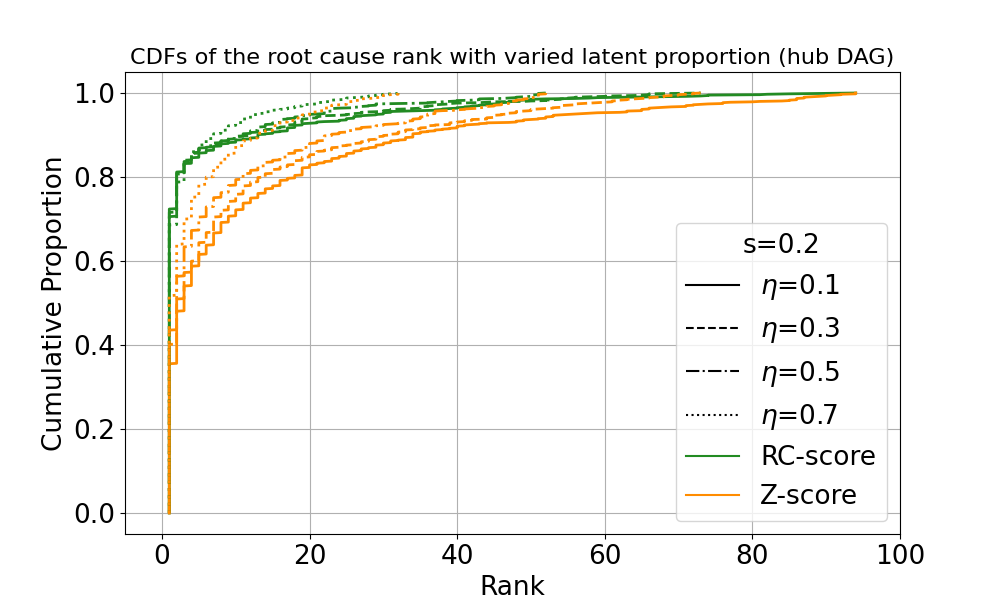}
	\end{minipage}
	\begin{minipage}[b]{0.48\textwidth}
		\includegraphics[width=\textwidth]{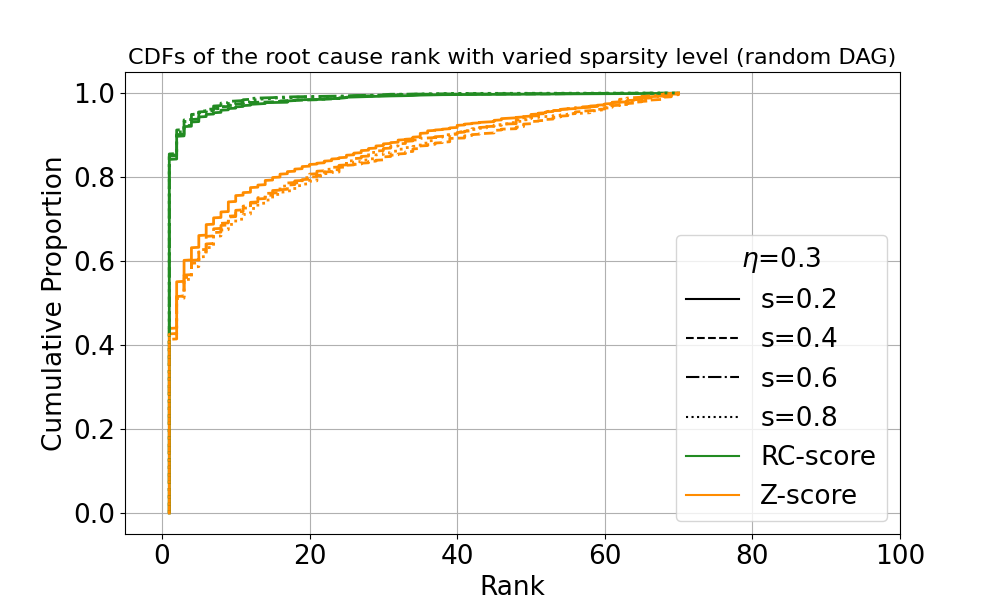}
	\end{minipage}
	\begin{minipage}[b]{0.48\textwidth}
		\includegraphics[width=\textwidth]{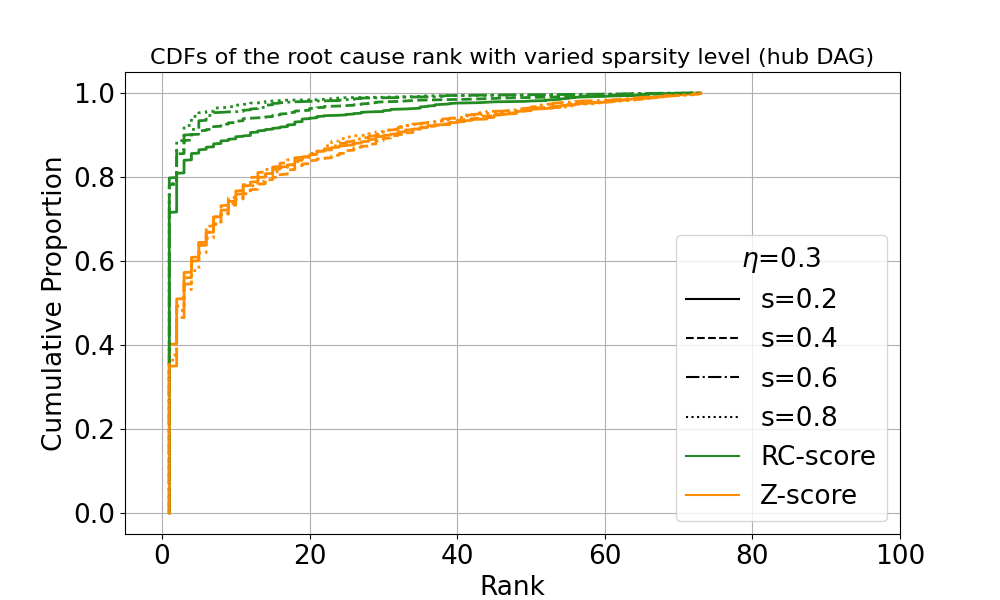}
	\end{minipage}
	\caption{CDFs of the root cause rank using the squared z-score (orange) and RC-score (green) in the presence of latent variables. The left and right columns display results for random and hub DAGs, respectively. The top and bottom plots show results for different latent proportions and sparsity levels, respectively. Different line types represent different simulation settings. }
	\label{Fig:MainSimuLatentP100}
\end{figure}

The results are shown in Figure~\ref{Fig:MainSimuLatentP100}, with different line types representing different simulation settings.
We observe that the RC-score outperforms the squared z-score across all settings, maintaining stable performance as the latent proportion $\eta$ and sparsity level $s$ vary, demonstrating its robustness to latent variables. 
The performance of the squared z-score seems to improve as $\eta$ increases, but this is an artifact caused by comparing the root cause to fewer variables.
The CDF of the RC-score remains relatively stable across different values of $\eta$ because the root cause already has a small rank.

\subsection{Simulations in high-dimensional settings} \label{appendix:SimuHighdim} 

We conduct simulations to evaluate the performance of Algorithm~\ref{Algo:RCDhighdim} in high-dimensional settings, alongside calculating the squared z-scores for comparison. 
We set $v=10$ random permutations for Algorithm~\ref{Algo:RCDhighdim}.
To reduce computational time, we only treat variables with a squared z-score larger than $1.5$ as the response in step 2 of Algorithm~\ref{Algo:RCDhighdim}.
Additionally,  as in previous simulations, we use thresholds $(0.1, 0.3, \dots, 5)$ in the first step of Algorithm~\ref{Algo:RCD} when it is called by Algorithm~\ref{Algo:RCDhighdim}.
We consider a hub graph for our simulations as it is inspired by genetic applications which involves high-dimensional data,
and we use Gaussian errors with the variances independently sampled from $U(1,5)$. 
The simulation setup follows the description in Section~\ref{sec:simu-setup}, but we now consider a larger hub graph.
Specifically, we consider 20 hub nodes. Each hub node has an upper node block of size 30, where the nodes point to the hub node, and a lower node block of size 20, to which the hub node points. Additionally, each hub node is pointed to by 9 nodes from other upper node blocks and points to 6 nodes from other lower node blocks. 
This results in a total of $p=1020$ variables.
Nodes within the upper and lower node blocks form a random DAG (as described in Section~\ref{sec:simu-setup}) with sparsity level $s=0.2$.
As in Section~\ref{sec:simu-setup}, we randomly generate $20$ matrices $B$.
For each $B$, we generate $n$ observational samples. In addition, we randomly choose $50$ root causes, and generate one interventional sample with intervention effect $\delta_r$ for each root cause.
In total, there are $m=1000$ interventional samples with possibly different root causes.

We consider the following scenarios to investigate the effects of sample size $n$ and intervention strength $\delta_r$:
\begin{itemize}
	\item Vary $n\in\{100, 200, 300, 400\}$ while fixing $\delta_r=12$.
	\item Vary $\delta_r\in\{8,12,16,20\}$ while fixing $n=200$.
\end{itemize}

\begin{figure}[ht]
	\centering
	\begin{minipage}[b]{0.48\textwidth}
		\includegraphics[width=\textwidth]{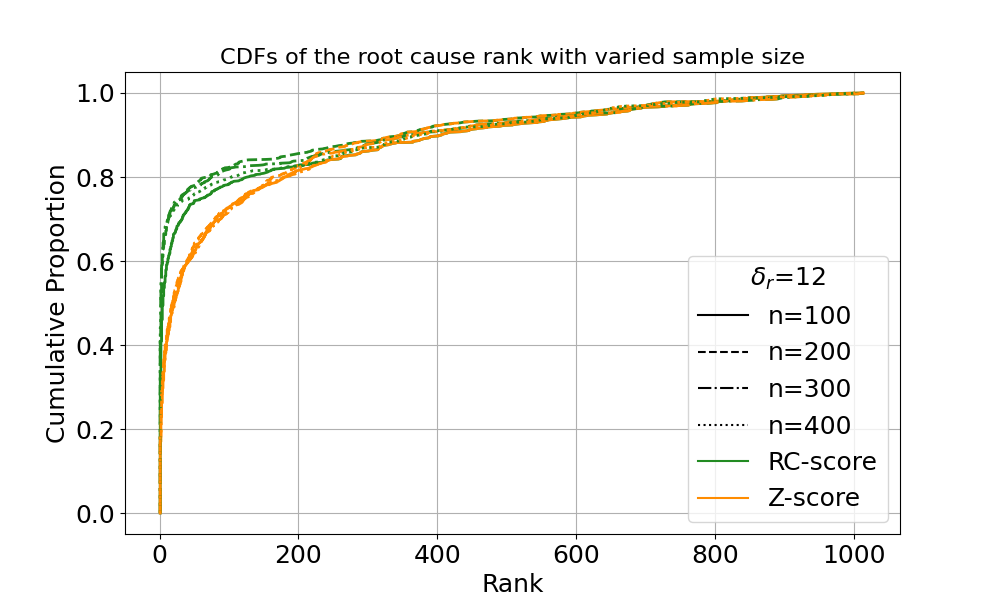}
	\end{minipage}
	\hfill
	\begin{minipage}[b]{0.48\textwidth}
		\includegraphics[width=\textwidth]{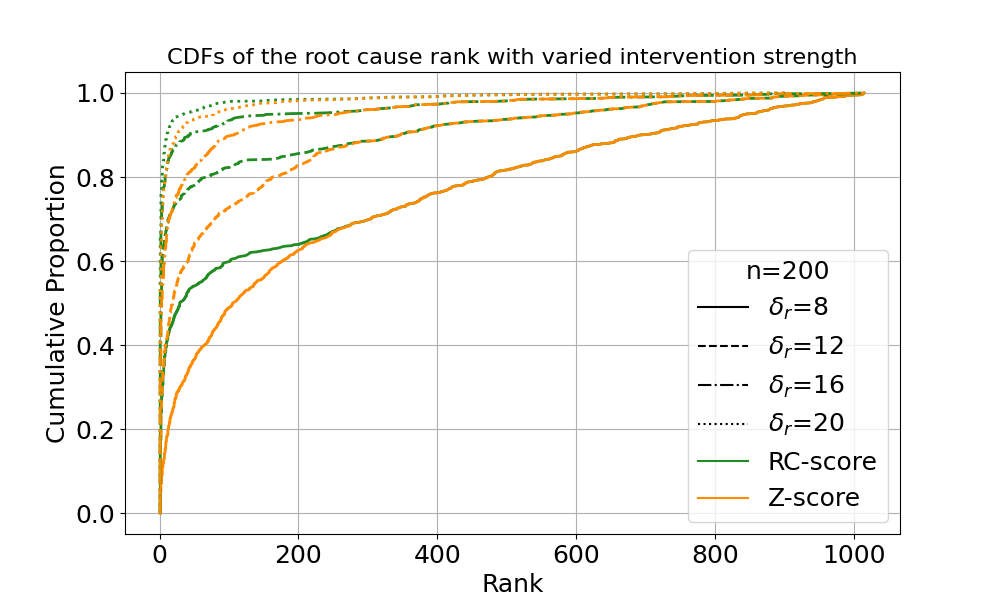}
	\end{minipage}
	\caption{CDFs of the root cause rank using the squared z-score (orange) and the RC-score for high-dimensional settings (Algorithm~\ref{Algo:RCDhighdim}; green). The left plot shows results for different sample sizes, while the right plot presents results for varying intervention strengths. Different line types represent different simulation settings. }
	\label{Fig:simuHighDim}
\end{figure}

The results are shown in Figure~\ref{Fig:simuHighDim}. 
We see that Algorithm~\ref{Algo:RCDhighdim} performs better than the squared z-scores in all settings.
As expected, with larger sample size and intervention strength, the performances become better.


\subsection{Comparison of results on separate and combined RNA-seq datasets in the real application} \label{appendix:realdataSSNSseparate}

\begin{figure}[h!]
	\centering
	\includegraphics[scale=0.3]{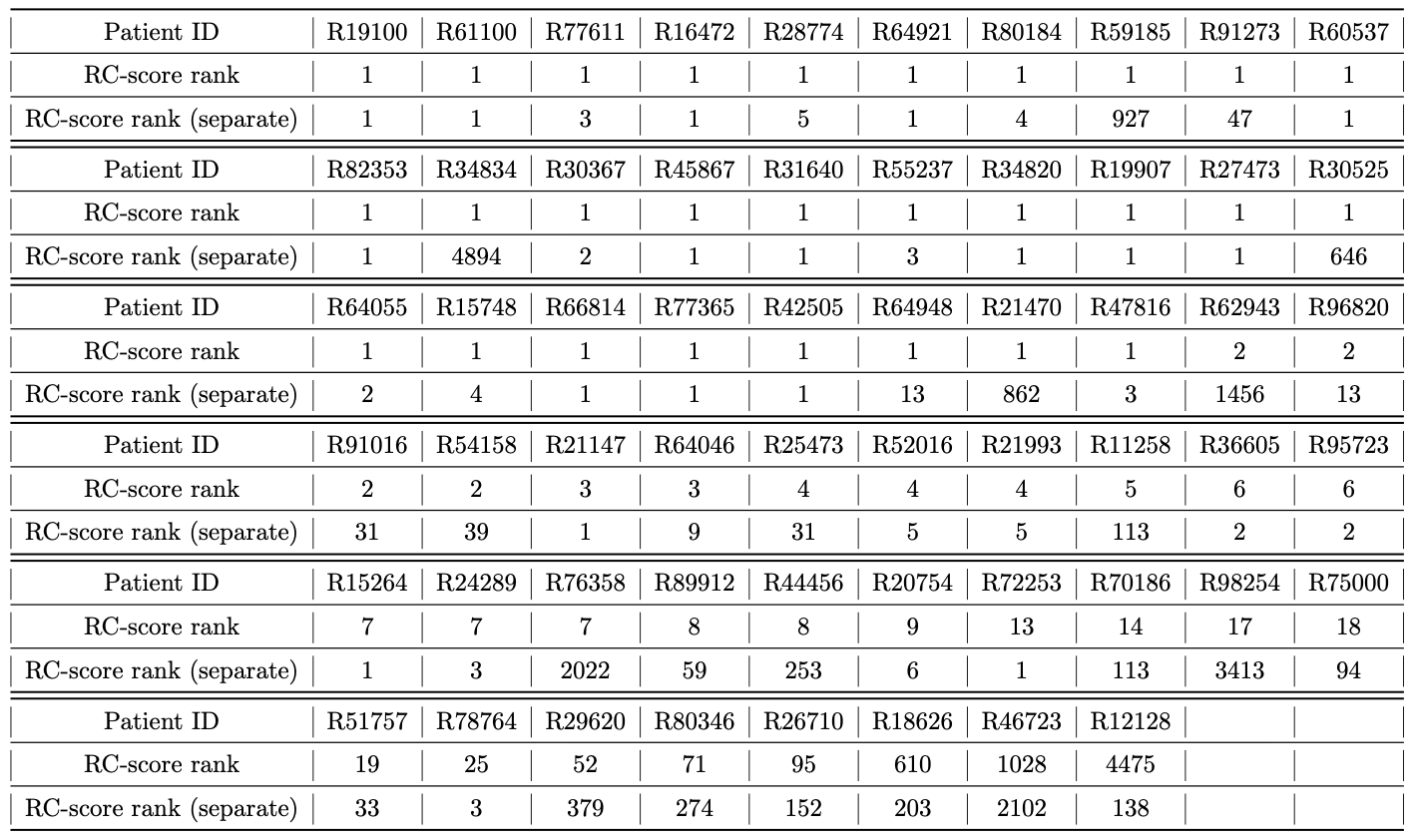}
	\caption{Table showing the rank of the root cause for $58$ patients based on the RC-score, applied to the non-strand-specific and strand-specific RNA-sequencing datasets separately (denoted by “RC-score rank (separate)”) and to the combined dataset (denoted by “RC-score rank”). A smaller rank is better, with rank 1 being the best.}
	\label{Fig:tab:realdataRankResultSSNSseparate}
\end{figure}

We ran the RC-score method on the strand-specific and non-strand-specific RNA-sequencing datasets separately, as well as on the combined dataset. Figure~\ref{Fig:tab:realdataRankResultSSNSseparate} shows the root cause ranks using the RC-score for all $58$ patients obtained using these two approaches. Note that each patient belongs either to the strand-specific or the non-strand-specific dataset. From the table, we can see that the results obtained by applying the RC-score method to the combined dataset are generally better than those obtained by applying it to the two datasets separately.

\subsection{Score plots for all patients in the real application} \label{appendix:realdata}
\begin{figure}[ht]
	\centering
	\begin{minipage}[b]{0.9\textwidth}
		\centering
		\includegraphics[scale=0.26]{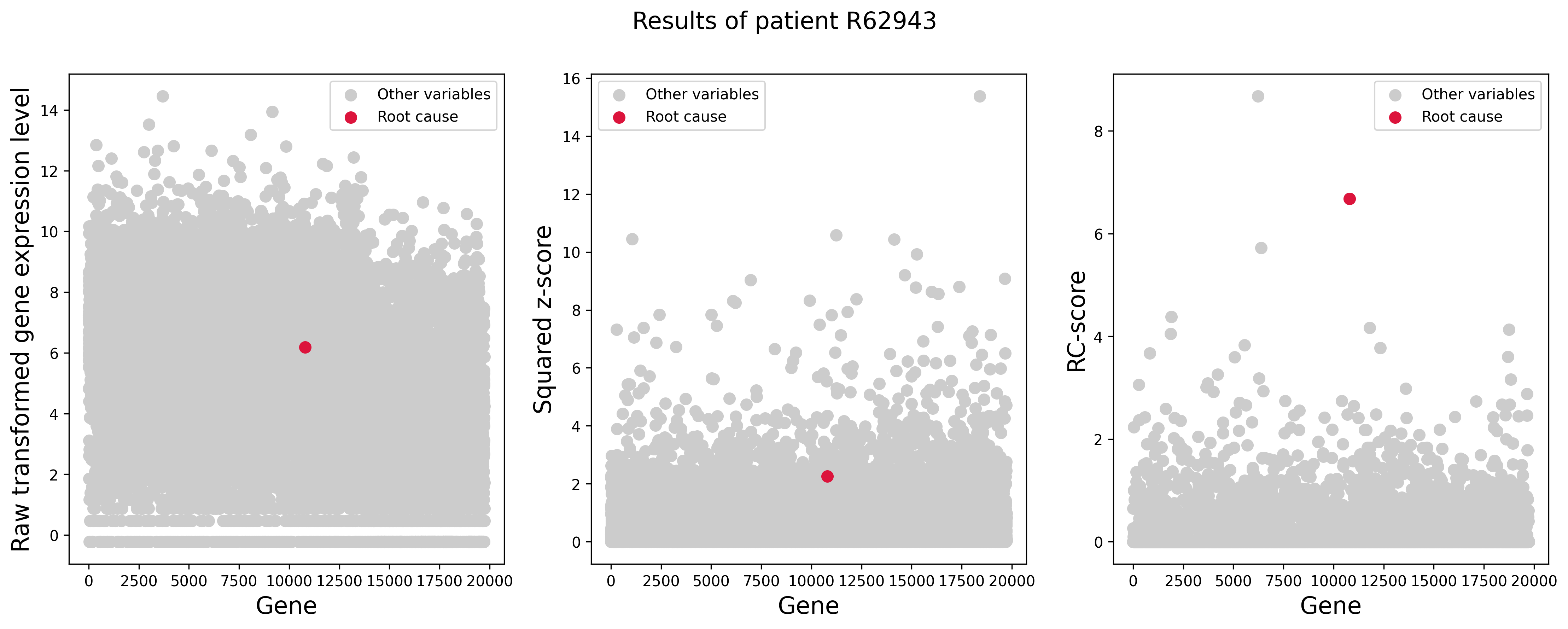}
	\end{minipage}
	\vspace{0.05cm}
	\begin{minipage}[b]{0.9\textwidth}
		\centering
		\includegraphics[scale=0.26]{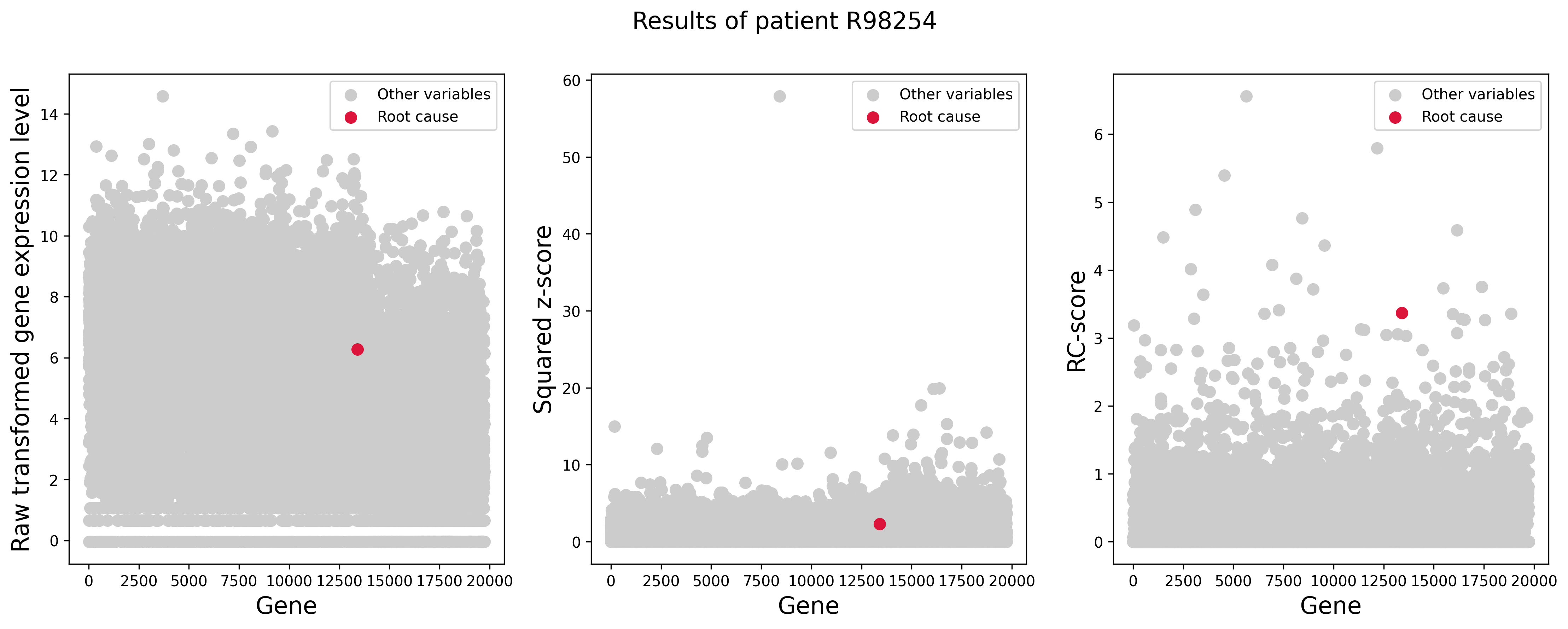}
	\end{minipage}
	\vspace{0.05cm}
	\begin{minipage}[b]{0.9\textwidth}
		\centering
		\includegraphics[scale=0.26]{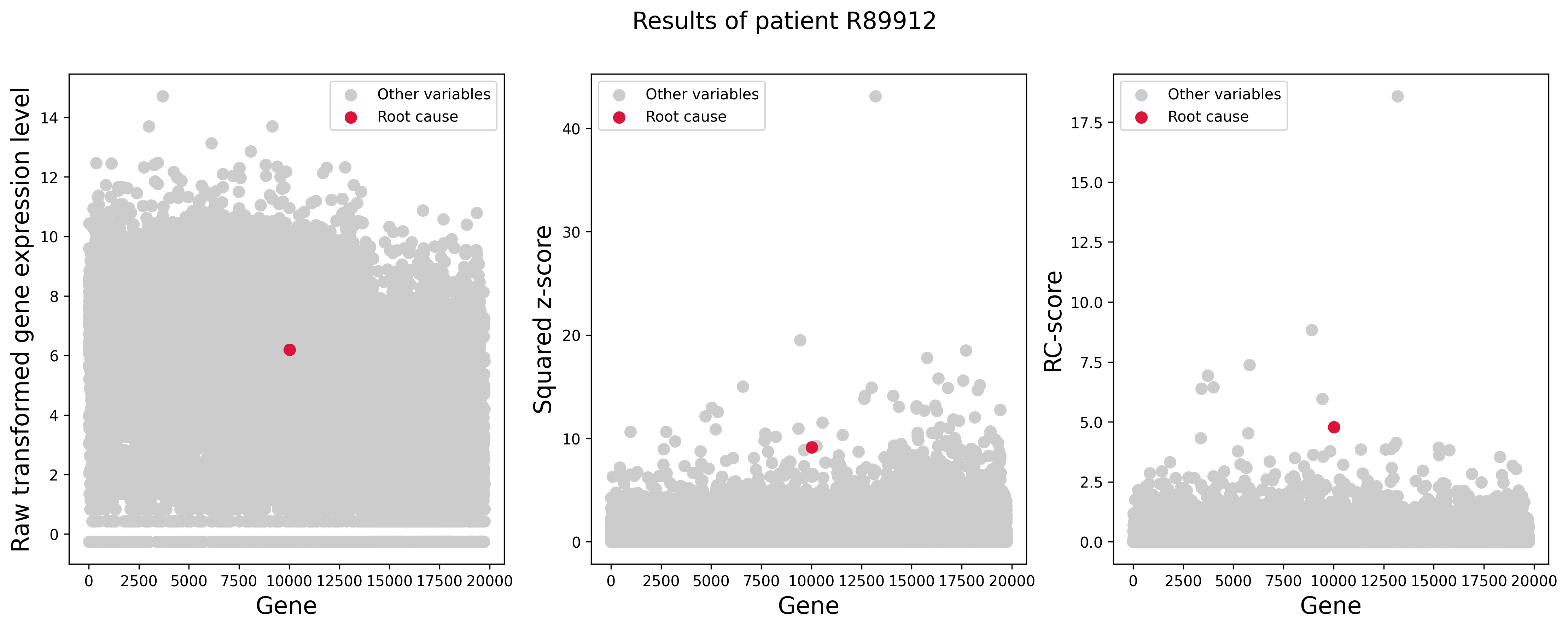}
	\end{minipage}
	\vspace{0.05cm}
	\begin{minipage}[b]{0.9\textwidth}
		\centering
		\includegraphics[scale=0.26]{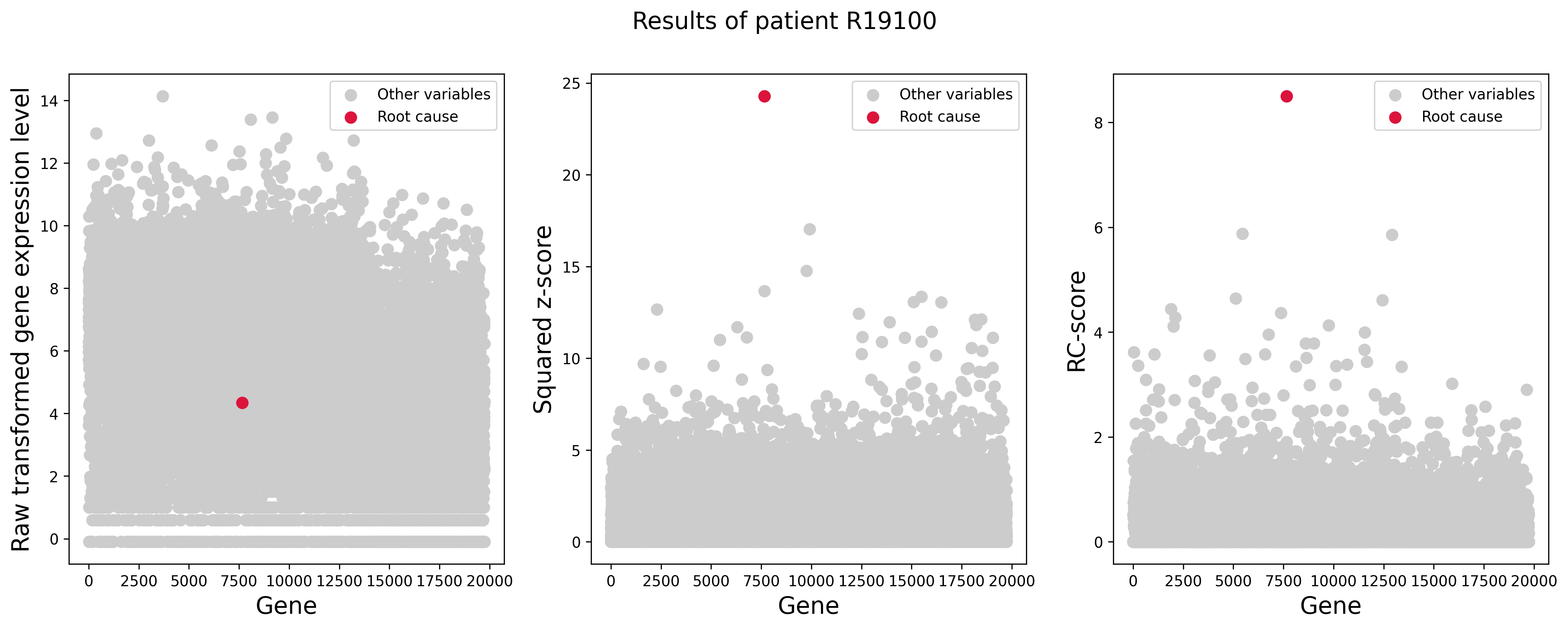}
	\end{minipage}
	\vspace{0.05cm}
	\begin{minipage}[b]{0.9\textwidth}
		\centering
		\includegraphics[scale=0.26]{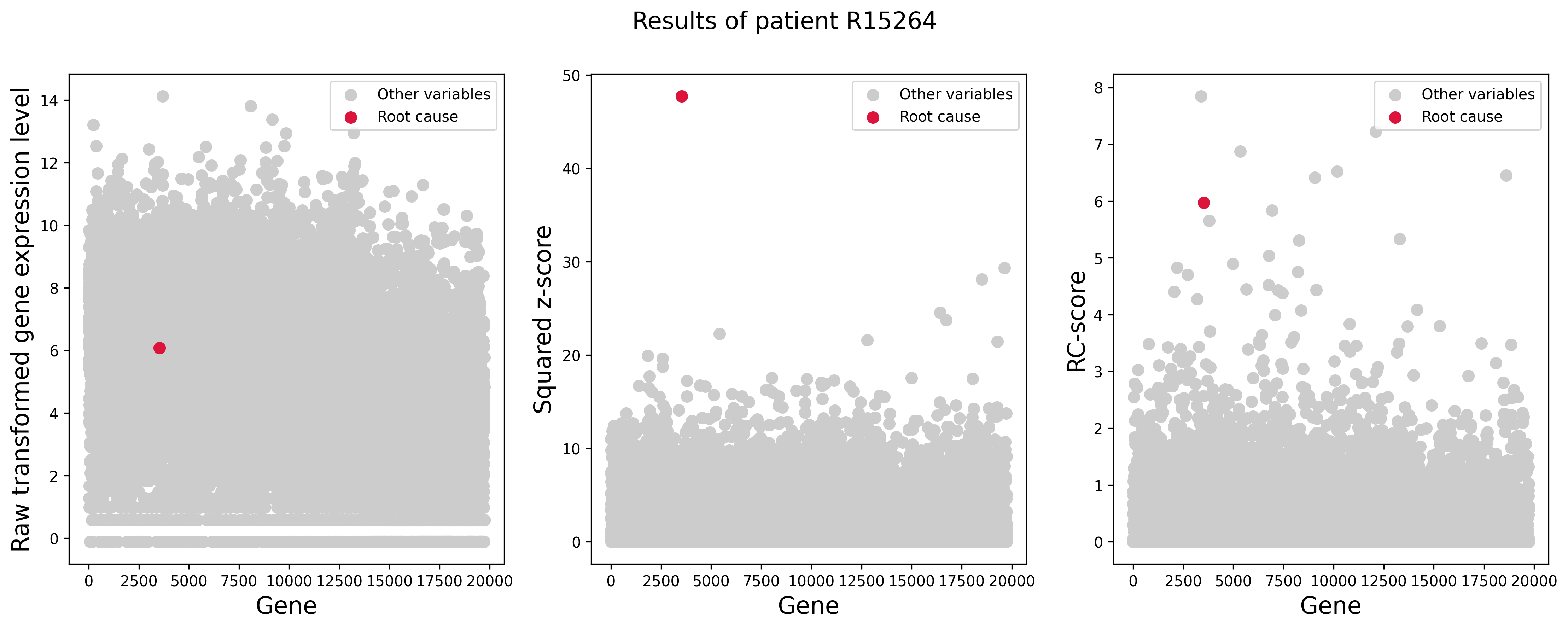}
	\end{minipage}
	\vspace{-0.5cm}
	\caption{The raw transformed gene expression levels, squared z-scores, and RC-scores of genes for patients $R62943$, $R98254$, $R89912$, $R19100$, and $R15264$.}
	\label{Fig:AppReal1}
\end{figure}

\begin{figure}[ht]
	\centering
	\begin{minipage}[b]{0.9\textwidth}
		\centering
		\includegraphics[scale=0.26]{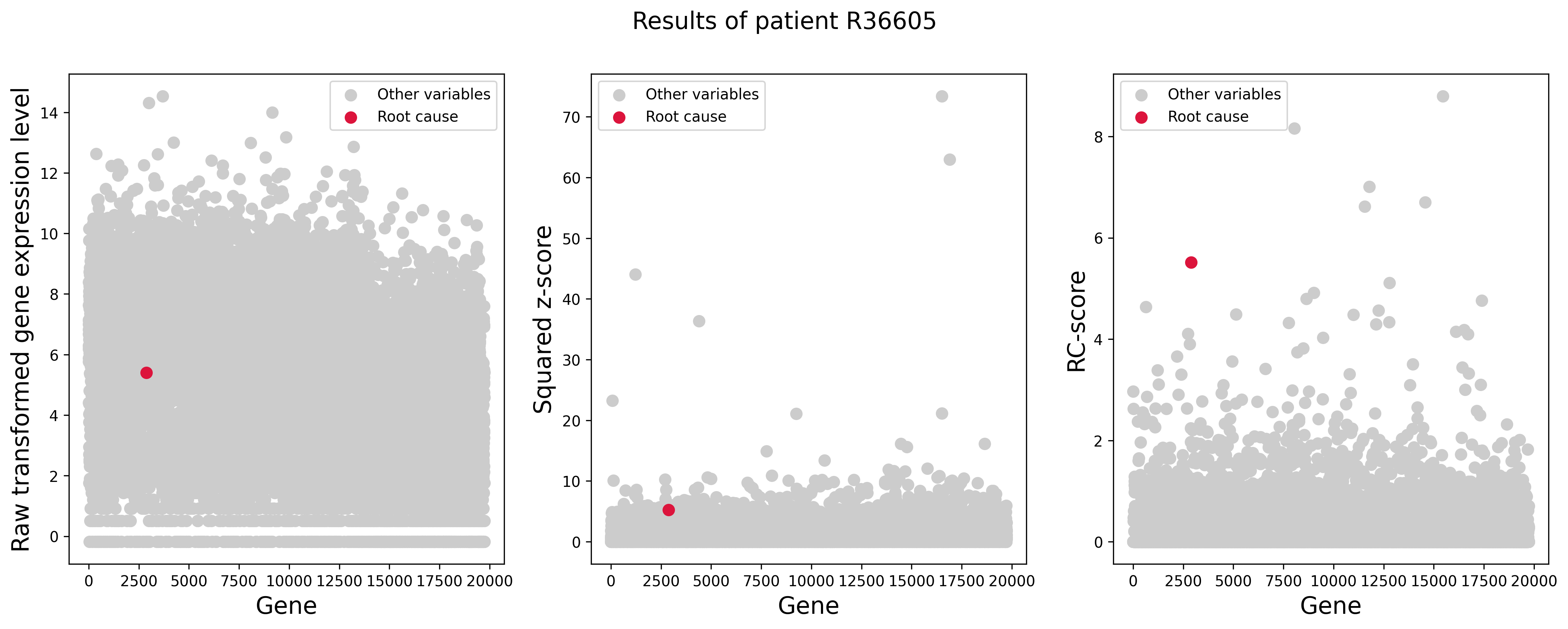}
	\end{minipage}
	\vspace{0.05cm}
	\begin{minipage}[b]{0.9\textwidth}
		\centering
		\includegraphics[scale=0.26]{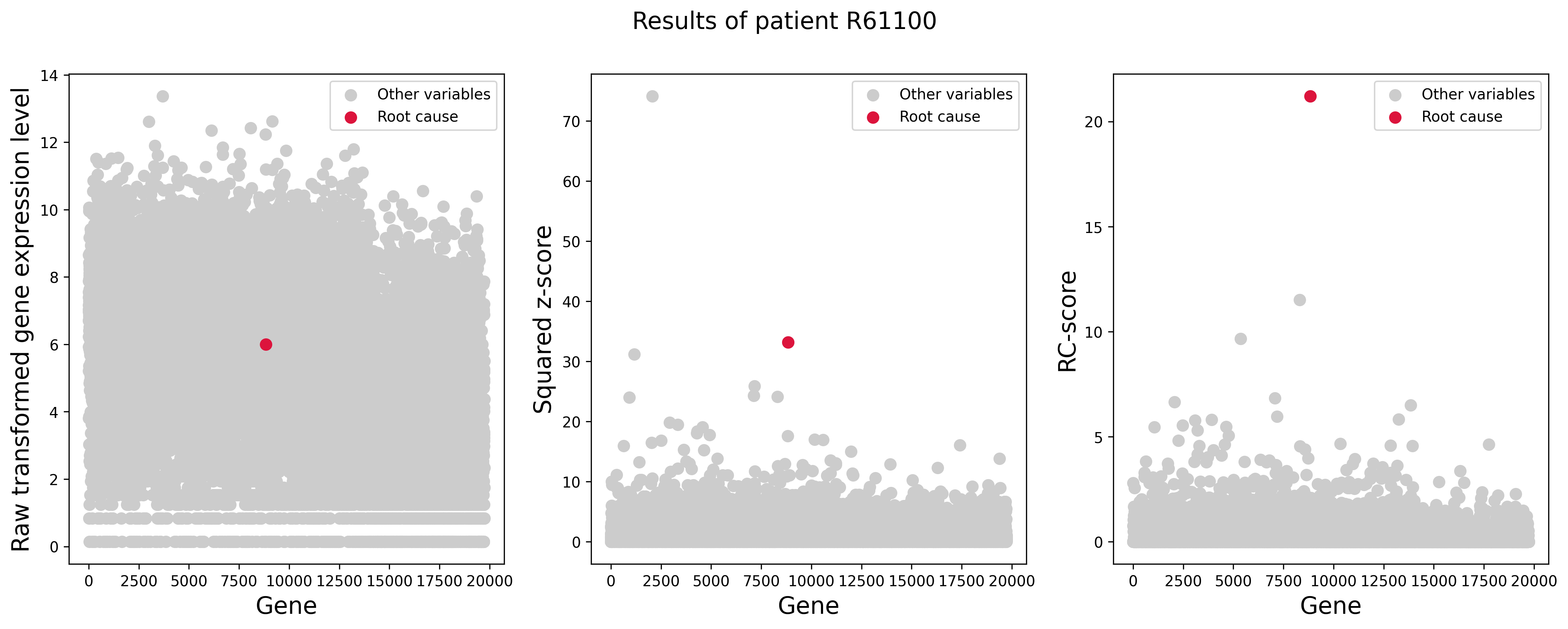}
	\end{minipage}
	\vspace{0.05cm}
	\begin{minipage}[b]{0.9\textwidth}
		\centering
		\includegraphics[scale=0.26]{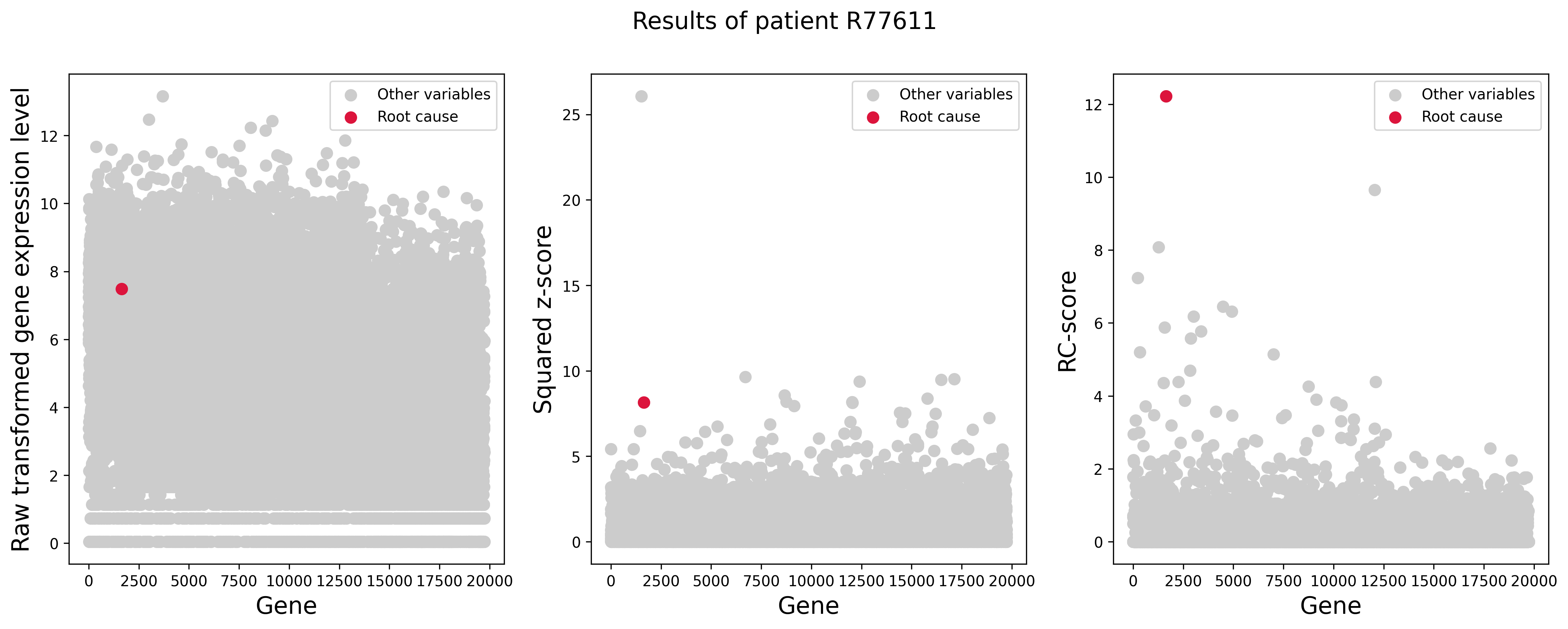}
	\end{minipage}
	\vspace{0.05cm}
	\begin{minipage}[b]{0.9\textwidth}
		\centering
		\includegraphics[scale=0.26]{plots/Supp/ThreeScoresPlots/ThreeScoresR16472.png}
	\end{minipage}
	\vspace{0.05cm}
	\begin{minipage}[b]{0.9\textwidth}
		\centering
		\includegraphics[scale=0.26]{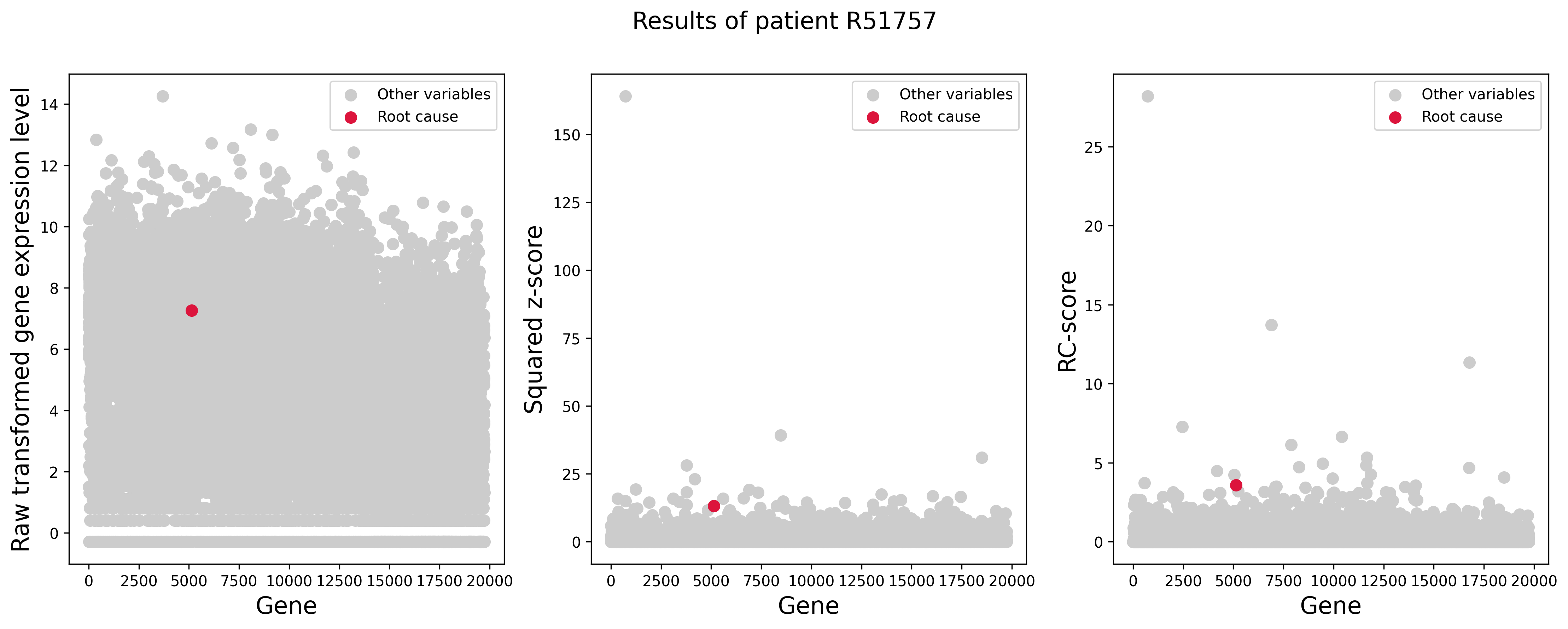}
	\end{minipage}
	\vspace{-0.5cm}
	\caption{The raw transformed gene expression levels, squared z-scores, and RC-scores of genes for patients $R36605$, $R61100$, $R77611$, $R16472$, and $R51757$.}
	\label{Fig:AppReal2}
\end{figure}

\begin{figure}[ht]
	\centering
	\begin{minipage}[b]{0.9\textwidth}
		\centering
		\includegraphics[scale=0.26]{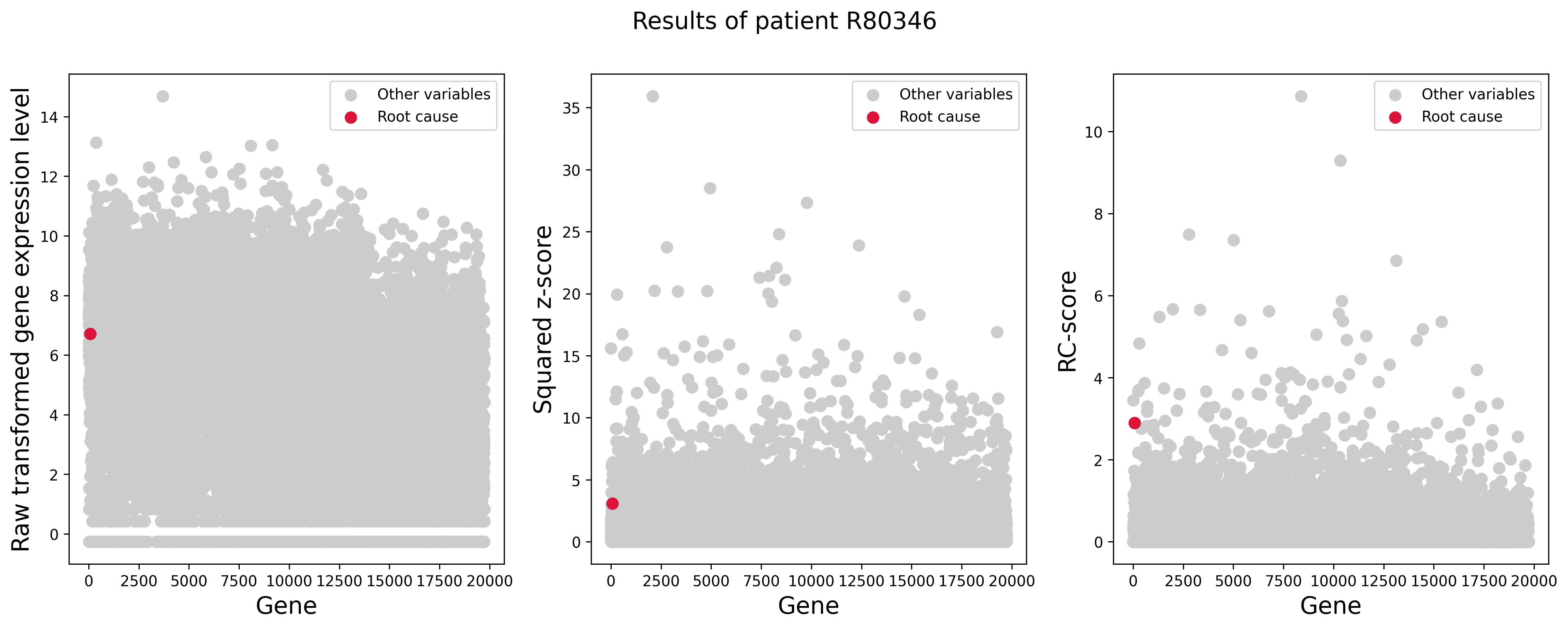}
	\end{minipage}
	\vspace{0.05cm}
	\begin{minipage}[b]{0.9\textwidth}
		\centering
		\includegraphics[scale=0.26]{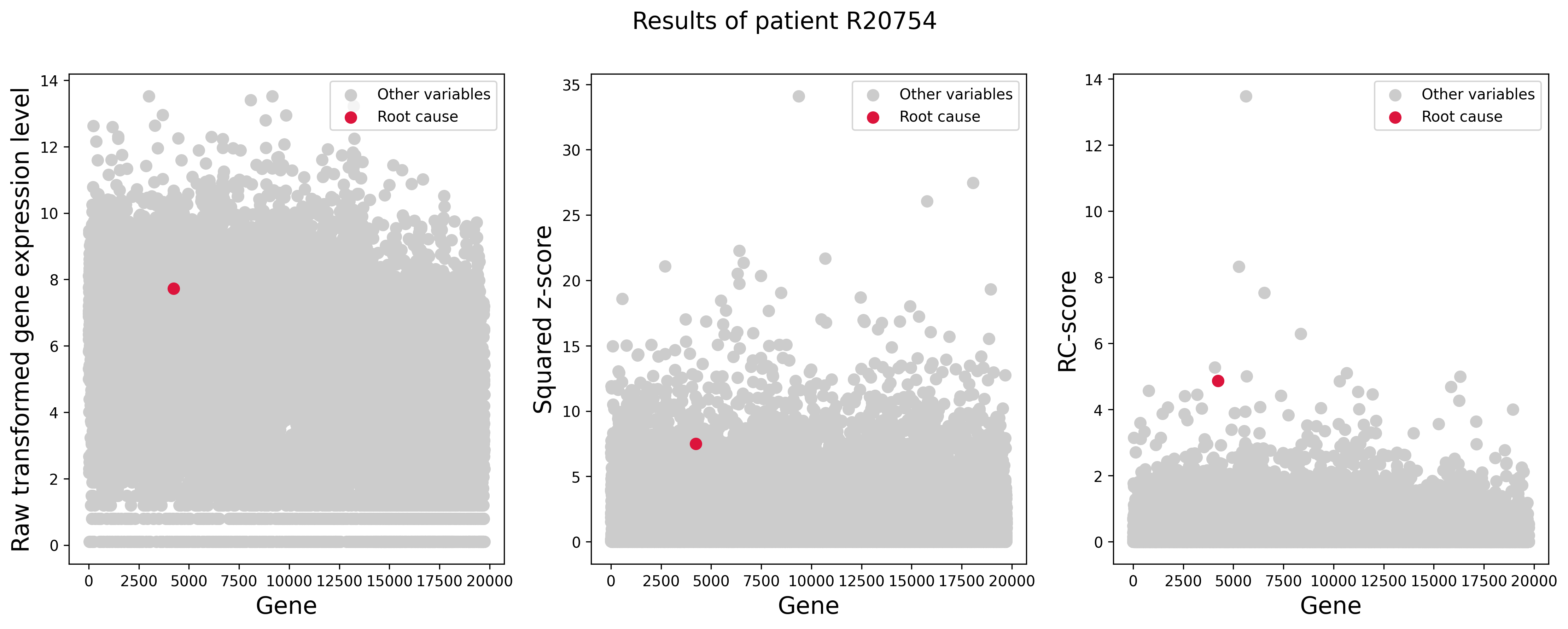}
	\end{minipage}
	\vspace{0.05cm}
	\begin{minipage}[b]{0.9\textwidth}
		\centering
		\includegraphics[scale=0.26]{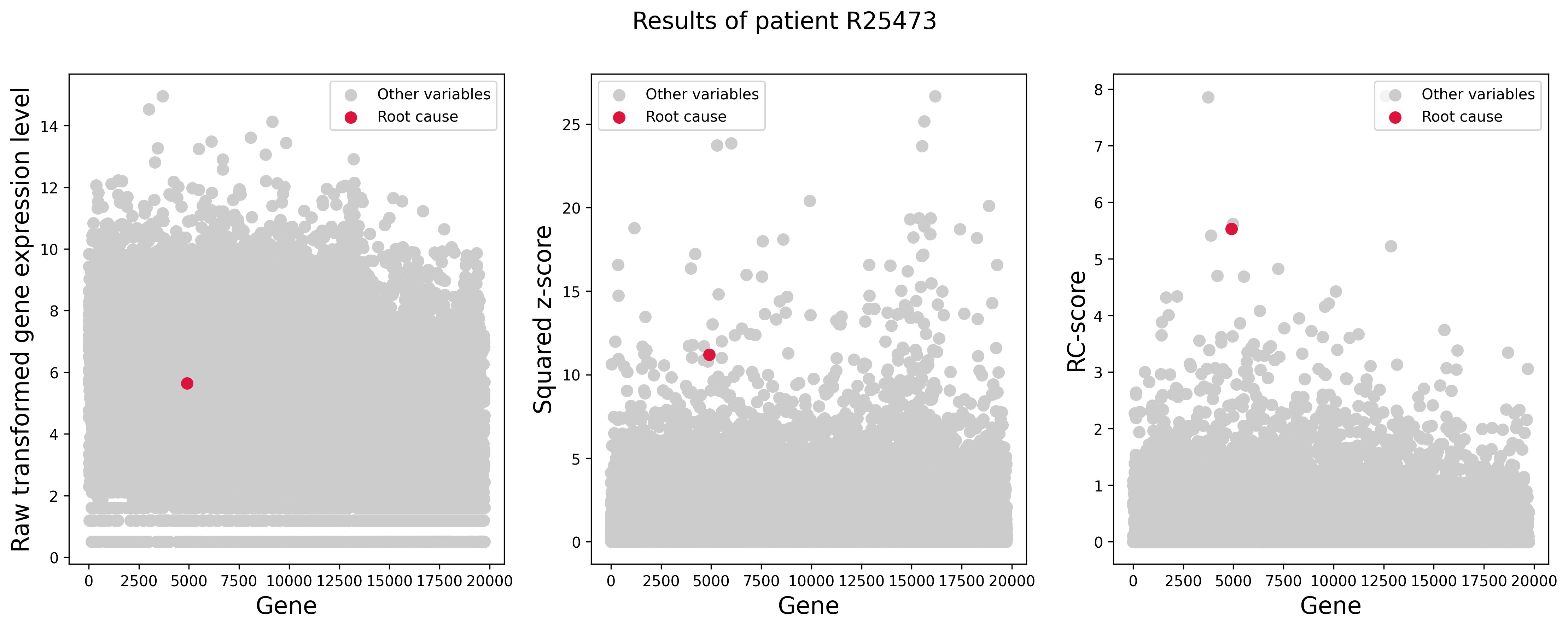}
	\end{minipage}
	\vspace{0.05cm}
	\begin{minipage}[b]{0.9\textwidth}
		\centering
		\includegraphics[scale=0.26]{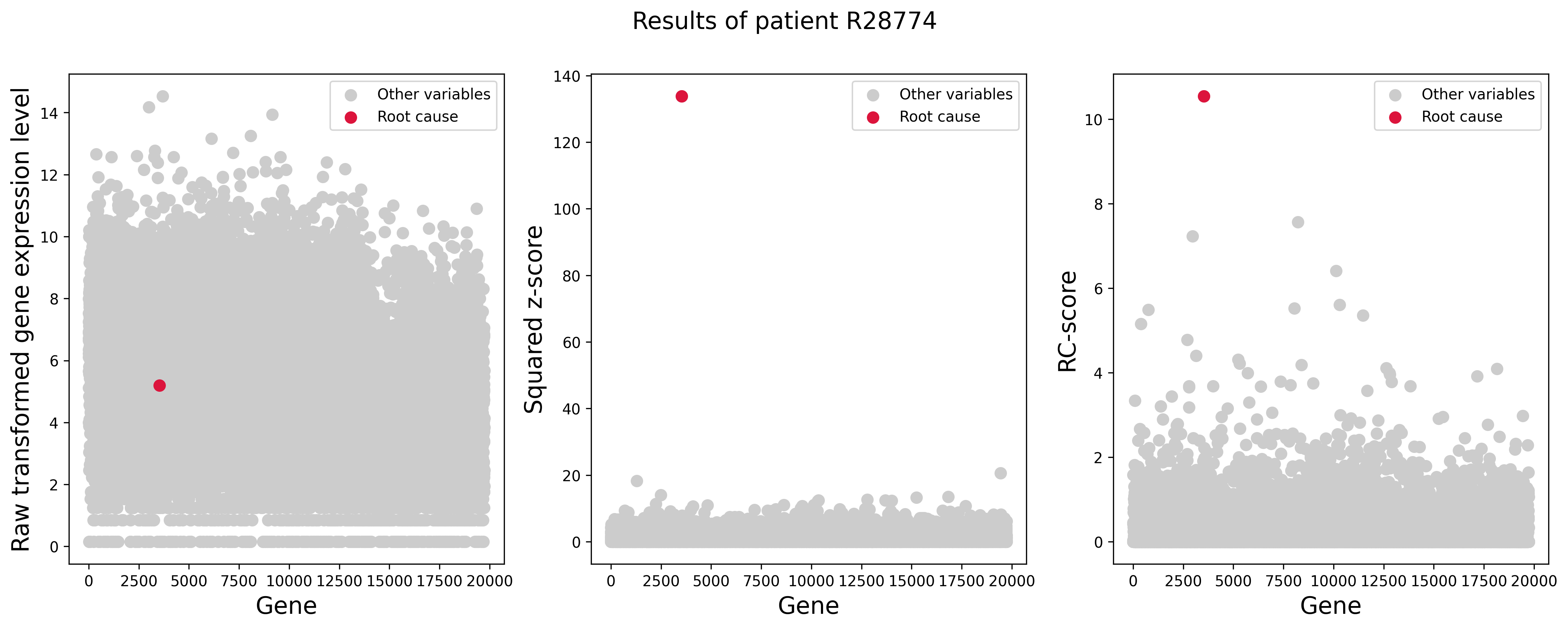}
	\end{minipage}
	\vspace{0.05cm}
	\begin{minipage}[b]{0.9\textwidth}
		\centering
		\includegraphics[scale=0.26]{plots/Supp/ThreeScoresPlots/ThreeScoresR96820.png}
	\end{minipage}
	\vspace{-0.5cm}
	\caption{The raw transformed gene expression levels, squared z-scores, and RC-scores of genes for patients $R80346$, $R20754$, $R25473$, $R28774$, and $R96820$.}
	\label{Fig:AppReal3}
\end{figure}

\begin{figure}[ht]
	\centering
	\begin{minipage}[b]{0.9\textwidth}
		\centering
		\includegraphics[scale=0.26]{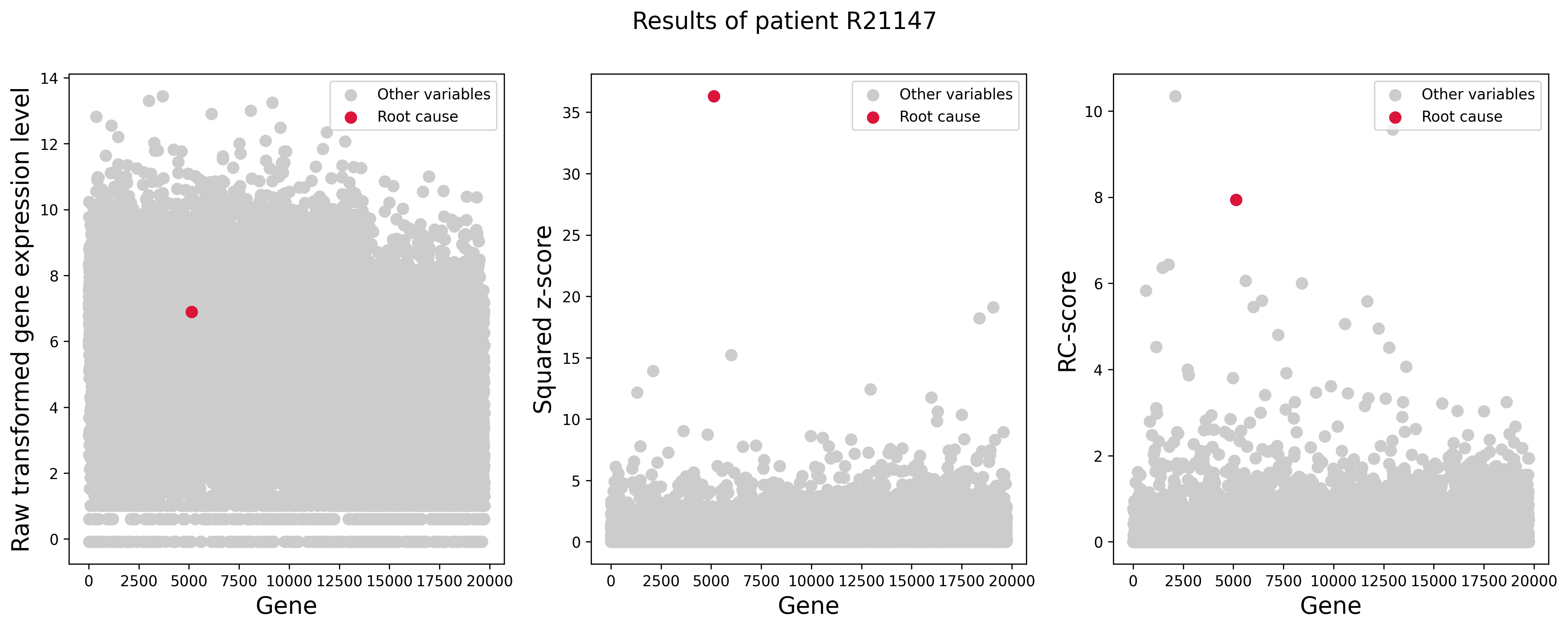}
	\end{minipage}
	\vspace{0.05cm}
	\begin{minipage}[b]{0.9\textwidth}
		\centering
		\includegraphics[scale=0.26]{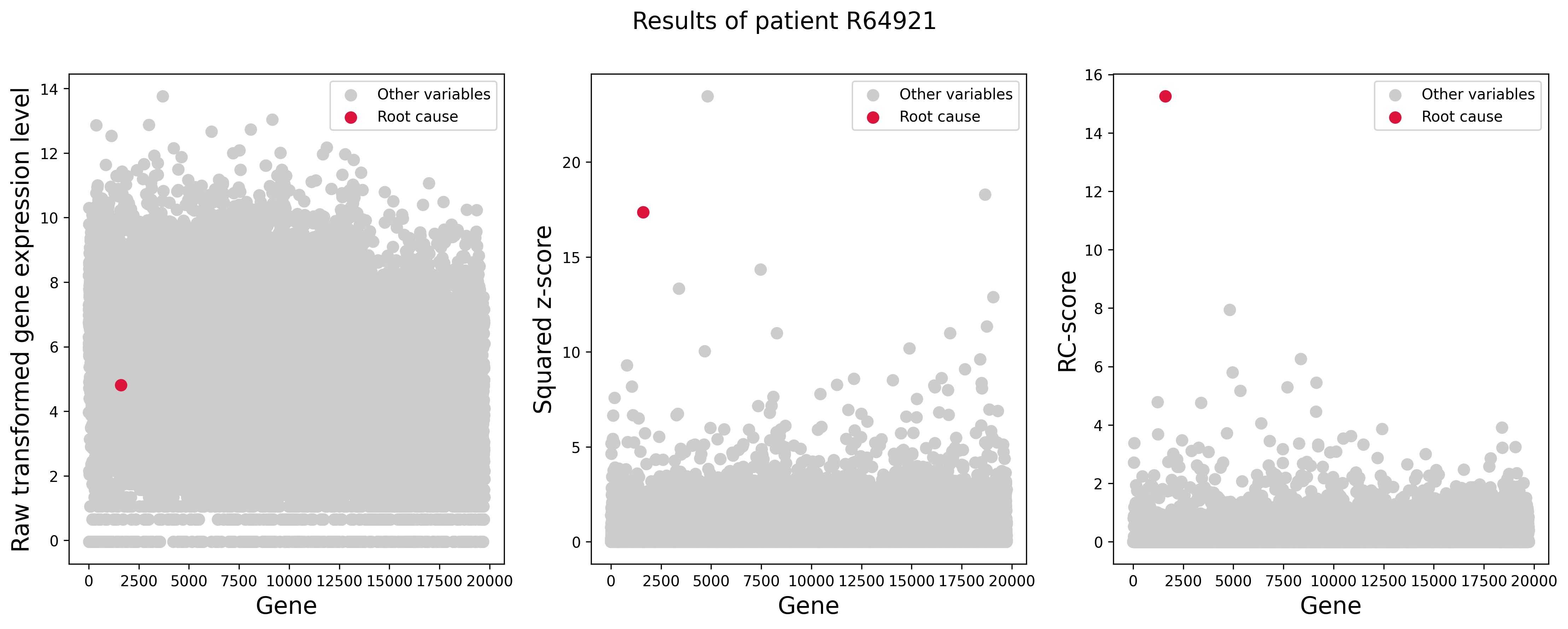}
	\end{minipage}
	\vspace{0.05cm}
	\begin{minipage}[b]{0.9\textwidth}
		\centering
		\includegraphics[scale=0.26]{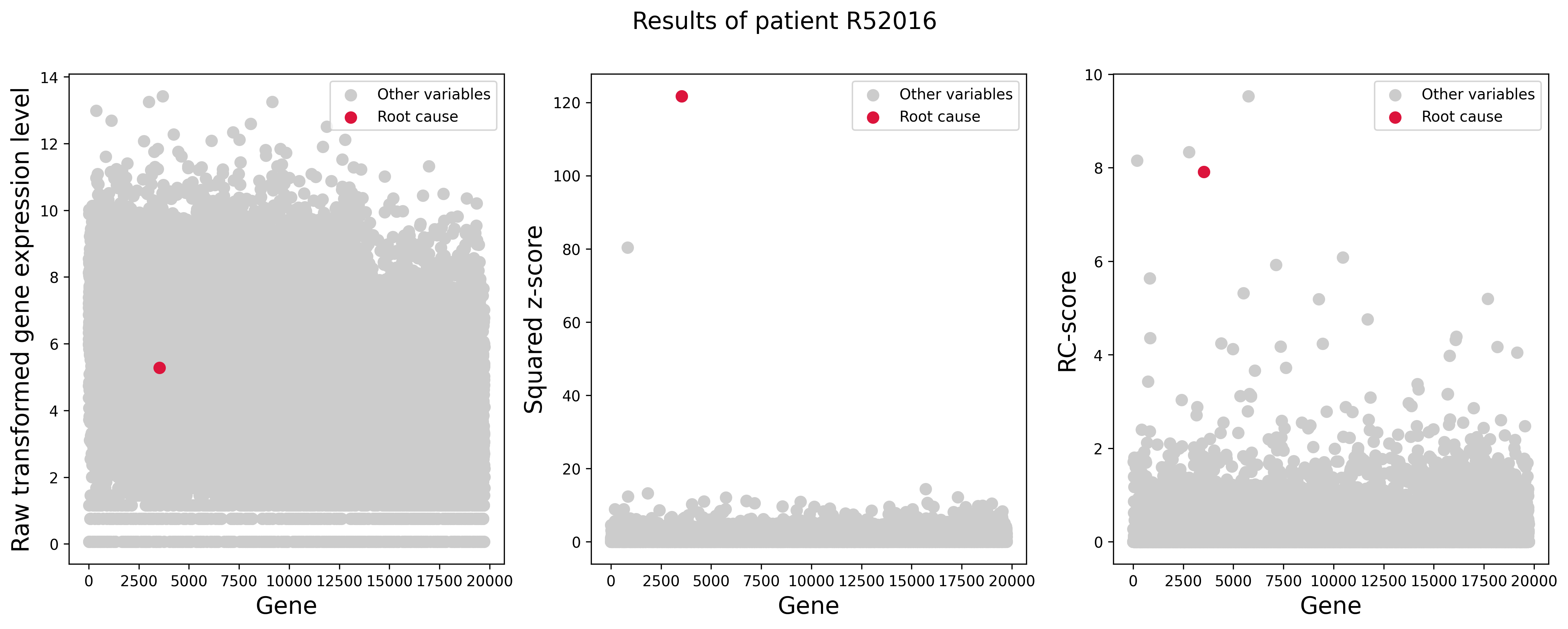}
	\end{minipage}
	\vspace{0.05cm}
	\begin{minipage}[b]{0.9\textwidth}
		\centering
		\includegraphics[scale=0.26]{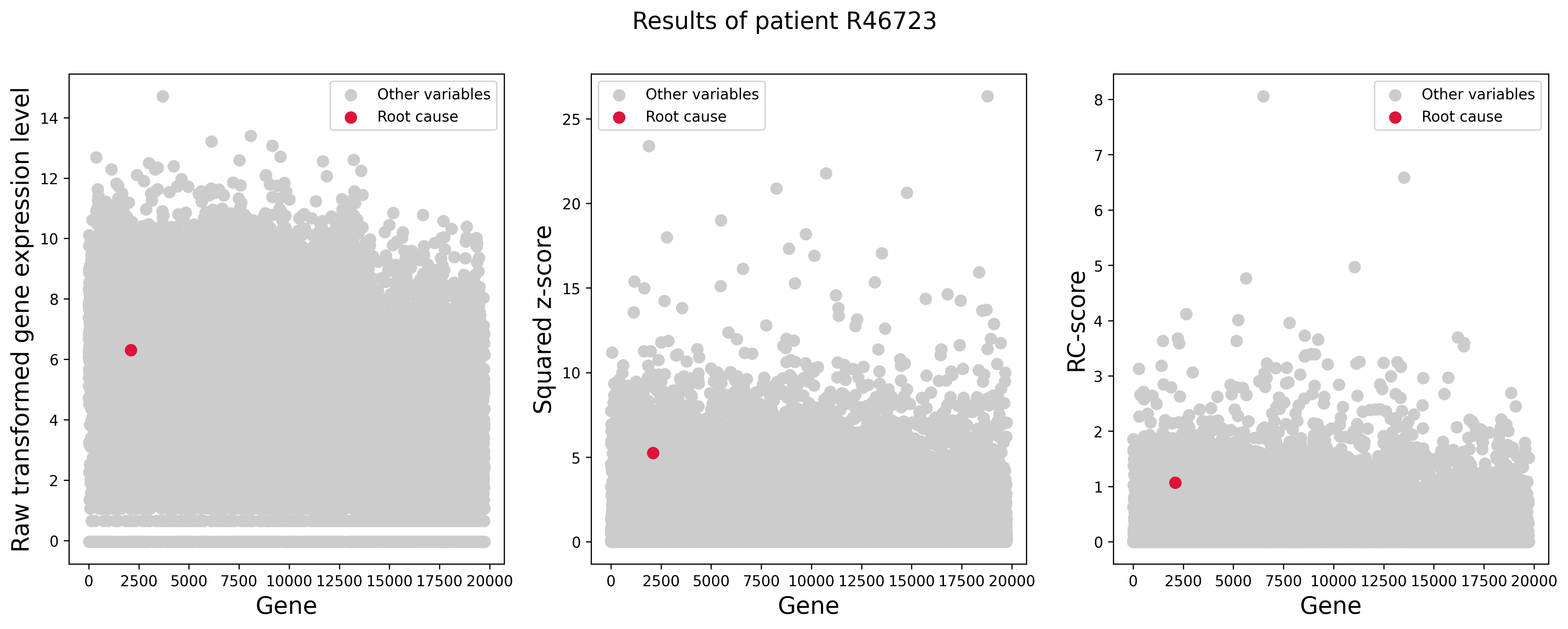}
	\end{minipage}
	\vspace{0.05cm}
	\begin{minipage}[b]{0.9\textwidth}
		\centering
		\includegraphics[scale=0.26]{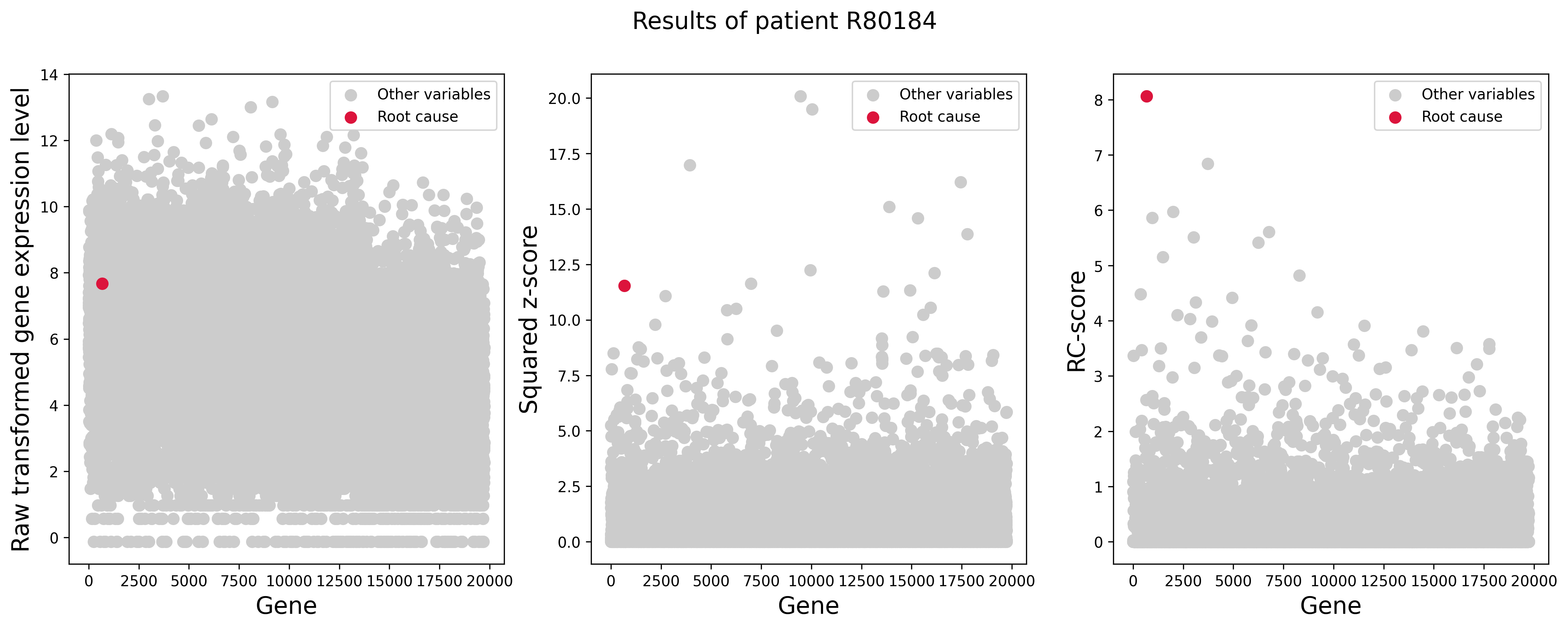}
	\end{minipage}
	\vspace{-0.5cm}
	\caption{The raw transformed gene expression levels, squared z-scores, and RC-scores of genes for patients $R21147$, $R64921$, $R52016$, $R46723$, and $R80184$.}
	\label{Fig:AppReal4}
\end{figure}

\begin{figure}[ht]
	\centering
	\begin{minipage}[b]{0.9\textwidth}
		\centering
		\includegraphics[scale=0.26]{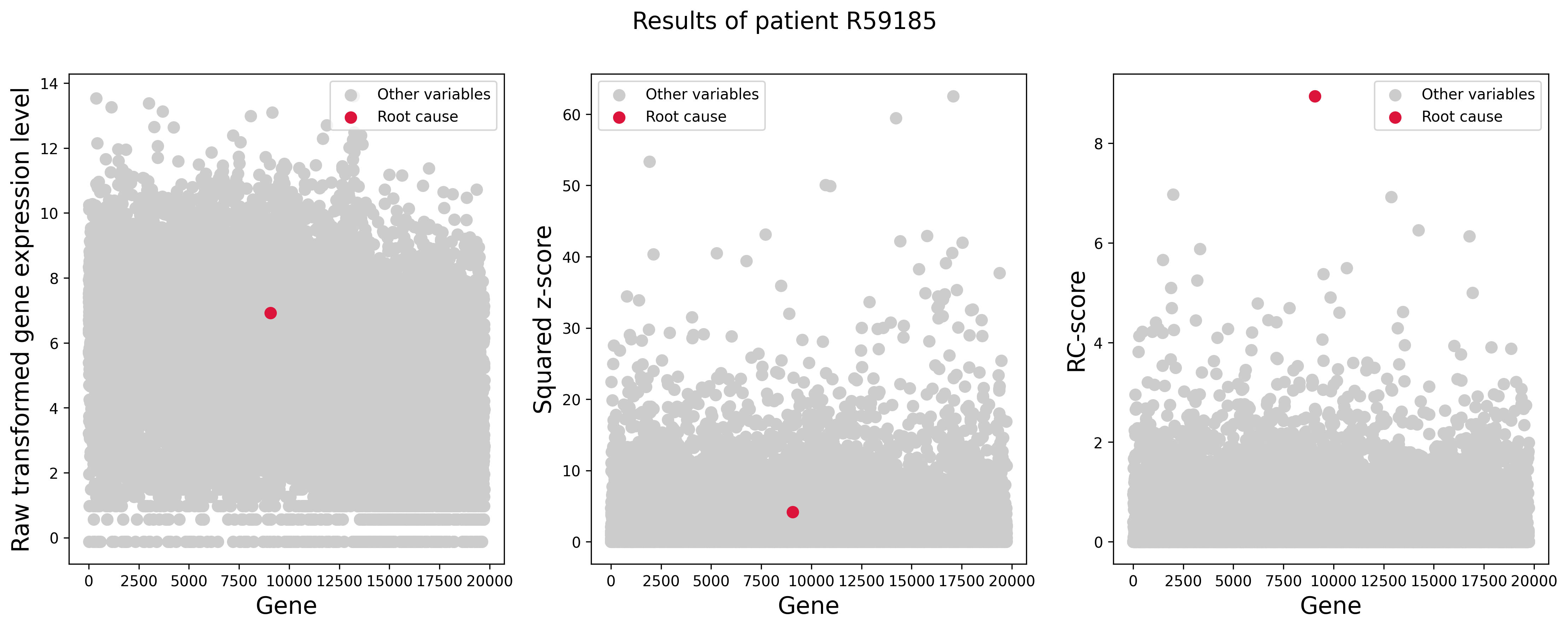}
	\end{minipage}
	\vspace{0.05cm}
	\begin{minipage}[b]{0.9\textwidth}
		\centering
		\includegraphics[scale=0.26]{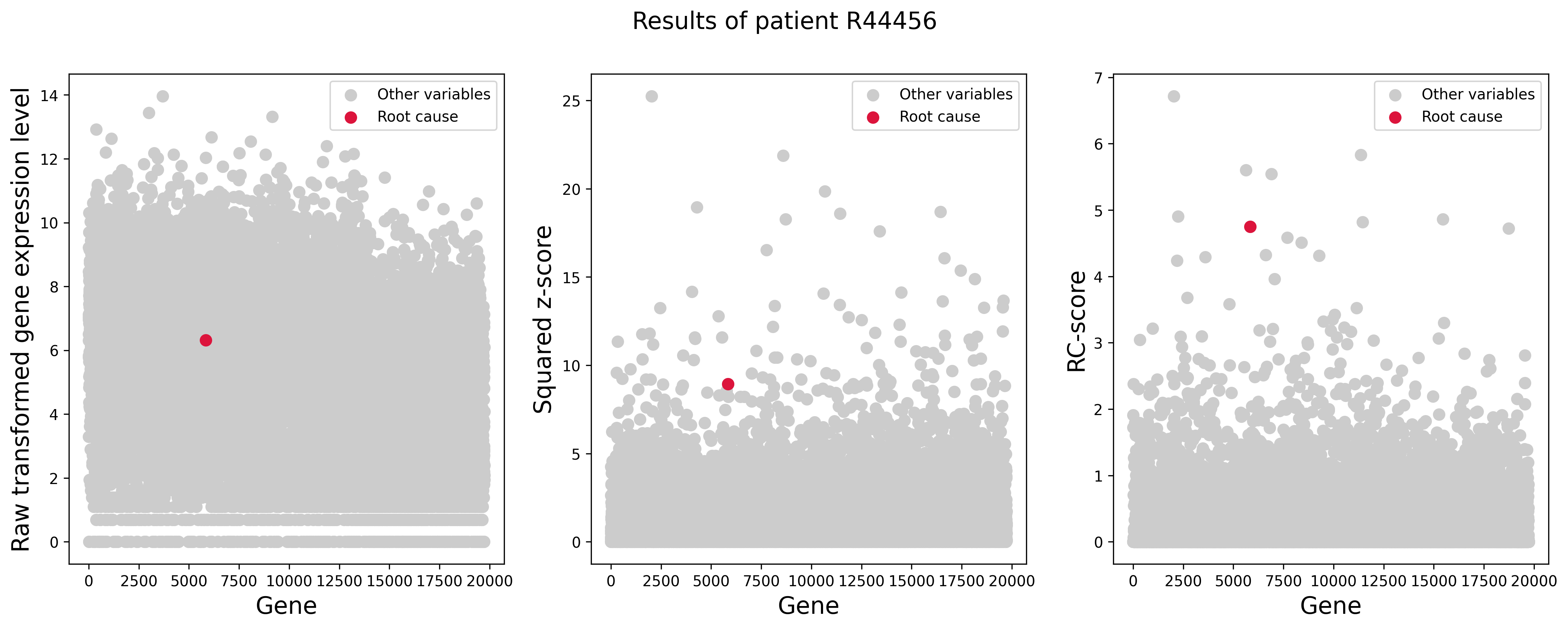}
	\end{minipage}
	\vspace{0.05cm}
	\begin{minipage}[b]{0.9\textwidth}
		\centering
		\includegraphics[scale=0.26]{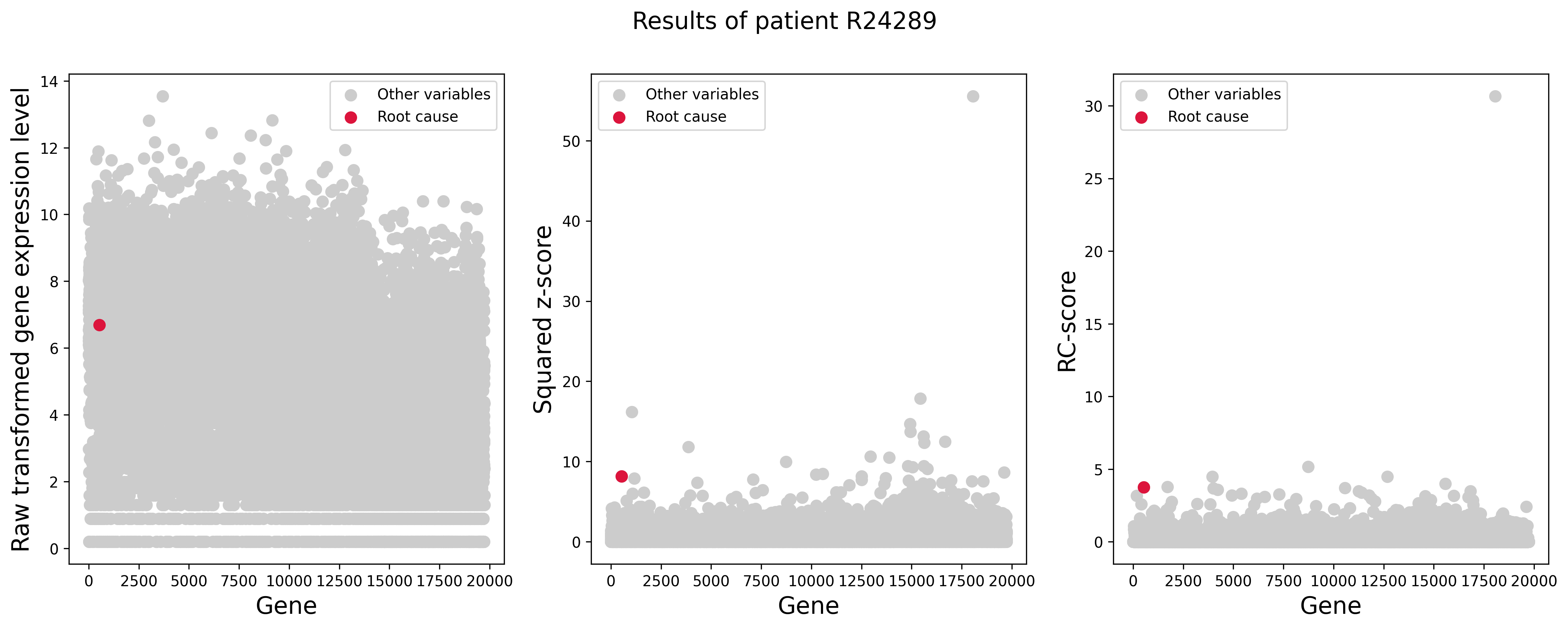}
	\end{minipage}
	\vspace{0.05cm}
	\begin{minipage}[b]{0.9\textwidth}
		\centering
		\includegraphics[scale=0.26]{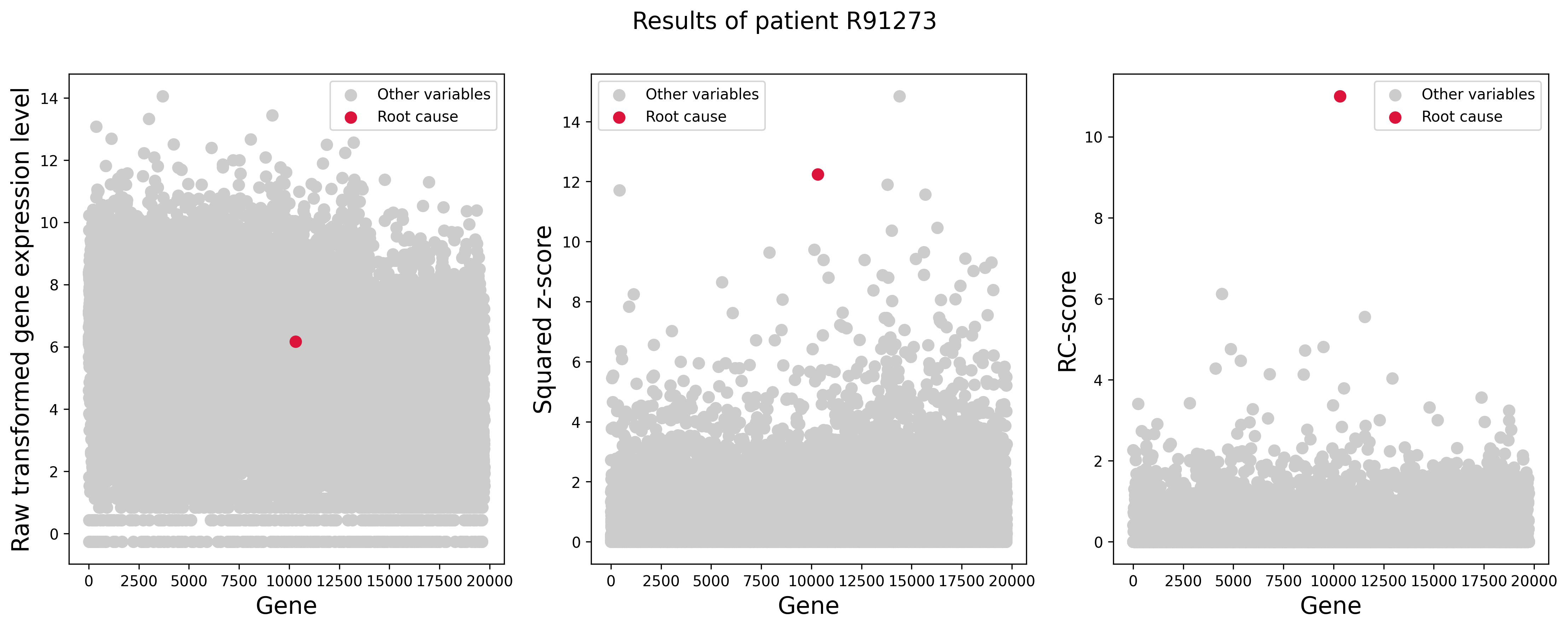}
	\end{minipage}
	\vspace{0.05cm}
	\begin{minipage}[b]{0.9\textwidth}
		\centering
		\includegraphics[scale=0.26]{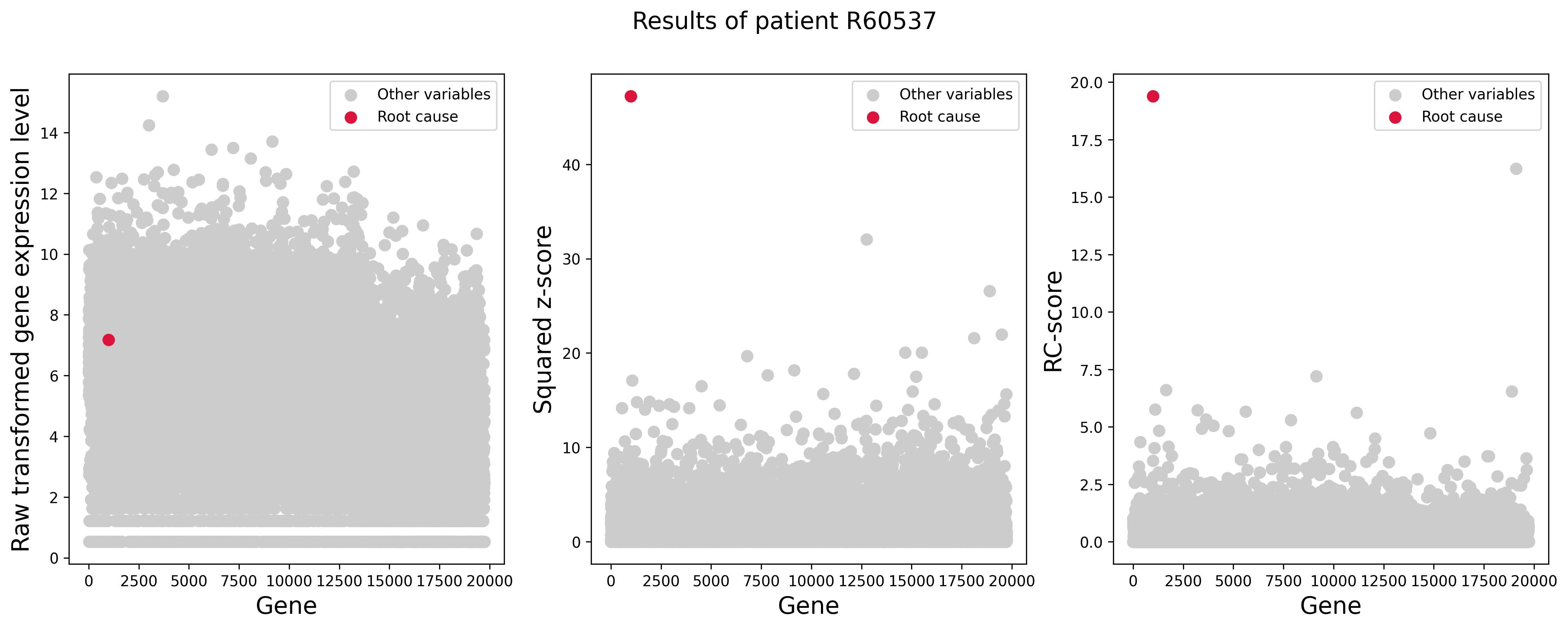}
	\end{minipage}
	\vspace{-0.5cm}
	\caption{The raw transformed gene expression levels, squared z-scores, and RC-scores of genes for patients $R59185$, $R44456$, $R24289$, $R91273$, and $R60537$.}
	\label{Fig:AppReal5}
\end{figure}

\begin{figure}[ht]
	\centering
	\begin{minipage}[b]{0.9\textwidth}
		\centering
		\includegraphics[scale=0.26]{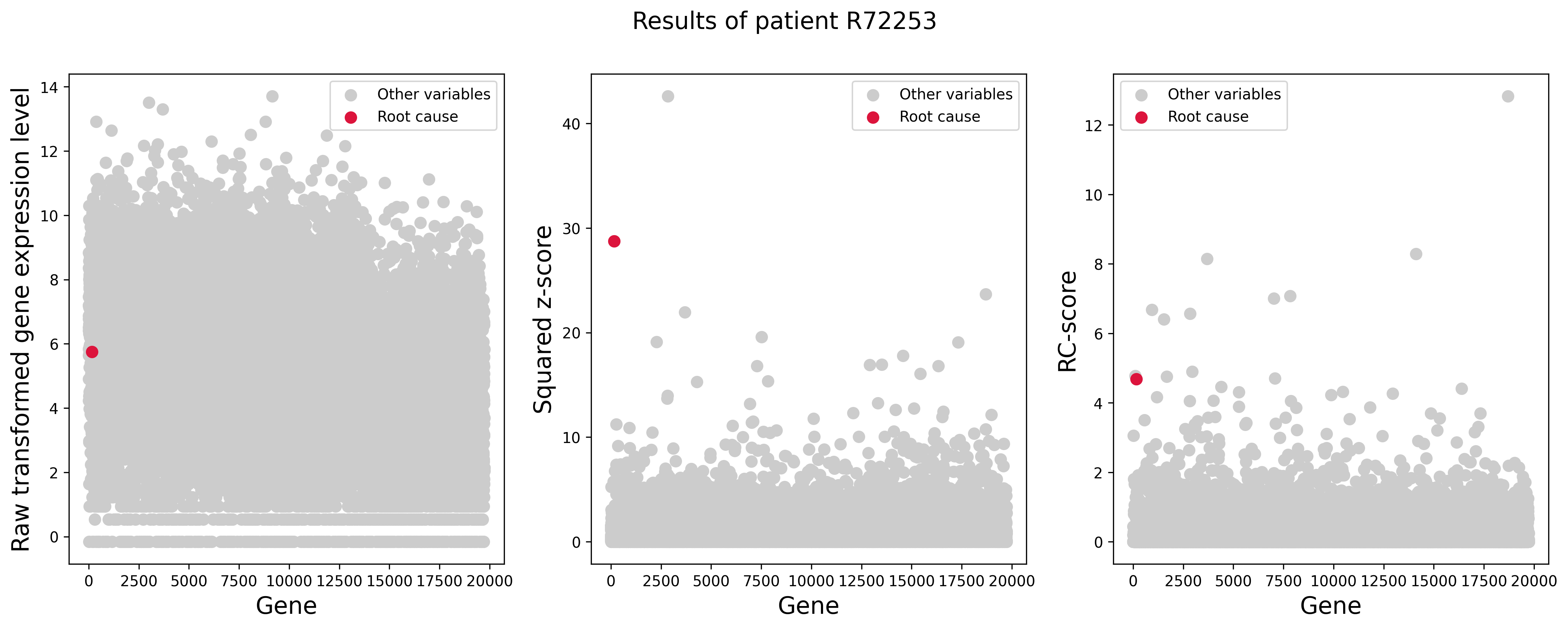}
	\end{minipage}
	\vspace{0.05cm}
	\begin{minipage}[b]{0.9\textwidth}
		\centering
		\includegraphics[scale=0.26]{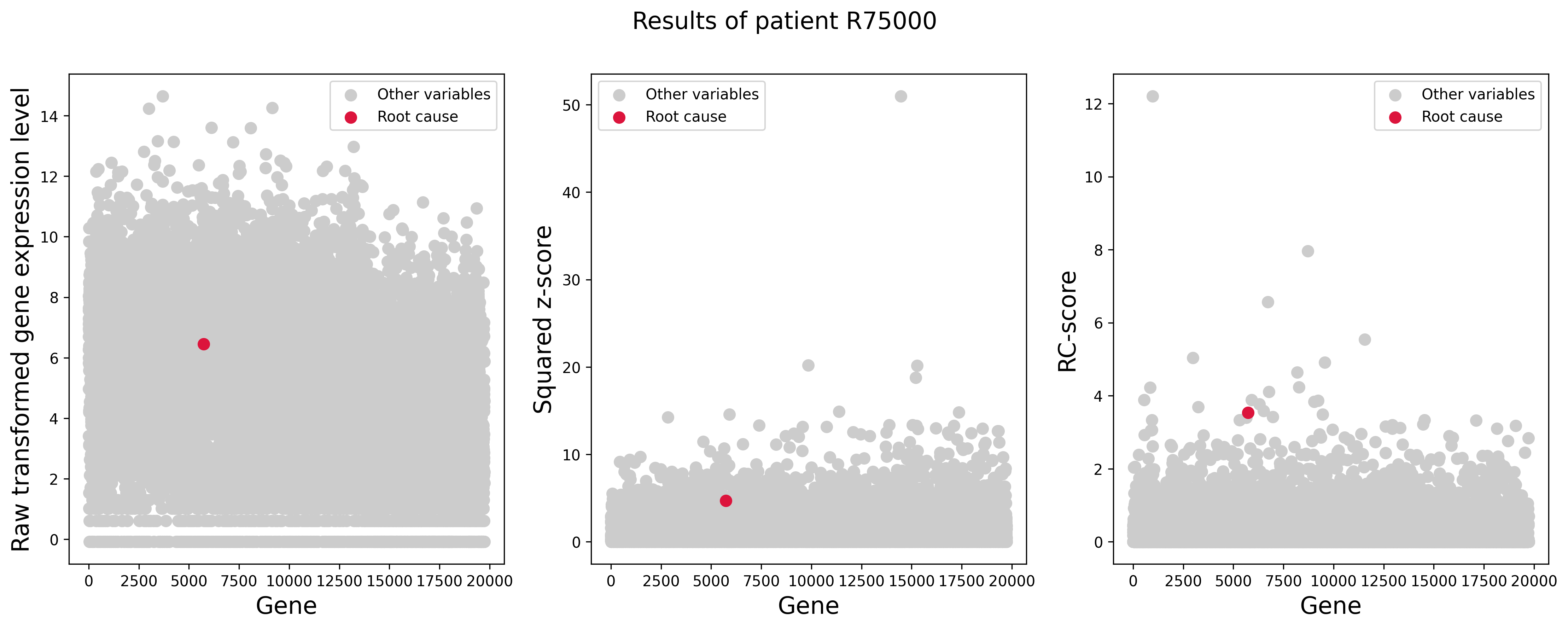}
	\end{minipage}
	\vspace{0.05cm}
	\begin{minipage}[b]{0.9\textwidth}
		\centering
		\includegraphics[scale=0.26]{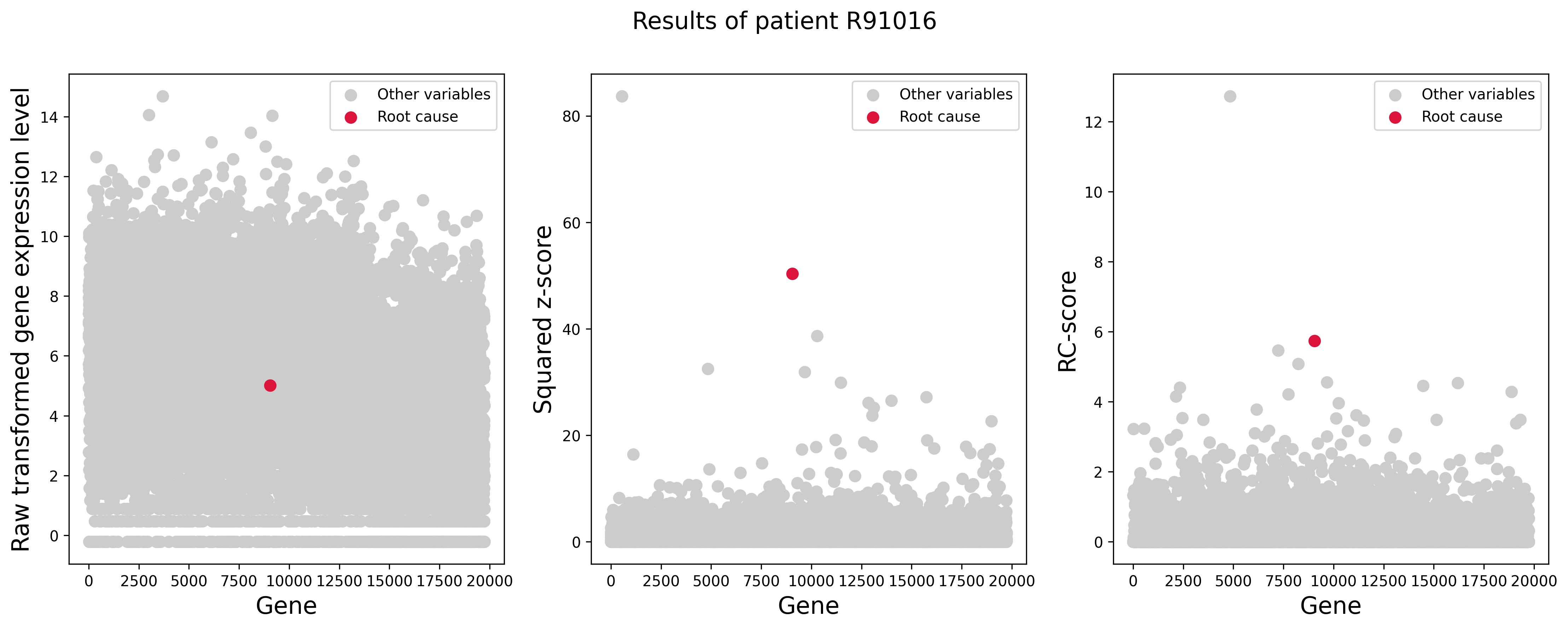}
	\end{minipage}
	\vspace{0.05cm}
	\begin{minipage}[b]{0.9\textwidth}
		\centering
		\includegraphics[scale=0.26]{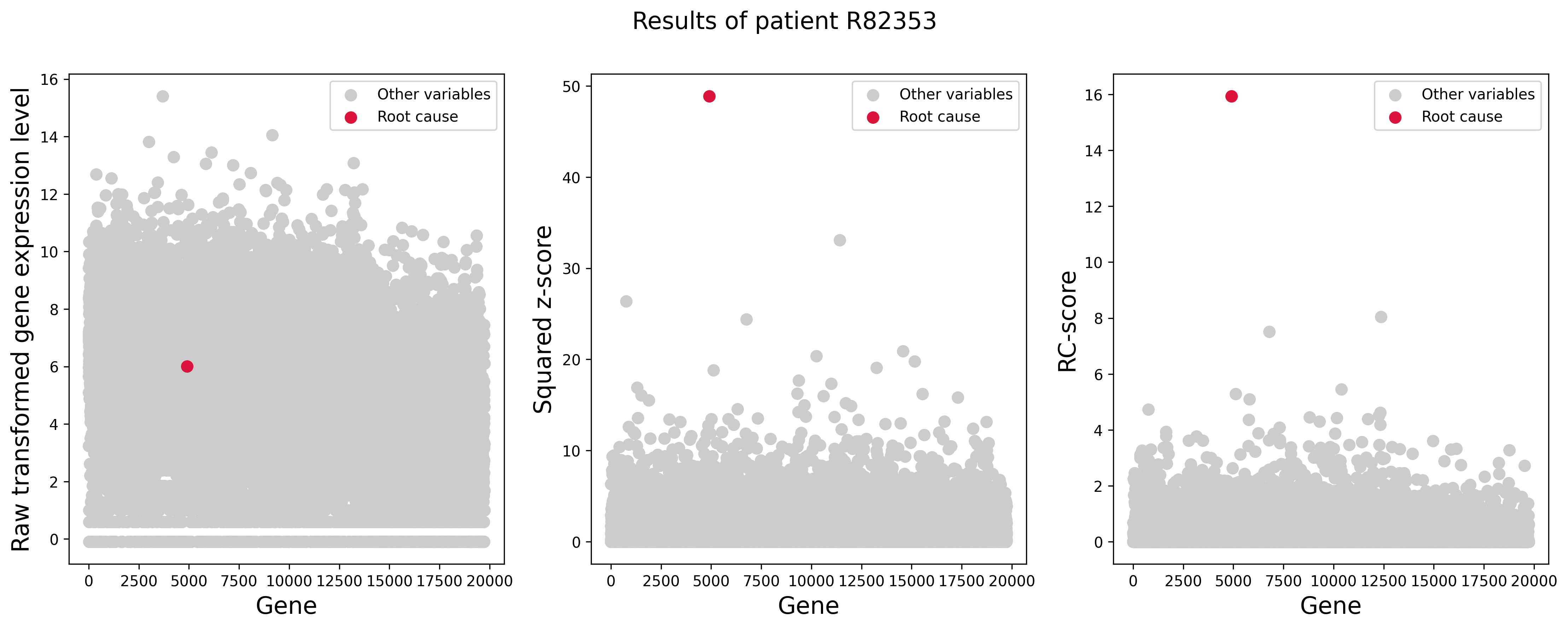}
	\end{minipage}
	\vspace{0.05cm}
	\begin{minipage}[b]{0.9\textwidth}
		\centering
		\includegraphics[scale=0.26]{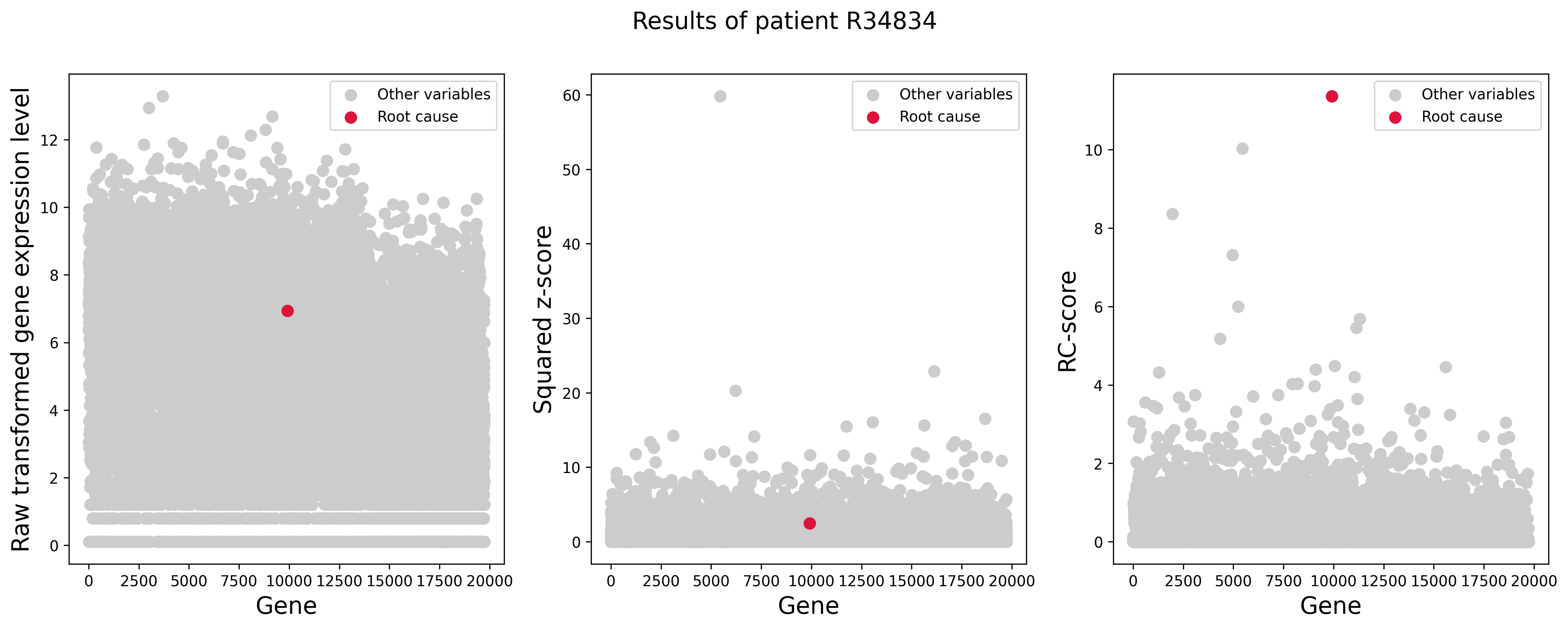}
	\end{minipage}
	\vspace{-0.5cm}
	\caption{The raw transformed gene expression levels, squared z-scores, and RC-scores of genes for patients $R72253$, $R75000$, $R91016$, $R82353$, and $R34834$.}
	\label{Fig:AppReal6}
\end{figure}

\begin{figure}[ht]
	\centering
	\begin{minipage}[b]{0.9\textwidth}
		\centering
		\includegraphics[scale=0.26]{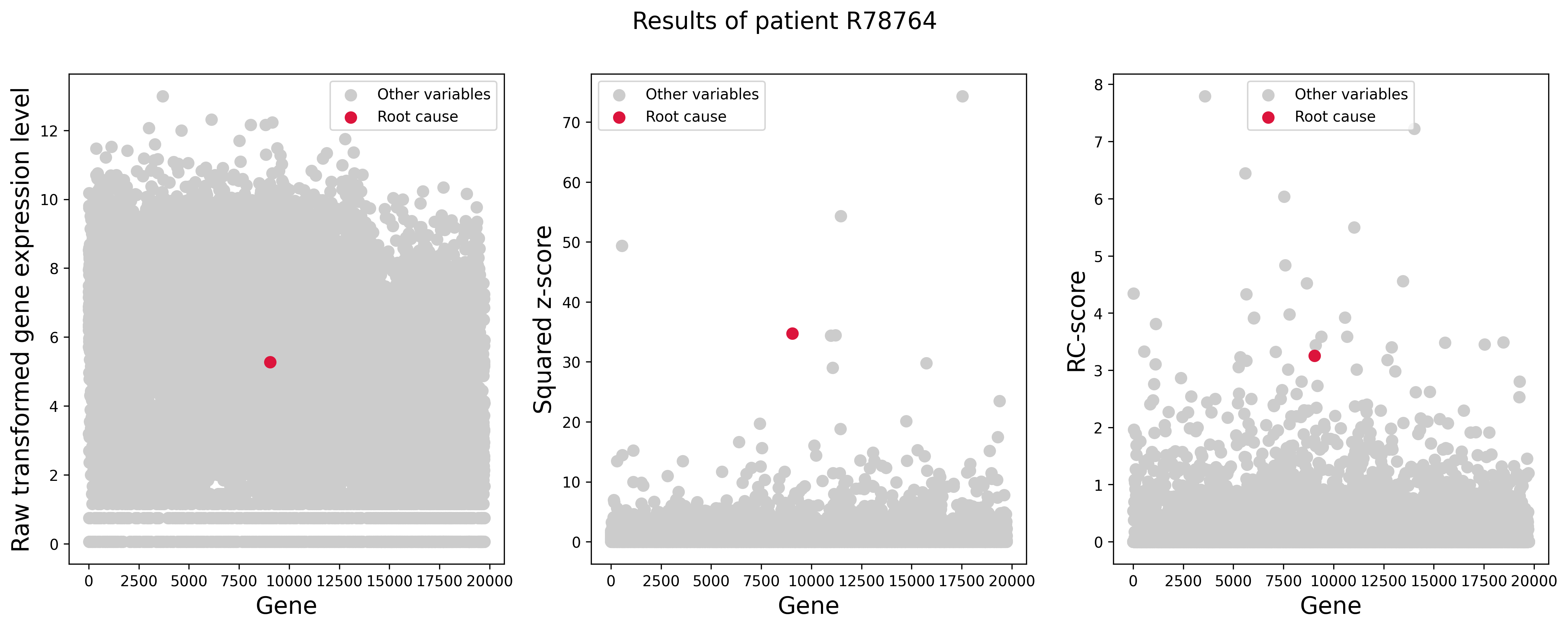}
	\end{minipage}
	\vspace{0.05cm}
	\begin{minipage}[b]{0.9\textwidth}
		\centering
		\includegraphics[scale=0.26]{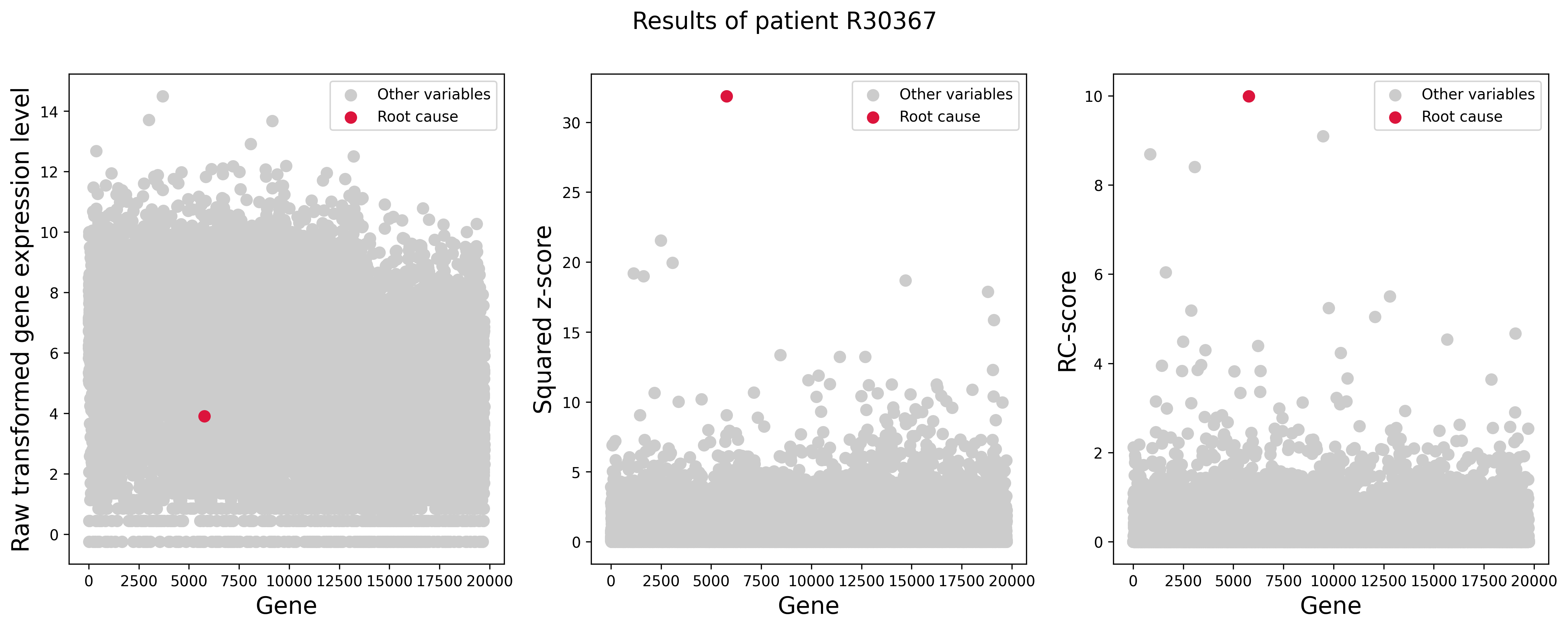}
	\end{minipage}
	\vspace{0.05cm}
	\begin{minipage}[b]{0.9\textwidth}
		\centering
		\includegraphics[scale=0.26]{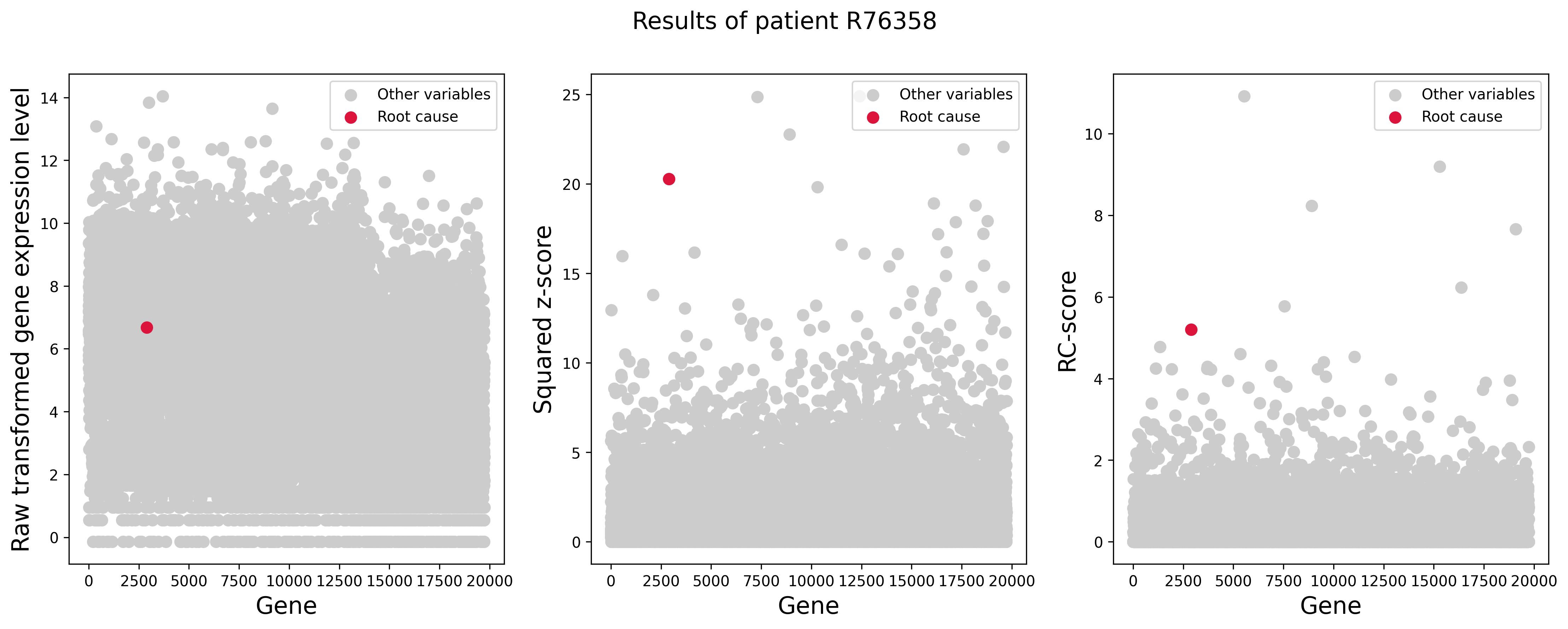}
	\end{minipage}
	\vspace{0.05cm}
	\begin{minipage}[b]{0.9\textwidth}
		\centering
		\includegraphics[scale=0.26]{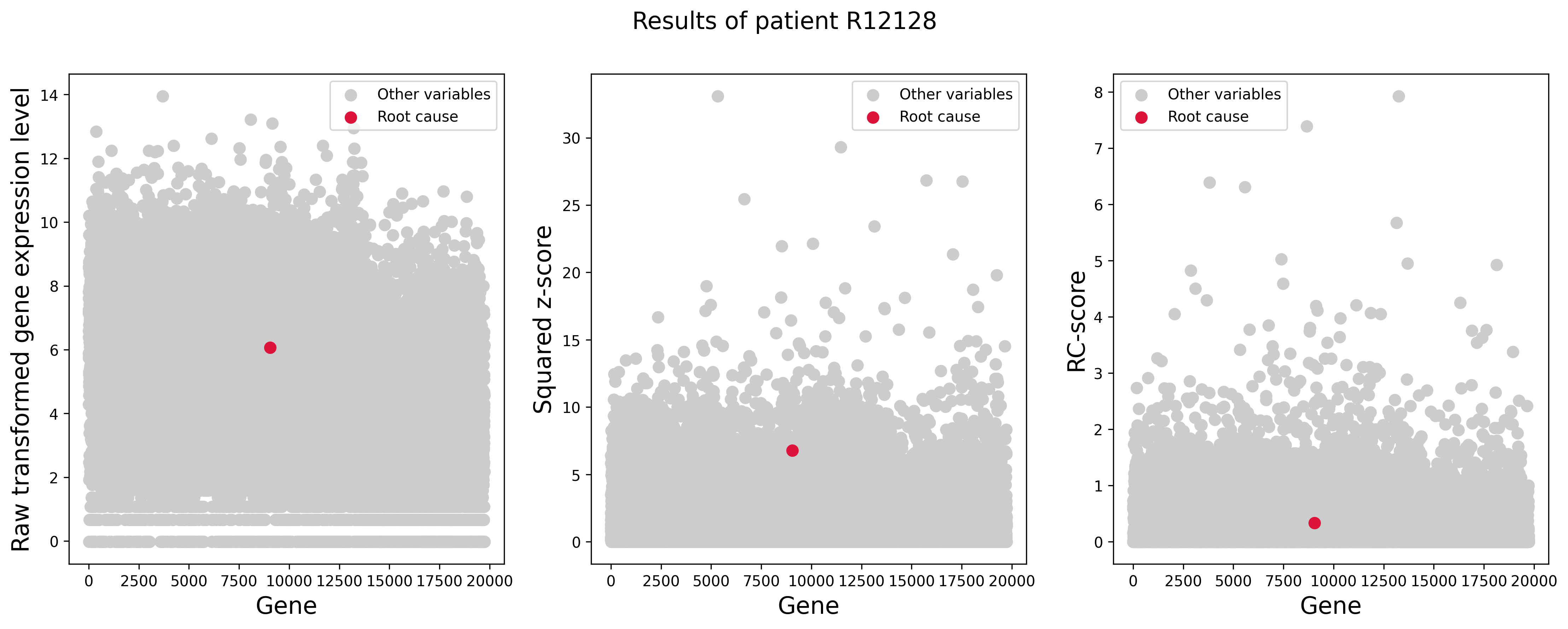}
	\end{minipage}
	\vspace{0.05cm}
	\begin{minipage}[b]{0.9\textwidth}
		\centering
		\includegraphics[scale=0.26]{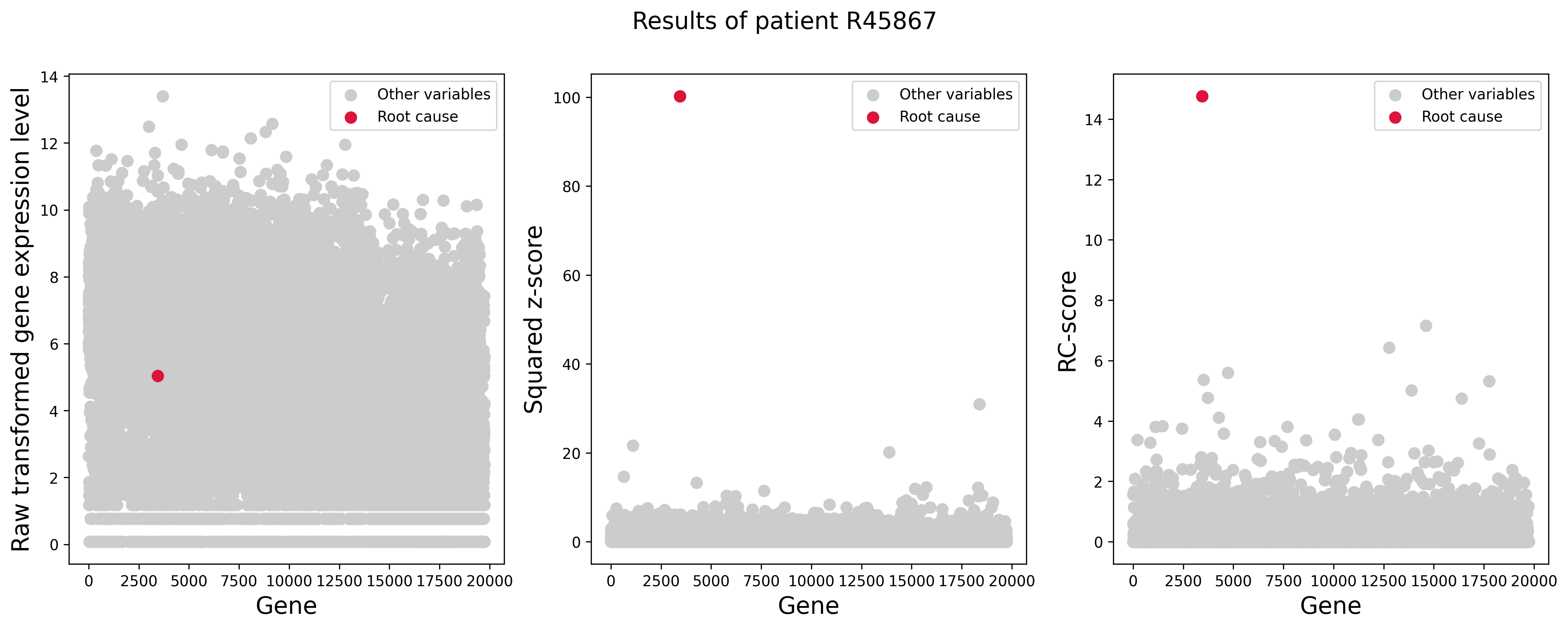}
	\end{minipage}
	\vspace{-0.5cm}
	\caption{The raw transformed gene expression levels, squared z-scores, and RC-scores of genes for patients $R78764$, $R30367$, $R76358$, $R12128$, and $R45867$.}
	\label{Fig:AppReal7}
\end{figure}

\begin{figure}[ht]
	\centering
	\begin{minipage}[b]{0.9\textwidth}
		\centering
		\includegraphics[scale=0.26]{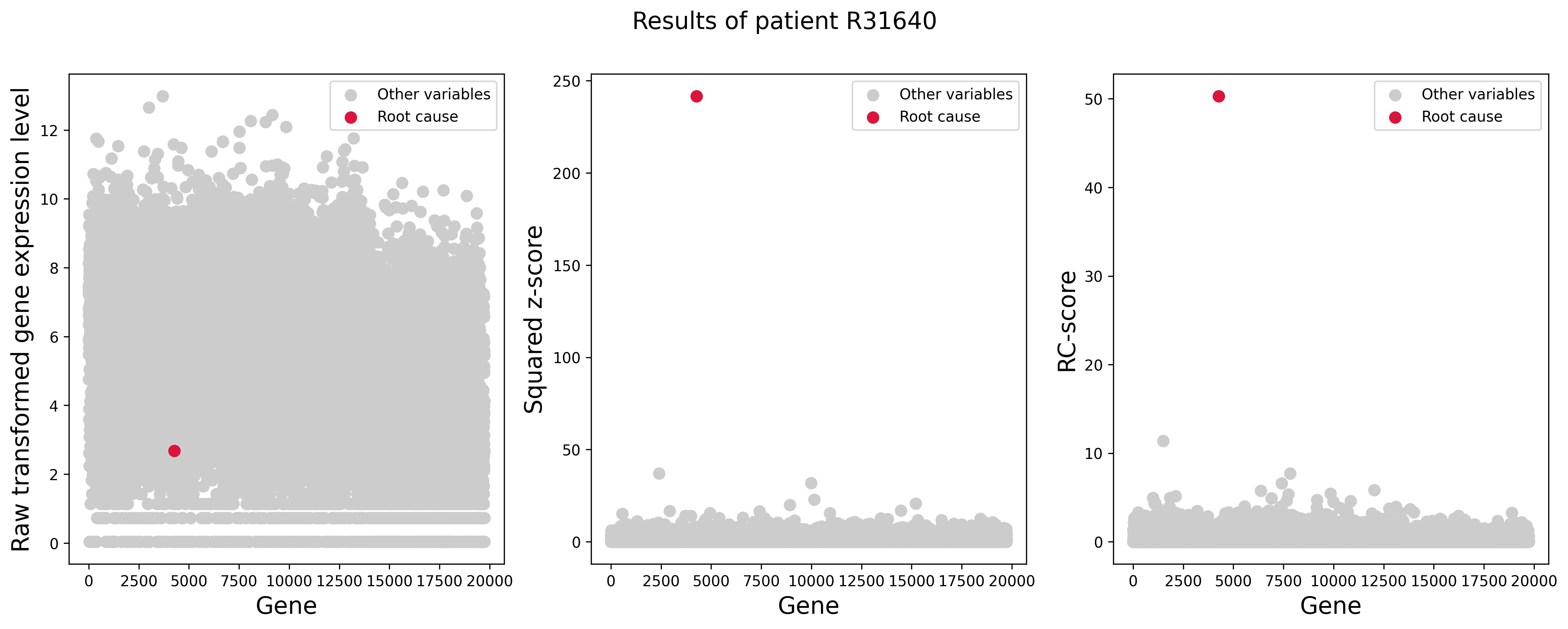}
	\end{minipage}
	\vspace{0.05cm}
	\begin{minipage}[b]{0.9\textwidth}
		\centering
		\includegraphics[scale=0.26]{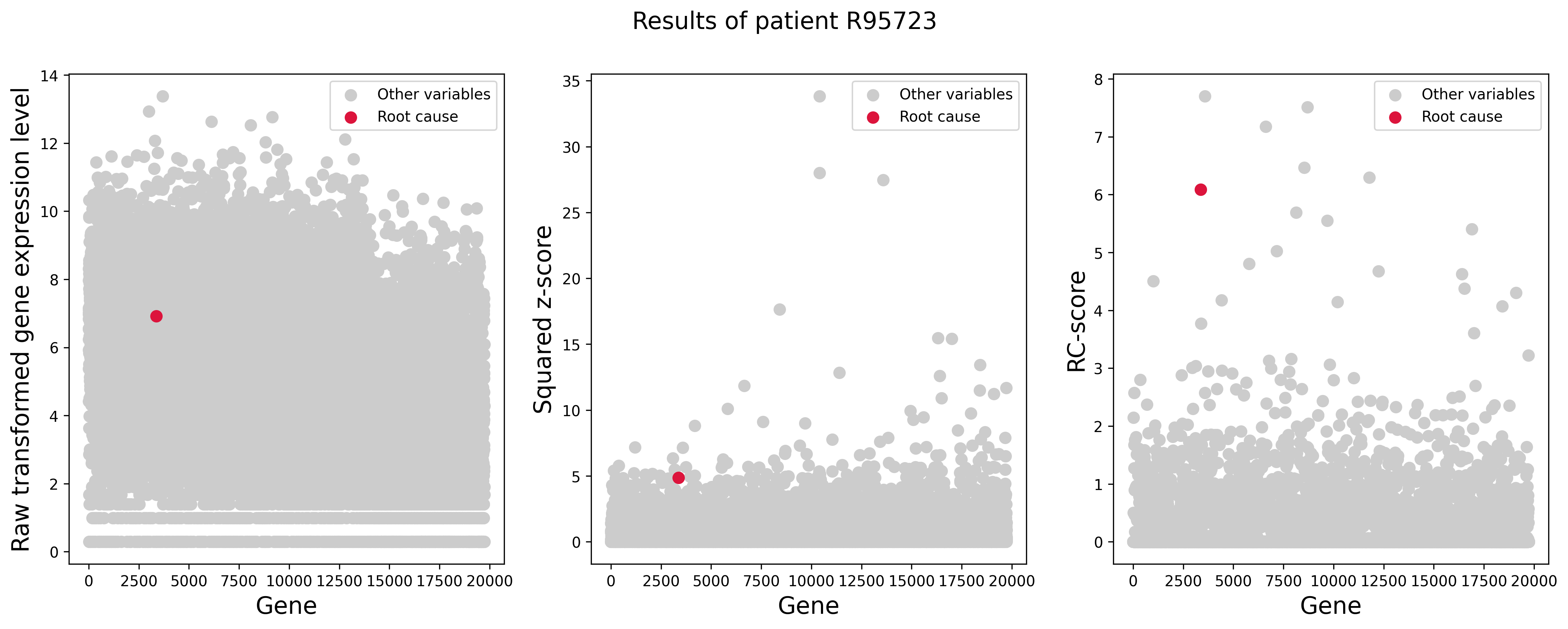}
	\end{minipage}
	\vspace{0.05cm}
	\begin{minipage}[b]{0.9\textwidth}
		\centering
		\includegraphics[scale=0.26]{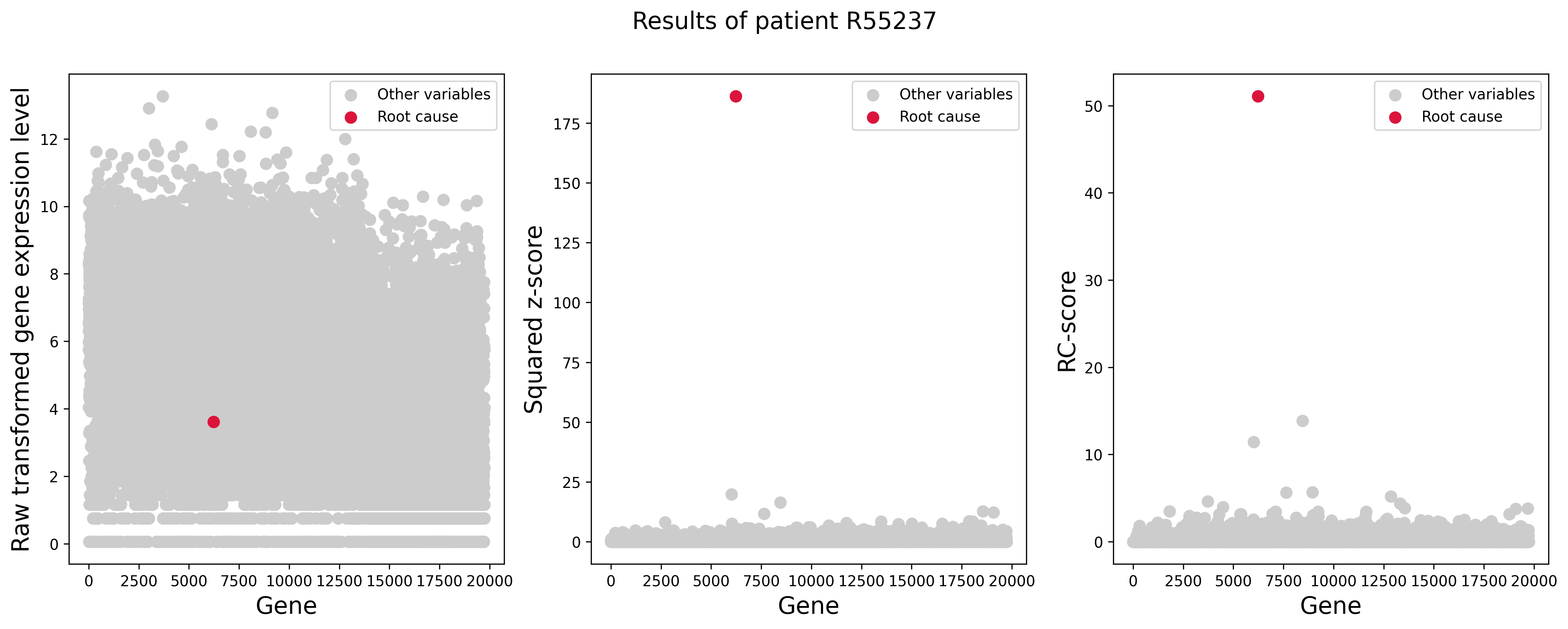}
	\end{minipage}
	\vspace{0.05cm}
	\begin{minipage}[b]{0.9\textwidth}
		\centering
		\includegraphics[scale=0.26]{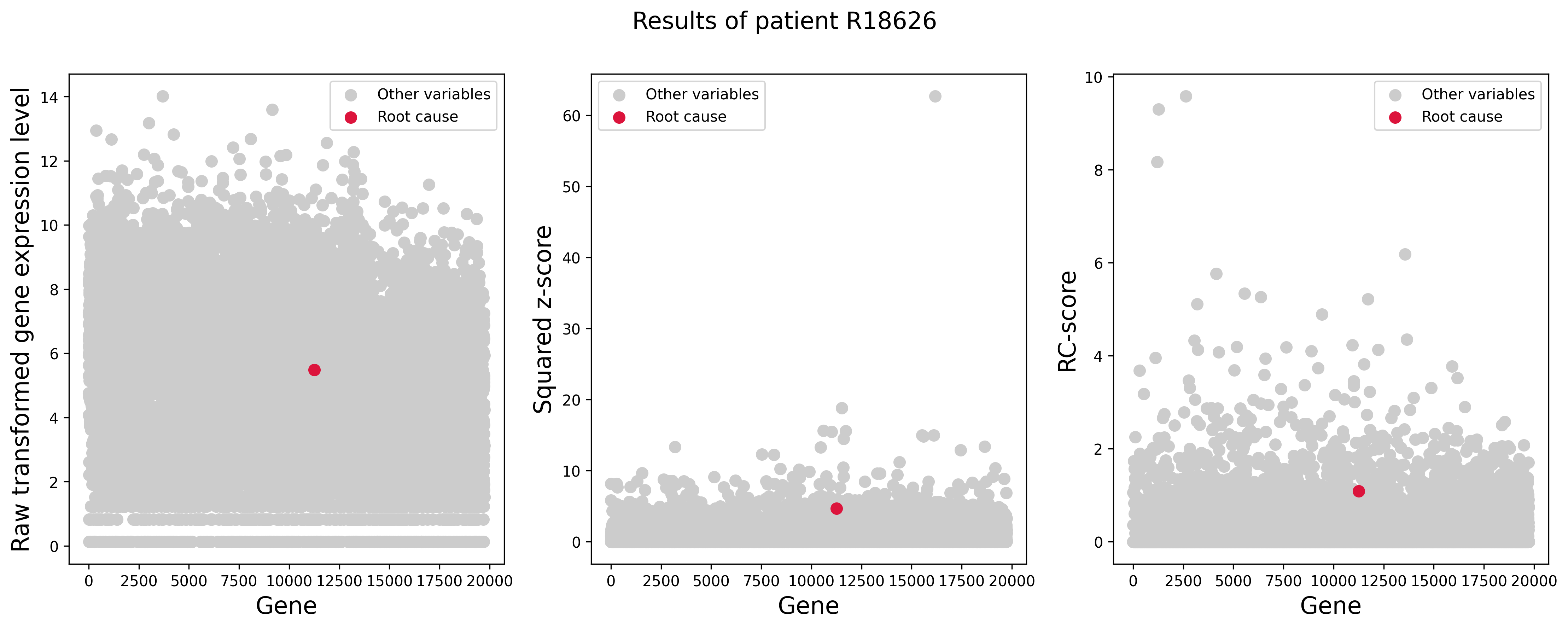}
	\end{minipage}
	\vspace{0.05cm}
	\begin{minipage}[b]{0.9\textwidth}
		\centering
		\includegraphics[scale=0.26]{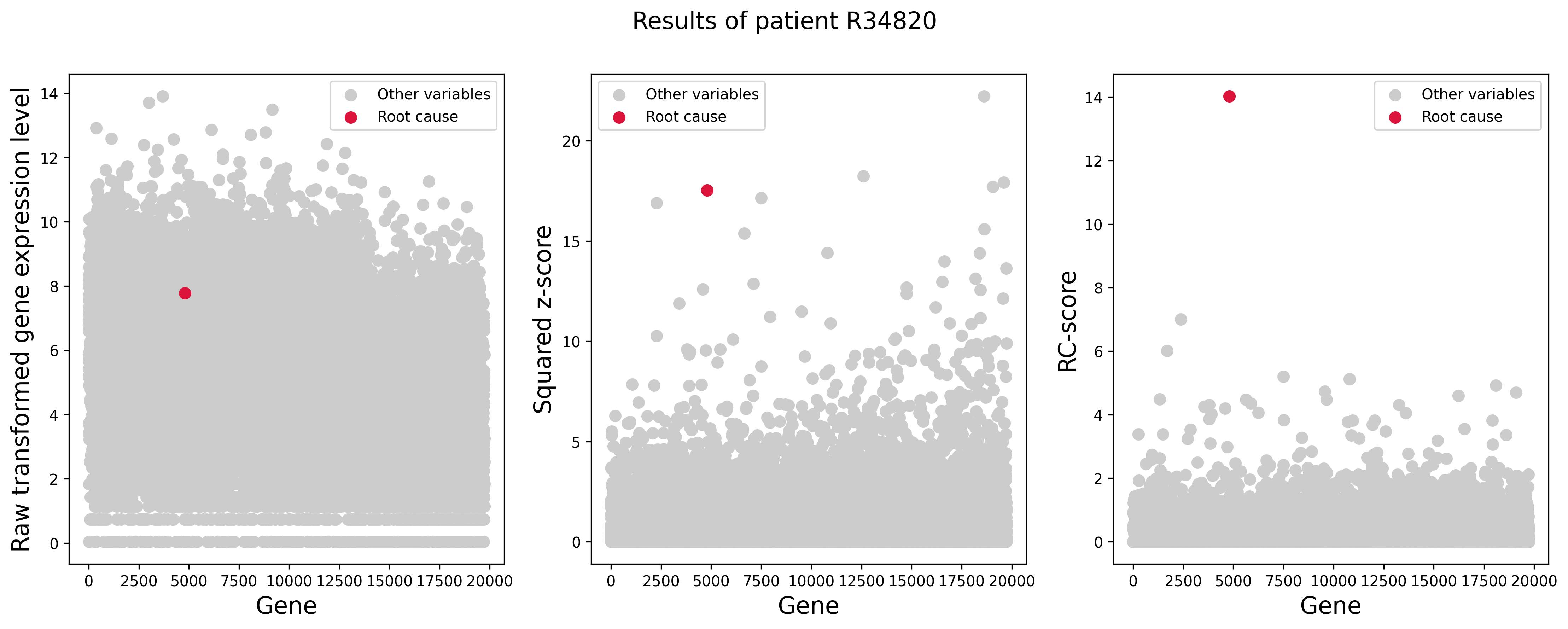}
	\end{minipage}
	\vspace{-0.5cm}
	\caption{The raw transformed gene expression levels, squared z-scores, and RC-scores of genes for patients $R31640$, $R95723$, $R55237$, $R18626$, and $R34820$.}
	\label{Fig:AppReal8}
\end{figure}

\begin{figure}[ht]
	\centering
	\begin{minipage}[b]{0.9\textwidth}
		\centering
		\includegraphics[scale=0.26]{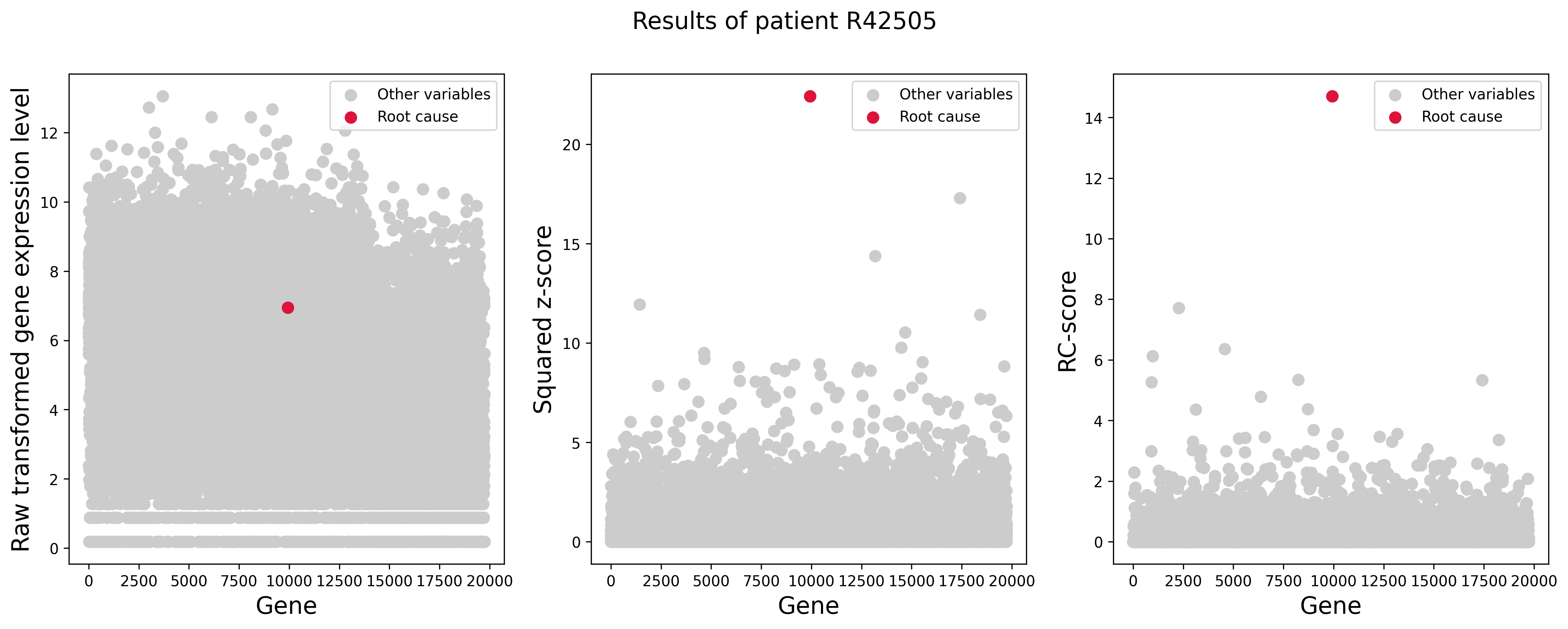}
	\end{minipage}
	\vspace{0.05cm}
	\begin{minipage}[b]{0.9\textwidth}
		\centering
		\includegraphics[scale=0.26]{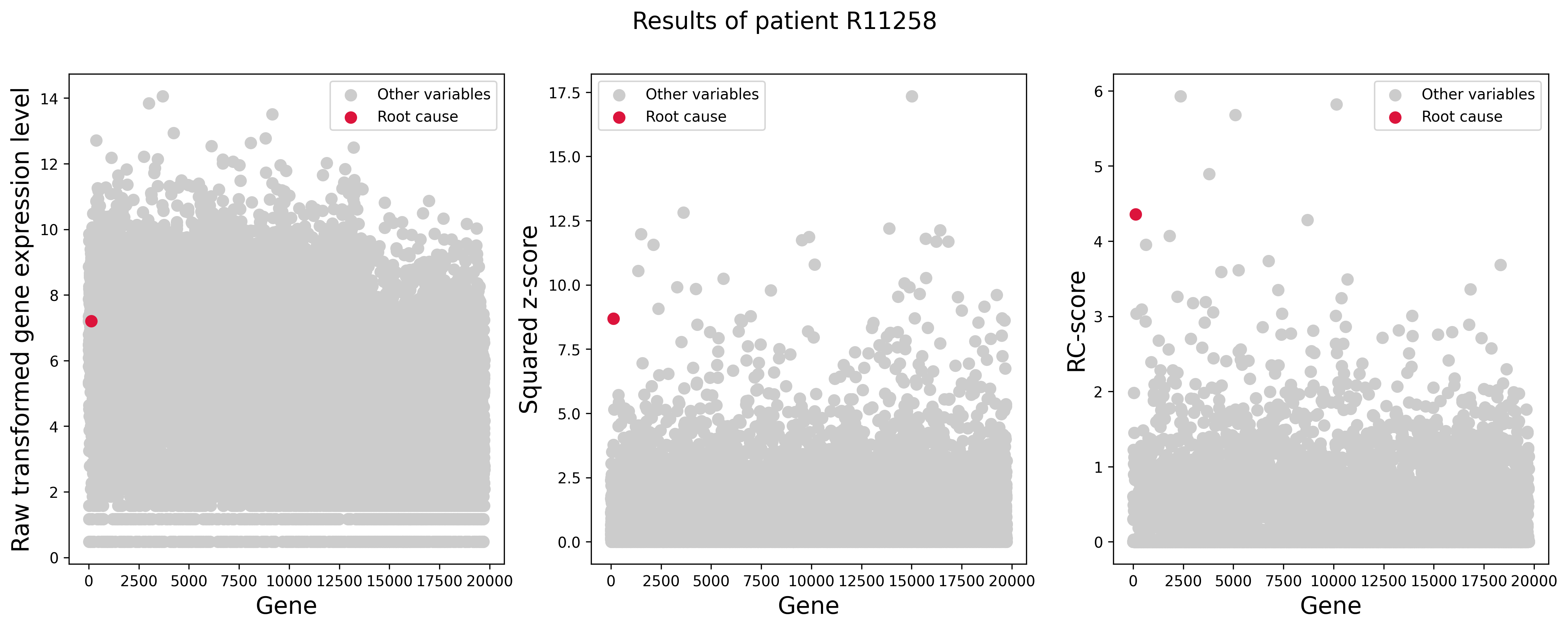}
	\end{minipage}
	\vspace{0.05cm}
	\begin{minipage}[b]{0.9\textwidth}
		\centering
		\includegraphics[scale=0.26]{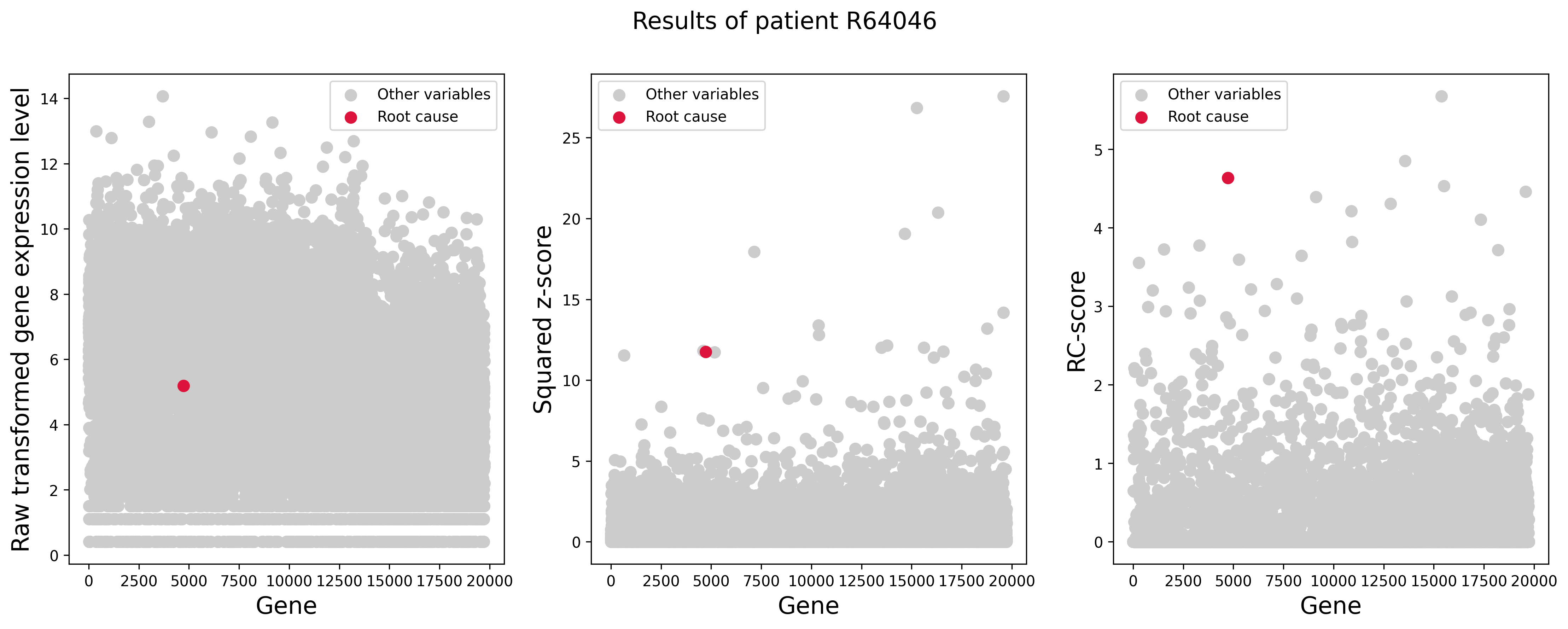}
	\end{minipage}
	\vspace{0.05cm}
	\begin{minipage}[b]{0.9\textwidth}
		\centering
		\includegraphics[scale=0.26]{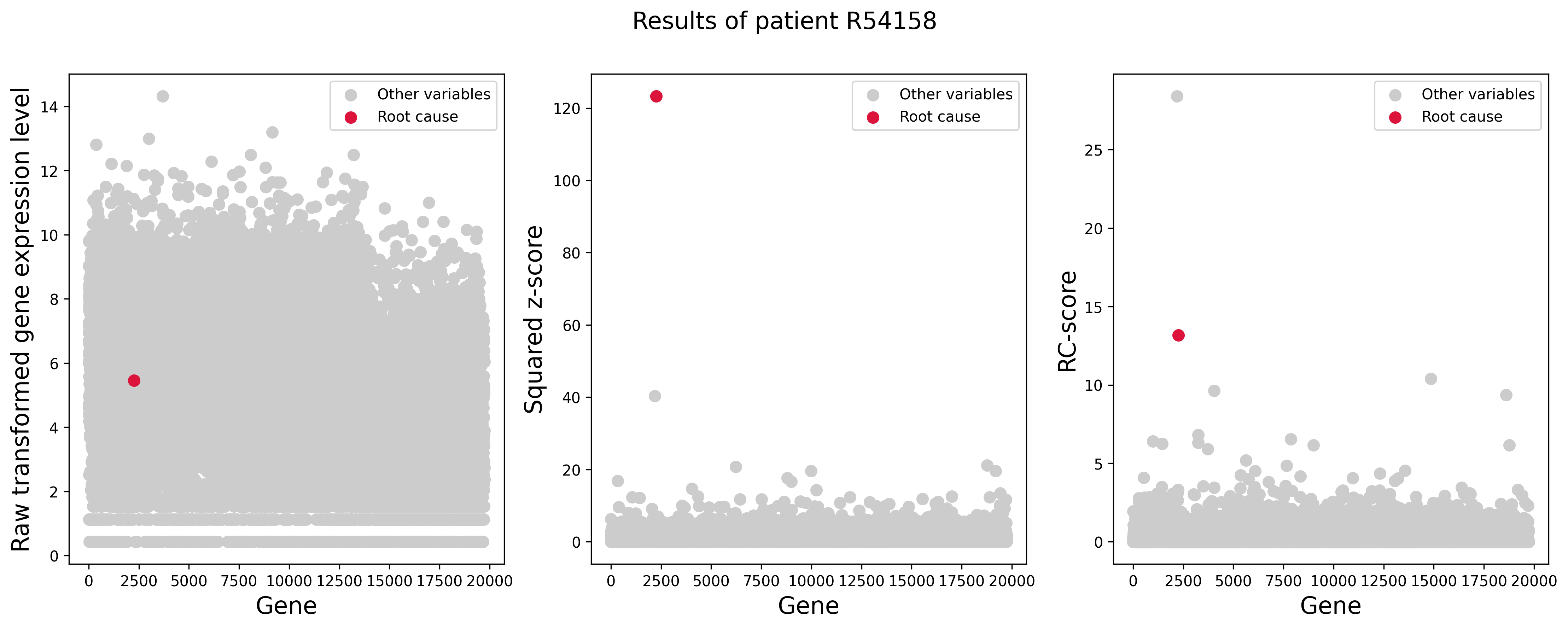}
	\end{minipage}
	\vspace{0.05cm}
	\begin{minipage}[b]{0.9\textwidth}
		\centering
		\includegraphics[scale=0.26]{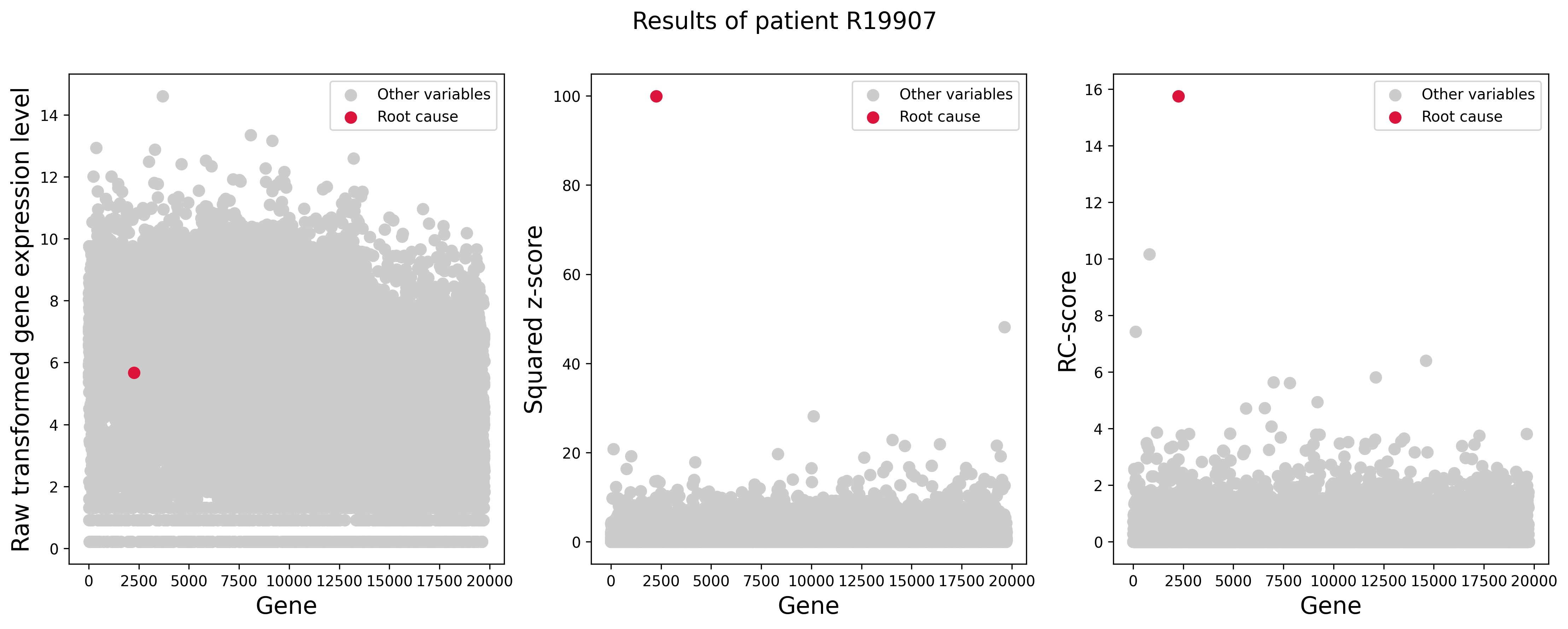}
	\end{minipage}
	\vspace{-0.5cm}
	\caption{The raw transformed gene expression levels, squared z-scores, and RC-scores of genes for patients $R42505$, $R11258$, $R64046$, $R54158$, and $R19907$.}
	\label{Fig:AppReal9}
\end{figure}

\begin{figure}[ht]
	\centering
	\begin{minipage}[b]{0.9\textwidth}
		\centering
		\includegraphics[scale=0.26]{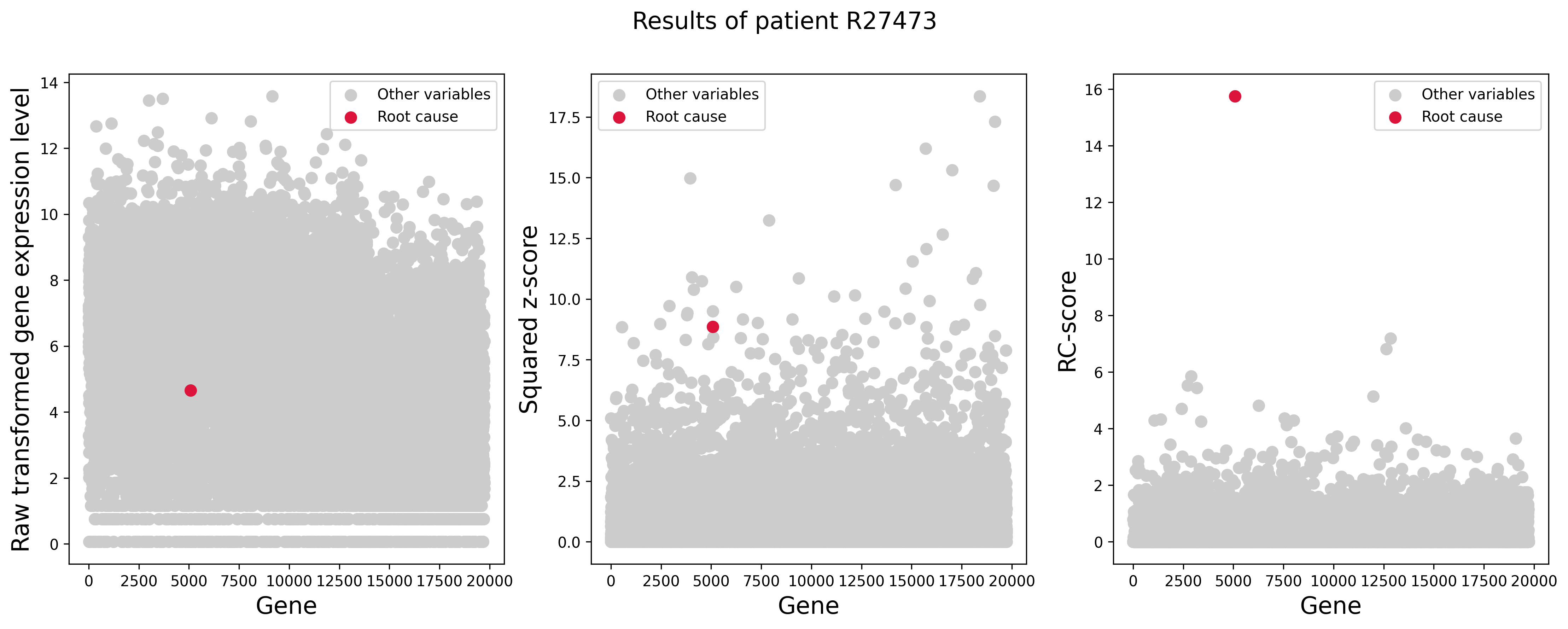}
	\end{minipage}
	\vspace{0.05cm}
	\begin{minipage}[b]{0.9\textwidth}
		\centering
		\includegraphics[scale=0.26]{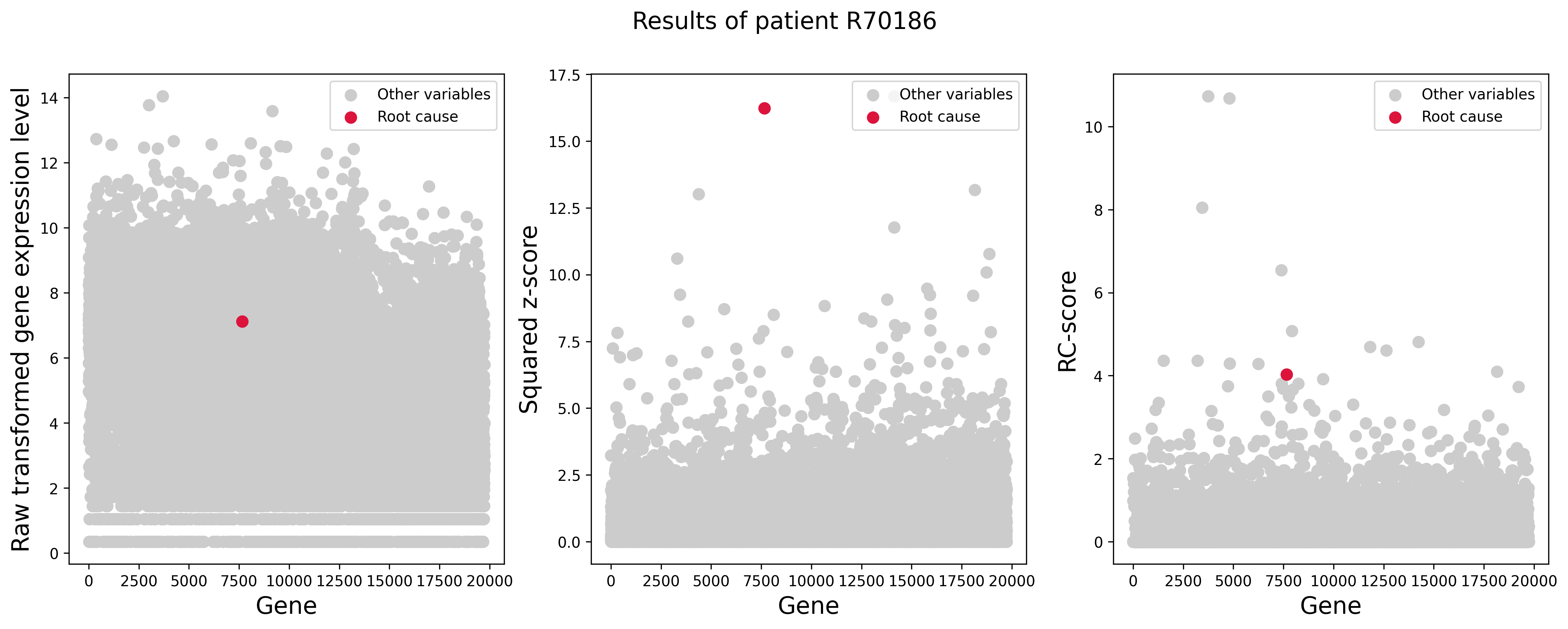}
	\end{minipage}
	\vspace{0.05cm}
	\begin{minipage}[b]{0.9\textwidth}
		\centering
		\includegraphics[scale=0.26]{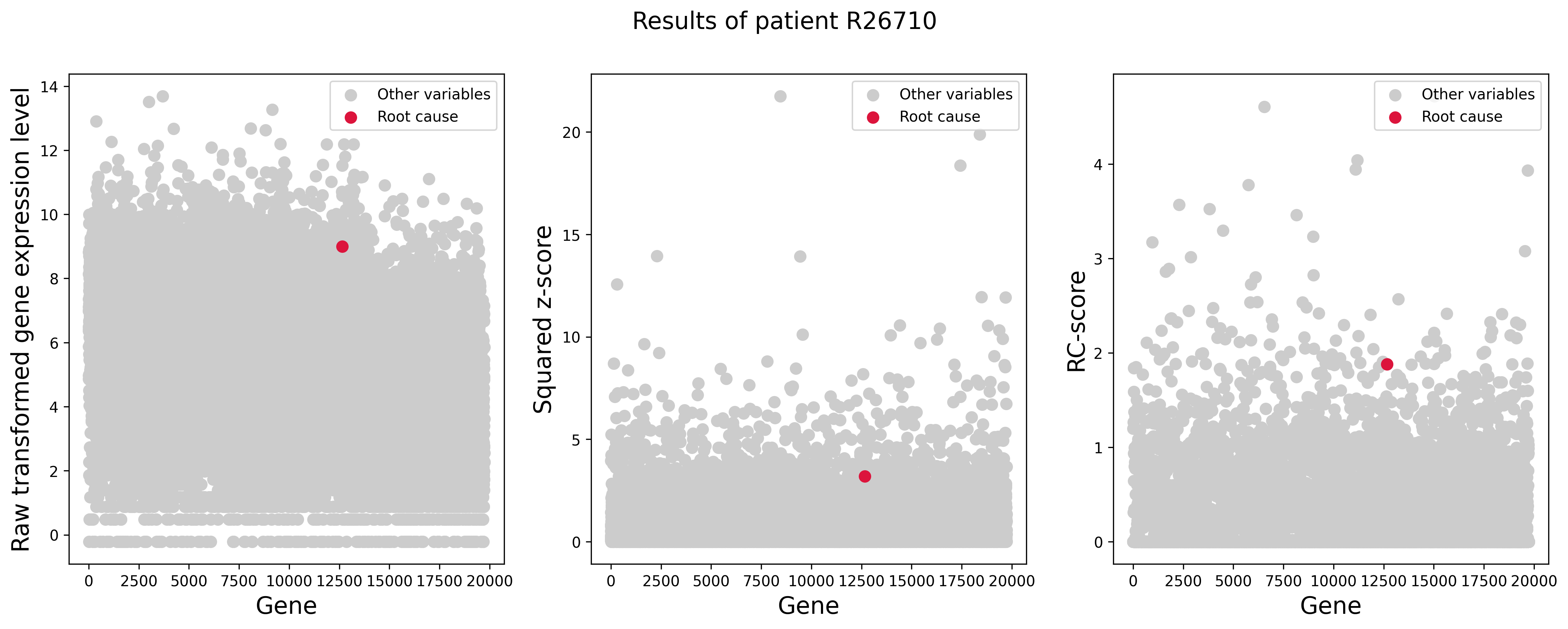}
	\end{minipage}
	\vspace{0.05cm}
	\begin{minipage}[b]{0.9\textwidth}
		\centering
		\includegraphics[scale=0.26]{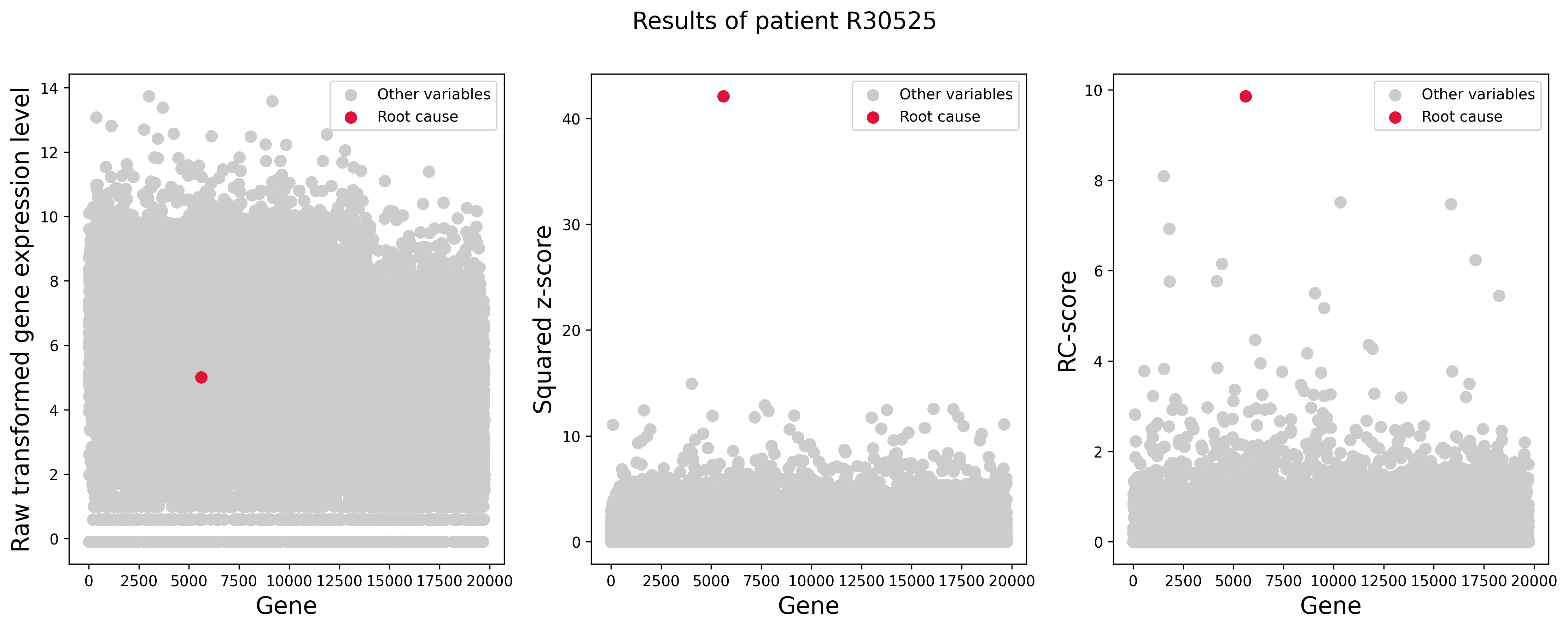}
	\end{minipage}
	\vspace{0.05cm}
	\begin{minipage}[b]{0.9\textwidth}
		\centering
		\includegraphics[scale=0.26]{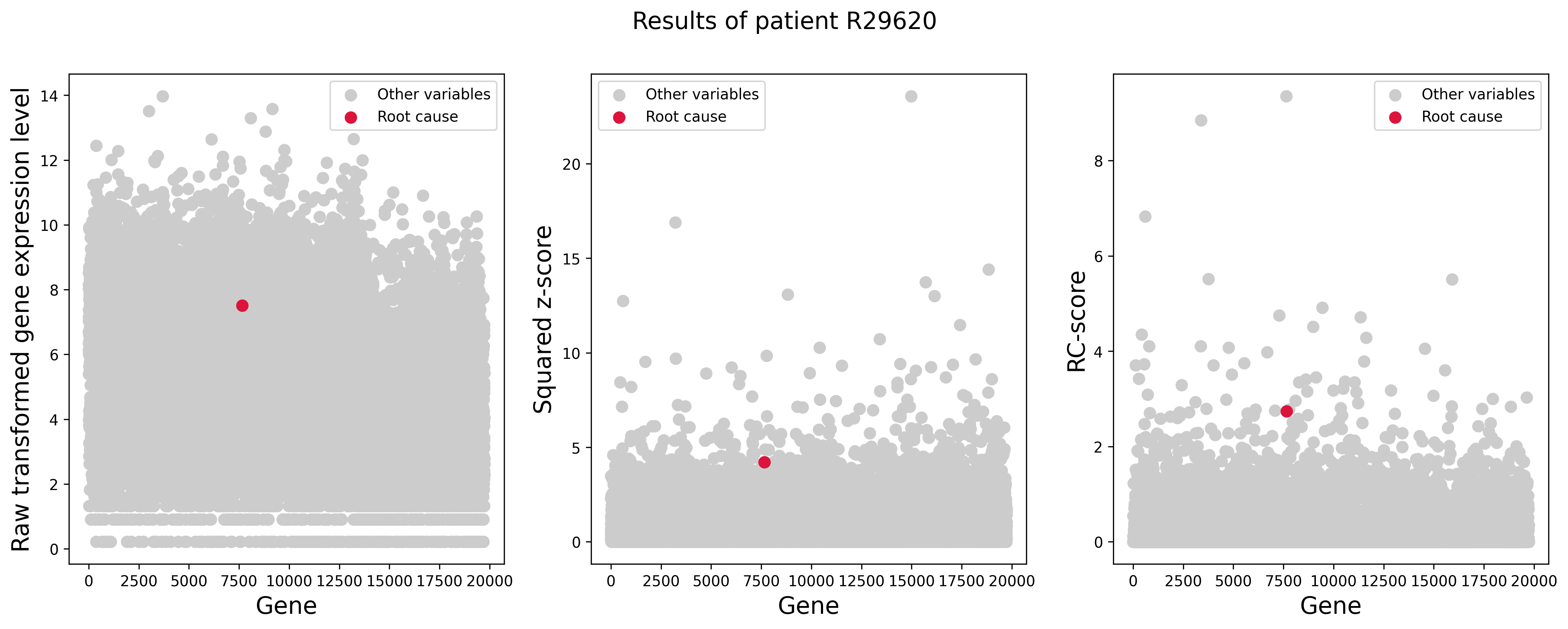}
	\end{minipage}
	\vspace{-0.5cm}
	\caption{The raw transformed gene expression levels, squared z-scores, and RC-scores of genes for patients $R27473$, $R70186$, $R26710$, $R30525$, and $R29620$.}
	\label{Fig:AppReal10}
\end{figure}

\begin{figure}[ht]
	\centering
	\begin{minipage}[b]{0.91
			\textwidth}
		\centering
		\includegraphics[scale=0.26]{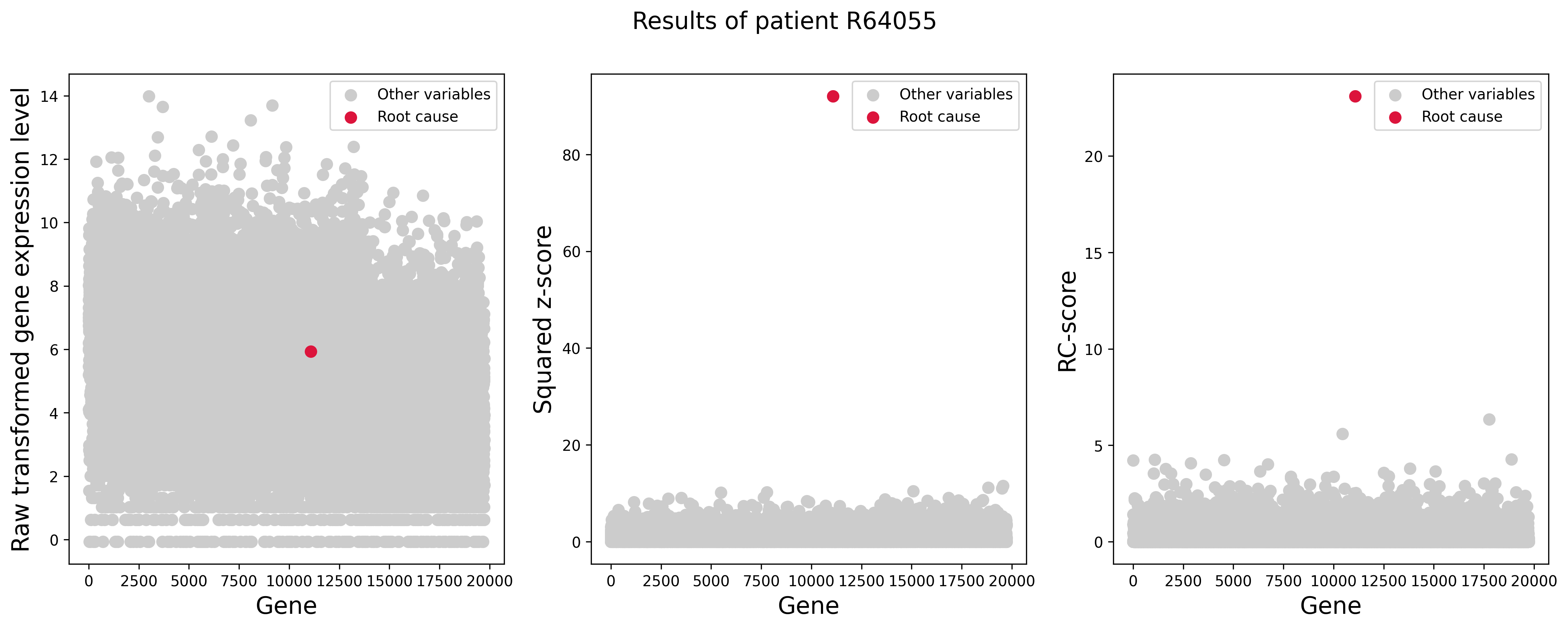}
	\end{minipage}
	\vspace{0.05cm}
	\begin{minipage}[b]{0.9\textwidth}
		\centering
		\includegraphics[scale=0.26]{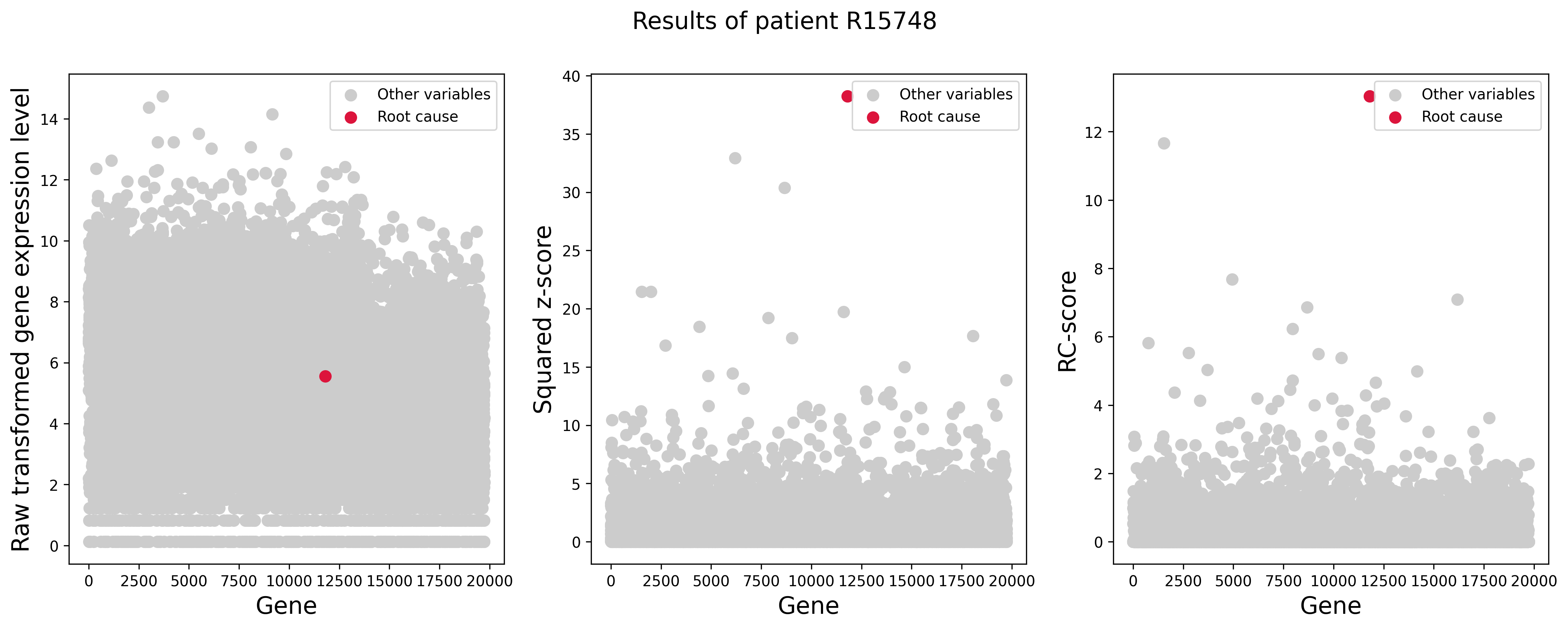}
	\end{minipage}
	\vspace{0.05cm}
	\begin{minipage}[b]{0.9\textwidth}
		\centering
		\includegraphics[scale=0.26]{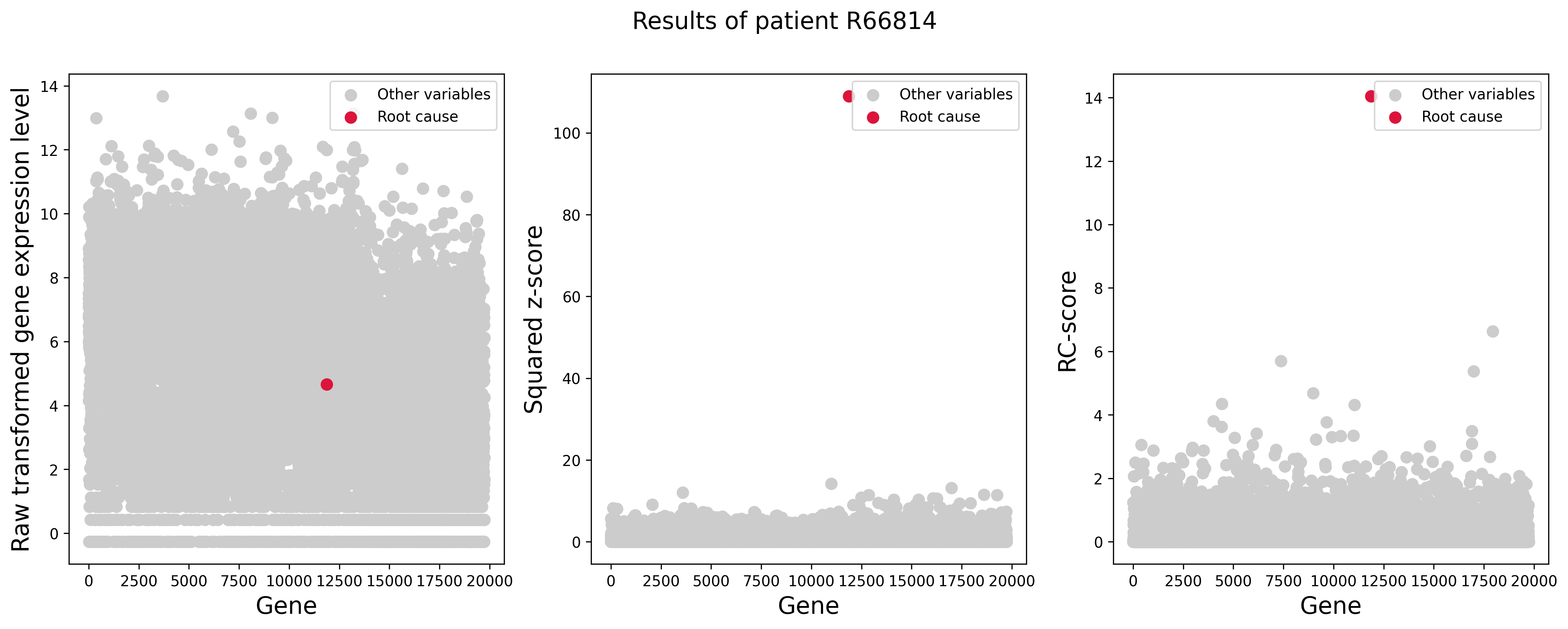}
	\end{minipage}
	\vspace{0.05cm}
	\begin{minipage}[b]{0.9\textwidth}
		\centering
		\includegraphics[scale=0.26]{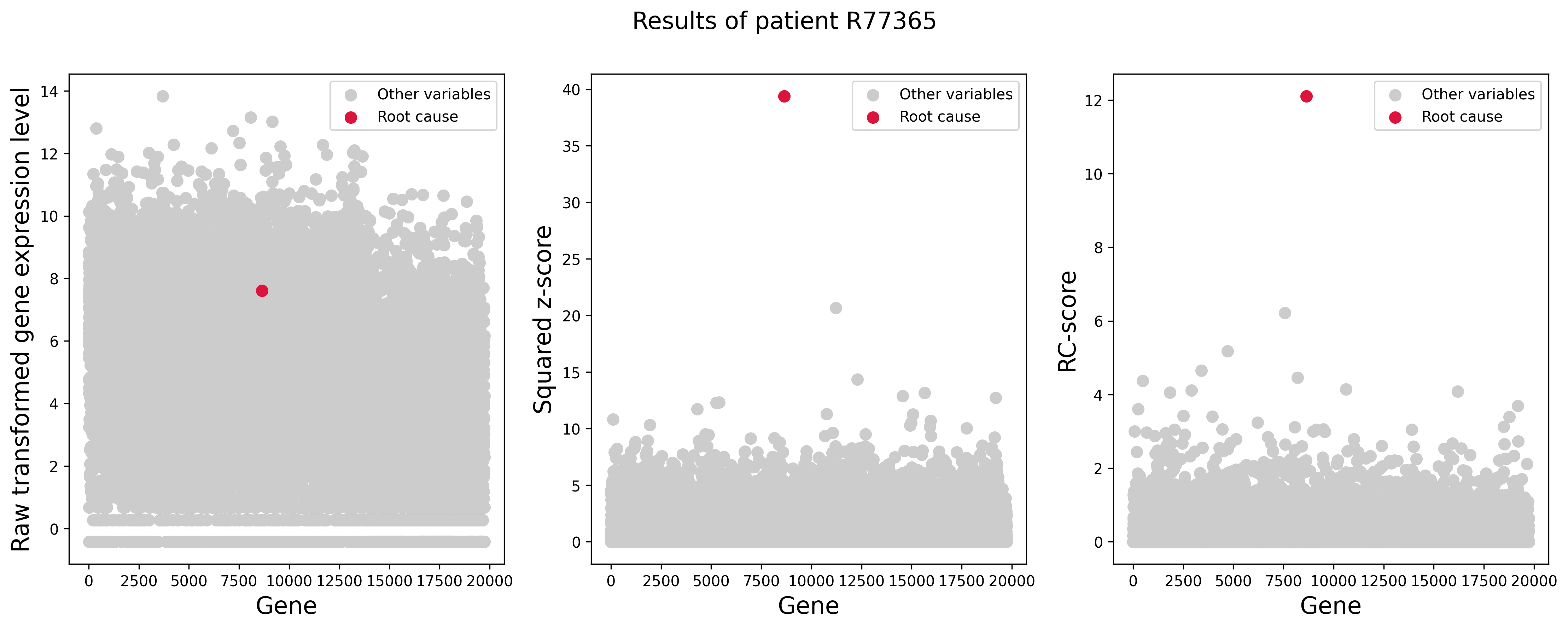}
	\end{minipage}
	\vspace{0.05cm}
	\begin{minipage}[b]{0.9\textwidth}
		\centering
		\includegraphics[scale=0.26]{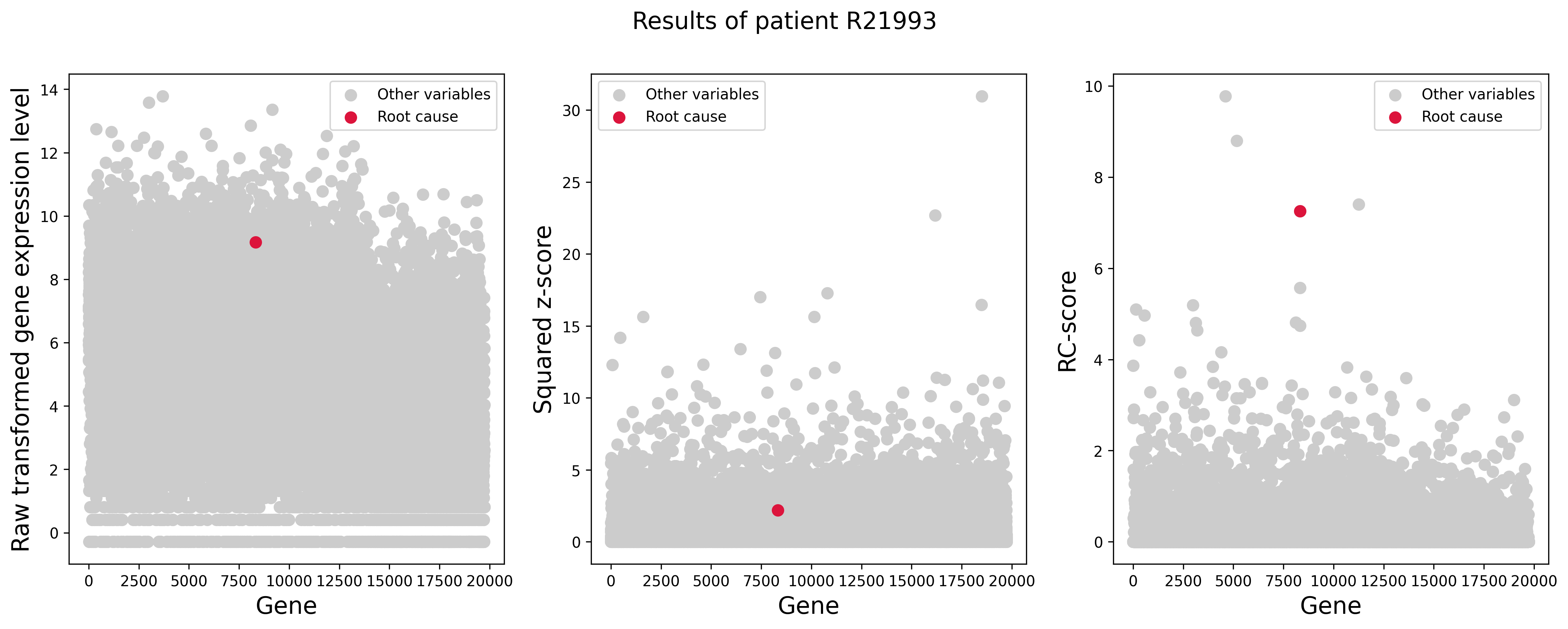}
	\end{minipage}
	\vspace{-0.5cm}
	\caption{The raw transformed gene expression levels, squared z-scores, and RC-scores of genes for patients $R64055$, $R15748$, $R66814$, $R77365$, and $R21993$.}
	\label{Fig:AppReal11}
\end{figure}

\begin{figure}[ht]
	\centering
	\begin{minipage}[b]{0.9\textwidth}
		\centering
		\includegraphics[scale=0.26]{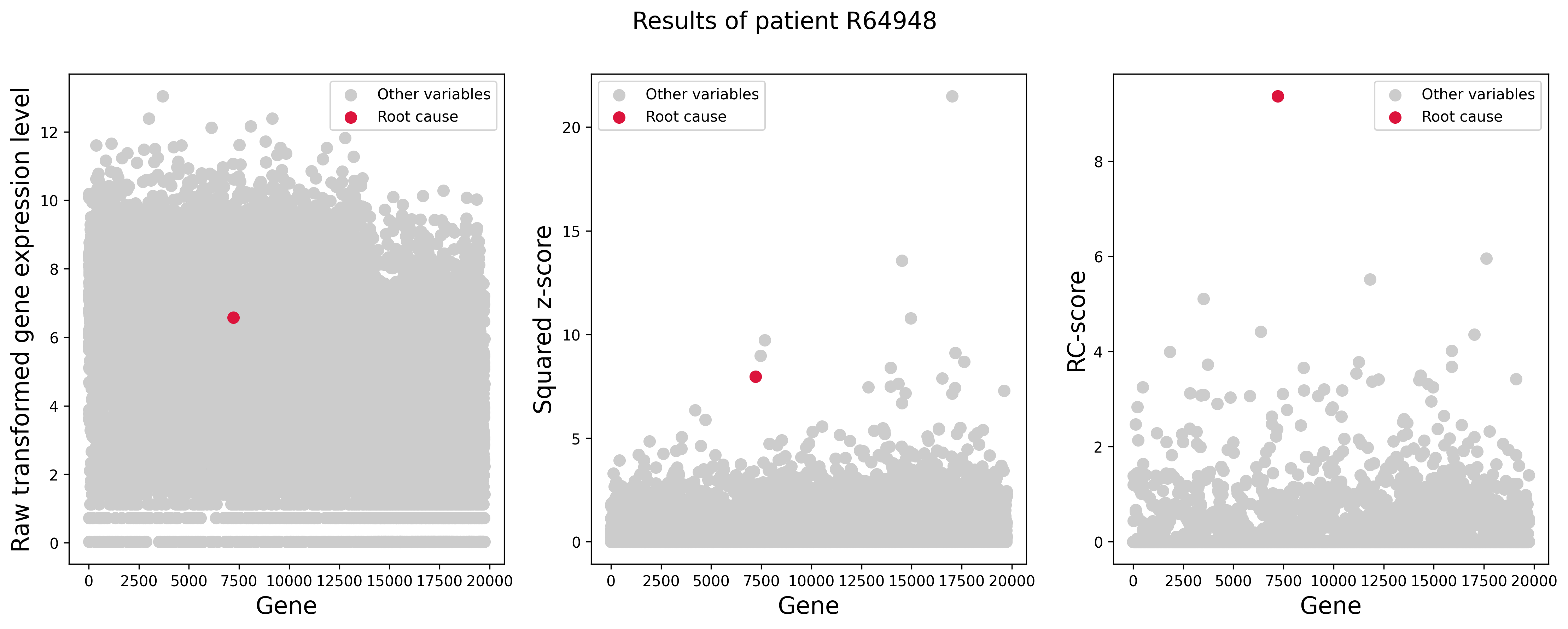}
	\end{minipage}
	\vspace{0.05cm}
	\begin{minipage}[b]{0.9\textwidth}
		\centering
		\includegraphics[scale=0.26]{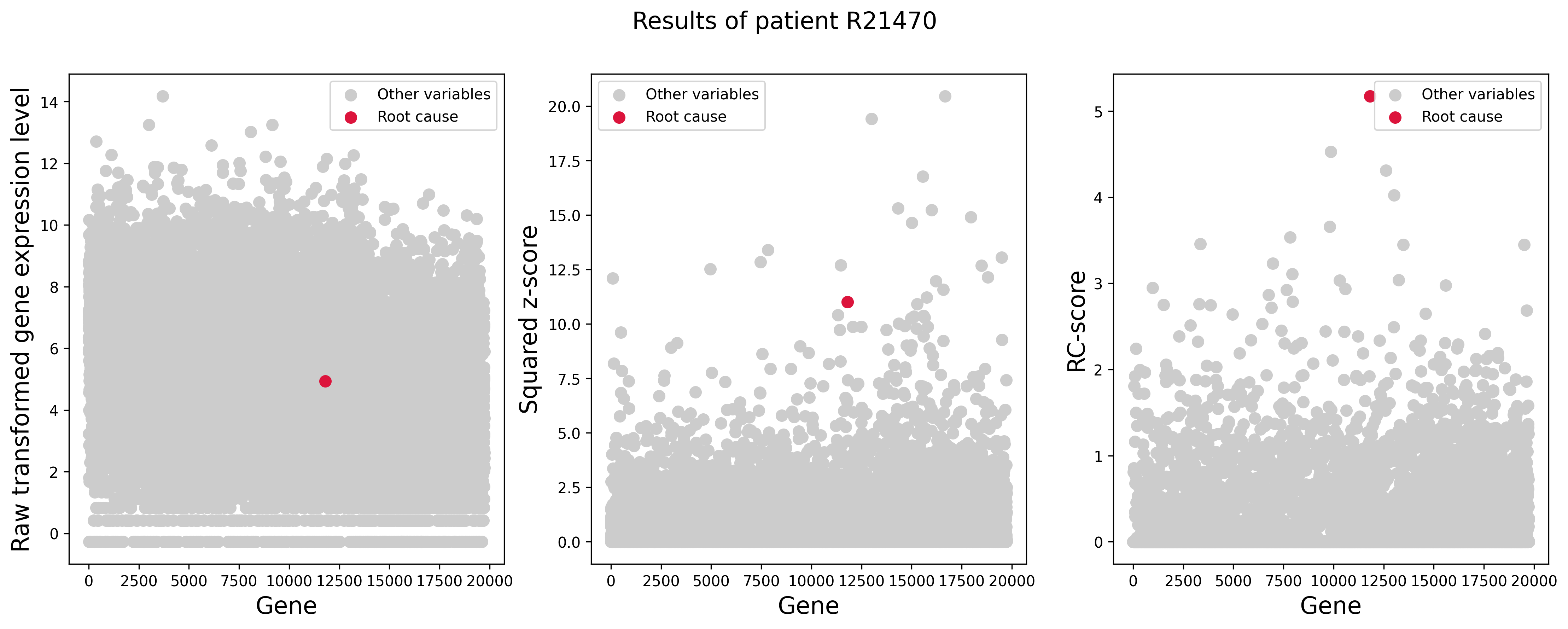}
	\end{minipage}
	\vspace{-0.5cm}
	\caption{The raw transformed gene expression levels, squared z-scores, and RC-scores of genes for patients $R64948$ and $R21470$.}
	\label{Fig:AppReal12}
\end{figure}

We show the raw transformed gene expression levels, squared z-scores, and RC-scores of genes for all $58$ patients
in Figures~\ref{Fig:AppReal1},~\ref{Fig:AppReal2},~\ref{Fig:AppReal3},~\ref{Fig:AppReal4},~\ref{Fig:AppReal5},~\ref{Fig:AppReal6},~\ref{Fig:AppReal7},
~\ref{Fig:AppReal8},~\ref{Fig:AppReal9},~\ref{Fig:AppReal10},~\ref{Fig:AppReal11}, and \ref{Fig:AppReal12}.

\end{document}